\documentclass[footinbib,a4paper,aps,superscriptaddress,reprint,twocolumn,preprintnumbers,amsmath,amssymb,nobalancelastpage,10pt]{revtex4-2}

\usepackage{bigints}
\usepackage{amsmath}
\usepackage{verbatim}
\usepackage{graphicx}
\usepackage{color}
\usepackage{braket}
\usepackage{yfonts}
\usepackage{soul}
\usepackage[normalem]{ulem}
\usepackage{dsfont}
\usepackage{xcolor}
\usepackage{amsmath}
\usepackage{verbatim}
\usepackage{graphicx}
\usepackage{color}
\usepackage{tikz}
\usepackage{subfigure}
\usepackage{float}
\usepackage{slashed}
\usepackage{mathptmx}
\usepackage{amssymb}
\usepackage{amsmath}
\usepackage{amsfonts}
\usepackage{array}

\usepackage{bm}
\usepackage{xcolor}
\graphicspath{{figures/}}

\usepackage{braket}

\definecolor{mygrey}{gray}{0.35}
\definecolor{myblue}{rgb}{0.2,0.2,0.8}
\definecolor{mygreen}{rgb}{0.2,0.8,0.5}
\definecolor{myzard}{cmyk}{0,0,0.05,0}
\definecolor{mywhite}{rgb}{1,1,1}
\definecolor{myred}{rgb}{1,0.,0.3}

\usepackage[colorlinks=true,citecolor=myblue,linkcolor=myred]{hyperref}

 \def\ee{\mathord{\rm e}}
 
 \def\ii{\mathord{\rm i}}

\def\half{\textstyle\frac{1}{2}}

\def\fourth{\textstyle\frac{1}{4}}

\renewcommand{\ii}{{\rm i}}
\renewcommand{\ee}{{\rm e}}
\def\beq{\begin{equation}}
\def\eeq{\end{equation}}
\def\barray{\begin{eqnarray}}
\def\earray{\end{eqnarray}}

\usepackage{latexsym,amsmath,amssymb,amstext}
\usepackage{graphicx}
 	\graphicspath{ {figuras/} }
\usepackage{dcolumn}
\usepackage{bm}
\usepackage[utf8]{inputenc}
\usepackage{amsmath, amsthm, amsfonts, amssymb}

\usepackage{physics}
\usepackage{dsfont}
\usepackage{graphicx}
\usepackage{cleveref} 
\usepackage{xfrac}
\usepackage{mathrsfs}
\usepackage[normalem]{ulem}
\usepackage{bbold}
\usepackage{soul}
\usepackage{mathtools}

\begin{document}


\title{ A  higher-order topological twist on cold-atom SO($5$) Dirac fields}


\author{A. Bermudez}
\affiliation{Instituto de Física Teórica, UAM-CSIC, Universidad Autónoma de Madrid, Cantoblanco, 28049 Madrid, Spain}

\author{D. González-Cuadra}
\affiliation{Institute for Theoretical Physics, University of Innsbruck, 6020 Innsbruck, Austria}
\affiliation{Institute for Quantum Optics and Quantum Information of the Austrian Academy of Sciences, 6020 Innsbruck, Austria}

\author{S. Hands}
\affiliation{Department of Mathematical Sciences, University of Liverpool, Liverpool L69 3BX, United Kingdom}

\begin{abstract}
     Ultracold Fermi gases of spin-3/2  atoms  provide a clean platform to realise SO($5$) models of  4-Fermi interactions in the laboratory. By confining the atoms in a two-dimensional Raman  lattice,  we show how  this system can be used as  a flexible quantum simulator of   Dirac quantum field theories (QFTs) that combine   Gross-Neveu and Thirring interactions  with a higher-order topological twist. We show that the lattice model corresponds to  a regularization of this  QFT with  an anisotropic twisted Wilson   mass. This allows us to access  higher-order topological states protected by a hidden SO($5$) symmetry, a remnant of the original  rotational symmetry of the 4-Fermi interactions that is not explicitly broken by the lattice discretization. Using  large-$N$ methods, we show that the 4-Fermi interactions lead to a rich phase diagram with various competing fermion condensates.   Our work opens a route for the implementation of correlated higher-order topological states with tunable interactions that has  interesting connections to non-trivial relativistic QFTs of Dirac fermions in $D=2+1$ dimensions. 
     
\end{abstract}
\maketitle

\setcounter{tocdepth}{2}
\begingroup
\hypersetup{linkcolor=black}
\tableofcontents
\endgroup

\section{\bf Introduction}

Symmetry plays a primary role in our most fundamental theories of Nature. So far, all forms of matter observed in the laboratory can be  ultimately described by the standard model~\cite{Peskin:1995ev}, a  relativistic quantum field theory (QFT) that contains  Dirac fermions locally coupled to both scalar and gauge bosons, and  is invariant under the Lorentz transformations. The role of symmetry goes beyond this relativistic invariance, as the specific form of the local fermion-boson couplings is dictated by the invariance of the QFT  under various groups of local, so-called gauge, symmetries~\cite{PhysRev.96.191}. Additionally, there are also  global symmetries that  leave the standard model invariant.  These symmetries can be spontaneously broken at a certain energy scale, such that the vacuum of this QFT displays a certain order parameter that is no longer invariant under the action of the symmetry. An example of great relevance in high-energy physics is chiral symmetry breaking~\cite{PhysRev.122.345}, which leads to a so-called chiral  condensate, and accounts for most of the mass of the matter  surrounding us. In the early days of quantum chromodynamics (QCD)~\cite{Peskin:1995ev}, various effective QFTs that  account for mass generation by chiral symmetry breaking were explored, including 4-Fermi QFTs such as the Nambu-Jona-Lasinio~\cite{PhysRev.122.345,PhysRev.124.246,RevModPhys.64.649,HATSUDA1994221} and Gross-Neveu~\cite{PhysRevD.10.3235,doi:10.1142/9789814412674_0004,ZAMOLODCHIKOV1978481,WITTEN1978110,BERG1978205} models. These QFTs describe self-interacting Dirac fermions with various forms of quartic interactions and, moreover, can be defined in different spacetime dimensions leading to clear analogies with higher-dimensional non-Abelian gauge theories~\cite{PhysRevLett.30.1343,PhysRevLett.30.1346}.  

Besides serving as effective models that capture  some of the phenomenology observed at particle colliders qualitatively, low-dimensional Dirac QFTs also emerge in condensed matter and atomic molecular and optical (AMO) physics, providing a playground to  test the QFT predictions quantitatively. Systems such as  graphene~\cite{Novoselov2005,Zhang2005}, topological insulators~\cite{Hsieh2008,Hsieh2009}, and ultracold atoms in optical lattices~\cite{Tarruell2012,Jotzu2014}, are clear examples leading to low-dimensional Dirac matter~\cite{doi:10.1080/00018732.2014.927109, Gonzalez_2020}. We remark that, away from   $D=1+1$ dimensions, it is not straightforward to find  experimental setups which, in spite of being  highly non relativistic,  are accurately described by Lorentz-invariant Dirac QFTs  at low energies. Moreover, in the presence of interactions, these QFTs can present a number of non-trivial properties for $D<4$~\cite{PhysRevD.7.2911,PhysRevD.21.2327,HANDS199329} that are still the subject of  active research. A particularly-relevant case is that of Dirac QFTs in $D=2+1$ dimensions  with a Gross-Neveu 4-Fermi term~\cite{PhysRevLett.62.1433,ROSENSTEIN199159} or variants thereof~\cite{PhysRevLett.63.2633}, which can give rise to strong correlations and novel critical phenomena. This type of 4-Fermi models  leads to different  phase transitions  that are no longer characterised by the above chiral condensate but, instead, require finding other symmetry-breaking channels with their associated order parameters. These generalised 4-Fermi models have been the subject of renewed interest in  recent years (see the  reviews~\cite{Boyack2021,herbut2023wilsonfisher} and references therein). 

In this work, we are  interested in 
 models of synthetic Dirac QFTs that can be implemented with  ultracold atoms~\cite{Bloch_2008}. A rather unique possibility of these systems is that,  in addition to the emerging Lorentz invariance, one can actually tailor the other symmetries, either local or global. These platforms can thus be used as analogue quantum simulators~\cite{Feynman_1982, Cirac2012,Bloch2012,Blatt2012,PRXQuantum.2.017003} for a specific QFT of interest, allowing us to test theoretical predictions in a controllable experimental environment, as well as to explore other regimes that cannot be accessed using current analytical or numerical methods, such as real-time dynamics or finite fermion densities~\cite{PhysRevLett.94.170201,NAGATA2022103991}.
In particular, we focus   on  spin-$3/2$ neutral  atoms at ultra-cold temperatures,  as their $s$-wave scattering leads to 4-Fermi terms that naturally yield a large symmetry group of  SO($5$) transformations~\cite{PhysRevLett.91.186402,doi:10.1142/S0217984906012213}. When these cold atoms are loaded on standard optical lattices, one obtains a non-relativistic SO($5$) Hubbard-type~\cite{PRSLSA_276_238} model that is interesting in its own right~\cite{PhysRevLett.91.186402,Eckert_2007,Szirmai_2011}, and leads to  a rich  phase diagram already  in $D=1+1$ dimensions~\cite{PhysRevLett.95.240402,PhysRevLett.95.266404,PhysRevLett.96.097205,PhysRevLett.105.050402}. In fact, the role of SO($5$) symmetry in condensed-matter systems is broader, as it also plays a key role in certain theories with competing magnetic and superconducting orders~\cite{doi:10.1126/science.275.5303.1089,RevModPhys.76.909}.

We show in our work that, by including additional Raman beams  that interfere with the standing wave underlying the above optical lattice, we can enter a new regime in which the SO($5$) 4-Fermi model corresponds to a specific discretization of a relativistic QFT of self-interacting Dirac fermions. More specifically, we demonstrate how this lattice regularization, which in general breaks explicitly the rotational symmetry, can actually preserve a  hidden  SO($5$) $\pi/2$ rotation, and provide a neat route to explore correlation effects in higher-order topological insulators (HOTIs)~\cite{doi:10.1126/sciadv.aat0346,doi:10.1126/science.aah6442,PhysRevB.96.245115}. In particular, this lattice discretization corresponds to an anisotropic version of twisted-mass Wilson fermions which, as we show, leads to flat bands and strictly-localised  zero-energy corner modes protected by the hidden SO($5$) symmetry. These anomalous corner modes are a boundary manifestation of a non-trivial topological invariant in the bulk~\cite{PhysRevB.97.205135}, which connects to the phenomenology explored for other lattice models of higher-order topological states. In these models, the study of the effects brought up by including interactions has seen an increase of activity very recently~\cite{You_2018, Dubinkin_2019, Kudo_2019, Laubscher_2019, Laubscher_2020, Sil_2020, Rasmussen_2020, bibo2020, Peng_2021, Guo_2022, Hackenbroich_2021, Otsuka_2021, gonzalez2022, Li_2022, Montorsi_2022, Wienand_2022, Aksenov_2023, Fraxanet_2023}, but still remains largely unexplored  in comparison to the situation in their first-order counterparts (see the reviews~\cite{Hohenadler_2013,doi:10.1142/S021797921330017X,Neupert_2015,Rachel_2018}).


We contribute to this line of research by exploring the effect that the SO($5$) 4-Fermi interactions have on the aforementioned Wilson-fermion HOTI. 
In connection to the  fermion condensate for chiral symmetry breaking in high-energy physics, we show that our model accounts for  a  competition between various possible condensation channels and that, at sufficiently strong interactions, the HOTI phase gives way to a pseudo-scalar fermion condensate where the hidden  SO($5$) rotational symmetry that protects the HOTI  gets spontaneously broken. We present a non-perturbative account of this phenomenon based on the large-$N$ limit of this 4-Fermi QFT, where one considers $N$ flavours of the Dirac fermions coupled by the quartic SO($5$) interactions. By resuming the leading-order Feynman diagrams for $N\to\infty$, we calculate the effective potential, and perform a  minimization that allows us to infer the values of various condensates. Moreover, this large-$N$ techniques can be readily used to obtain an estimate for the many-body topological invariant, allowing us to chart the entire phase diagram of the model showing, in addition to the aforementioned condensates, correlated HOTIs and trivial band insulators. Since the model studied can be realised in experiments of spin-$3/2$ neutral atoms in Raman optical lattices, possible future experiments could test these predictions and their connection to non-trivial strongly-coupled QFTs.

\section{\bf Euclidean 4-Fermi  field theories with a twist} 

In this section, we introduce  our model of interacting  Dirac fields in $D=2+1$ dimensions, which is motivated by a specific Kaluza-Klein-like dimensional reduction. We also discuss  a non-standard lattice regularization that will allow us to study the  non-perturbative phenomena  induced by fermion-fermion  interactions on higher-order topological groundstates. 

\subsection{ Dimensional reduction and SO($5$)$\,\,\mapsto\,\,$SO($3$)  }

 Our model of self-interacting Dirac matter is built from a relativistic QFT of  fermions with rotationally-invariant 4-Fermi interactions. As discussed in Appendix~\ref{appA}, the partition function of this QFT can be written as a path integral~\cite{negele_orland_2019} over two independent Grasmmann spinors  $\psi(x),\overline{\psi}(x)$~\cite{0201304503}, which  represent the  Dirac fermions in  a $3$-dimensional Euclidean spacetime with imaginary time $x=(\tau,\boldsymbol{x})$, where $\boldsymbol{x}=(x_1,x_2)$. The path integral is expressed in terms of an  Euclidean  action that contains two terms  $S=S_0+S_{\rm int}$. The first one describes  free Dirac fermions with two possible mass terms
\begin{equation}
\label{eq:action_free}
    S_{\rm 0}=\!\int\!\!{\rm d}^{3}x\,\overline{\psi}(x)\bigg( \gamma^\mu\partial_\mu+ \ii m_{1}\gamma^{3}+ \ii m_2\gamma^5\bigg)\psi(x),
\end{equation}
where $\mu\in\{0,1,2\}$ labels the spacetime coordinates, $\partial_\mu=\partial/\partial x^{\mu}$, and $m_1, m_2$ are the corresponding bare masses, which will be latter connected to a mass twisting. Here,  we have used Einstein's convention of repeated index summation, and  natural units $\hbar=c=1$. The set of 5 gamma matrices $\{\gamma^0,\gamma^1,\gamma^2,\gamma^{3},\gamma^5\}$  fulfill $\{\gamma^a,\gamma^b\}=\gamma^a\gamma^b+\gamma^b\gamma^a=2\delta^{a,b}\mathbb{1}_{d_s}$, which can only be satisfied by considering  Grassmann fields with $d_s=4$ spinor components. We recall that, in an Euclidean metric, these gamma matrices  are all Hermitian, and can be defined by various possible  choices of tensor products of operators within the Pauli basis $\{\mathbb{1},\sigma^x,\sigma^y,\sigma^z\}$~\cite{vanproeyen2016tools,Freedman:2012zz}. Although the  specific choice of gamma matrices is arbitrary at the level of the QFT~\eqref{eq:action_free}, 
the implementation based on spin-$3/2$ cold-atom gases   discussed  in Sec.~\ref{sec:ham_int_atoms} fixes  our  choice of the spacetime gamma matrices to
\beq
\label{eq:gamma_spatial}
\gamma^0=\sigma^y\otimes\mathbb{1}_2,\,\,\,\,\gamma^1=\sigma^x\otimes\sigma^x,\,\,\,\,\gamma^2=\sigma^x\otimes\sigma^y.
\eeq
In addition,  the remaining gamma matrices are also fixed as
\beq
\label{eq:gamma_masses}
\gamma^3=\sigma^x\otimes\sigma^z, \,\,\,\, \gamma^5=-\sigma^z\otimes\mathbb{1}_2.
\eeq

As  discussed in Appendix~\ref{appA}, this set of matrices  actually forms a reducible representation of the Clifford algebra in the underlying 3-dimensional  spacetime. These gamma matrices can  be used to define the  generators of   Lorentz transformations, which correspond to a spinor representation of the  SO($3$)  rotations in the Euclidean metric. We remark that such Lorentz transformations could also be generated using  an irreducible representation of the Clifford algebra, which would only require using two-component spinors~\cite{vanproeyen2016tools,Freedman:2012zz}. However, this  would not allow for the      two independent mass terms  in Eq.~\eqref{eq:action_free}, as the spacetime gamma matrices would already exhaust all the possible mutually anti-commuting Hermitian matrices, e.g. $\gamma^0=\sigma^z,\gamma^1=\sigma^x,\gamma^2=\sigma^y$. In this case, there can only be a single mass term $m_0\overline{\psi}\psi$ that breaks parity~\cite{ROSENSTEIN199159}. As we will see, having two anti-commuting   masses~\eqref{eq:action_free}   plays a key role in the phenomena discussed in our work.

As discussed in detail in Appendix~\ref{appB}, a different perspective motivating the choice of the action~\eqref{eq:action_free} is that the reducible representation~\eqref{eq:gamma_spatial}-\eqref{eq:gamma_masses} is the result of a Kaluza-Klein-type compactification. In the original Kaluza-Klein context~\cite{D_Bailin_1987}, gravity and electrodynamics in $D=3+1$ dimensions were shown to result from the compactification of an extra dimension in a   5-dimensional theory of pure  gravity. In the current context~\eqref{eq:action_free}, the situation is much simpler, as one only needs to consider a    QFT of Dirac fermions in a higher 5-dimensional spacetime coupled to background fields that will induce the two mass terms~\cite{Freedman:2012zz,Ryu_2010}. In the larger spacetime,  SO($5$) symmetry  is manifest in the action, and can also be used to understand the structure  of the  Lorentz-invariant  4-Fermi self-interactions, as we now discuss. 
Considering the reducible representation of the Clifford algebra,    we  can define   SO($5$)-invariant 4-Fermi interactions that can be written as  follows
\begin{equation}
\label{eq:action_qurtic}
S_{\rm int}\!=\!\!\int\!\!{\rm d}^{3}x\,\frac{g^2}{2}\!\!\left({\color{blue}-}J_\mu J^\mu+(\overline{\psi}\gamma^3\psi)^2+(\overline{\psi}\gamma^5\psi)^2-(\overline{\psi}\psi)^2\!\right)\!,
\end{equation}
where we have introduced  the Euclidean fermion current as $J^\mu={\ii}\overline{\psi}\gamma^\mu\psi$. This action can
  again be interpreted from the perspective of   dimensional reduction, where the first five terms can be written as the squared norm of a higher-dimensional vector  whose components are fermion bilinears. Therefore, the norm of the vector is conserved under the   SO($5$) Lorentz transformations (see Appendix~\ref{appB}). In addition, the last term in Eq.~\eqref{eq:action_qurtic} is a scalar under these SO($5$) spatial rotations, and thus also remains invariant. From the perspective of the higher-dimensional parent  theory, these quartic terms     correspond to  a linear combination of the standard  Gross-Neveu and Thirring~\cite{THIRRING195891,PhysRevD.11.2088} interactions. This is very different from the interactions allowed by an irreducible representation, which can all be reduced to  a single Gross-Neveu quartic interactions. On the other hand, for a reducible representation,  the physics of  Gross-Neveu and Thirring QFTs in $D=2+1$  turns out to be very different~\cite{hep-lat/9706018}.
Therefore, working with  reducible gamma matrices  leads to a richer  dimensionally-reduced QFT with more interaction channels than those allowed by an irreducible representation. 
The  structure of Eq.~\eqref{eq:action_qurtic} will yield a competition of various  fermion condensates  with the aforementioned topological  groundstates.

\subsection{Anisotropic mass twisting of Wilson  fermions and the explicit breakdown of SO($3$) symmetry}
\label{sec:an_twisted}

\begin{figure*}[t]
	\centering
	\includegraphics[width=2\columnwidth]{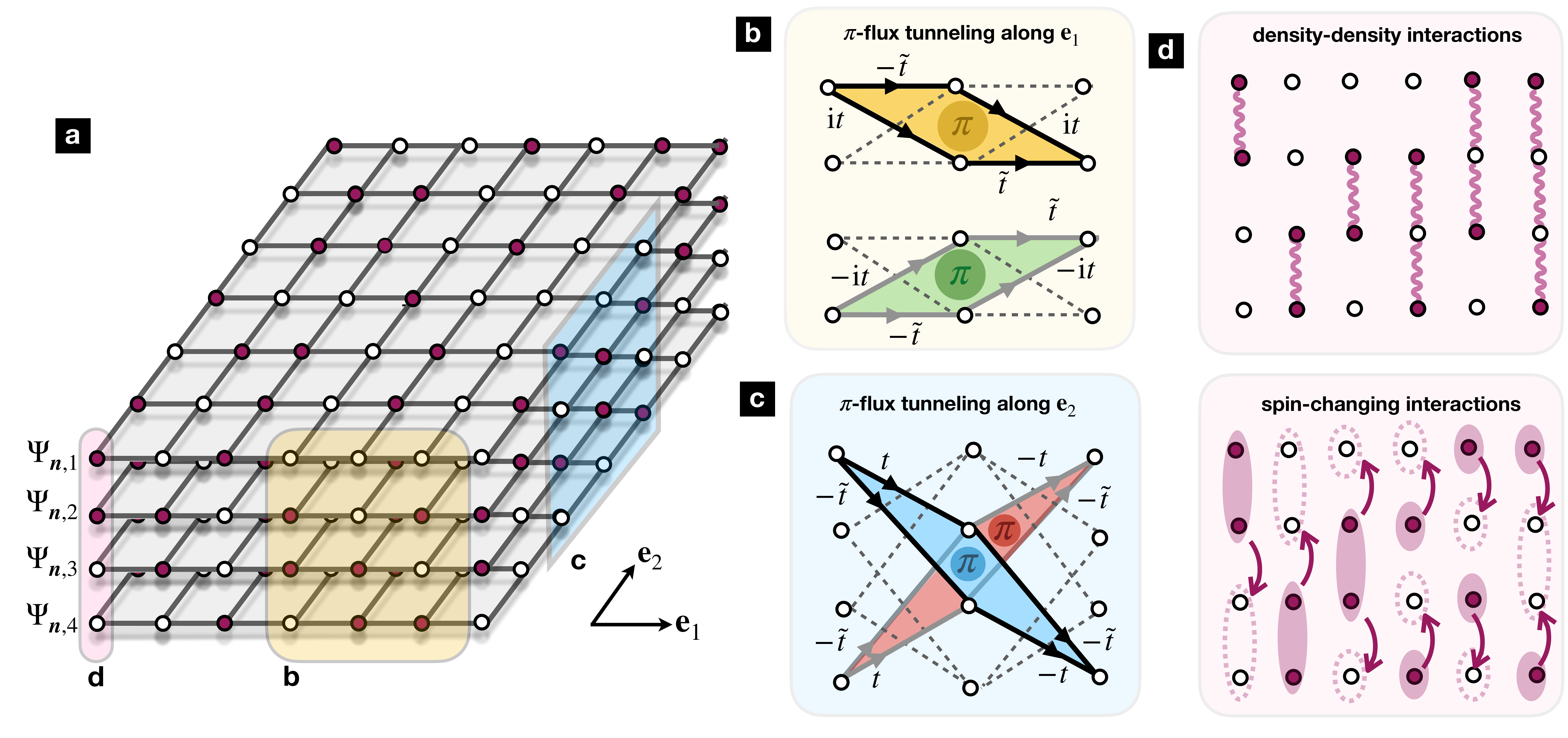}
	\caption{{\bf Multi-layer scheme for the lattice regularization of the SO($5$) 4-Fermi QFT:} {\bf (a)} In the Hamiltonian formulation, the spatial directions are discretized on a square lattice, and the  spinor components can be represented as four-layer model. The yellow region in the inset {\bf (b)} describes the discretization of the kinetic and twisted Wilson mass along the $x$ axis. The layers only get coupled in pairs, and can be interpreted as a distribution of rhomboid plaquettes pierced by $\pi$-fluxes corresponding to a pair of decouple Creutz ladders. The blue region   in the inset {\bf (b)} describes the discretization of the kinetic and twisted Wilson mass along the $y$ axis. This tunnelling couples all four layers, and also has  underlying $\pi$-fluxes which, together with those of {bf (b)} lead to flat bands. The purple region in the inset {\bf (d)} represents the SO($5$) 4-Fermi interactions, which include Hubbard density-density interactions among all possible inter-layer pairings (upper panel), and also spin-changing collisions that can be interpreted as inter-layer pair tunnelling processes.    }
	\label{fig:scheme_lattice}
\end{figure*}

After setting up the continuum QFT~\eqref{eq:action_free}-\eqref{eq:action_qurtic}, we now discuss how  higher-order topological~\cite{doi:10.1126/sciadv.aat0346,doi:10.1126/science.aah6442,PhysRevB.96.245115} and trivial groundstates may arise as the result of a non-standard lattice regularization. We  introduce this regularization   by the use of Wilson fermions~\cite{Wilson1977} for the more standard first-order topological insulators~\cite{RevModPhys.82.3045,RevModPhys.83.1057}, and then highlight the ``twist'' that is required to connect to higher-order topology.  

Let us start by noting that all of these  regularizations 
are related to the  problem of  fermion doubling~\cite{NIELSEN198120,NIELSEN1981173} in lattice field theories~\cite{gattringer_lang_2010}. In  particular,  we  shall consider a non-zero lattice spacing $a$ along the spatial directions, such that  $\boldsymbol{x}=\sum_ja(n_j-N_j/2){\bf e}_j=a(\boldsymbol{n}-\boldsymbol{N}/2)$ with $\boldsymbol{n}\in\Lambda_{\rm s}=\mathbb{Z}_{N_1}\times\mathbb{Z}_{N_2}$ and $\boldsymbol{N}=(N_1,N_2)$ contains an even number of lattice sites per axis, leading to $N_{\rm s}=N_1N_2$ as the total number of lattice sites (see Fig.~\ref{fig:scheme_lattice}{\bf (a)}). The spatial derivatives  in Eq.~\eqref{eq:action_free} are  substituted by finite differences,  while the Euclidean time remains continuous $\tau\in\mathbb{R}$.  This asymmetric treatment of the spacetime will allow us to make a direct connection with the Hamiltonian approach to field theories in  section~\ref{sec:HOTI}. The fermion doubling can   be understood by writing  the free action~\eqref{eq:action_free} after this regularization which,  in momentum space, reads 
 \begin{equation}
 \label{eq:free_action_naive}
    S_{\rm 0}=\!\!\int_k\overline{\psi}(k) \big(\gamma^\mu \hat{k}_\mu+\ii m_1\gamma^3+\ii m_2\gamma^5\big)\psi(k),
\end{equation}
where $ k=(k^0,\boldsymbol{k})$ is the three-momentum, an we have introduced the short-hand notation $\!\int_k:=\frac{1}{(N_{\rm s}a)^2}\!\!\sum_{\boldsymbol{k}\in{\rm BZ}}\int\!\!\frac{{\rm d}k^0}{2\pi}$. In the expression above,  we have defined $\hat{k}_0=k_0\in\mathbb{R}$ using the zero-temperature limit of the Matsubara frequencies~\cite{negele_orland_2019}. Additionally, the spatial components of the momentum $\hat{\boldsymbol{k}}$ are related to the corresponding crystal momenta $\boldsymbol{k}$  via 
\beq 
\label{eq:discrete_momentum}
\hat{k}_1=\frac{1}{a}\sin(k_1a),\hspace{2ex}\hat{k}_2=\frac{1}{a}\sin(k_2a).
\eeq
 We recall that, as a consequence of the lattice regularization,  the crystal momenta  are quantised within the first Brillouin zone $\boldsymbol{k}\in{\rm BZ}$, namely   $k_j=-{\pi}/{a}+{2\pi n_j}/{aN_j}$ with  $n_j\in\mathbb{Z}_{N_j}$. By a Taylor expansion $\boldsymbol{k}\mapsto\boldsymbol{k}_{{\rm D},\boldsymbol{\ell}}+ \boldsymbol{k}$ for $|\boldsymbol{k}|\ll\Lambda_{\rm c}=\pi/a$, around any of the four Dirac points 
 \beq
 \label{eq:Dirac_points}
 \boldsymbol{k}_{{\rm D},\boldsymbol{\ell}}=\frac{\pi}{a}(\ell_1,\ell_2),\hspace{2ex}\boldsymbol{\ell}\in\mathbb{Z}_2\times \mathbb{Z}_2,
 \eeq
 one finds  that
 the long-wavelength action stemming from Eq.~\eqref{eq:free_action_naive} coincides with that of a massless Dirac fermion~\eqref{eq:action_free}. Altogether, the long-wavelength infra-red (IR) features are governed by $N_{\rm D}=4$ Dirac fermions instead of one.
 Typically, one refers to $\boldsymbol{\ell}\in\{(0,1),(1,0),(1,1)\}$ as the spurious fermion doublers, each of which has a different emergent chirality $\gamma^5_{\boldsymbol{\ell}}=(-1)^{\ell_1+\ell_2}\gamma^5$. The presence of these spurious  doublers can change the physics considerably, specially when adding further interactions such as those in Eq.~\eqref{eq:action_qurtic}.

The idea of K. Wilson to cope with fermion doubling~\cite{Wilson1977} was to add a momentum-dependent {\em Wilson mass}, which sends these doublers to the ultra-violet (UV) cutoff scale of the QFT, such that they become very massive and are not expected to interfere with the low-energy physics of the remaining Dirac fermion  $\boldsymbol{\ell}=(0,0)$. In Appendix~\ref{app:Chern}, we show that the standard Wilson regularization amounts to setting $m_2\mapsto\overline{m}_2=0$ and $m_1\mapsto \overline{m}_1(\boldsymbol{k})$ in Eq.~\eqref{eq:wilson_mass}, which leads to two copies of a Chern insulator~\cite{PhysRevLett.61.2015,PhysRevB.74.085308,PhysRevB.78.195424,PhysRevResearch.4.L042012,ZIEGLER2022168763}. In this sense, 
 the use of a reducible representation of the gamma matrices is rather trivial, and we could have obtained  the same topological features using an irreducible representation and two-component spinors.
 
Note, however, that we have  only  used  one of  the mass terms in Eq.~\eqref{eq:free_action_naive}, and followed the  Wilson regularisation scheme verbatim. In this work, we explore a different  non-standard regularization which, although having a similar effect on the doublers, leads to very different manifestations of topology. We consider an {\it anisotropic   twisted Wilson mass} regularization
$S=S_0^{\rm TM}+S_{\rm int}$, in which the free part is expressed in momentum space as
 \begin{equation}
\label{eq:twisted_mass_S_0}
    S_{\rm 0}^{\rm TM}=\int_k\overline{\psi}(k)\bigg( \gamma^\mu \hat{k}_\mu+\ii m_1\!(\boldsymbol{k})\gamma^3+\ii m_2(\boldsymbol{k})\gamma^5\bigg)\psi(k).
\end{equation}
This contains the above  anisotropic twisted Wilson mass
\begin{equation}
 \label{eq:twisted_mass}
 \begin{split}
   m_1\!(\boldsymbol{k})=m_1+\frac{r}{a}\bigg(1-\cos(k_1a)\bigg),\\
   m_2(\boldsymbol{k})=m_2+\frac{r}{a}\bigg(1-\cos(k_2a)\bigg),
\end{split}
\end{equation}
where $r\in\mathbb{R}$ is the analogue of the Wilson parameter in the standard regularization of App.~\ref{app:Chern}.
The interacting part of the action $S_{\rm int}$
is defined  by considering  the  4-Fermi terms in Eq.~\eqref{eq:action_qurtic}, but discretizing the spatial coordinates such that  the integral becomes a shorthand for $\!\int\!\!{\rm d}^{3}x:= {a^2}\sum_{\boldsymbol{n}\in\Lambda_{\rm s}}\int{\rm d}\tau$.

Twisted-mass Wilson fermions have been previously considered in lattice gauge theories~\cite{ALPHA_Collaboration_2001,Roberto_Frezzotti_2001,Carlos_Pena_2004,Frezzotti_2004,PhysRevD.78.014509,PhysRevD.95.094515,PhysRevLett.127.252001,PhysRevLett.130.241901,Saez:2022ptd} (see~\cite{SHINDLER200837} for a detailed account), although the twisting procedure is very different from the one considered in our work. In order to understand the differences, let us describe mass twisting in a broader context. We start by focusing on the standard situation, which involves even-dimensional spacetimes such as $D=3+1$. As discussed  in Appendix~\ref{appA}, the usual mass twisting follows from an axial rotation of angle $\theta$ on the standard Dirac mass term $m\overline{\psi}\psi$, leading to a pair of anti-commuting mass matrices proportional to $m_1=m\cos\theta, m_2=m\sin\theta$~\eqref{eq:Euc_action_twisted}. The new mass term  for $\theta\neq 0$   breaks parity~\eqref{eq:parity} explicitly, unless the  rotation/twisting angle is promoted to a pseudo-scalar  axion field $\theta\mapsto\theta(\tau,\boldsymbol{x})$~\cite{PhysRevLett.38.1440,PhysRevLett.40.279,PhysRevLett.40.223}. Let us now consider this mass twisting for a Wilson-fermion lattice regularization, which  upgrades the parity-invariant mass to a momentum-dependent one $m_1\mapsto \overline{m}_1(\boldsymbol{k})$ similar to Eq.~\eqref{eq:wilson_mass}, but accounting for the larger spatial dimensions. This Wilson term leads again to a different mass for each of the Dirac points,  and can lea to  a non-zero topological invariant. From the perspective of  first-order topological insulators~\cite{PhysRevB.78.195424}, a non-zero    twisted  mass $m_2>0$ breaks both parity and time-reversal,  making  the  topological invariant no longer quantised~\cite{PhysRevLett.105.190404} which    to connects to  axion electrodynamics~\cite{PhysRevLett.58.1799}.

All of these effects require having an odd number of Dirac points with negative Wilson masses which, in the context of gauge theories such as QCD, would shift  the value of the  vacuum theta angle. When the goal is to recover  the continuum limit of QCD,  it is sensible to avoid this possibility unless one is using  a discretization based on  domain-wall fermions with an extra spatial dimension~\cite{KAPLAN1992342,GOLTERMAN1993219,PhysRevLett.108.181807}. Accordingly,  the works on lattice QCD with Wilson fermions typically  work in regions of parameter space that are close to a critical line in order to recover a continuum limit, but making sure that   these topological contributions to the theta angle are zero. The motivation to include  a mass twisting is then completely different from the above discussion. When trying to improve  the lattice regularization to achieve a faster convergence to the continuum~\cite{SYMANZIK1983187,LUSCHER1985250, JANSEN1996275}, the approaches based on Wilson fermions ~\cite{JANSEN1998185,DELLAMORTE2005117} can take advantage of the mass twisting~\cite{Roberto_Frezzotti_2001,Carlos_Pena_2004}. In fact, as first shown in~\cite{Frezzotti_2004}, by working with a maximal twist angle $\theta=\pi/2$, one can find an automatic $\mathcal{O}(a)$ improvement that only requires controlling  a single parameter of the model. 

If we now move back  to our $D=2+1$ dimensions, as noted above, an irreducible representation of the gamma matrices would forbid considering a mass twisting. On the other hand,  our reducible representation~\eqref{eq:gamma_spatial}-\eqref{eq:gamma_masses} permits  additional mass terms~\eqref{eq:action_free}, yielding a a 3-dimensional version of the  mass twisting. Still, we remark that this twisted Wilson mass  would be very different from  our anisotropic mass twisting in Eq.~\eqref{eq:twisted_mass}. As discussed in~\cite{SHINDLER200837}, the usual  mass twisting for Wilson fermions can be transformed to a  physical basis where, rather than the bare masses, one rotates the Wilson term responsible for the momentum dependence of the mass. Exploring  angles different from the full twist, e.g. $\theta=\pi/4$, would bring us closer to the type of Wilson mass twisting of Eq.~\eqref{eq:twisted_mass}. Yet, there is a fundamental difference, the Wilson mass twisting considered in lattice gauge theories is always isotropic. We are instead considering that the dependence on momentum of the twisted Wilson masses~\eqref{eq:twisted_mass} is highly anisotropic: $m_1(\boldsymbol{k})$ only depends on $k_1$, and $m_2(\boldsymbol{k})$   on $k_2$. In turn, as will become clear below, this means that the real-space Wilson term  is anisotropic; it is different when the fermions tunnel to neighbouring sites along each of the two spatial directions. The other important difference is that, instead of considering a mass twisting that combines $\overline{\psi}\psi$ and $\overline{\psi}\ii\gamma^5\psi$, we  are admixing $\overline{\psi}\ii\gamma^3\psi$ and $\overline{\psi}\ii\gamma^5\psi$  terms, both of which are parity invariant in $D=2+1$ dimensions. This will be very important to get a model with a hidden SO($5$) rotational symmetry, which will  protect the higher-order topological state.

In the following section, we will show in detail how the  groundstate of the free lattice action can  be characterised by  a topological invariant which is, however,   distinct from that~\eqref{eq:Chern_number} of the Chern insulators discussed above. Indeed, as shown in Sec.~\ref{sec:HOTI}, the groundstate  in this case  corresponds to a higher-order topological insulator (HOTI). The  bulk-boundary correspondence~\cite{https://doi.org/10.1002/pssb.202000090} leads to a boundary manifestation that differs from the edge states of  Chern insulator, as we find zero-energy  states that are only localised in the corners of the spatial lattice. To make this connection clearer, we  start by introducing a Hamiltonian  version of this QFT.

\section{\bf Cold-atom Hamiltonian field theory}
\label{sec:HOTI}
In this section, we present the Hamiltonian of the above Euclidean field theory~\eqref{eq:twisted_mass_S_0}, which will  be useful when discussing the HOTI, and a possible cold-atom implementation. 

\subsection{The Creutz-Hubbard multi-layer model}

Since our discretization keeps the imaginary time continuous, one can also describe the system through a Hamiltonian lattice field theory by rotating back to real time $tau\mapsto -\ii t$. In the Hamiltonian formulation~\cite{Peskin:1995ev,PhysRevD.11.395}, one works with field operators instead of Grassmann variables. We thus define $\Psi_{\boldsymbol{n}}=(\Psi_{\boldsymbol{n},0},\Psi_{\boldsymbol{n},1},\Psi_{\boldsymbol{n},2},\Psi_{\boldsymbol{n},3})^{\rm t}$  and $\Psi^{\dagger}_{\boldsymbol{n}}=(\Psi^\dagger_{\boldsymbol{n},0},\Psi^\dagger_{\boldsymbol{n},1},\Psi^\dagger_{\boldsymbol{n},2},\Psi^\dagger_{\boldsymbol{n},3})$ in terms of fermionic creation-annihilation operators defined on the lattice sites (see Fig.~\ref{fig:scheme_lattice} {\bf (a)}), which are supplemented with  the following equal-time anti-commutation relations 
\beq
\label{eq:car}
\left\{\Psi^{\phantom{\dagger}}_{\boldsymbol{n_1},\sigma_1},\Psi^\dagger_{\boldsymbol{n}_2,\sigma_2}\right\}=\frac{1}{a^2}\,\delta_{\boldsymbol{n}_1,\boldsymbol{n}_2}\delta_{\sigma_1,\sigma_2}.
\eeq
In a Minkowski spacetime, the adjoint is no longer independent but, instead,  related to the creation operators  $\overline{\Psi}_{\boldsymbol{n}}^{\phantom{\dagger}}={\Psi}^\dagger_{\boldsymbol{n}}\gamma^0$. The  Hamiltonian operator $H$ governing the dynamics of these fields can be found from the partition function discussed in Appendix~\ref{appA}, recalling that the  basis of fermionic coherent states   $\Psi_{\boldsymbol{n}}\ket{\psi_{\boldsymbol{n}}}={\psi_{\boldsymbol{n}}}\ket{\psi_{\boldsymbol{n}}}$, $\bra{\psi_{\boldsymbol{n}}}\overline{\Psi}_{\boldsymbol{n}}=\bra{\psi_{\boldsymbol{n}}}\overline{\psi}_{\boldsymbol{n}}$ is used to write the partition function $Z={\rm Tr}\{\ee^{-\beta H}\}$ as a path integral over the Grassmann  fields $\overline{\psi}_{\boldsymbol{n}},{{\psi}}_{\boldsymbol{n}}$~\cite{negele_orland_2019}. The identified operator  can be written as the sum of two terms $H=H_0^{\rm TM}+H_{\rm int}$. 
The  free term $H_0^{\rm TM}$ is obtained by the discretization of the spatial derivatives of Eq.~\eqref{eq:action_free} in terms of finite differences, which leads to  tunnelling terms between nearest neighbours. Additionally,  the   anisotropic twisted Wilson mass is also realized by including tunnelling terms that give momentum-dependence to the local masses in~\eqref{eq:action_free} according to Eq.~\eqref{eq:twisted_mass}. Altogether, this leads to the quadratic lattice Hamiltonian
\beq
\label{eq:tm_wilson_H}
H_0^{\rm TM}=a^2\!\!\!\sum_{\boldsymbol{n}\in\Lambda_s}\sum_{j=1,2}\!\bigg(\!\!\left(\Psi^\dagger_{\boldsymbol{n}}\mathbb{T}_j\Psi_{\boldsymbol{n}+{\bf e}_j}+{\rm H.c.}\right)+\Psi^\dagger_{\boldsymbol{n}}\mathbb{M}_j\Psi_{\boldsymbol{n}}\!\bigg),
\eeq
where  we have introduced  the  tunnelling matrices
\beq
\begin{split}
\label{eq:T_matrix}
\mathbb{T}_1=-\ii t\alpha^1- \tilde{t}\alpha^3,\hspace{2ex}
\mathbb{T}_2=-\ii t\alpha^2- \tilde{t}\alpha^5.
\end{split}
\eeq
These depend on the following tunnelling strengths
\beq
t=\frac{1}{2a},\hspace{1ex} \tilde{t}=\frac{r}{2a}, 
\eeq
as well as on the Dirac $\alpha$-matrices defined as
\beq
\label{eq:alphas}
\begin{split}
\alpha^1=\ii\gamma^0\gamma^1=\sigma^z\otimes\sigma^x,\hspace{2ex}\alpha^2=\ii\gamma^0\gamma^2=\sigma^z\otimes\sigma^y,\\
\alpha^3=\ii\gamma^0\gamma^3=\sigma^z\otimes\sigma^z,\hspace{2ex}\alpha^5=\ii\gamma^0\gamma^5=\sigma^x\otimes\mathbb{1}_2.
\end{split}
\eeq
Note that the above matrices are still expressed in terms of products of the Euclidean gamma matrices in Eqs.~\eqref{eq:gamma_spatial}-\eqref{eq:gamma_masses}. In Minkowski spacetime, it is customary to work with $\{\hat{\gamma}^a,\hat{\gamma}^b\}=2\eta^{a,b}\mathbb{1}_{d_s}$, where $\eta={\rm diag}\{1,-1,\cdots,-1\}$ is the  metric. Using the  prescription $\hat{\gamma}^0=\gamma^0,\hat{\gamma}^1=\ii\gamma^1,\hat{\gamma}^2=\ii\gamma^2,\hat{\gamma}^3=\ii\gamma^3,\hat{\gamma}^5=\gamma^5$, one  recovers the standard conventions for the Hamiltonian lattice field theory of Dirac fermions~\cite{Peskin:1995ev}. Using the standard definitions of the Dirac $\alpha$ and $\beta$ matrices, we also find 
\beq
\label{eq:beta}
\beta=\gamma^0=\sigma^y\otimes\mathbb{1}_2,\hspace{2ex} \alpha^j=\beta\hat{\gamma}^j,
\eeq
where the later coincide with the expressions  in Eq.~\eqref{eq:alphas}. 

In the above lattice Hamiltonian~\eqref{eq:tm_wilson_H}, we have also introduced the  mass matrices 
\beq
\label{eq:tunnelling_mass_matrices}
\mathbb{M}_1= \tilde{m}_1\alpha^3,\hspace{2ex}\mathbb{M}_2= \tilde{m}_2\alpha^5,
\eeq
which are expressed in terms of the following parameters
\beq
 \tilde{m}_1=m_1+2\tilde{t}, \hspace{1ex} \tilde{m}_2=m_2+2\tilde{t}.
\eeq
In addition to this quadratic Hamiltonian~\eqref{eq:tm_wilson_H}, the 4-Fermi terms  in Eq.~\eqref{eq:action_qurtic} lead directly to the  quartic interactions
\beq
\label{eq:hubbard_ints}
H_{\rm int}=a^2\!\!\!\sum_{\boldsymbol{n}\in\Lambda_s}\!\!\!\frac{g^2}{2}\!\Bigg(({\Psi}^\dagger_{\boldsymbol{n}}\Psi_{\boldsymbol{n}})^2-({\Psi}^\dagger_{\boldsymbol{n}}\boldsymbol{\alpha}\Psi_{\boldsymbol{n}})^2-({\Psi}^\dagger_{\boldsymbol{n}}\beta\Psi_{\boldsymbol{n}})^2\Bigg).
\eeq
In the next section, we will see how this specific interaction emerges naturally when considering spin-$3/2$ Fermi gases tightly-confined by optical potentials~\cite{PhysRevLett.91.186402,doi:10.1142/S0217984906012213}. We will argue that this connection fixes the choice of the $\alpha$ and $\beta$ matrices to those in Eqs.~\eqref{eq:alphas}-~\eqref{eq:beta}, and thus forces the choice of gamma matrices~\eqref{eq:gamma_spatial}-\eqref{eq:gamma_masses} in our 4-Fermi QFT in Eqs.~\eqref{eq:action_free} and~\eqref{eq:action_qurtic}. 

In the Hamiltonian formulation, the discussion of the SO($5$) invariance must be revisited   in light of the definition of the adjoint operator below Eq.~\eqref{eq:car}. In fact, the Euclidean SO($5$) Lorentz invariance   must now be described in terms of  SO($1,4$) Lorentz transformations, where the boosts do not admit a unitary spinor representation~\cite{Peskin:1995ev}. One could define a completely analogous Kaluza-Klein compactification, where the above  Hamiltonians arise from a 5-dimensional parent model  
regularised on a lattice. In analogy to the Euclidean action, the continuum limit of the dimensionally-reduced model is expected to recover the lower-dimensional SO(1,2) invariance. On the other hand, we are not only interested in the continuum limit, but also in the HOTI phases of the theory where one can go beyond this continuum emergent symmetry.  From this perspective, we should look for other transformations, including   discrete  spatial transformations, which correspond to exact symmetries of the full lattice model in~\eqref{eq:tm_wilson_H} and~\eqref{eq:hubbard_ints}. 


As depicted in Fig.~\ref{fig:scheme_lattice} {\bf (a)}, the non-interacting Hamiltonian~\eqref{eq:tm_wilson_H} can be interpreted as a multi-layer fermionic model with both intra- and inter-layer tunnellings.  An aspect that will be important in our analysis below is that there are certain background $\pi$-fluxes that dress the tunnelling along certain plaquettes involving the inter-layer synthetic dimensions (see Fig.~\ref{fig:scheme_lattice} {\bf (b)-(c)}). These fluxes lead to flat-band regimes, and a generalization of the so-called Aharonov-Bohm cages~\cite{PhysRevLett.81.5888}, which become very useful to understand the bulk-boundary correspondence.  In this regard, our model can be considered as a  higher-dimensional multi-layer version of the Creutz ladder~\cite{PhysRevLett.83.2636,RevModPhys.73.119,PhysRevLett.102.135702,PhysRevB.86.155140,PhysRevLett.112.130401,PhysRevB.96.035139,PhysRevB.99.054302}. Moreover, the quartic interactions~\eqref{eq:hubbard_ints} are purely local, and can thus be interpreted as a  Hubbard-like interaction. For $D=1+1$ dimensions, the Hubbard interaction maps exactly onto a Gross-Neveu quartic term~\cite{Junemann_2017,BERMUDEZ2018149,PhysRevB.99.125106,PhysRevB.106.045147}, although one could also use a bosonic species to mimic an auxiliary field that carries the Gross-Neveu interactions~\cite{PhysRevLett.105.190403}. For $D=2+1$ dimensions, when working with an irreducible representation and two-component spinors, the Hubbard interaction maps again into a Gross-Neveu quartic term~\cite{PhysRevResearch.4.L042012,ZIEGLER2022168763}.  In the current reducible case, where we have four-component spinors,  the Hubbard-type interaction is richer and contains both inter-layer density-density interactions among all pairs of spinor states, as well as spin-changing collisions that involve effective inter-layer  pair tunnellings (see Fig.~\ref{fig:scheme_lattice} {\bf (d)}). 

\subsection{Spin-$3/2$ atoms and  4-Fermi interactions }
\label{sec:ham_int_atoms}

In this section, we present the details of how spin-3/2 fermionic atoms can naturally lead to 
the 4-Fermi interactions in Eq.~\eqref{eq:hubbard_ints}. This is an example of the  unique opportunity emphasised in the introduction: the possibility of tailoring local and global symmetries that  connect to interesting models of high-energy physics.  This brings us closer to the field of quantum simulators~\cite{Cirac2012,Bloch2012,Blatt2012,PRXQuantum.2.017003}; controllable  quantum many-body systems that  behave according to a specific model of interest~\cite{Feynman_1982}. In the context of quantum simulators for high-energy physics (see the reviews~\cite{https://doi.org/10.1002/andp.201300104,Zohar_2016,doi:10.1080/00107514.2016.1151199,Banuls2020,Carmen_Banuls_2020,doi:10.1098/rsta.2021.0064,Klco_2022, https://doi.org/10.48550/arxiv.2204.03381,dimeglio_2023}), there have been  several proof-of-principle experiments showing the quantum simulation of relativistic QFTs~\cite{Gerritsma2010,PhysRevLett.106.060503,PhysRevLett.107.240401,Tarruell2012,LeBlanc_2013,Uehlinger2013,PhysRevLett.111.185307,Jotzu2014,doi:10.1126/science.1259052,doi:10.1126/science.aad5812,Martinez2016,Dai2017,PhysRevA.98.032331,Schweizer2019,Kokail2019,Pineiro_2019,HuertaAlderete2020,PhysRevResearch.5.L012006,PhysRevX.10.021041,PhysRevD.101.074512,Mil1128,Yang2020,doi:10.1126/science.abl6277,Atas2021,PhysRevLett.127.212001,PRXQuantum.3.020324,PhysRevD.103.094501,PhysRevD.104.034501,PhysRevD.105.074504,PhysRevLett.129.040402,PhysRevResearch.4.L022060,https://doi.org/10.48550/arxiv.2207.03473,https://doi.org/10.48550/arxiv.2207.01731,mildenberger2022probing,https://doi.org/10.48550/arxiv.2209.10781,Frolian2022,PhysRevResearch.5.023010,charles2023simulating,zhang2023observation}, including  lattice gauge theories~\cite{PhysRevD.10.2445,PhysRevD.11.395}. The case of gauge theories is particularly demanding in terms of the required resources, as the tailored symmetries must be local and, ultimately, non-Abelian,  requiring the introduction of additional gauge degrees to allow for this local symmetries. On the other hand, for synthetic  Dirac matter with quartic interactions, the requirements are in principle milder, as one  restricts to global and spacetime symmetries, including  non-Abelian ones, but one can dispense with the  extra  gauge degrees of freedom. 

Let us consider a gas of fermionic neutral atoms that can be tightly confined by optical potentials in a square lattice $\boldsymbol{x}=\sum_j\frac{\lambda_{\rm L}}{2}(n_j-N_j/2){\bf e}_j$ with $\boldsymbol{n}\in\Lambda_{\rm s}=\mathbb{Z}_{N_1}\times\mathbb{Z}_{N_2}$. We emphasise that the physical lattice spacing is set by  half the wavelength $\lambda_{\rm L}/2$ of the laser that leads to the optical-lattice potential, which is kept fixed in the experiment. This physical lattice spacing  will not be mapped onto the lattice-field-theory spacing $a$, which must be sent to $a\to0$ to recover the continuum limit. Another difference is that, in second quantization~\cite{Jaksch_2005, Lewenstein_2007, Gross_2017}, the atoms are described by  dimensionless   operators $f^\dagger_{\boldsymbol{n}, \sigma}$ ($f^{\vphantom{\dagger}}_{\boldsymbol{n}, \sigma}$)  that create (annihilate) an atom in the position specified by $\boldsymbol{n}$,  and in the  internal electronic internal state given by $\sigma\in \mathcal{S}_\sigma$. The set $\mathcal{S}_\sigma$ generally depends on the particular type of atom, and its specific isotope, which can also control the bosonic/fermionic nature of the operators. We will be interested in the fermionic case, where 
\beq
\label{eq:car}
\left\{f^{\phantom{\dagger}}_{\boldsymbol{n_1},\sigma_1},f^\dagger_{\boldsymbol{n}_2,\sigma_2}\right\}=\delta_{\boldsymbol{n}_1,\boldsymbol{n}_2}\delta_{\sigma_1,\sigma_2}.
\eeq
In the tight-binding regime, the system is thus described by a spin-conserving Hamiltonian, and its non-interacting part can be written~\cite{Jaksch_2005, Lewenstein_2007, Gross_2017} in second quantization as
\begin{equation}
\label{eq:sc_H}
    H_{\rm sc} = \sum_{\boldsymbol{n}}\sum_{\sigma, j} \left(-\tilde{t}_j f^\dagger_{\boldsymbol{n}, \sigma}f^{\vphantom{\dagger}}_{\boldsymbol{n} + \mathbf{e}_j, \sigma} + \text{H.c.}\right),
\end{equation}
where $\tilde{t}_j$ is the standard nearest-neighbour tunnelling coupling along the $j \in \{1, 2\}$ axis, and we set $\tilde{t}_1 = \tilde{t}_2 =: \tilde{t}$.

In order to find  the set of internal states $\mathcal{S}_{\sigma}$, we need to consider the atomic energy level structure, focusing in particular in the groundstate manifold. For instance, in the case of $^{6}$Li Alkali atoms, we have principal quantum number $n=2$, total orbital angular momentum $L=0$ and spin $S=1/2$ which, in spectroscopic notation leads to the $2^{2}S_{1/2}$ groundstate manifold. The total nuclear spin is $I=1$, which leads to a couple of hyperfine levels with total angular momentum $F\in\{1/2,3/2\}$. If we focus on the lower-energy state $F=1/2$, the set of internal states is given by the two possible Zeeman sub-levels $M_F=\pm\half$, namely  $\mathcal{S}_\sigma=\{-\half,+\half\}=:\{0,1\}$. At sufficiently cold temperatures, the scattering of the dilute Fermi gas is dominated by $s$-wave collisions between pairs of  $^{6}$Li atoms, which mostly contribute~\cite{Jaksch_2005}  to  a Hubbard density-density interactions that can be written as
\begin{equation}
\label{eq:sH_hubb}
    H_{\rm int} = \sum_{\boldsymbol{n}} \sum_{\sigma_1 \neq \sigma_2}\half{U_{\sigma_1\sigma_2}} n_{\boldsymbol{n}, \sigma_1} n_{\boldsymbol{n}, \sigma_2},
\end{equation}
where $n_{\boldsymbol{n}, \sigma}=f^{{\dagger}}_{\boldsymbol{n},\sigma}f^{\phantom{\dagger}}_{\boldsymbol{n},\sigma}$ is the  number operator, and $
U_{\sigma_1\sigma_2}=U_0=\sqrt{{8}/{\pi}}ka_0E_{\rm R}\left({V_{0,x}V_{0,y}V_{0,z}}/{E_{\rm R}^3}\right)^{\!{1}/{4}}$ is the Hubbard coupling strength~\cite{Bloch_2008}. Here,  $k_{\rm L}=2\pi/\lambda_{\rm L}$ is the laser wavevector, and $E_{\rm R}=k_{\rm L}^2/2m_a$ is the recoil energy of the $^{6}$Li atoms of mass $m_{a}$. In this expression, we have  introduced the optical-potential depths along the different axes, which will be constrained to $V_{0,z}\gg V_{0,x},V_{0,y}$ such that the dynamics takes place within   the $xy$ plane. Finally, a key quantity in the Hubbard coupling is
 the s-wave scattering length  $a_0$, which only depends on singlet scattering channel~\cite{Bloch_2008}. At cold temperatures and within the lowest hyperfine multiplet,  the inter-atomic potential is rotationally invariant within the total angular momentum of the colliding pair  $\boldsymbol{F}_{\rm t}=\boldsymbol{F}_{\rm 1}+\boldsymbol{F}_{\rm 2}$, which in this case leads to a couple of channels with $F_{\rm t}\in\{0,1\}$. Due to Pauli exclusion principle and the effective contact interactions for the $s$-wave channel, only the singlet case $F_{\rm t}=0$ is allowed, which is described by the above $s$-wave scattering length $a_0$. We note that the Hubbard interaction in Eq.~\eqref{eq:sH_hubb} has a global SU($2$) symmetry and, moreover, its strength can be controlled via Feshbach resonances by e.g.  applying a magnetic field~\cite{RevModPhys.82.1225}. 

This type of interaction~\eqref{eq:sH_hubb} would suffice to make connections to Gross-Neveu interactions for an irreducible representation of the gamma matrices~\cite{BERMUDEZ2018149,PhysRevResearch.4.L042012,ZIEGLER2022168763}, since the spinor components are only two. In this work, however, we are interested in reducible representations with a larger number spinor components $\mathcal{S}_\sigma=\{0,1,2,3\}$, where a larger non-Abelian symmetry appears in the interactions. A well-known example of large non-Abelian global symmetries in the scattering appears for other atomic species, such as $^{87}$Sr Alkaline-earth atoms. In this case,  there are two valence electrons, and  the  groundstate manifold has principal number $n=5$, and vanishing  total  spin and orbital angular momentum $S=L=0$, leading to the manifold $5^1S_0$. For vanishing $J=0$, there is no hyperfine splitting due to the nuclear spin, such that $F=I=9/2$, and we get a single multiplet with $N=10$ Zeeman sub-levels $M_F\in\{-9/2,-7/2,\cdots,9/2\}$. Since there is no hyperfine coupling, the atoms all interact with an $s$-wave scattering length that is independent of the nuclear features and, thus, equal for all of the $N=10$ sub-levels $U_{\sigma_1\sigma_2}=U_0$ $\forall\sigma_1\neq\sigma_2$. Hence, Eq.~\eqref{eq:sH_hubb} has an exact SU($N$) symmetry ~\cite{Cazalilla_2014,CAPPONI201650}. When considering also the  long-lived $5^3P_0$ level, one gets more flexibility, leading to the so-called two-orbital SU($N$) Hubbard models~\cite{Gorshkov2010}. When considering a mixed-species Fermi gas with a couple of alkaline-earth atoms, the inter- and intra-orbital scattering preserve the SU($N$) symmetry. Provided that one can control their corresponding scattering lengths  via Feshbach resonances, there are specific conditions where the two-orbital $SU(N)$ interactions would connect to  the Gross-Neveu term between $N$  fermion flavours.

In this article, however, we are interested in a specific type of 4-Fermi term that goes beyond the Gross-Neveu couplings~\eqref{eq:hubbard_ints}. These interactions, even for a single fermion flavor, have a non-Abelian SO($5$) symmetry. As realised in~\cite{PhysRevLett.91.186402,doi:10.1142/S0217984906012213}, it turns out that  there is an exact SO($5$) symmetry in the theory of $s$-wave scattering when working with spin-$3/2$ alkali gases, similar to  the case of $^{6}$Li  for the $F=3/2$ hyperfine multiplet. The only caveat is that we should consider other  atomic species in which the $F=3/2$ multiplet $\mathcal{S}_\sigma\in\{-3/2,-1/2,1/2,3/2\}=:\{0,1,2,3\}$ corresponds to the lowest-energy level, as  the scattering of the higher-energy hyperfine levels can otherwise lead to processes that bring the atoms into the lower hyperfine multiplet~\cite{PhysRevLett.81.742}. There are various possible Alkaline-earth atoms, such as the fermionic isotope $^{132}$Cs, which fulfill this condition and have an $F=3/2$ low-energy multiplet. To the best of our knowledge, experiments with Cesium have been reported only for  the bosonic $^{133}$Cs isotope, e.g.~\cite{Weber2003,doi:10.1126/science.1175850}. Other possibilities would be to work with the atomic species $^{9}$Be, $^{135}$Ba, $^{137}$Ba, and $^{201}$Hg.

Following our discussion above, the total angular momentum of a colliding pair could be $F_{\rm t}\in\{0,1,2,3\}$ in this case, where Pauli exclusion principle forbids the odd-momentum channels. We thus  have a pair of scattering lengths for the singlet $a_0$ and quintet $a_2$ channels.   The contact interactions can be written in terms of projection operators on these  two total angular momenta $P_{F_{{\rm t}}}(\boldsymbol{n})=\sum_{M_F}\sum_{\sigma_1,\sigma_2}\langle \sigma_1\sigma_2|F_{\rm t},M_{F_{\rm t}}\rangle f_{\boldsymbol{n},\sigma_1}f_{\boldsymbol{n},\sigma_2}$ ~\cite{PhysRevLett.81.742}, namely
\beq
 H_{\rm int} =\sum_{\boldsymbol{n}}\sum_{F_{\rm t}=0,2}U_{F_{\rm t}}P^\dagger_{F_{{\rm t}}}(\boldsymbol{n})P^{\phantom{\dagger}}_{F_{{\rm t}}}(\boldsymbol{n})
\eeq
which can be controlled by the individual  coupling strengths $U_{F_{\rm t}}=\sqrt{{8}/{\pi}}ka_{F_{\rm t}}E_{\rm R}\left({V_{0,x}V_{0,y}V_{0,z}}/{E_{\rm R}^3}\right)^{\!{1}/{4}}$. As shown in~\cite{PhysRevLett.91.186402}, these interactions can be rewritten as a linear combination of 4-Fermi terms  with a certain definition of gamma matrices. In order to connect to our previous discussion, we first need to define field operators with the right units, such that the spinor operator is $\Psi_{\boldsymbol{n}}=(f_{\boldsymbol{n},0},f_{\boldsymbol{n},1},f_{\boldsymbol{n},2},f_{\boldsymbol{n},3})^{\rm t}/a$, where $a$ is an effective lattice spacing that still needs to be connected to the microscopic cold-atom parameters. Using the corresponding Clebsch-Gordan coefficients $\langle \sigma_1\sigma_2|F_{\rm t},M_{F_{\rm t}}$, and the specific $\alpha$-$\beta$ Dirac matrices~\eqref{eq:alphas}-\eqref{eq:beta}, we find
\beq
\label{eq:hubbard_ints_so5}
\begin{split}
H_{\rm int}=a^2\!\!\!\sum_{\boldsymbol{n}\in\Lambda_s}\!\!\!\frac{1}{2}\!\Bigg(\!\tilde{g}^2({\Psi}^\dagger_{\boldsymbol{n}}\Psi_{\boldsymbol{n}})^2-g^2\Big(({\Psi}^\dagger_{\boldsymbol{n}}\boldsymbol{\alpha}\Psi_{\boldsymbol{n}})^2+({\Psi}^\dagger_{\boldsymbol{n}}\beta\Psi_{\boldsymbol{n}})^2\Big)\!\!\Bigg)\!\!,
\end{split}
\eeq
where we have introduced the vector $\boldsymbol{\alpha}=(\alpha^1,\alpha^2,\alpha^3,\alpha^5)$, and the individual coupling constants 
\beq
\label{eq:int_stregths}
\frac{g^2}{a^2}=\frac{U_2-U_0}{2},\hspace{2ex} \frac{\tilde{g}^2}{a^2}=\frac{3U_0+5U_2}{8}.
\eeq
 This brings us already really close to the desired 4-Fermi term in Eq.~\eqref{eq:hubbard_ints}, which would require tuning $g^2=\tilde{g}^{2}$, which would require setting $a_0=-13a_2/11$, we would recover exactly the interaction term that combines Thirring, Gross-Neveu and squared mass terms. Let us emphasise, however, that the coupling proportional to $\tilde{g}^2$, which is proportional to the temporal component of the fermion current $J_0^2$, shall not play any role in the phase diagram of the fermionic QFT.  Therefore, even if one cannot adjust the scattering lengths of the singlet and quintet channels, the physics will be completely equivalent, at least under the half-filling conditions explored in our work. In fact, as will also be clearer in our discussion below after discussing the cold-atom implementation of the quadratic part,  there is a hidden SO($5$) symmetry corresponding to a $90^{\rm o}$ rotation, regardless of the relative value of these two couplings. This symmetry is responsible for protecting the higher-order topological state  discussed in Sec.~\ref{sec:hiddern_SO5}.


\subsection{Raman lattices and twisted Wilson fermions}
\label{sec:Raman_lattices}
\begin{figure*}[t]
	\centering
	\includegraphics[width=\textwidth]{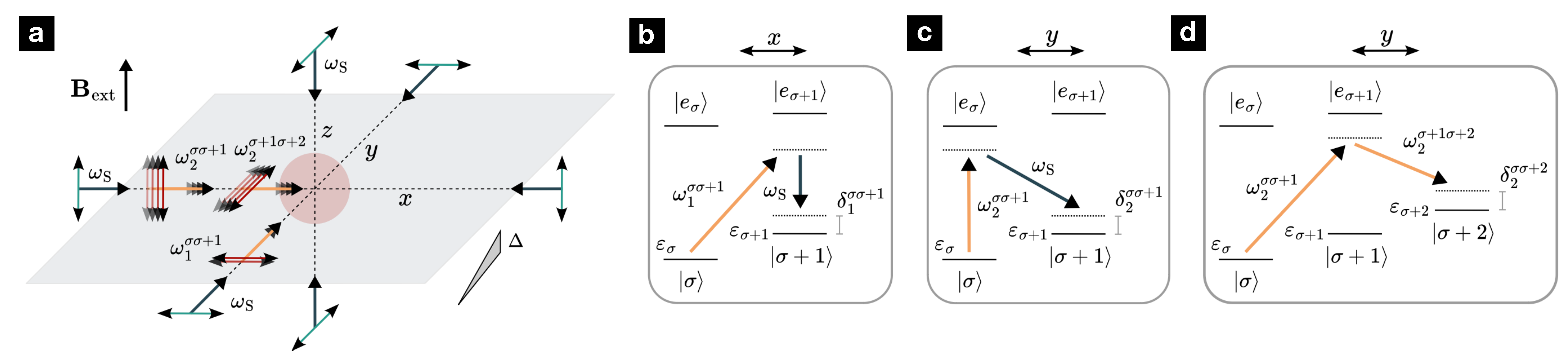}
	\caption{{\bf Raman optical lattice:}  {\bf (a)} The atom cloud (light red sphere) is subjected to a 3D optical lattice, created using three pairs of counter-propagating laser beams (blue arrows) with mutually orthogonal polarizations (green arrows) and frequencies $\omega_{\rm S}$. The atoms can be then confined into a 2D plane by increasing the potential depth in the $z$ direction. Additionally, travelling Raman beams (orange arrows) with frequencies $\omega^{\sigma\sigma^\prime}_j$ and appropriate polarizations (red arrows) generate spin-changing processes, as discussed in the main text. Finally, a magnetic field $\mathbf{B}_{\rm ext}$ is applied in the $z$ direction to split the hyperfine atomic energy levels, and a gradient obtained by lattice acceleration creates and energy difference $\Delta$ between nearest-neighbour sites in the $y$ direction. The panels {\bf (b)}, {\bf (c)} and {\bf (d)} depict the two-photon Raman transitions giving rise to spin-changing processes, where $\varepsilon_\sigma$ is the electronic energy of $\ket{\sigma}$, and $\delta^{\sigma\sigma^\prime}_j$  the corresponding detunings.}
	\label{fig:Raman_lattice}
\end{figure*}


Let us now discuss how to realise the twisted Wilson mass regularization in the cold-atom system. In order to obtain the tunnelling structure required by Eq.~\eqref{eq:tm_wilson_H}, we need to extend Eq.~\eqref{eq:sc_H} by considering additional Raman beams that also assist spin-changing tunnelling processes against certain energy offsets provided by  the Zeeman effect of an external magnetic field. This experimental scheme falls within the so-called Raman optical lattices~\cite{PhysRevLett.112.086401,PhysRevLett.113.059901,Wu83,PhysRevLett.121.150401,Songeaao4748,PhysRevResearch.5.L012006}, which have been exploited as quantum simulators of   synthetic spin-orbit coupling~\cite{Galitski2013,zhai_2015,book_soc}. In previous works~\cite{PhysRevResearch.4.L042012,ZIEGLER2022168763,PhysRevB.106.045147,FulgadoClaudio2023fermionproduction}, we highlighted the potential of Raman optical lattices for the quantum simulation of  Gross-Neveu-type QFTs with a standard Wilson discretization . For square lattices, these proposals connect to the recent realization of Chern insulators in~\cite{PhysRevResearch.5.L012006}. The goal of this section is to present a Raman-lattice scheme for four spinor components that could serve as a quantum simulator of our SO($5$) Dirac field theory regularised with the anisotropic twisted Wilson mass.

Let us first focus on tunnellings that change the atomic states corresponding to the  spinor components $\mathcal{S}_\sigma\in\{0,1,2,3\}$ by one unit, i.e. $\sigma\mapsto\sigma^\prime = \sigma + 1$, where we follow the conventions of  Ref.~\cite{ZIEGLER2022168763}. Along the $x$ ($y$) axis, these tunnellings can be assited by adding Raman beams along the $y$ ($x$) axis polarized in the $x$ ($z$) direction [Fig.~\ref{fig:Raman_lattice}{\bf (a)}]. Due to the difference in polarization, the latter, together with the $z$ ($x$)-polarized standing wave responsible for the standard optical lattice along the $x$ ($y$) axis, give rise to two-photon spin-changing Raman processes. The key observation is that, due to the different spatial periodicity of the standard lattice and these Raman process,  this spin-changing terms  cannot contribute with on-site terms, but drive instead  an assisted tunnelling. It can be shown that, in the tight-binding limit, this configuration gives rise to  spin-changing tunnellings~\cite{PhysRevLett.112.086401,PhysRevLett.113.059901,Wu83,PhysRevLett.121.150401} that read
\begin{equation}
\label{eq:sf_H}
\begin{aligned}
H^{\sigma}_{{\rm sf}, j} = -\sum_{\boldsymbol{n}} \Big[&\ii t^{\sigma\sigma+1}_j \ee^{\ii(\delta^{\sigma\sigma+1}_j t - \phi^{\sigma\sigma+1}_{j,\boldsymbol{n}})}f^\dagger_{\boldsymbol{n}, \sigma^{\vphantom{\prime}}}f^{\vphantom{\dagger}}_{\boldsymbol{n} +\mathbf{e}_j, \sigma+1}  \\
-&\ii t^{\sigma\sigma+1}_j \ee^{\ii(\delta^{\sigma\sigma+1}_j t - \phi^{\sigma\sigma+1}_{j,\boldsymbol{n}})}f^\dagger_{\boldsymbol{n}, \sigma^{\vphantom{\prime}}}f^{\vphantom{\dagger}}_{\boldsymbol{n} -\mathbf{e}_j, \sigma+1}  + \text{H.c.}\Big],
\end{aligned}
\end{equation}
where $t^{\sigma\sigma+1}_j$ is the corresponding  Raman-assisted tunnelling along the $j$-th axis. Here, we have introduced  $\phi^{\sigma\sigma+1}_{j,\boldsymbol{n}} = \phi^{\sigma\sigma+1}_j - \pi (n_1 + n_2)$, where $\phi^{\sigma\sigma+1}_j$ is the relative phase between the standing wave and the Raman beam. We have also introduced   $\delta^{\sigma\sigma+1}_j = \omega_{\rm S} - \omega^{\sigma\sigma+1}_j - (\epsilon_\sigma - \epsilon_{\sigma+1})$, which is the corresponding detuning for the two-photon Raman transition, with $\omega_{\rm S}$ and $\omega^{\sigma\sigma+1}_j$ the frequencies of the standing wave and the Raman beam, respectively, and $\epsilon_\sigma$ is the electronic energy for the level $\sigma$, which are controlled by the external magnetic field [see the two-photon transitions  in Fig.~\ref{fig:Raman_lattice}{\bf (b)} and {\bf (c)}].

In order to realize the lattice field theory described by Eq.~\eqref{eq:tm_wilson_H}, we need to combine these spin-changing processes. In particular, we need to connect the states $(0,1)$ and $(2,3)$ both in the $x$ and $y$ directions, and choose the proper phases $\phi^{\sigma\sigma^\prime}_j$, which can be checked by inspecting the structure of the $\mathbb{T}_1$ and $\mathbb{T}_2$ matrices in Eq.~\eqref{eq:T_matrix}. Additionally, we need processes that flip the spinor twice in the $y$ direction, connecting the states $(0,2)$ and $(1,3)$. These can be obtained in a similar fashion by using instead $y$-polarized Raman beams in the $x$ direction [Fig.~\ref{fig:Raman_lattice}{\bf (a)}], leading to the same expression as in Eq.~\eqref{eq:sf_H} for the case $\sigma\mapsto \sigma^\prime = \sigma + 2$, where $\phi^{\sigma\sigma+2}_2$ is now the relative phase between two Raman beams and $\delta^{\sigma\sigma+2}_2 = \omega^{\sigma+1\sigma+2}_2 - \omega^{\sigma\sigma+1}_2 - (\epsilon_\sigma - \epsilon_{\sigma+2})$, with $\omega^{\sigma+1\sigma+2}_2$ the frequency of the y-polarized Raman beam in the $x$ direction [Fig.~\ref{fig:Raman_lattice}{\bf (d)}]. Let us now detail how the relative phases need to be tuned for each different tunnelling process.

We first focus on $\mathbb{T}_1$~\eqref{eq:T_matrix}, this is, the tunnelling processes along the $x$ axis. The corresponding spin-changing terms can be obtained by using two Raman beams with the same polarization, as explained above, connecting the pairs $(0,1)$ and $(2,3)$. The latter can be produced from the same laser source using acusto-optical modulators to generate beams with different detunings $\delta^{\sigma\sigma^\prime}_j$ and phases $\phi^{\sigma\sigma^\prime}_j$. Here, we choose in particular $\delta^{01}_x = - \delta^{23}_x =: \delta_x$ and $\phi^{01}_x = \phi^{23}_x = \pi$. After performing the following gauge transformation and rescaling 
\begin{equation}
\label{eq:psi_f}
\begin{aligned}
\Psi_{\boldsymbol{n}, 0} &= \ee^{\ii\frac{\delta_x}{2}t}\frac{f_{\boldsymbol{n}, 0}}{a}, & \Psi_{\boldsymbol{n}, 2} &= \ee^{-\ii\frac{\delta_x}{2}t}\ee^{\ii\pi(n_1+n_2)}\frac{f_{\boldsymbol{n}, 2}}{a}, \\
\Psi_{\boldsymbol{n}, 1} &= \ee^{-\ii\frac{\delta_x}{2}t}\ee^{\ii\pi(n_1+n_2)}\frac{f_{\boldsymbol{n}, 1}}{a}, & \Psi_{\boldsymbol{n}, 3} &= \ee^{\ii\frac{\delta_x}{2}t}\frac{f_{\boldsymbol{n}, 3}}{a},
\end{aligned}
\end{equation}
it can be easily checked that the above configuration gives rise to the terms
$a^2\sum_{\boldsymbol{n}}\left[\left(\Psi^\dagger_{\bf{n}}\mathbb{T}_1\Psi_{\bf{n}+{\bf e}_1}+{\rm H.c.}\right)+\Psi^\dagger_{\bf{n}}\mathbb{M}_1\Psi_{\bf{n}}\right]$ in Eq.~\eqref{eq:tm_wilson_H}, with the following parameters,
$\tilde{m}_1 = {\delta_x}/{2},  t = {1}/{2a},  r = {t}/{\tilde{t}}$,
where we take $t := t^{01}_x = t^{23}_x$. We remark that the required $\mathbb{M}_1$ mass matrix~\eqref{eq:tunnelling_mass_matrices} is  generated  by the Raman-assisted tunnelling once we move to the above rotating frame.

Let us now consider the spin-changing processes associated to $\mathbb{T}_2$~\eqref{eq:T_matrix}, for which we also add a spin-independent gradient along the $y$ axis,
\begin{equation}
\label{eq:H_Delta}
    H_{\rm grad} = \Delta \sum_{n_x,n_y} n_y \sum_\sigma f^\dagger_{\boldsymbol{n}, \sigma}f^{\vphantom{\dagger}}_{\boldsymbol{n}, \sigma},
\end{equation}
which can be implemented e.g. by accelerating the optical lattice in that direction~\cite{Struck_2012}. This gradient serves a two-fold purpose. First, for $\Delta \gg \tilde{t}$, it suppresses the spin-conserving tunnelling in the $y$ direction, which is absent in Eq.~\eqref{eq:tm_wilson_H}. Additionally, it allows us to tune the relative phase between the two terms in Eq.~\eqref{eq:sf_H}, as required by $\mathbb{T}_2$. Specifically, for each of the four spin-changing pairs of terms involved in  $\mathbb{T}_2$, we employ two Raman beams, and choose the values of the detunings as follows: $\delta^{01}_{y} - \delta_x = \delta^{23}_{y} + \delta_x = - \tilde{\delta}^{01}_y + \delta_x = - \tilde{\delta}^{23}_y - \delta_x = \Delta$, where $\tilde{\delta}^{\sigma\sigma^\prime}_y$ denotes the second Raman beam, as well as $\delta^{01}_{y} = \delta^{02}_{y}$, $\delta^{23}_{y} = \delta^{13}_{y}$, $\tilde{\delta}^{01}_{y} = \tilde{\delta}^{02}_{y}$ and $\tilde{\delta}^{23}_{y} = \tilde{\delta}^{13}_{y}$. For every pair of levels, this allows each Raman beam to independently assists one single spin-changing process in Eq.~\eqref{eq:sf_H}. The phases of these processes can now be chosen freely, which can be seen by transforming first to the interaction picture with respect to the gradient term in Eq.~\eqref{eq:H_Delta}, and then applying the rotating-wave approximation in the limit of large detunings. In particular, if we take $\phi^{01}_y = \tilde{\phi}^{01}_y = \phi^{23}_y = \tilde{\phi}^{23}_y = - \pi/2$ and $\phi^{02}_y = - \tilde{\phi}^{02}_y = - \phi^{13}_y = \tilde{\phi}^{13}_y = - \pi / 2$, this configuration generates the terms $a^2\sum_{\boldsymbol{n}}\left(\Psi^\dagger_{\bf{n}}\mathbb{T}_2\Psi_{\bf{n}+{\bf e}_2}+{\rm H.c.}\right)$ in Eq.~\eqref{eq:tm_wilson_H} after applying again the transformations in Eq.~\eqref{eq:psi_f}, where we take $t^{01}_y = t^{23}_y = t$ and $t^{02}_y = t^{13}_y = \tilde{t}$. Finally, the mass term $a^2\sum_{\boldsymbol{n}}\Psi^\dagger_{\bf{n}}\mathbb{M}_2\Psi_{\bf{n}}$ can be  obtained by driving transitions between the $(0,1)$ and $(2,3)$ spinor pairs using microwave drivings with a Rabi frequency that gives the remaining micorsocopic parameter $\tilde{m}_2=\Omega_y/2$.

Let us finally note that the relevant dimensionless parameter that appear in the phase diagrams to be discussed in the rest of the article correspond to
\beq
m_1a=\frac{\delta_x}{4t}-r,\hspace{2ex}m_2a=\frac{\Omega_y}{4t}-r, \hspace{2ex}\frac{g^2}{a}=\frac{U_2-U_0}{4t},
\eeq
where we recall that the Wilson parameter is controlled by the ratio of the tunnellings $r=t/\tilde{t}$, and the fact that we have neglect the $\tilde{g}^2$ interaction as it will play no role (see the discussion below Eq.~\eqref{eq:int_stregths}). The important feature of this mapping is that all of the  relevant parameters can be tuned independently in the experiments. As noted previously, the continuum limit does not require sending $\lambda_{\rm L}\to 0$, but actually working in the vicinity of possible critical lines in parameter space $(m_1a,m_2a,g^2/a)$, which we start to explore below.

\section{\bf Higher-order topological insulators (HOTIs) and hidden SO($5$) symmetry}

In this section, we discuss the regions of parameter space where the free twisted-mass Wilson regularization can lead to HOTIs. In addition of presenting exact expressions for the zero-energy corner modes and the associated topological invariants, we will also discuss the symmetry  responsible for protecting these topological states.   
Although the original SO($5$) invariance of the 4-Fermi interactions  is explicitly broken by the lattice discretization, 
we find a hidden discrete SO($5$) rotation that protects  the   groundstate topology. 

\subsection{Flat bands and zero-energy corner modes}
\label{sec:hiddern_SO5}

We now discuss the topological features of the half-filled groundstate in the absence of interactions $g^2=0$. 
By performing a Fourier transform 
$\Psi_{\boldsymbol{n}}=\frac{1}{a\sqrt{N_{\rm s}}}\sum_{\boldsymbol{k}\in{\rm BZ}}\ee^{\ii \boldsymbol{k}\cdot\boldsymbol{n}a}\Psi_{\boldsymbol{k}}$, 
the lattice model~\eqref{eq:tm_wilson_H} becomes  $H_0^{\rm TM}=\sum_{\boldsymbol{k}\in{\rm BZ}}\Psi^{{\dagger}}_{\boldsymbol{k}}\mathbb{H}_0(\boldsymbol{k})\Psi^{\phantom{\dagger}}_{\boldsymbol{k}}$, where the single-particle Hamiltonian can be written as
\begin{widetext}
\beq
\label{eq:h_0}
\mathbb{H}_0(\boldsymbol{k})=\begin{pmatrix}
-2t\sin(k_1a)\sigma^x-2t\sin(k_2a)\sigma^y+(\tilde{m}_1-2\tilde{t}\cos(k_1a))\sigma^z & (\tilde{m}_2-2\tilde{t}\cos(k_2a))\mathbb{1}_2  \\
(\tilde{m}_2-2\tilde{t}\cos(k_2a))\mathbb{1}_2 & 2t\sin(k_1a)\sigma^x+2t\sin(k_2a)\sigma^y-(\tilde{m}_1-2\tilde{t}\cos(k_1a))\sigma^z  
\end{pmatrix}.
\eeq
\end{widetext}

\begin{figure}[t]
	\centering
	\includegraphics[width=1\columnwidth]{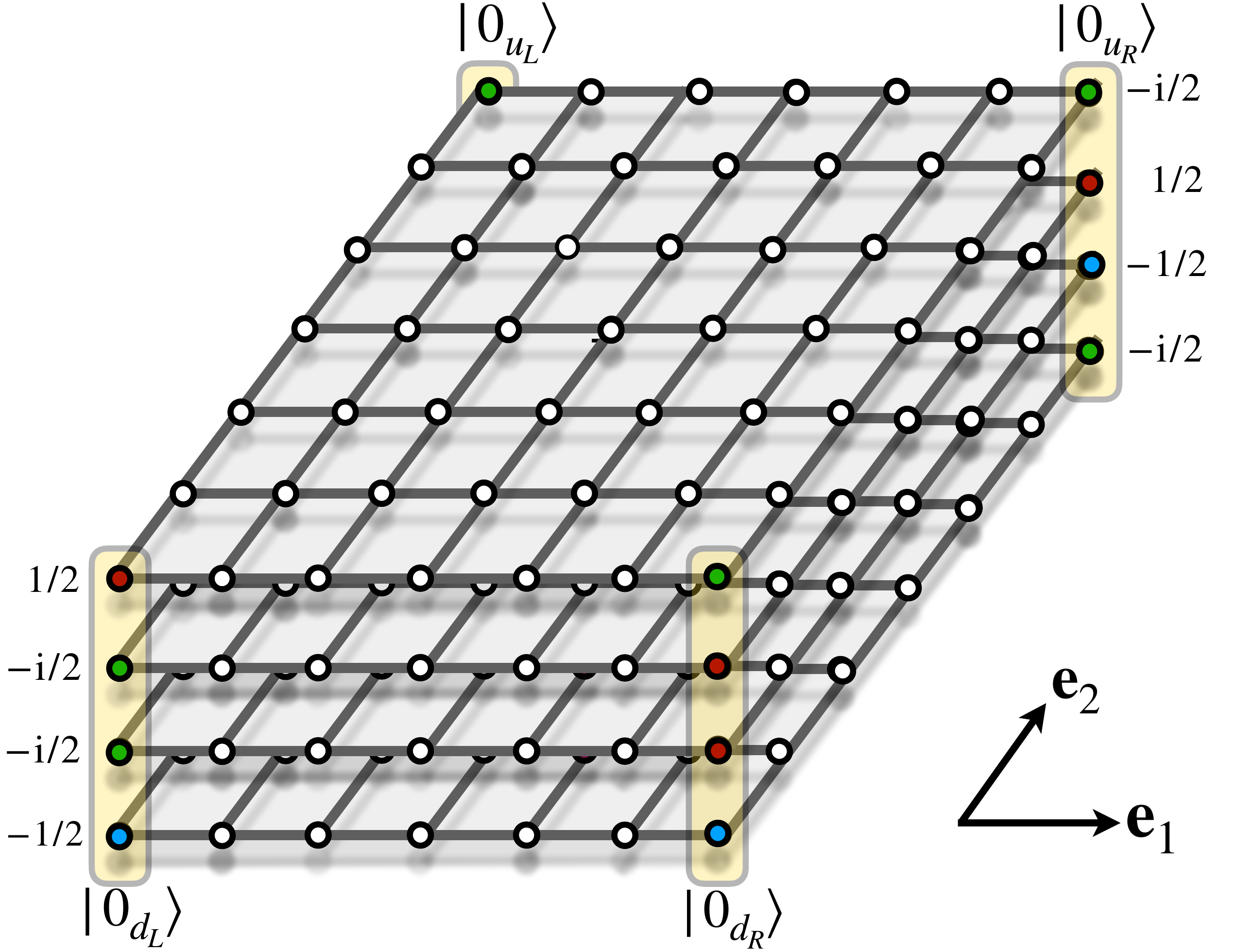}
	\caption{{\bf Anomalous corner states in the flat-band regime:}  For a finite multi-layer with parameters $m_1=m_2=r/a$ and $r=1$, the Hamiltonian displays two-fold degenerate flat bands $
\epsilon_{q,\pm}(\boldsymbol{k})=\pm2t$ and four zero-energy modes~\eqref{eq:corner_states} that are strictly localised to the corners of the lattice. The colouring of the sites at the corners corresponds to the complex phases of Eq.~\eqref{eq:corner_states}. }
	\label{fig:corner_free}
\end{figure}

Matrix diagonalization $\mathbb{H}_0(\boldsymbol{k})\ket{\epsilon_{q,\pm}(\boldsymbol{k})}=\epsilon_{q,\pm}(\boldsymbol{k})\ket{\epsilon_{q,\pm}(\boldsymbol{k})}$ with  $q\in\{1,2\}$, yields a band structure with four energy  bands $\epsilon_{q,\pm}(\boldsymbol{k})=\pm \epsilon(\boldsymbol{k})$ that display a  two-fold  degeneracy

\beq
\label{energy_bands}
\epsilon(\boldsymbol{k})=\sqrt{\sum_{j=1,2}\!\left(\!4t^2\sin^2(k_ja)\!+\!\big(\tilde{m}_j-2\tilde{t}\cos(k_ja)\big)^{\!2}\right)}.
\eeq 
This  
expression allows one to  realise that, by setting $\tilde{m}_1=\tilde{m}_2=0$ and $t=\tilde{t}$ (i.e. fixing $r=1$ and $m_1a=m_2a=-1$), the energy bands become totally flat $
\epsilon_{q,\pm}(\boldsymbol{k})=\pm2t$. This is the consequence of the  Aharonov-Bohm phases depicted in Fig.~\ref{fig:scheme_lattice}, which lead to a  destructive interference  forbidding  the tunnelling of bulk fermions  more than two sites apart along
any of the spatial directions (see, for instance, the vanishing amplitude of the two black-grey paths in Figs.~\ref{fig:scheme_lattice} {\bf (b)} and {\bf (c)}). This can be understood as the phenomenon of Aharonov-Bohm caging~\cite{PhysRevLett.81.5888}, and finds its minimal manifestation at the corners of the multi-layer, where a single fermion with certain amplitudes over the various spinor components must remain localised. Considering that we have a total of $N_{\rm s}=N_1N_2$ lattice sites, with $N_j$ sites per spatial direction, we find that the states corresponding to such  localised  solutions on the  corners
$\boldsymbol{d}_L=(1,1)$, $\boldsymbol{d}_R=(N_1,1)$, $\boldsymbol{u}_L=(1,N_2)$, and $\boldsymbol{u}_R=(N_1,N_2)$  have zero energy (see Fig.~\ref{fig:corner_free}). These zero modes can be expressed as follows 
\beq
\label{eq:corner_states}
\begin{split}
\ket{0_{\boldsymbol{d}_L}}&=\textstyle{\frac{1}{{2}}}\left(\Psi^\dagger_{\boldsymbol{d}_L,0}-\ii\Psi^\dagger_{\boldsymbol{d}_L,1}-\ii\Psi^\dagger_{\boldsymbol{d}_L,2}-\Psi^\dagger_{\boldsymbol{d}_L,3}\right)\ket{\rm vac},\\
\ket{0_{\boldsymbol{d}_R}}&=\textstyle{\frac{1}{{2}}}\left(\Psi^\dagger_{\boldsymbol{d}_R,0}+\ii\Psi^\dagger_{\boldsymbol{d}_R,1}+\ii\Psi^\dagger_{\boldsymbol{d}_R,2}-\Psi^\dagger_{\boldsymbol{d}_R,3}\right)\ket{\rm vac},\\
\ket{0_{\boldsymbol{u}_L}}&=\textstyle{\frac{1}{{2}}}\left(\Psi^\dagger_{\boldsymbol{u}_L,0}-\ii\Psi^\dagger_{\boldsymbol{u}_L,1}+\ii\Psi^\dagger_{\boldsymbol{u}_L,2}+\Psi^\dagger_{\boldsymbol{u}_L,3}\right)\ket{\rm vac},\\
\ket{0_{\boldsymbol{u}_R}}&=\textstyle{\frac{1}{{2}}}\left(\Psi^\dagger_{\boldsymbol{u}_R,0}+\ii\Psi^\dagger_{\boldsymbol{u}_R,1}-\ii\Psi^\dagger_{\boldsymbol{u}_R,2}+\Psi^\dagger_{\boldsymbol{u}_R,3}\right)\ket{\rm vac}.
\end{split}
\eeq

As argued in the following section, these corner states are the boundary manifestation of certain topological invariants~\cite{Hayashi2018,Hayashi2019} in the bulk bands of the system, which can lead to a topological quadrupole in analogy to the Bernevig-Benalcazar-Hughes (BBH) model~\cite{doi:10.1126/science.aah6442,PhysRevB.96.245115} of HOTIs, in particular to second-order topological insulators. When moving away from  the flat-band limit, e.g. by increasing the masses  $\tilde{m}_1=\tilde{m}_2>0$ or by switching on the quartic interactions $g^2>0$, the perfect Aharonov-Bohm interference will disappear, and  these zero modes will no longer be perfectly localised at the corners but, instead, start to spread within the bulk of the system. In particular, a certain localization length will emerge, which  characterises the exponential decay of the corner-state amplitude as one moves towards the bulk. These corner states, which  remain pinned to zero energy until the bulk energy gap is closed, are an example of  anomalous boundary states, which  only exist in the presence of the bulk. Moreover, they  are protected by a certain hidden  symmetry. 

We previously argued that, in the long-wavelength limit around  one of the Dirac points~\eqref{eq:Dirac_points}, the Euclidean action~\eqref{eq:twisted_mass_S_0} recovers the    SO($3$) rotational symmetry of the Lorentz group. In the Hamiltonian formulation, this corresponds to the  SO($1,2$) group of  boosts, and two-dimensional spatial rotations of angle $\theta$ within the $xy$ plane. Let us emphasise that this is a property of the long-wavelength limit, since the   lattice regularised action~\eqref{eq:twisted_mass_S_0} is not  invariant under arbitrary  rotations $k_1\mapsto k_1\cos\theta-k_2\sin\theta,k_2\mapsto k_2\cos\theta+k_1\sin\theta$. On the other hand, there may exist other  spatial symmetries that are exact for the  full lattice model, including the 4-Fermi interactions~\eqref{eq:action_qurtic}. For instance, as discussed in more detail in Appendix~\ref{appB}, there are two  parity symmetries which, at the level of the  fermionic field operators, act   as follows
\beq
\label{eq:parity_main}
\begin{split}
\Psi_{\boldsymbol{n}}\mapsto\mathcal{P}_1\Psi_{\boldsymbol{n}}&=\gamma^{1}\Psi_{(-n_1,n_{2})},\\
\Psi_{\boldsymbol{n}}\mapsto\mathcal{P}_2\Psi_{\boldsymbol{n}}&=\gamma^{2}\Psi_{(n_1,-n_{2})}.
\end{split}
\eeq
These transformations  correspond to mirror symmetries that take either $(k_1,k_2)\mapsto(-k_1,k_2)$ or $(k_1,k_2)\mapsto(k_1,-k_2)$, and clearly commute with the single-particle  Hamiltonian in Eq.~\eqref{eq:h_0}:  $\gamma^1\mathbb{H}_0(-k_1,k_2)\gamma^1=\gamma^2\mathbb{H}_0(k_1,-k_2)\gamma^2=\mathbb{H}_0(k_1,k_2)$. The composition  of the two parities corresponds to the lattice inversion, which is  a symmetry of the full  Hamiltonian corresponding to  a specific SO($1,2$)   rotation with angle $\pi$.

 We remark that the above symmetries need not   exhaust all possibilities, as there may be additional spatial symmetries that are not connected to the SO($1,2$) group. The possibility of finding such hidden symmetries becomes clear when inspecting the form of the interacting term~\eqref{eq:hubbard_ints} which, in fact, allows for generic SO($5$) rotations. In the Hamiltonian formulation, these admit a unitary representation, and do not include boosts but only spatial rotations. By simple inspection, it is clear that the first quartic-term $\sum_{\boldsymbol{n}}(\Psi_{\boldsymbol{n}}^\dagger\Psi_{\boldsymbol{n}}^{\phantom{\dagger}})^2$ that appears in Eq.~\eqref{eq:hubbard_ints} will be a scalar under any  such unitary transformation. In contrast,  the remaining terms can be rewritten as $\sum_{\boldsymbol{n}}\boldsymbol{N}^2_{\boldsymbol{n}}$, which involves the norm  of a 5-component vector of fermion bilinears
 \beq
 \boldsymbol{N}_{\boldsymbol{n}}=\Psi_{\boldsymbol{n}}^\dagger\big(\beta,\alpha^1,\alpha^2,\alpha^3,\alpha^5\big)\Psi_{\boldsymbol{n}}.
 \eeq 
 Accordingly, this part of the 4-Fermi interaction  is  also invariant under $ SO(5)$ rotations $\boldsymbol{N}_{\boldsymbol{n}}\mapsto R\boldsymbol{N}_{\boldsymbol{n}'}$, where $R$ is an orthogonal matrix $R^{\rm t}R=\mathbb{1}$ with ${\rm det}R=1$, and  $\boldsymbol{n}'$ is the corresponding two-dimensional rotation of the lattice. 
 
 Although  the twisted-mass Wilson regularization~\eqref{eq:tm_wilson_H}  breaks explicitly this arbitrary SO($5$) symmetry, there can exist specific rotation angles $\theta$ and matrices $R$ that correspond to an exact  discrete symmetry of the full model.  In particular, let us focus on  a discrete $\pi/2$-rotation that transforms the spatial coordinates as $\boldsymbol{n}=(n^1,n^2)\mapsto\boldsymbol{n}'=(-n^2,n^1)$. It is important to emphasise that the action of this  rotation  on the Dirac spinor $\Psi_{\boldsymbol{n}}\mapsto S_R\Psi_{\boldsymbol{n}'}$ will {\em not} be the same as that of the corresponding SO($1,2$) Lorentz rotation $\Lambda$, namely  $\Psi_{\boldsymbol{n}}\mapsto S_\Lambda\Psi_{\boldsymbol{n}'}$. As discussed in Appendix~\ref{appB},  such a Lorentz rotation is $S_\Lambda={\rm exp}\{\half\theta\gamma^{1}\gamma^2\}=\mathbb{1}_2\otimes{\rm exp}\{\frac{\ii}{4}\pi\sigma^z\}$, 
  and one can then check that invariance of the  lattice model   $H_0^{\rm TM}\mapsto H_0^{\rm TM}$   would  require a momentum-independent mass, which is no longer the case with the Wilson-mass  regularisation~\eqref{eq:twisted_mass}. In order to find a  $\pi/2$-rotation that leaves the model invariant we can, instead, look for a different rotation $S_R$ within the above  SO($5$) group of rotations. This  SO($5$) invariance requires that   
the Dirac $\alpha$ and $\beta$ matrices in Eqs.~\eqref{eq:alphas}-\eqref{eq:beta} transform as 
\beq
\label{sp:trans_alpha_beta}
\begin{split}
S_R^\dagger\alpha^1S_R^{\phantom{\dagger}}&=+\alpha^2,\hspace{2ex}  S_R^\dagger\alpha^2S_R^{\phantom{\dagger}}=-\alpha^1,\\ S_R^\dagger\alpha^3S_R^{\phantom{\dagger}}&=+\alpha^5,\hspace{2ex}    S_R^\dagger\alpha^5S_R^{\phantom{\dagger}}=+\alpha^3, \hspace{2ex}
 S_R^\dagger\beta^{} S_R^{\phantom{\dagger}}=-\beta.
\end{split}
\eeq
We note that the first row of Eq.~\eqref{sp:trans_alpha_beta} coincides with the transformations that would be obtained from the $\pi/2$-rotation of the  SO($1,2$) Lorentz group. On the other hand, the second row describes   different transformation laws that are crucial  to attain invariance of the twisted Wilson mass~\eqref{eq:tm_wilson_H}. One can check that Eq.~\eqref{sp:trans_alpha_beta} has the following solution 
\beq
\label{eq:c4}
 \Psi_{\boldsymbol{n}}\mapsto S_R\Psi_{\boldsymbol{n}'},\hspace{2ex} S_R=\frac{1}{\sqrt{2}}\left(\begin{matrix}
S^\dagger& S\\
S& -S^\dagger
\end{matrix}\right), \hspace{2ex}
\eeq
 where 
$S={\rm exp}\{\ii\frac{\pi}{4}(1-\sigma^z)\}$ 
is known as the phase-gate in quantum computing~\cite{nielsen00}, which maps the eigenvectors of $\sigma^x$ onto those of $\sigma^y$, namely $S\ket{\pm_x}=\ket{\pm_y}$.
 Combining these transformations with the action of the rotation on the crystal momenta, we find that the Bloch Hamiltonian~\eqref{eq:h_0} is indeed invariant when   $m_1=m_2$:  $S_R^{\phantom{\dagger}}\mathbb{H}_0(k_2,-k_1)S_R^\dagger=\mathbb{H}_0(k_1,k_2)$. At the level of the zero modes~\eqref{eq:corner_states}, the spatial part of this hidden SO($5$) transformation respects the set of corners, and one says that these anomalous boundary states are protected by this hidden spatial symmetry. Since they only have support on a region of codimension 2, they are the boundary manifestation of symmetry-protected HOTI groundstates.  We note that this   symmetry can be interpreted as the multi-layer counterpart  of the  $\mathcal{C}_4$ symmetry of the BBH model~\cite{doi:10.1126/science.aah6442,PhysRevB.96.245115}.

Before closing this subsection, we remark that the unitary transformation on the spinors $\Psi_{\boldsymbol{n}}\mapsto S_R\Psi_{\boldsymbol{n}'}$, together with the transformation of the Dirac $\alpha$ and $\beta$ matrix~\eqref{sp:trans_alpha_beta}, can be used to show that the vector of bilinears simply transforms as
 \beq
 \label{eq:so_5_hidden}
 \boldsymbol{N}_{\boldsymbol{n}}\mapsto \boldsymbol{N}_{\boldsymbol{n'}}=\Psi_{\boldsymbol{n}'}^\dagger\big(-\beta,\alpha^2,-\alpha^1,\alpha^5,\alpha^3\big)\Psi_{\boldsymbol{n'}}.
 \eeq 
  It is then clear that its norm is conserved and, thus, the quartic interactions~\eqref{eq:hubbard_ints} are also left invariant under this hidden SO($5$) rotation $H_{\rm int}\mapsto H_{\rm int}$.
 Altogether, this proves that the full lattice model is invariant $H=H^{\rm TM}_0+H_{\rm int}\mapsto H$. Since the   set of all four corners  is also invariant under a $\pi/2$ rotation, we expect that the anomalous corner states will be protected by this symmetry and, thus, robust when varying the microscopic parameters of the  model and switching on  interactions. The only  possibility to get rid of them is by closing the bulk energy gap, which would signal a quantum phase transition to a trivial band insulator or to a groundstate with symmetry-broken long-range order. We will explore these possibilities in the sections below but, first, let us provide a bulk perspective of the higher-order topology associated to these corner modes.


\subsection{ Higher-order topological invariants }

According to our current understanding of the bulk-boundary correspondence of symmetry-protected topological phases~\cite{https://doi.org/10.1002/pssb.202000090}, the anomalous zero modes are a boundary manifestation of a topological band structure in the bulk. For the present model, this phase should be a HOTI od second order with a certain non-vanishing topological invariant. As shown in~\cite{Hayashi2018,Hayashi2019}, this invariant can be expressed as the product of two winding numbers, which will allow us to find the  phase transitions between the HOTI and a  trivial insulator for  $g^2=0$.

At the level of the twisted-mass free Hamiltonian~\eqref{eq:tm_wilson_H}, one sees that  $\beta \mathbb{H}_0(\boldsymbol{k})\beta=-\mathbb{H}_0(\boldsymbol{k})$. This  corresponds to the $\mathsf{AIII}$  class in the classification of topological insulators under global symmetries, and shows that the band structure~\eqref{energy_bands} always comes in pairs of  positive-negative energies. The idea now is to find an alternative unitarily-equivalent representation of the Dirac matrices that intertwines the effect of this $\mathsf{AIII}$ symmetry with the dimensionality of the problem. This is, once more, only allowed by the fact that we are working with a reducible representation of the  Clifford algebra.  By applying the following unitary transformation $U$, built again from the $S$-gate, one finds that the  $\beta$-matrix transforms as
\beq
U=\frac{1}{\sqrt{2}}\left(\begin{matrix}
\ii S^\dagger& S\\
-\ii S& S^\dagger
\end{matrix}\right),\hspace{2ex} U^\dagger\beta U=\sigma^y\otimes\sigma^z=:\beta_2\otimes\beta_1,
\eeq
where  $\beta_1=\sigma^z$ and $\beta_2=\sigma^y$. On the other hand,    the Dirac $\alpha$ matrices transform according to the following expressions
\beq
\begin{split}
U^\dagger\alpha^1U&=+\mathbb{1}_2\otimes\sigma^x,\hspace{2ex}  U^\dagger\alpha^2U=+\sigma^x\otimes\sigma^z,\\ U^\dagger\alpha^3U&=-\mathbb{1}_2\otimes\sigma^y,\hspace{2ex}    U^\dagger\alpha^5U=-\sigma^z\otimes\sigma^z,
\end{split}
\eeq
which can be used to show that the transformed Bloch Hamiltonian~\eqref{eq:h_0} has the following tensor-product  structure
\beq
\label{eq:AIII_HOTI}
U^\dagger \mathbb{H}_0(\boldsymbol{k})U=\mathbb{1}_2\otimes \mathbb{h}_{1}(k_1)+\mathbb{h}_{2}(k_2)\otimes\beta_1.
\eeq
Here, we have defined the following single-particle Hamiltonians that only depend on the kinetic energy and Wilson mass along a specific direction in momentum space
\beq
\label{eq:ind_hams}
\begin{split}
\mathbb{h}_1(k_1)&=t\sin(k_1a)\sigma^x-(\tilde{m}_1-2\tilde{t}\cos(k_1a))\sigma^y,\\
\mathbb{h}_2(k_2)&=t\sin(k_2a)\sigma^x-(\tilde{m}_2-2\tilde{t}\cos(k_2a))\sigma^z.
\end{split}
\eeq

\begin{figure}[t]
	\centering
	\includegraphics[width=1\columnwidth]{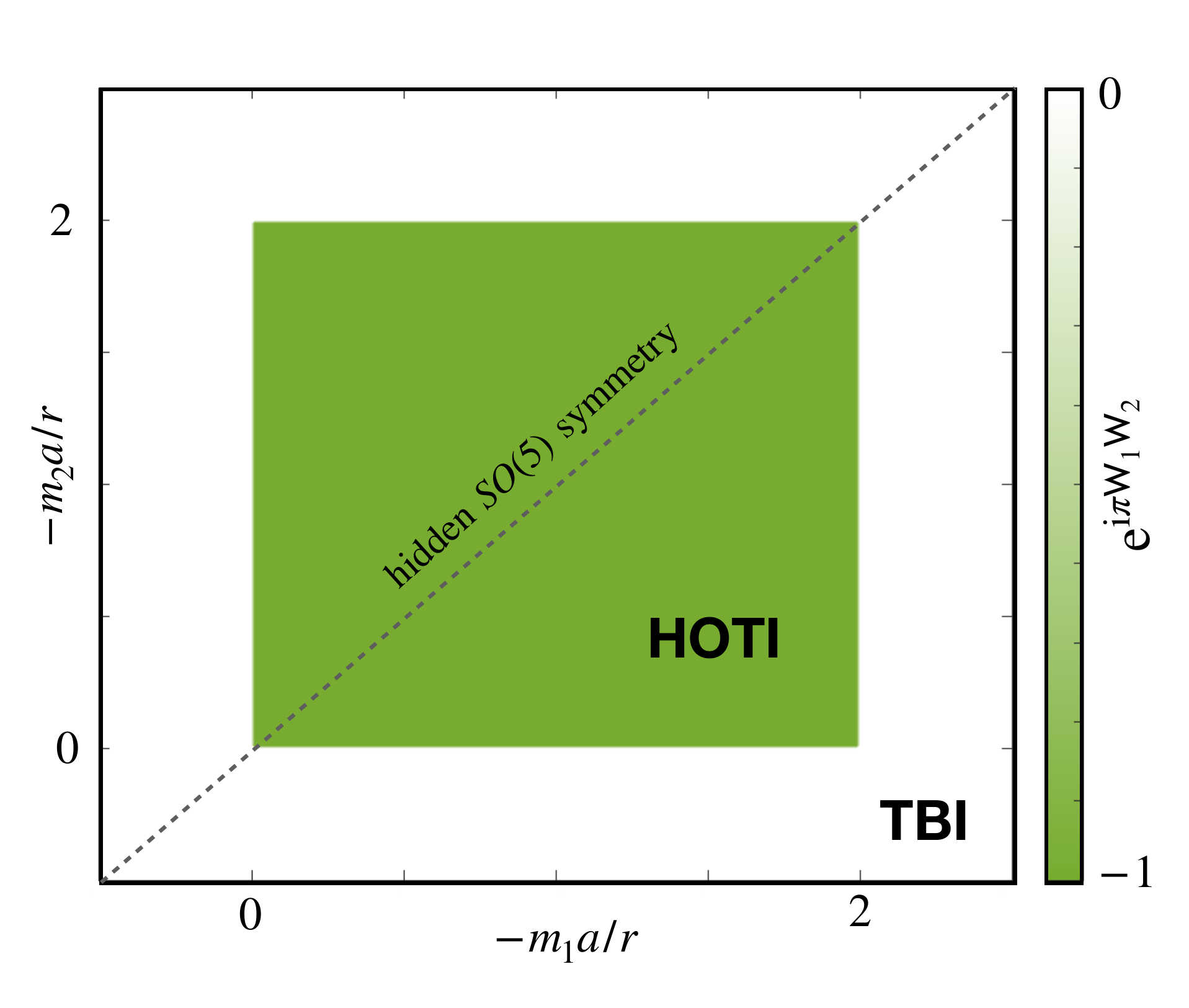}
	\caption{{\bf Non-interacting higher-order topological phase diagram:} We represent the topological invariant in Eq.~\eqref{eq:top_inv} as a function of the bare twisted masses $m_1$ and $m_2$. In the green inner square $m_ja\in(-2r,0)$, the topological invariant is non-trivial $\ee^{\ii\pi\mathsf{W}_1\mathsf{W}_2}=-1$, and the groundstate of the twisted-Wilson lattice model corresponds to a higher-order topological insulator (HOTI). The shaded line with $m_1=m_2$ represents the regime where the model has a hidden SO($5$) symmetry that protects the  corner modes. The white region where $m_ja\notin(-2r,0)$ for one or both masses corresponds to a trivial band insulator (TBI) with trivial topological invariant $\ee^{\ii\pi\mathsf{W}_1\mathsf{W}_2}=+1$.   }
	\label{fig:HOTI_free}
\end{figure}

One can  check that  each of these Hamiltonians has an individual $\mathsf{AIII}$ symmetry $\beta_1\mathbb{h}_1(k_1)\beta_1=-\mathbb{h}_1(k_1)$, and $\beta_2\mathbb{h}_2(k_2)\beta_2=-\mathbb{h}_2(k_2)$, which guarantees that the full Hamiltonian fulfills  the desired transformation  $\beta \mathbb{H}_0(\boldsymbol{k})\beta=-\mathbb{H}_0(\boldsymbol{k})$. Each of these terms~\eqref{eq:ind_hams} can be understood as a two-band model  corresponding  to an un-twisted Wilson regularization of a $(1+1)$-dimensional Dirac-fermion QFT~\cite{BERMUDEZ2018149}. This tensor-product construction also highlights that the corner states are not equivalent to  the edge states of   $\mathsf{AIII}$ topological insulators of  one-dimensional chains, arranged along the boundaries of a square lattice. It is really the two-dimensional bulk that is required to  host and protect these corner modes.

The lower-dimensional band structures of $\mathbb{h}_j(k_j)$ have a pair of Dirac points corresponding to the projection of the previous Dirac points~\eqref{eq:Dirac_points} onto the respective axis $k_{{\rm D},\ell_j}=\boldsymbol{k}_{{\rm D},\boldsymbol{\ell}}\cdot{\bf e}_j$, each of which presents  a different  mass
\beq
k_{{\rm D},j}=\frac{\pi}{a}\ell_j,\hspace{1ex}\ell_j\in\{0,1\},\hspace{2ex} m_j(\boldsymbol{k}_{{\rm D},\boldsymbol{\ell}})=\tilde{m}_j-2\tilde{t}(-1)^{\ell_j}.
\eeq
In analogy to our discussion of the Chern insulator~\eqref{eq:Chern_number} an the mass matrix~\eqref{eq:mass_matrix}, we can now define two mass matrices
\beq
M_{{\rm W},1}=\!\!\sum_{l_1=0,1}\!\!m_1(\boldsymbol{k}_{{\rm D},\boldsymbol{\ell}})\ket{\ell_1}\!\bra{\ell_1},\hspace{1ex}M_{{\rm W},2}=\!\!\sum_{l_2=0,1}\!\!m_2(\boldsymbol{k}_{{\rm D},\boldsymbol{\ell}})\ket{\ell_2}\!\bra{\ell_2}\!,
\eeq
each of which contains the information about each of the twisted Wilson masses at the corresponding projection of the Dirac point.
The Berry connection  for each of these projections is  $\mathcal{A}_i(k_i)=-\ii \bra{\epsilon_{-}({k}_i)}\partial_{k_i}\ket{\epsilon_{-}({k}_i)}$, where $\ket{\epsilon_{-}({k}_i)}$ are the  negative-energy modes that would be filled in the corresponding lower-dimensional groundstate. One can define a Chern-Simons form~\cite{Ryu_2010} associated to this Berry connection or, equivalently, a 
 so-called Zak's phase~\cite{PhysRevLett.62.2747}, which plays the role of the above Chern number~\eqref{eq:Chern_number} in this reduced dimensionality. We find that each of these  invariants 
 \beq
 \label{eq:Chern-Simons_forms_def}
 \mathsf{CS}_j=\frac{1}{2\pi}\int\!\!{\rm d}k_j\mathcal{A}_j(k_j)=\frac{1}{2\pi}{\rm arg}\left\{{\rm Det}\big({M}_{{\rm W},j}\big)\right\}.
 \eeq
 is again non-trivial when an odd number of the projected Dirac points have a negative twisted Wilson mass
  \beq
 \label{eq:Chern-Simons_forms}
 \mathsf{CS}_j=\frac{1}{4}\sum_{\ell_j=0,1}(-1)^{\ell_j}{\rm sgn}\{m_j(\boldsymbol{k}_{{\rm D},\boldsymbol{\ell}})\}.
 \eeq
As is well-known for standard first-order  topological insulators~\cite{bernevig_hughes_2013}, these topological invariants $ \mathsf{CS}_j=\mathsf{W}_j/2$ are proportional to the winding number $\mathsf{W}_j$  of the mappings $\hat{\boldsymbol{d}}_j(k_j)={\boldsymbol{d}}_j(k_j)/||{\boldsymbol{d}}_j(k_j)||: {\rm BZ}\mapsto U(1)$, where ${\boldsymbol{d}}_j(k_j)$ is the vector of coefficients of the individual Hamiltonians~\eqref{eq:ind_hams} in the Pauli basis $h_j(k_j)={\boldsymbol{d}}(k_j)\cdot\boldsymbol{\sigma}$. As discussed in Reference~\cite{Ryu_2010}, one can define a  topological invariant that is also gauge invariant by considering the Wilson loop associated to such winding number. In this way, by simply multiplying the  winding numbers together and exponentiating them, we obtain a topological invariant for HOTIs with $\mathsf{AIII}$ symmetry~\eqref{eq:AIII_HOTI}, which reads
\beq
\label{eq:top_inv}
\ee^{\ii\pi\mathsf{W}_1\mathsf{W}_2}=\left\{\begin{array}{ll} 
-1 &\,\,\,\, {\rm if}\,\,\,\, -2r<m_ja<0, \,\,\,\,\forall j\in\{1,2\},\\
+1 &\,\,\,\, {\rm else}\,\,\,\, 
\end{array}\right.
\eeq

A non-vanishing invariant $\ee^{\ii\pi\mathsf{W}_1\mathsf{W}_2}=-1$ signals the non-trivial topology of the bulk, and must have a boundary manifestation in the form of  corner states. Indeed, for $m_1a=m_2a=r$, which corresponds to the previous flat-band limit $\tilde{m}_1=\tilde{m}_2=0$ when $r=1$, we find that $\ee^{\ii\pi\mathsf{W}_1\mathsf{W}_2}=-1$ in the bulk, while the zero-energy corner states are those of  Eq.~\eqref{eq:corner_states}. Away from this flat-band limit, and while $m_ja\in(-2r,0)$ in both directions, the groundstate is still a HOTI with $\ee^{\ii\pi\mathsf{W}_1\mathsf{W}_2}=-1$, and the anomalous boundary states  remain exponentially localised to the corners. The non-interacting HOTI in parameter space corresponds to  the square  displayed in Fig.~\ref{fig:HOTI_free}. For mass parameters in the inner square, the groundstate is a HOTI whereas, outside this region, it is  trivial. 

In the next section, we will explore the fate of this HOTI as the fermion self-interactions $g^2$ are increased. Let us note that, in our $D=3$-dimensional spacetime, considering the role of the  quartic interactions by simple power counting can be misleading. Indeed, the dimensions of the Dirac field is $[\Psi_{\boldsymbol{n}}]=\mathsf{L}^{-1}$, such that   $[t_j]=[\tilde{t}_j]=[m_1]=[m_2]=\mathsf{L}^{-1}$, whereas the coupling strength has units of length $[g^2]=\mathsf{L}$ and, thus, a negative energy dimension.  Naive power-counting arguments would then suggest the 4-Fermi interaction  is perturbatively {\em irrelevant}, such that the QFT in $D=2+1$ would be non-renormalizable. On the other hand, it is well-known that the relevant/irrelevant nature of the couplings  can be modified after resummation in the large-$N$ limit~\cite{coleman_1985}, whereupon the Thirring-like interaction $(\Psi^\dagger_{\boldsymbol{n}}\Psi_{\boldsymbol{n}})^2+\sum_{j=1,2}(\Psi_{\boldsymbol{n}}^\dagger\alpha^j\Psi_{\boldsymbol{n}})^2$ becomes {\em marginal}, whereas the Gross-Neveu-like interactions $(\Psi^\dagger_{\boldsymbol{n}}\beta\Psi_{\boldsymbol{n}})^2$ become {\em relevant}. Even if  the corresponding  4-Fermi  field theories are not perturbatively renormalizable, they become $1/N$ renormalizable~\cite{doi:10.1142/9789814412674_0004,PhysRevD.21.2327,HANDS199329,PhysRevLett.62.1433,ROSENSTEIN199159}. The goal of the following section is to analyse the effect of the 4-Fermi interactions on the HOTI, including also the  mass-squared interactions $\sum_{j=3,5}(\Psi^\dagger_{\boldsymbol{n}}\alpha^j\Psi_{\boldsymbol{n}})^2$, using the large-$N$ expansion.

\section{\bf Correlated HOTI\lowercase{s},  fermion condensates and  quantum phase transitions}

In this subsection, we discuss the effect of the 4-Fermi interactions in detail. We argued previously that the HOTI groundstate is protected by a hidden SO($5$) symmetry and, thus, should be robust under symmetric perturbations unless those are sufficiently strong such that the bulk energy gap closes allowing for a change of the topological invariant,  or if a certain symmetry-breaking phase transition takes place. We also showed in Eq.~\eqref{eq:so_5_hidden} that the 4-Fermi interactions preserve this hidden symmetry, such that one expects  the HOTI phase to become a correlated HOTI  as one increases the coupling strength $g^2>0$. Eventually, when the interactions are sufficiently strong, there may be a symmetry-breaking phase transition at some $g^2_c$, which paves the way for the appearance of new phases of matter  typically referred to as fermion condensates in the QFT literature. The goal of this section is to explore the possible condensates allowed by the rich SO($5$) structure of the self-interactions, and provide a quantitative account about which of the fermion condensates is expected to form at which point in parameter space $(m_1a,m_2a,g^2/a)$.

\subsection{Auxiliary fields and the hidden  symmetry}
To accomplish this goal, we shall return to the Euclidean formulation of our SO($5$) Dirac matter in Eqs.~\eqref{eq:action_free} and~\eqref{eq:action_qurtic}, where we can make use of controlled approximations such as the large-$N$ expansion~\cite{coleman_1985}. To present a  non-perturbative, yet tractable, account of the 4-Fermi interactions, one can generalise the Euclidean action in Eqs.~\eqref{eq:action_free} and~\eqref{eq:action_qurtic} to $N$ flavours 
\beq
\begin{split}
\psi(x)\mapsto\psi(x)&=(\psi_1(x),\psi_2(x),\cdots, \psi_N(x))^{\rm t}\\
\overline{\psi}(x)\mapsto\overline{\psi}(x)&=(\overline{\psi}_1(x),\overline{\psi}_2(x),\cdots, \overline{\psi}_N(x)).
\end{split}
\eeq
The free part of the regularised Wilson twisted-mass  action $S_0^{\rm TM}$~\eqref{eq:twisted_mass_S_0} simply becomes a sum of the corresponding actions for each of the fermion flavours. In contrast, the  4-Fermi term does  couple the different flavours where, to have a consistent $N\to\infty$ limit, one must rescale the coupling strength as
\begin{equation}
\label{eq:action_qurtic_app}
S_{\rm int}\!=\!\!\int\!{\rm d}^{3}x\frac{g^2}{2N}\!\!\left({\color{blue}-}J_\mu J^\mu+(\overline{\psi}\gamma^3\psi)^2+(\overline{\psi}\gamma^5\psi)^2-(\overline{\psi}\psi)^2\!\right)\!,
\end{equation}
where the gamma matrices for $N$ flavours appearing the in the above bilinears  should be understood as $\gamma^a\mapsto\mathbb{1}_N\otimes\gamma^a$. We will assume this in all the expressions below, which leads to an additional $U(N)$ symmetry in  flavour space.

The first step of the large-$N$ approximation is to introduce auxiliary fields via a Hubbard-Stratonovich transformation~\cite{startonovich,PhysRevLett.3.77}, such that the partition function becomes quadratic in the Grassmann spinors. We need to introduce  six real  bosonic fields $a_\mu(x),\sigma_1(x),\sigma_2(x)$, and $\pi(x)$, such that the partition function can be exactly rewritten as
\beq
\label{eq:z_N_flavours}
Z=\int[{\rm D}\overline{\psi}{\rm D}{\psi}{\rm D}a_\mu{\rm D}\sigma_j{\rm D}\pi]{\rm e}^{-S^{\rm TM}_0-S_{\rm int}'-\int\!\!{\rm d}^3x\frac{N}{2g^2}\left(a_\mu a^\mu+\sigma_j\sigma^j+\pi^2\right)}.
\eeq
As a consequence of this transformation, we get the following action coupling between the fermionic and  bosonic fields
\beq
\label{eq:aux_field_int}
S_{\rm int}'=\!\int\!\!{\rm d}^3x\overline{\psi}(x)\!\left(\!\ii \gamma^\mu \!a_{\mu}(x)+\ii\gamma^3\!\sigma_1(x)+\ii\gamma^5\!\sigma_2(x)+\pi(x)\!\right)\!\!\psi(x)\!.
\eeq
Except for the $\pi$-field, the rest  can be incorporated in the free action $S_0^{\rm TM}$ by the following substitution
\beq
\partial_\mu\mapsto \partial_\mu+\ii a_\mu(x),\hspace{2ex}
m_j\mapsto m_j+\ii\sigma_j(x),
\eeq
which  shows that  the auxiliary fields $a_\mu(x)$ act as an effective gauge-like potential, the components of which admix by the Lorentz transformations, and couple to the fermions minimally. We stress, however, that the free action~\eqref{eq:z_N_flavours} of these auxiliary fields is not gauge invariant, but simply a mass-like term that becomes very heavy in the large-$N$ limit. On the other hand, the  $\sigma_j(x)$ are scalar fields that couple to the fermions via a pair of twisted  Yukawa-type couplings. To be more accurate, we should consider the Euclidean formulation of the two parity transformations~\eqref{eq:parity_main}, which amount to 
\beq
\label{eq:parity_transf}
\begin{split}
\mathcal{P}_1\psi(x)= \gamma^1\psi(\tau,-x_1,x_2),\hspace{1ex} \mathcal{P}_1\overline{\psi}(x)= -\overline{\psi}(\tau,-x_1,x_2)\gamma^1,\\
\mathcal{P}_2\psi(x)= \gamma^2\psi(\tau,x_1,-x_2),\hspace{1ex} \mathcal{P}_2\overline{\psi}(x)= -\overline{\psi}(\tau,x_1,-x_2)\gamma^2.
\end{split}
\eeq
Since the 4-Fermi interactions are invariant under these parity transformations, we know that $S_{\rm int}'\mapsto S_{\rm int}'$, which require the  auxiliary gauge-like fields to transform as
\beq
\begin{split}
&\mathcal{P}_1a_\mu(x)= -(1-2\delta_{\mu, 0})a_\mu(\tau,-x_1,x_2),\\
&\mathcal{P}_2a_\mu(x)= -(1-2\delta_{\mu, 0})a_\mu(\tau,x_1,-x_2), 
\end{split}
\eeq
whereas the auxiliary $\sigma$ fields transform as
\beq
\begin{split}
\hspace{1ex}\mathcal{P}_1\sigma_j(x)=\sigma_j(\tau,-x_1,x_2),\\
\hspace{1ex}\mathcal{P}_2\sigma_j(x)=\sigma_j(\tau,x_1,-x_2).
\end{split}
\eeq
Therefore, we see that the $\sigma$ fields  are all parity even, the $a_0(x)$ component of the gauge-like  field is parity even, while the spatial components $\boldsymbol{a}(x)$ are parity odd.  

We have so far left out  the discussion of the $\pi(x)$ field, as it gives rise to a new term that was not present in $S_0^{\rm TM}$. Considering the parity transformations in Eq.~\eqref{eq:parity_transf}, we see that in order to recover parity invariance, the $\pi(x)$ field should transform as  
\beq
\begin{split}
&\mathcal{P}_1\pi(x)\mapsto-\pi(\tau,-x_1,x_2),\\
&\mathcal{P}_2\pi(x)\mapsto-\pi(\tau,x_1,-x_2),
\end{split}
\eeq
and, thus, corresponds to a pseudo-scalar field that is parity odd. According to Eq.~\eqref{eq:aux_field_int}, this pseudo-sclalar auxiliary field couples to the fermions via the standard  Yukawa coupling.

Let us finally discuss the hidden  SO($5$) symmetry~\eqref{eq:c4} which, for the Euclidean Grassmann fields,  corresponds to 
\beq
\label{eq:C4_grassmann}
\psi(\tau,\boldsymbol{x})\mapsto  S_R\psi(\tau,x_2,-x_1),\hspace{1ex}\overline{\psi}(\tau,\boldsymbol{x})\mapsto  -\overline{\psi}(\tau,x_2,-x_1)S_R^\dagger,
\eeq
where we recall that $S_R$ is the unitary matrix given by Eq.~\eqref{eq:c4}. Considering that the Euclidean gamma matrices transform as 
\beq
\begin{split}
S_R^\dagger\gamma^1S_R&=-\gamma^2,\hspace{2ex}  S_R^\dagger\gamma^2S_R=+\gamma^1,\\ S_R^\dagger\gamma^3S_R&=-\gamma^5,\hspace{2ex}    S_R^\dagger\gamma^5S_R=-\gamma^3, \hspace{2ex}    S_R^\dagger\gamma^0S_R=-\gamma^0,
\end{split}
\eeq
one can readily see that the original  Euclidean action Eqs.~\eqref{eq:action_free} and~\eqref{eq:action_qurtic} is also invariant under this hidden rotational symmetry~\eqref{eq:C4_grassmann} when $m_1=m_2$. Once the auxiliary fields are introduced, this hidden symmetry implies that the vector field should  transform as 
\beq
\label{eq:c4_vector}
\begin{split}
&a_0(x)\mapsto +a_0(\tau,x_2,-x_1),\\
&a_1(x)\mapsto +a_2(\tau,x_2,-x_1),\\
&a_2(x)\mapsto -a_1(\tau,x_2,-x_1),
\end{split}
\eeq
whereas the scalar and pseudo-scalar fields must fulfill
\beq
\begin{split}
\label{eq:c4_scalars}
\sigma_1(x)&\mapsto+\sigma_2(\tau,x_2,-x_1),\\ \sigma_2(x)&\mapsto+\sigma_1(\tau,x_2,-x_1),\\
\pi(x)&\mapsto-\pi(\tau,x_2,-x_1).
\end{split}
\eeq
One can check that these pair of equations~\eqref{eq:c4_vector} and~\eqref{eq:c4_scalars} define a transformation on a vector of auxiliary fields $\boldsymbol{\phi}(x):=(a_{0}(x),a_{1}(x),a_{2}(x),\sigma_1(x),\sigma_2(x),\pi(x))\mapsto O\boldsymbol{\phi}(\tau,x_2,-x_1)$, with $O^{\rm t}O=\mathbb{1}$ and ${\rm det}O=+1$. We could then say that, within the Euclidean formulation where the field and the adjoint are independent Grassmann fields,  the hidden symmetry~\eqref{eq:C4_grassmann} can be interpret as a specific SO($6$) rotation of the auxiliary fields and fermion bilinears. We note that the interacting part of the Euclidean action has indeed a larger symmetry under arbitrary $SO(6)$ rotations when expressed in term of auxiliary fields, which gets broken down by the lattice regularization of the free fermions. It is only for the specific $\pi/2$ rotation above that one  recovers invariance of the full Euclidean action. However, it must be noted that this  transformation is a rotation of the auxiliary fields about the axis $a_0(x)\mapsto a_0(x')$. If one considers the   adjoint definition in the Hamiltonian approach,  $a_0(x)$  should always remain invariant, such that  the hidden symmetry reduces to the previous SO($5$) rotation.

The  idea of the large-$N$ method in the present context is to assume that these scalar fields will be homogeneous in the groundstate of the interacting theory $\boldsymbol{\phi}(x)=\boldsymbol{\Phi},  \,\,\,\forall x$, and try to determine the  regime in parameter space $(m_1a,m_2a,g^2/a)$ where some of them achieve a non-zero vacuum expectation value. This is the region of parameter space  where the groundstate supports a specific  combination of fermion condensates \beq
A_\mu=\langle a_\mu(x)\rangle,\hspace{1ex} \Sigma_j=\langle \sigma_j(x)\rangle,   \hspace{1ex} \Pi=\langle \pi(x)\rangle.
\eeq
Each of these fermion condensates is responsible for breaking a particular symmetry, and may even change completely the QFT that governs the continuum limit in the vicinity of such a symmetry-breaking phase transition. This is the case of the vector condensate, which is proportional to a fermion current  $A_\mu\propto\langle J_\mu\rangle$, and would thus forbid recovering Lorentz invariance even in the continuum limit. Such condensates have been identified in related two-band models~\cite{PhysRevResearch.4.L042012, Ziegler_2022}.

The vacuum expectation values of the scalar and pseudo-scalar condensates play a different role. In fact, the scalar condensates $\Sigma_1\ \propto\langle \overline{\psi}\ii\gamma^3\psi\rangle$ and $\Sigma_2\propto\langle \overline{\psi}\ii\gamma^5\psi\rangle$ are generally non-zero except for a particular line in parameter space $m_1a=m_2a=-r$, which is a consequence of the  twisted Wilson mass regularization. These scalar condensates do not break any of the parity symmetries~\eqref{eq:parity_main}, but contribute with a renormalization of the  bare twisted masses $m_j\mapsto m_j+\Sigma_j$ which, as will be discussed below, can change abruptly  the value of the topological invariant~\eqref{eq:top_inv}. When the bare masses $m_1=m_2=:m$, the two scalar condensates  take equal values $\Sigma_1=\Sigma_2=:\Sigma$, such that the hidden SO($5$) protecting symmetry~\eqref{eq:c4_scalars} is not broken, and one can still talk about the symmetry-protected HOTI. By increasing the interaction $g^2$,  as discussed below, one of the possibilities is that  the values of $\Sigma$ will change and lead to an  interaction-induced quantum phase transitions between the correlated HOTI, and a trivial band insulator with no corner states and a vanishing many-body topological invariant.

Before closing this section, we comment on the remaining fermion condensate $\Pi\propto\langle\overline{\psi}\psi \rangle$. Although in even spacetime dimensions, this condensate is parity even and associated to the breakdown of chiral symmetry, in our even-dimensional QFT it plays a  different role. As discussed above, this condensate is odd under any of the parities~\eqref{eq:parity_main}, and a finite vacuum expectation value would imply the spontaneous breakdown of parity. The possibility of finding such condensates in the standard Wilsonian lattice regularization of Dirac QFTs was initially considered by S. Aoki~\cite{PhysRevD.30.2653}, and it is typically referred to as an Aoki condensate in lattice gauge theories. In our current model of HOTIs,  rather than parity breaking, it is more important to consider  the hidden SO($5$) symmetry, which is  spontaneously broken  by a non-zero value of  this $\pi$ condensate~\eqref{eq:c4_scalars}. We note that the formation of vector condensates $\boldsymbol{A}\neq\boldsymbol{0}$ would also break spontaneously the protecting symmetry in light of Eq.~\eqref{eq:c4_scalars}. As shown below, understanding the competition of the different condensation channels is the key to understand the phase diagram of our HOTI.

\subsection{Large-$N$ condensates and the effective potential}

In this subsection, we describe the results of the aforementioned large-$N$ technique to chart the phase diagram of the interacting HOTI. As advanced previously, there are various  possible fermion condensates characterised by different vacuum expectation values, which could be obtained in the $N\to \infty$ limit by solving a set of gap equations. As discussed in the context of lattice gauge theories~\cite{Eguchi:1983gq}, these gap equations  are, however, only valid  for non-vanishing values of the vacuum expectation values $\Phi_a\not=0$. 
On the other hand,  we are also interested in the competition of the HOTI with a trivial band insulator, where the symmetry-breaking fermion condensates vanish and there is no spontaneous symmetry breaking. In order to explore the whole phase diagram,  we need to go beyond the gap equations and obtain   the large-$N$ effective potential $V_{\rm eff}(\boldsymbol{\Phi})$, the minimum of which will provide the values of the auxiliary fields in  $\boldsymbol{\Phi}$ for any specific point in parameter space determining, in particular, which  of the possible symmetry-breaking condensates prevail. 

The large-$N$ effective potential can be obtained diagrammatically by considering the Feynman diagrams with a  single fermion loop and an increasing number of external lines describing the auxiliary fields. Any other one-particle irreducible diagram with more fermion loops and internal auxiliary lines contributes with a higher order in $1/N$, and can thus be neglected when $N\to\infty$. In the standard calculation  of the chiral-invariant Gross-Neveu QFT, one can  show that only diagrams with an even number of external auxiliary legs can give a non-zero contribution~\cite{coleman_1985}. On the other hand, for our twisted Wilson mass regularization, one cannot apply the same arguments, and must also take into account the diagrams with an odd number of external legs. A similar situation can also be found in Chern-insulator models with a standard Wilson discretization although, there, one finds that these odd terms are zero due to the vanishing of the corresponding integrals~\cite{PhysRevResearch.4.L042012,ZIEGLER2022168763}. For the present model, this is not the case, and both the even and odd Feynman diagrams  have a non-zero contribution that must be  accounted for (see Appendix~\ref{app:eff_potential}).

The resummation  of these Feynman diagrams yields    the leading-order quantum radiative corrections $\delta V_{\rm eff}(\boldsymbol{\Phi})$ to the classical potential of the auxiliary fields
\beq
\label{eq:eff_V}
V_{\rm eff}(\boldsymbol{\Phi})=\frac{N}{2g^2}\boldsymbol{\Phi}^2+\delta V_{\rm eff}(\boldsymbol{\Phi}).
\eeq
As we have assumed that these auxiliary fiels are homogeneous, all the relevant information is included in this effective potential, which 
 plays a crucial role in determining the parameter regimes where the groundstate displays    non-zero  vacuum expectation values  $\boldsymbol{\Phi}\neq 0$. In fact, the classical part of the potential predicts a zero vacuum expectation value $\boldsymbol{\Phi}= 0$, and it is the contribution of radiative corrections $\delta V_{\rm eff}(\boldsymbol{\Phi})$ for $g^2>0$ that can change the minimum of Eq.~\eqref{eq:eff_V} allowing for condensation. As discussed in Appendix~\ref{app:eff_potential}, this resummation can be accomplished to all orders of $g^2$, which allows us to address non-perturbative effects in the phase diagram of the model. We find that the quantum correction can be expressed as
\beq
\label{eq:quant_corr}
\frac{\delta V_{\rm eff}}{2N}=-\!\!\int_{k}\log\!\left(\frac{k_0^2+(\hat{\boldsymbol{k}}+\boldsymbol{A})^2+\boldsymbol{m}^2(\boldsymbol{k},\boldsymbol{\Sigma})+\Pi^2}{k_0^2+\hat{\boldsymbol{k}}{\phantom{k}}^{\!\!\!\!\!2}+\boldsymbol{m}^2(\boldsymbol{k},\boldsymbol{0})}\right)\!\!,
\eeq
where we recall that $\hat{\boldsymbol{k}}$ is the regularised spatial momentum in Eq.~\eqref{eq:discrete_momentum}, and  we have introduced the shifts in the twisted Wilson masses~\eqref{eq:twisted_mass} stemming from the additive renormalisations that one finds for a non-zero values by  the  scalar condensates
\beq
\label{eq:renormalised_twisted_mass}
\boldsymbol{m}(\boldsymbol{k},\boldsymbol{\Sigma})= \big(m_1(\boldsymbol{k})+\Sigma_1,m_2(\boldsymbol{k})+\Sigma_2\big).
\eeq
We also recall that the integral symbol is a short-hand notation~\eqref{eq:free_action_naive} for the spatial mode sum and the zero-temperature limit of the Matsubara sum. 
As discussed in  Appendix~\ref{app:eff_potential}, the temporal component of the pseudo-vector field $A_0$ does not contribute to the effective potential and, thus, cannot condense. The occurrence of other condensation channels will depend  on the  minimum of  $V_{\rm eff}(\boldsymbol{A},\boldsymbol{\Sigma},\Pi)$. For $m_1=m_2$, this effective potential can be easily seen to be invariant under the hidden SO($5$) rotation that takes $k_1\mapsto k_2$, $k_2\mapsto -k_1$, together with  $A_1\mapsto A_2$, $A_2\mapsto -A_1$, $\Sigma_1\mapsto\Sigma_2$, $\Sigma_2\mapsto\Sigma_1$, and $\Pi\mapsto-\Pi$. 

Let us note that the above expressions with the specific lattice regularization are in fact more general, and would also apply to continuum QFTs with the same 4-Fermi interactions. This would simply require  substituting $\hat{\boldsymbol{k}}\mapsto\boldsymbol{k}$ and $m_j(\boldsymbol{k})\mapsto m_j$, recovering in this way the underlying Lorentz invariance. On the other hand, the expressions could also be used with a discretized Euclidean time by substituting $k_0\mapsto\sin(k_0a)/a$ with $k_0=-{\pi}/{a}+{2\pi (n_0+1/2)}/{aN_0}$ with  $n_0\in\mathbb{Z}_{N_0}$. From the perspective of the cold-atom quantum simulator, one is interested in the continuum-time limit and the Hamiltonian field theory. In this case,  one can actually perform the integral over  $k_0\in\mathbb{R}$, and   express the radiative corrections as follows   
\beq
\label{eq:large_n_correction}
\delta V_{\rm eff}(\boldsymbol{\Phi})=2N\bigintsss_{\boldsymbol{k}}\!\!\left(\!\epsilon(\hat{\boldsymbol{k}}+\boldsymbol{A},\boldsymbol{m}(\boldsymbol{k},\boldsymbol{\Sigma}),\Pi)-\epsilon(\hat{\boldsymbol{k}},\boldsymbol{m}(\boldsymbol{k},\boldsymbol{0}),0)\!\right)\!\!,
\eeq
where we have introduced a short-hand  for the spatial-momenta sum within the Brillouin zone $\int_{\boldsymbol{k}}=\frac{1}{(N_{\rm s}a)^2}\!\!\sum_{\boldsymbol{k}\in{\rm BZ}}$. In addition, we have generalised the energy dispersion relation in Eq.~\eqref{energy_bands} to account for the possible non-zero vacuum expectation values of the fermion condensates 
\beq
\epsilon\big(\hat{\boldsymbol{k}}+\boldsymbol{A},\boldsymbol{m}(\boldsymbol{k},\boldsymbol{\Sigma}),\Pi\big)=\sqrt{(\hat{\boldsymbol{k}}+\boldsymbol{A})^2+\boldsymbol{m}^2\!(\boldsymbol{k},\boldsymbol{\Sigma})+\Pi^2}.
\eeq
Expression~\eqref{eq:large_n_correction} has a very simple interpretation, the large-$N$ radiative corrections are given by the total shift of single-particle energy levels in the filled bands, i.e. those with negative energies  forming the Dirac sea, once some of  the condensates form $\boldsymbol{\Phi}\neq \boldsymbol{0}$. For the scalar condensates $\Sigma_j$, such corrections appear  as soon as the interactions are switched on $g^2>0$. The only exception is the straight line at $m_1a=m_2a=-r$, in which the scalar condensates vanish by symmetry arguments. For the vector $\boldsymbol{A}$ and pseudo-scalar $\Pi$ condensates, the situation is completely different. A non-zero value of these condensates would break the hidden SO($5$) symmetry in Eqs.~\eqref{eq:c4_vector}-\eqref{eq:c4_scalars}, which can only happen spontaneously for a sufficiently-strong coupling $g^2>g^2_{\rm c}(m_1,m_2)$.  In order to find out which of the condensates prevails, we need to  minimize the full  effective potential
\beq
\label{eq:large_N_solution}
\boldsymbol{\Phi}^\star={\rm argmin}\left\{\frac{N}{2g^2}\boldsymbol{\Phi}^2+\delta V_{\rm eff}(\boldsymbol{\Phi})\right\},
\eeq
which is  an unconstrained  multi-parameter non-linear minimization problem that must be addressed numerically, as we detail in the following subsection. 

\subsection{Self-energy and correlated HOTIs}

In the previous subsection, we have discussed the procedure to find the values of $\boldsymbol{\Phi}^\star$, which will allow us to detect the symmetry-breaking condensates and localise the critical lines that separate them from the HOTI and the trivial band insulator. On the other hand, we would also like to predict the flow of the critical lines separating the HOTI from the trivial band insulator as the coupling increases $g^2>0$. Since  topological phases cannot be distinguished by a local order parameter, we also need to calculate the topological invariant~\eqref{eq:top_inv} away from the non-interacting free theory. An approximation that  has already been used for the many-body topological invariants of standard topological insulators~\cite{PhysRevLett.105.256803,PhysRevB.83.085426,PhysRevX.2.031008,PhysRevB.86.165116,Wang_2013}, deals with the so-called topological Hamiltonian.

Many-body topological invariants can be defined via  the two-point  Green's functions $G(x_1- x_2 ) = \langle\mathcal{T} \{\Psi^\dagger (x_1 )\Psi(x_2 )\}\rangle$~\cite{doi:10.1143/JPSJ.75.123601,PhysRevLett.122.146601}, where the expectation value is calculated over the groundstate with non-zero interactions. Following the prescriptions of quantum-many body physics within condensed matter~\cite{negele_orland_2019}, by going to momentum space, the inverse of the Green's function can be expressed  as
\beq
G^{-1}(\ii k_0,\boldsymbol{k})=\ii k_0- \mathbb{H}_0(\boldsymbol{k})+\mathsf{\Sigma}_{\rm s}(\ii k_0,\boldsymbol{k}),
\eeq
where $\mathbb{H}_0(\boldsymbol{k})$ is the single-particle Hamiltonian, the $N$-flavour version of Eq.~\eqref{eq:h_0} in  our case, and $\mathsf{\Sigma}_{\rm s}(\ii k_0,\boldsymbol{k})$ is the so-called self-energy. This self-energy contains all the one-particle irreducible ``tadpole'' contributions  to the fermion propagator arising from intermediate scattering processes in which particle-antiparticle pairs are virtually created from the groundstate. Within our large-$N$ theory, we have precisely calculated those at leading order in $N$ by introducing the auxiliary fields. In fact, the above condensates can be readily used to approximate this self-energy as
\beq
\mathsf{\Sigma}_{\rm s}(\ii k_0,\boldsymbol{k})=\mathbb{1}_N\otimes\left(\gamma^0\gamma^\mu A_\mu+\ii\gamma^0\gamma^3\Sigma_1+\ii\gamma^0\gamma^5\Sigma_2+\gamma^0\Pi\right),
\eeq
which has no momentum dependence $\mathsf{\Sigma}_{\rm s}(\ii k_0,\boldsymbol{k})=\mathsf{\Sigma}_{\rm s}(0,\boldsymbol{0})$ since we have assumed the condensates to be homogeneous.

As discussed in~\cite{PhysRevLett.105.256803,PhysRevB.83.085426,PhysRevX.2.031008,PhysRevB.86.165116,Wang_2013}, the static contributions to the self-energy $\mathsf{\Sigma}_{\rm s}(0,\boldsymbol{k})$ can be used to define the so-called topological Hamiltonian. To consider the same symmetry class as the non-interacting one, we set $\boldsymbol{A}=\boldsymbol{0}$ and $\Pi=0$, such that the hidden SO($5$) rotation is preserved. In this case, the only contribution to the topological Hamiltonian stems from the scalar condensates
\beq
\mathbb{H}_{\rm t}(\boldsymbol{k})=\mathbb{H}_0(\boldsymbol{k})+\mathsf{\Sigma}_{\rm s}(0,\boldsymbol{k})=\mathbb{H}_0(\boldsymbol{k})+\ii\gamma^0\gamma^3\Sigma_1+\ii\gamma^0\gamma^5\Sigma_2.
\eeq
Within this large-$N$ approximation, the many-body topological invariant can be expressed in terms of the two Chern-Simons forms~\eqref{eq:Chern-Simons_forms} with the  twisted 
Wilson masses renormalised by the two scalar condensates~\eqref{eq:renormalised_twisted_mass}, namely
 \beq
 \label{eq:Chern-Simons_forms}
 \mathsf{W}_j(\boldsymbol{\Sigma})=\frac{N}{2}\sum_{\ell_j=0,1}(-1)^{\ell_j}{\rm sgn}\{m_j(\boldsymbol{k}_{{\rm D}},\boldsymbol{\Sigma})\},
 \eeq
The full topological invariant of the correlated HOTI can be approximated, within the large-$N$ limit, as follows
\beq
\label{eq:top_inv_int}
\ee^{\ii\pi\mathsf{W}_1(\boldsymbol{\Sigma})\mathsf{W}_2(\boldsymbol{\Sigma})}=\left\{\begin{array}{ll} 
(-1)^N &\,\, {\rm if}\,\,\, -\frac{2r}{a}<m_j+\Sigma_j<0 \,\,\,\,\forall j.\\
(+1)^N &\,\, {\rm else}\,\,\, 
\end{array}\right.
\eeq
Accordingly, the large-$N$ solution obtained by minimizing the effective potential~\eqref{eq:large_N_solution} can be readily used to extract $\Sigma_1,\Sigma_2$ in the parameter regime where the hidden SO($5$) symmetry is still preserved, and localise the critical surface that separates the correlated HOTI from the trivial band insulator. We recall again that, in order to have the symmetry protection, we need to consider the regime where $m_1=m_2$ and $\Sigma_1=\Sigma_2$, which will cut this  critical surface determining a critical line.

\subsection{Large-$N$ results and phase diagram}

 We start by making the assumption that only the   pseudo-scalar auxiliary field  condenses $\Pi=\langle\pi\rangle\not=0$,
which we expect describes the leading symmetry-breaking channel among all other condensates, and should set in at a certain  value of the interaction strength $g^2>0$. This assumption will allow us to derive a set of simple gap equations that can be later used as a reference in the minimization of the full effective potential~\eqref{eq:large_N_solution} that contains  all possible condensation channels~\eqref{eq:large_n_correction}. As with gap equations in other models~\cite{PhysRevResearch.4.L042012,ZIEGLER2022168763,PhysRevB.106.045147}, this formalism is applicable   
whenever the order parameter is non-zero $\Pi>0$~\cite{Eguchi:1983gq}. 
As advanced in the previous sections, the $\sigma$ condensates typically attain  non-zero values for any point in parameter space $(m_1a,m_2a,g^2/a)$, except for the line at fixed twisted mass $m_1a=m_2a=r$.  We hence consider that 
$\boldsymbol{\Sigma}=\langle\boldsymbol{\sigma}\rangle\not=\boldsymbol{0}$. 
Therefore, the main assumption in the following gap equations is that the vector condensate vanishes $\boldsymbol{A}
=\langle\boldsymbol{a}\rangle=\boldsymbol{0}$ (this will be justified in Fig.~\ref{fig:conds} below). The gap equations are derived by 
solving the set of non-linear equations given by the stationary point  $\partial_\Pi V_{\rm eff}|_{A_\mu=0}=\partial_{\Sigma_1} V_{\rm eff}|_{A_\mu=0}=\partial_{\Sigma_2} V_{\rm eff}|_{A_\mu=0}=0$. Setting the Wilson parameter to $r=1$ henceforth, we find
\begin{equation}
{1\over g^2}=4\!\!\bigintsss\!\!\!\frac{{\rm d}k_0}{2\pi}\!\!\bigintsss_{\boldsymbol{k}}\,\,\frac{1}{k_0^2+\hat{\boldsymbol{k}}{\phantom{k}}^{\!\!\!\!\!2}+\boldsymbol{m}^2(\boldsymbol{k},\boldsymbol{\Sigma})+\Pi^2},
\label{eq:gap}
\end{equation}
where we have assumed that $\Pi>0$ to simplify the equation. In addition, 
we get the following two equations   from the derivatives with respect to the scalar condensates
\begin{equation}
{m_j a\over g^2}=-4\!\!\bigintsss\!\!\!\frac{{\rm d}k_0}{2\pi}\!\!\bigintsss_{\boldsymbol{k}}\,\,\frac{1-\cos k_ja}{k_0^2+\hat{\boldsymbol{k}}{\phantom{k}}^{\!\!\!\!\!2}+\boldsymbol{m}^2(\boldsymbol{k},\boldsymbol{\Sigma})+\Pi^2},
\label{eq:gapm}
\end{equation}
where we have used Eq.~\eqref{eq:gap} to simplify them, getting an expression that relates to the twisted masses $m_1,m_2$.

\begin{figure}[t]
	\centering
	\includegraphics[width=1\columnwidth]{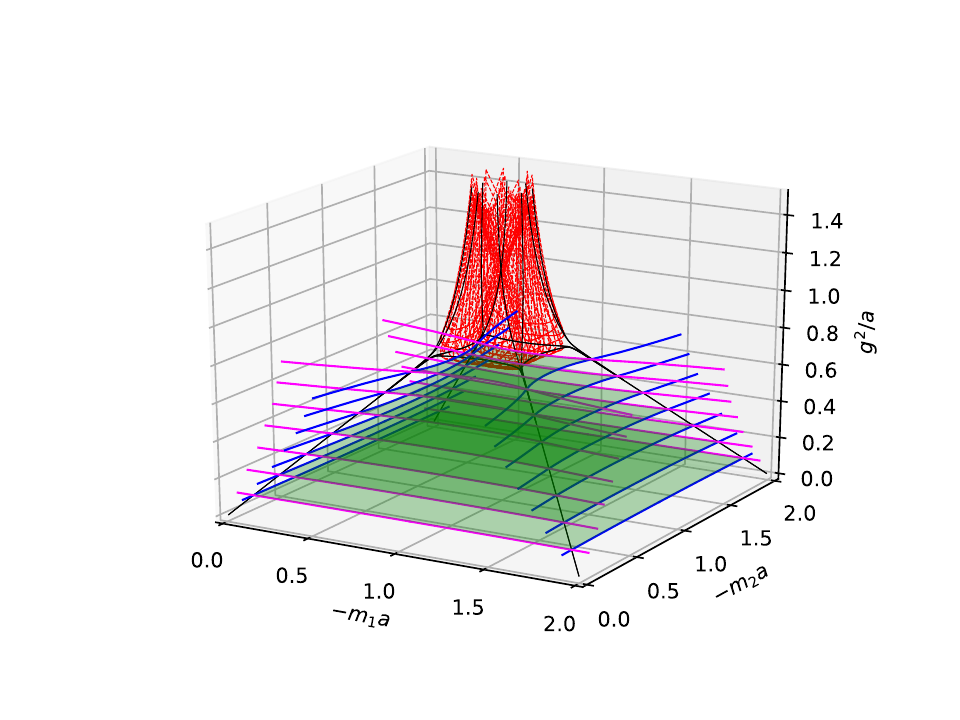}
	\caption{{\bf Interacting higher-order topological phase diagram I:}  We consider parameter space  $(m_1a,m_2a,g^2/a)$, and represent the different phases in the twisted-Wilson lattice model with 4-Fermi interactions. The blue and magenta curves~\eqref{eq:gapless} delimit the inner green area~\eqref{eq:HOTI_region} at fixed-$g^2$ slices that has a non-trivial many-body topological invariant $\ee^{\ii\pi\mathsf{W}_1(\boldsymbol{\Sigma})\mathsf{W}_2(\boldsymbol{\Sigma})}=-1$ for an odd number of fermion flavours.  The red lines are obtained by solving the gap equations to localise the boundary of the parity-breaking $\Pi$ condensate~\eqref{eq:gap}-\eqref{eq:gapm}. The black lines demonstrate that this pseudo-scalar condensate actually grows from the four corners of the non-interacting HOTI phase at $g^2=0$ (see Fig. \ref{fig:HOTI_free}), forming the shape of an Eiffel tower that rests on the correlated HOTI. All numerical evaluations employed a spatial lattice with $N_{\rm s}=32^2$ sites.    }
	\label{fig:tour}
\end{figure}

Equations (\ref{eq:gap})-(\ref{eq:gapm}) can be solved with $\Pi=0^+$ to map out the
boundary of the phase hosting a pseudo-scalar condensate. This critical region  is a surface in  parameter space
$(m_1a,m_2a,g^2/
a)$, and is  shown in red in Fig.~\ref{fig:tour}.
This surface is spanned by a number of trajectories that represent solutions to the gap equations for  $\Pi=0^+$,  for which  we fix a different  value of the  twisted masses renormalised by the scalar condensates $(m_j+\Sigma_j)$. These trajectories are   plotted
 in Fig.~\ref{fig:tour} as a collection of red dashed curves; the resulting network 
visualises a closed surface which descends from strong coupling down to $g^2/a\approx0.7$. There are a couple of interesting comments: {\it (i)} The volume inside the red surface describes the spontaneous symmetry-broken phase with a pseudo-scalar condensate. This condensate  breaks any of the parities, and corresponds to the so-called Aoki phase found in other lattice field theories ~\cite{PhysRevD.30.2653}. As emphasised above, the more important thing is that this pseudo-scalar condensate also breaks spontaneously the hidden SO($5$) symmetry responsible for the protection of the HOTI. Hence the higher-order topological invariant an the corner modes cannot coexist with this pesudo-scalar condensate.  {\it (ii)} The volume that contains this SO($5$) breaking condensate is centered around the symmetry line $m_1a=m_2a=-1$ in which  the scalar condensates vanish $\boldsymbol{\Sigma}=\boldsymbol{0}$, and gets more and more compressed as the interaction strength becomes  large  $g^2/a\gg 1$.  On the other hand,   for $g^2=0$, the gap equation (\ref{eq:gap}) becomes singular at the four corners
$(m_1,m_2)\in\{(0,0),(0,-2),(-2,0),(-2,-2)\}$, and we have found that  this phase extends all the way down to weak
coupling in four very sharp spikes. Let us note that this is a large-$N$ prediction, and a different method should be used to determine the extent of this phase for finite $N$.

\begin{figure}[t]
	\centering
	\includegraphics[width=0.8\columnwidth]{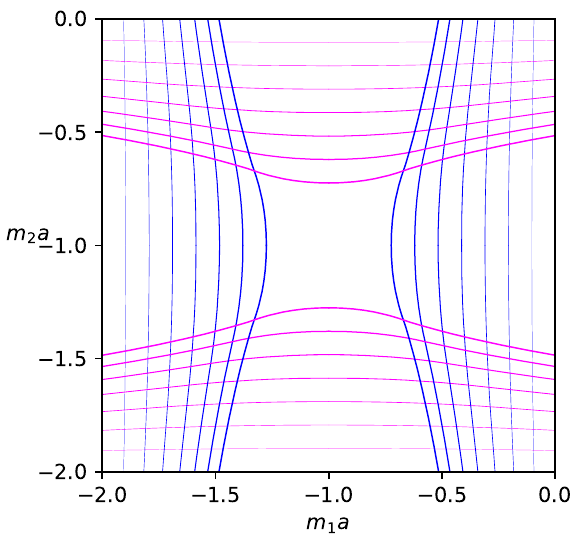}
	\caption{{\bf Interacting higher-order topological phase diagram II:} 
 Here the lines where the gap functions $f_1=0$ (blue) and $f_2=0$ (magenta) are plotted
 for increasing values of the coupling strength $g^2/a\in\{0.1,0.2,0.3,0.4,0.5,0.6,0.7\}$, which grow as one moves towards the center of the square. The HOTI phase is contained within a star-shaped region that is delimited by groups of four of these lines, corresponding to the same value of the coupling strength $g^2/a$.}
	\label{fig:star}
\end{figure}
As already remarked above,  the  gap equations~\eqref{eq:gap}-\eqref{eq:gapm} are only valid for $\Pi>0$. On the other hand, the original question that motivated the study was to see the extent of the HOTI as one increases interactions, which would require exploring the region with $g^2>0$ for which $\Pi=0$, more particularly, the weaker-coupling  regime  beneath the Aoki phase. To explore it, we will directly minimize the effective potential $V_{\rm eff}(\boldsymbol{0},\boldsymbol{\Sigma},0)$
at a specific $(m_1a,m_2a,g^2/a)$. As discussed in Eq.~\eqref{eq:top_inv_int}, the many-body topological invariant for the HOTI in the large-$N$ limit is non-trivial   $\ee^{\ii\pi\mathsf{W}_1(\boldsymbol{\Sigma})\mathsf{W}_2(\boldsymbol{\Sigma})}=-1$ when
\begin{equation}
-2<m_1a+\Sigma_1a<0,\hspace{2ex}-2<m_2a+\Sigma_2a<0,
\label{eq:HOTI_region}
\end{equation}
and we have an odd number of flavours. 
The correlated
HOTI phase will then be contained in a 
region bounded by contours along which a gap function $f_j(k_ja,\Sigma_ja)=m_ja+(1-\cos
k_ja)+\Sigma_ja$ (with $j=1,2$) vanishes at either the origin $k_j=0$, or at the zone edge $k_j=\pi/a$.
The procedure is then to search for solutions of the non-linear equations
\begin{equation}
m_j+\Sigma_j=n_j,\hspace{2ex}n_j\in\{0,-2\},
\label{eq:gapless}
\end{equation}
using the value of the scalar condensates $\Sigma_1,\Sigma_2$ obtained by numerically minimisation of $V_{\rm eff}(\boldsymbol{0},\boldsymbol{\Sigma},0)$. In practice, we define a circle centred at the line of
symmetry where $\boldsymbol{\Sigma}=\boldsymbol{0}$ by defining the twisted masses as
$(m_1,m_2)=(-1/a,-1/a)+m(\cos\theta,\sin\theta)$, for $m>0$ and $\theta\in[0,2\pi)$. We then search for the roots of Eq.~(\ref{eq:gapless}) by scanning first in $\theta$ and, subsequently,  in $m$. The resulting
contours along which a gap function  vanishes $f_j(k_j,\Sigma_j)=0$ are plotted as blue (magenta) lines in
Figs.~\ref{fig:tour} and\ref{fig:star} for various values of the coupling strength $g^2$. Each pair of blue (magenta) lines for a fixed coupling strength corresponds to  the renormalised twisted mass proportional to $\ii\gamma^3$
($\ii\gamma^5$) satisfying $f_1(0,\Sigma_1)=0$ or $f_1(\pi/a,\Sigma_1)=0$ ($f_2(0,\Sigma_2)=0$ or $f_2(\pi/a,\Sigma_2)=0$). The region of the correlated HOTI phase is
enclosed within the areas inside the four intersecting lines, and is depicted in Fig.~\ref{fig:tour} by a  shaded  green area that connects to the green square in the non-interacting limit (Cf. Fig.~\ref{fig:HOTI_free}), and projected onto the $(m_1,m_2)$-plane in Fig.~\ref{fig:star}.
As $g^2$ increases, the borders of this region curve inwards, and the  HOTI shrinks until it roughly coincides with the lateral extent of the
Aoki phase at the critical coupling $g^2/a\simeq0.7$ (note the lower surface of the Aoki
phase has convex curvature, confirmed in Fig.~\ref{fig:conds}). We remark that the hidden SO(5) symmetry responsible for the protection of the anomalous corner modes takes place within the $m_1=m_2$ area of the green volume of Fig.~\ref{fig:tour}. On the other hand, all the empty region surrounding both the green and red volumes corresponds to a trivial band insulator, in which the topological invariant is trivial and the pseudo-scalar condensate vanishes. All the critical surfaces predicted within our large-$N$ methods correspond to higher-order quantum phase transitions, either topological or symmetry-breaking ones.

Once we have discussed the phase diagram of Fig.~\ref{fig:tour} in detail, we should check for the consistency of the assumptions we made about the competing condensates by  
minimising the full effective potential $V_{\rm eff}(\boldsymbol{A},\boldsymbol{\Sigma},\Pi)$. We recall again that   one can set $A_0=0$ as discussed in Appendix~\ref{app:eff_potential}, but must consider the other competing condensation channels
along lines of fixed $(m_1,m_2)$. Fig.~\ref{fig:conds} shows the
resulting condensates for two choices of $(m_1,m_2)$ at different distances from the line of
symmetry $m_1=m_2=-1/a$. The $\Pi$ condensate signalling the SO($5$)-breaking Aoki phase rises from zero at a critical
coupling $g_c^2/a\approx 0.7$, and the opposite signs of the scalar condensates  $\Sigma_1=-\Sigma_2$ correspond to
$(m_1,m_2)$ lying on opposite sides of the symmetric line $(-1/a,-1/a)$. Crucially,
the current condensate $A_1$ remains zero throughout,  justifying  our
previous assumption where we set $\boldsymbol{A}=\boldsymbol{ 0}$. Solutions with $\boldsymbol{ A}\not=\boldsymbol{0}$, as found for an interacting  Chern-insulator two-band model in~\cite{ZIEGLER2022168763,PhysRevResearch.4.L042012}, could only be found in our lattice model by artificially constraining $\Pi=0$. Otherwise, we find that the  SO($5$)-breaking Aoki condensate always describes the leading condensation channel  $\Pi>0$ as one increases the coupling strength. Further from the line of symmetry, as shown by the dashed lines in Fig.~\ref{fig:conds}
the picture remains qualitatively the same, but with values of the scalar condensates $\Sigma_i$  that become larger in
magnitude; this time at strong-enough coupling the trajectory actually re-enters the
SO($5$)-symmetric phase, reflected by the renewed vanishing of $\Pi$ and the kinks in
$\Sigma_i(g^2)$. however, rather than re-entering into the HOTI phase, one goes into a trivial band insulator where, even if the SO($5$) symmetry is preserved, the topological invariant is trivial and the zero states are no longer localised at the corners of the system.   

 \begin{figure}[t]
	\centering
	\includegraphics[width=1\columnwidth]{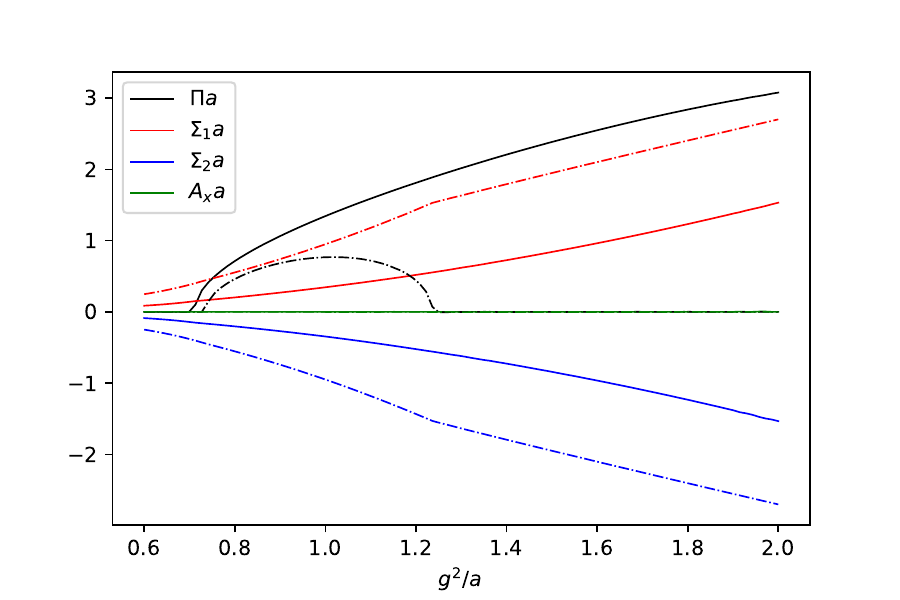}
	\caption{{\bf Competing condensation channels:}  Value of the various fermion condensates as a function of the coupling strength $g^2/a$ for the twisted masses $(m_1a,m_2a)=(-0.95,-1.05)$ (full lines) and $(m_1a,m_2a)=(-0.85,-1.15)$  (dot-dashed lines). The opposite signs of $\Sigma_1,\Sigma_2$ mirror the signs of $-1-m_{1,2}a§$. Note that the vector condensate vanishes throughout $\boldsymbol{A}=\boldsymbol{0}$  (green lines).   }
	\label{fig:conds}
\end{figure}


\section{\bf Conclusion and Outlook}

In this work, we have presented a non-standard lattice regularization of Dirac QFTs based on a new type of Wilson fermions that have an anisotropic twisted Wilson mass. We have shown that the anisotropic twisted Wilson mass is responsible for the occurrence of HOTIs that display zero-energy corner modes, and a non-vanishing topological invariant in the bulk. We have discuss a cold-atom implementation of this lattice field theory that exploits Raman optical lattices, and spin-$3/3$ Fermi gases of alkaline-earth atoms. Interestingly, the $s$-wave scattering of this atoms leads to a SO($5$) invariant 4-Fermi interaction, which leads to an interesting competition of HOTI phases and various fermion condensates. We explored the full phase diagram of the model using large-$N$ techniques, and argued that the correlated HOTI phase eventually gives way to a parity-breaking fermion condensate.  

Since the microscopic parameters of the model can be independently tuned in the proposed cold-atom experiment, it woudl be very ineteresting that the predictions presented  in this work could be tested in future experiments. We note that the recent experimental work on the quantum simulator of Chern insulators using $^{87} $Sr Fermi gases in Raman optical lattices~\cite{PhysRevResearch.5.L012006} is very promising in this direction. If one can use these methods for a different alkaline-earth atoms such as $^{132}$Cs, controlling the four spin states as discussed in our work, the experiment would directly probe the SO($5$) self-interacting Dirac QFT that has been the subject of our work. This experiment, together with related theory works~\cite{PhysRevLett.121.250403,PhysRevB.107.125132}, have also shown that it is possible to perform certain measurements to infer the value of a topological invariant similar to the Chern number. It would be interesting to explore if such methods can also be adapted to the spin-$3/2$ Fermi gas, and used to infer the value of our higher-order topological invariant. Other than that, further studies are required in order to propose other measurement schemes. For instance, in order to retrieve the order parameter associated to the fermion condensates discussed in this work, some sort of spin-resolved imaging by either illuminating the gas and processing its shadow, or using quantum gas microscopes,  should be required. These methods should be combined with microwave transition and spin-selective techniques  to infer the atomic densities corresponding to the order parameters.

From a more theoretical perspective, it would be very interesting to explore finite-temperature and finite-density phases in this model of SO($5$) interacting Dirac matter. In particular, by moving away from half filling, one should consider other possible condensation channels that likely include superconducting and inhomogeneous orders, or even new exotic orders that go beyond the Landau symmetry breaking paradigm. The study of those drawing further connections between high-energy physics, condesed matter, and AMO physics,    will contribute to the growing interest in this interdisciplinary line of research~\cite{PhysRevB.99.064105,PhysRevLett.124.131601,PhysRevD.101.094512,PhysRevD.104.094515,PhysRevD.106.114515,Roose2021,Czajka2022,PhysRevD.107.014509,sym14020265}.


\acknowledgements
 A. B. acknowledges support
from  PID2021-127726NB-I00 (MCIU/AEI/FEDER, UE), from
the Grant IFT Centro de
Excelencia Severo Ochoa  CEX2020-001007-S,
funded by MCIN/AEI/10.13039/501100011033, and from the CSIC Research Platform
on Quantum Technologies PTI-001. S.H. was supported by the STFC Consolidated Grant ST/T000813/1. D.G.-C. is supported by the Simons Collaboration on Ultra-Quantum Matter, which is a grant from the Simons Foundation (651440, P.Z.).

\appendix
\section{ Clifford algebra and SO($D$)  Dirac fermions}
\label{appA}
In this appendix, we review several aspects that appear in the description of relativistic QFTs of Dirac fermions in arbitrary spatial dimensions and Euclidean imaginary time. We discuss  the role of the  special orthogonal group of transformations in the description of both  spacetime and internal symmetries. We start from the partition function of a Dirac field of mass $m$ which, in  natural units $\hbar=c=1$,  can be written as a functional integral $Z=\int[{\rm D}\overline{\psi}\,{\rm D}{\psi}]\ee^{-S_0}$, provided that the fermionic field $\psi(x)$ and the adjoint $\overline{\psi}(x)$ are described by mutually anti-commuting Grassmann fields with  anti-periodic boundary conditions in the imaginary-time direction~\cite{negele_orland_2019,0201304503}.   The partition function thus depends on the Euclidean action   
\begin{equation}
\label{eq:Euc_action}
    S_{\rm 0}=\int\!{\rm d}^Dx\overline{\psi}(x)\big(\gamma^\mu\partial_\mu+m\big)\psi(x),
\end{equation}
where we have introduced $\mu\in\{0,\cdots, D-1\}$ to label the Euclidean spacetime coordinates $x=(\tau,\boldsymbol{x})$ and derivatives  $\partial_\mu=\partial/\partial x^\mu$, using Einstein's criterion of repeated index summation. We have also introduced the  gamma matrices $\{\gamma^\mu\}$, which  are the generators of the Clifford algebra with Euclidean metric $\mathsf{Cl}(0, D)$x~\cite{vanproeyen2016tools,Freedman:2012zz},  and must thus fulfill 
\beq
\label{eq:clifford}
\big\{\gamma^{\mu},\gamma^{\nu}\big\}=\gamma^{\mu}\gamma^{\nu}+\gamma^{\nu}\gamma^{\mu}=2g^{\mu\nu}\mathbb{1}_{d_s},
\eeq
where the Euclidean metric  $g^{\mu\nu}=\delta^{\mu,\nu}=\delta^{\mu}_{\nu}=\delta_{\mu,\nu}$ is defined through the Kronecker delta, and $\mathbb{1}_{d_s}$ is the identity matrix. The Euclidean gamma matrices are thus mutually anti-commuting, all square to the identity, and do not have any distinction between upper and lower indexes $\gamma^\mu=\gamma_\mu$. 

In an irreducible representation, these generators can be expressed by   square Hermitian  matrices $\gamma_\mu\in\mathsf{Herm}(\mathbb{C}^{d_s})$ acting on a vector space of dimension $d_s=2^{\lfloor{D/2}\rfloor}$, where $ \lfloor{x}\rfloor$ stands for the greatest integer less than or equal to $x$. Accordingly, for spacetime dimensions $D=2n$ or $D=2n+1$ with $n\in\mathbb{Z}^+$, they are $2^n\times 2^n$ Hermitian matrices that can be built from specific tensor products within the orthogonal Pauli basis $\gamma^\mu\in\mathcal{B}_n=\{\mathbb{1}_2,\sigma^x,\sigma^y,\sigma^z\}^{\otimes^n}$. As discussed  for instance in reference~\cite{vanproeyen2016tools}, there are simple recipes to construct these gamma matrices, and the only difference between  odd $D=2n+1$, and even $D=2n$ spacetime dimensions is that the latter has an additional   $\gamma^{2n}=\gamma_\star$,  obtained by multiplying all the   matrices of $D=2n$ together $\gamma_\star=(-\ii)^{n}\gamma^0\gamma^1\cdots\gamma^{2n-1}$. 

Once the gamma matrices are known, we can obtain the $2^D$ elements of the Clifford algebra using products. In particular, the anti-symmetric products   leads to elements of the Clifford algebra  being ordered according their rank as 
\beq
\label{eq:clifford}
\begin{split}
\mathsf{Cl}(0, 2n)=\big\{\mathbb{1}_{d_{s}},\gamma^{\mu_1},\gamma^{\mu_1\mu_2},\gamma^{\mu_1\mu_2\mu_3},\cdots,\gamma^{\mu_1\mu_2\cdots \mu_{2n}}\big\}_{\mu_j=0}^{2n-1},\\
\mathsf{Cl}(0, 2n+1)=\big\{\mathbb{1}_{d_{s}},\gamma^{\mu_1},\gamma^{\mu_1\mu_2},\gamma^{\mu_1\mu_2\mu_3},\cdots,\gamma^{\mu_1\mu_2\cdots \mu_{2n}}\big\}_{\mu_j=0}^{2n},
\end{split}
\eeq
where $\gamma^{\mu_1\mu_2}=\half[\gamma^{\mu_1},\gamma^{\mu_2}]$, and the remaining  are defined recursively  $\gamma^{\mu_1\mu_2\cdots \mu_j}=\half[\gamma^{\mu_1},\gamma^{\mu_2\cdots \mu_j}],$ for $j\in\{3,\cdots,D\}$.  

In the context of the Dirac QFT, the elements of the Clifford group with  rank 1, 2 play  a key role. The rank-1 elements, namely the aforementioned gamma matrices, enter in the definition of the action~\eqref{eq:Euc_action}. The rank-2 elements $\{\gamma^{\mu_1\mu_2}\}$, which correspond to the $D(D-1)/2$ anti-symmetric products of the gamma matrices, serve as the generators of the spacetime rotations $x^\mu\mapsto\Lambda^{\mu}_{\nu}x^\nu$ with $\Lambda\in SO(D)$, which would correspond to  Lorentz transformations SO($1,D-1$) if we rotated back to real time $\tau\to\ii t$~\cite{Peskin:1995ev}. Indeed, any such transformation $\Lambda$ has a representation in terms of a rotation of angle $\theta_{\mu\nu}$ within the $(\mu\nu)$-plane  and the   infinitesimal generator  $\gamma^{\mu\nu}$. This so-called spinor representation reads $S_\Lambda={\rm exp}\{\fourth\omega_{ab}\gamma^{ab}\}$  where $\omega_{ab}=\theta_{ab}(\delta^{a}_{\mu}\delta^{b}_{\nu}-\delta^{a}_{\nu}\delta^{b}_{\mu})$, which can be easily checked to yield a unitary representation of the SO($D$) group $S^{-1}_\Lambda=(S_\Lambda)^\dagger$. This leads to a crucial difference with respect to Minkowski spacetime, and forbids defining  the adjoint as $\overline{\psi}(x)=\psi^\dagger\!(x)\gamma_0$. The field and its adjoint are mutually anti-commuting Grassmann spinors with an even number  of components $d_s$, and one postulates that they transform under spacetime rotations as 
\beq
\label{eq:spinor_lorentz}
\psi(x)\mapsto S_\Lambda\psi\left(\Lambda^{-1}x\right), \overline{\psi}(x)\mapsto \overline{\psi}\left(\Lambda^{-1}x\right)S^{-1}_\Lambda.
\eeq
  One then  finds that  the Euclidean action~\eqref{eq:Euc_action} is invariant under  SO($D$), which also requires  using the transformations of the rank $j<D$ Clifford elements as tensors under  SO($D$) 
\beq
\label{eq:so_d_transformations_clifford}
 S^{-1}_\Lambda\gamma^{\mu_1\mu_2\cdots \mu_j}S_\Lambda=\Lambda^{\mu_1}_{\nu_1}\Lambda^{\mu_2}_{\nu_2}\cdots\Lambda^{\mu_j}_{\nu_j}\gamma^{\nu_1\nu_2\cdots \nu_j}.
\eeq
This shows that the gamma matrices transform as a vector under SO($D$), such that $\overline{\psi}\psi$ and $\overline{\psi}\gamma^a\partial_a\psi$ are  scalars, and the Euclidean action is invariant under SO($D$). 

For even spacetime dimensions $D=2n$, the highest-rank  element of the Clifford algebra~\eqref{eq:clifford} also plays an important role. It can be used to define an additional Hermitian matrix that anti-commutes with all the spacetime gamma matrices, and is thus left invariant under any SO($2n$) rotation 
\beq
\label{eq:gamma_star}
\gamma_{\star}=(-\ii)^{n}\gamma^0\gamma^1\cdots\gamma^{2n-1}\mapsto S^{-1}\!(\Lambda)\gamma_\star S(\Lambda)=\gamma_\star.
\eeq
In $D=4$, this is typically called the chiral gamma matrix $\gamma^5=-\gamma^0\gamma^1\gamma^2\gamma^3$, which can be used to decompose the Dirac spinor into  left- and right-handed two-component spinors, the so-called chiral Weyl fermions~\cite{Peskin:1995ev}. Alternatively, in any even spacetime dimension, this gamma matrix  can  serve to propose a twisting of the scalar mass 
\begin{equation}
\label{eq:Euc_action_twisted}
    S_{\rm 0}=\int\!{\rm d}^Dx\overline{\psi}(x)\left(\gamma^\mu\partial_\mu+m_1+\ii m_2\gamma_\star\right)\psi(x),
\end{equation}
where $m_1=m\cos\theta$, and $m_2=m\sin\theta$. The  anti-commuting mass terms can be expressed as 
\beq
m\ee^{\ii\theta\gamma_\star}\overline{\psi}\psi= m\cos\theta\overline{\psi}\psi+\ii m\sin\theta\overline{\psi}\gamma_\star\psi,
\eeq
which   respects  Lorentz SO($2n$) invariance according to Eqs.~\eqref{eq:spinor_lorentz} and ~\eqref{eq:gamma_star}, and can be seen as the result of an axial rotation $\psi\mapsto{\rm exp}\{\ii\frac{\theta}{2}\gamma_\star\}\psi$,  $\overline{\psi}\mapsto \overline{\psi}{\rm exp}\{\ii\frac{\theta}{2}\gamma_\star\}$. On the other hand, the second one breaks explicitly the parity symmetry since 
\beq
\label{eq:parity}
\left.	\begin{matrix}
\mathcal{P}\psi(\tau,\boldsymbol{x})=\gamma^0\psi(\tau,-\boldsymbol{x}) \\
\mathcal{P}\overline{\psi}(\tau,\boldsymbol{x})=\overline{\psi}(\tau,-\boldsymbol{x})\gamma^0
\end{matrix}
\right\}\implies
\left.\begin{matrix}\overline{\psi}\psi\mapsto\overline{\psi}\psi
\\
\overline{\psi}\gamma_{\star}\psi\mapsto-\overline{\psi}\gamma_\star\psi
\end{matrix}\right\}
\eeq

For odd spacetime dimensions $D=2n+1$, this $\gamma_\star$ matrix plays the role of the gamma matrix  for the new  spatial direction $\gamma^{2n}=\gamma_\star$. Therefore,  the product of all spacetime gamma matrices is trivial $\gamma^0\gamma^1\cdots\gamma^{2n}\propto\mathbb{1}_{d_s}$, and the SO($2n+1$)-invariant Dirac action for free Dirac fields can only take the form of Eq.~\eqref{eq:Euc_action}. Therefore, only the standard mass term $m\bar{\psi}\psi$ can be considered which, as discussed in the following section would break the invariance under the corresponding parity transformation. In the following subsection, we will explain  how to go beyond these limitations when considering a reducible representation of the Clifford algebra.

\section{  Dimensional reduction    and  4-Fermi interactions}
\label{appB}

Motivated by the experimental situations discussed in the main text, we can also consider reducible representations of the Clifford algebra for a specific spacetime dimension. In this appendix, we will consider odd  dimension $D=2n+1$, and understand a reducible representation of the Clifford algebra as a consequence of an effective dimensional reduction. We  consider  $2n+3$ dimensions initially, where the spinor dimension is doubled with respect to the irreducible one of $D=2n+1$, and we get two additional gamma matrices $\gamma^{2n+1}$, and $\gamma_\star=\gamma^{2n+2}$. We will label the higher-dimensional spacetime coordinates with latin indexes $a\in\{0,1,\cdots, 2n+2\}$. The 
SO($2n+3$)-invariant action is that of Eq.~\eqref{eq:Euc_action}, and we will focus in the massless case $m=0$, namely
\begin{equation}
    S_{\rm 0}=\int\!\!{\rm d}^Dx\overline{\psi}(x)\gamma^a\partial_a\psi(x).
\end{equation}
 We can rewrite this action by separating the contribution of the two extra spatial dimensions
\begin{equation}
    S_{\rm 0}=\!\int\!\!{\rm d}^{2n+3}x\overline{\psi}(x)\big(\gamma^\mu\partial_\mu+\gamma^{2n+1}\partial_{2n+1}+\gamma_{\star}\partial_{2n+2}\big)\psi(x),
\end{equation}
where the index $\mu\in\{0,\cdots, 2n\}$ is restricted to the lower number of  dimensions $D=2n+1$ in the reduced  spacetime. 

As discussed in~\cite{Ryu_2010}, the 
dimensional reduction is inspired by the  Kaluza-Klein compactification~\cite{D_Bailin_1987}, and  proceeds  in two steps. In the first one, the $x_{2n+2}$ spatial direction is compactified to a circle   $x_{2n+2}+r=x_{2n+2}$ with a  very small radius $r\to 0$. Considering that the Grassmann fields are periodic in the spatial direction, the corresponding momentum $p_{2n+2}=-\ii\partial_{2n+2}\propto \ell_{2n+2}/r$ gets quantised in terms of the integers $\ell_{2n+2}\in\mathbb{Z}_{N_{2n+2}}$, one readily sees that  only  the quantum number $\ell_{2n+2}=0$ plays a role in the low-energy physics as $r\to 0$. From the perspective of the non-compact  dimensions, one gets a tower of very heavy Dirac fields, and focusing on low energies amounts to a truncation of such high-energy modes~\cite{Freedman:2012zz}.   In the presence of  an additional scalar field $\sigma_{\star}(x)$ that is minimally coupled to the fermions $\partial_{2n+2}\to\partial_{2n+2}+\ii\sigma_\star(x)$, and assuming that this scalar field is homogeneous $\sigma_\star(x)=m_\star$, the dimensional reduction leads to an effective low-energy action action that reads 
\begin{equation}
\label{eq:S_E_dim_reduced}
    S_{\rm 0}=\!\int\!\!{\rm d}^{2n+2}x\overline{\psi}(x)\big(\gamma^\mu\partial_\mu+\gamma^{2n+1}\partial_{2n+1}+ \ii m_\star \gamma_\star\big)\psi(x).
\end{equation}
which is now invariant under Lorentz transformations  in the reduced spacetime, such that SO($2n+3$)$\,\,\mapsto\,\,$SO($2n+2$).

In a second step, we compactify the  $x_{2n+1}$ direction,  introducing also  a minimally-coupled scalar field $\sigma_{2n+1}(x)$. Projecting again onto the low-energy physics when $r\to 0$, and thus considering only the $\ell_{2n+1}=0$ quantised momentum for a homogeneous field $\sigma_{2n+1}(x)=m_{2n+1}$, we are led to 
\begin{equation}
\label{eq:dim_reduced_action}
    S_{\rm 0}=\!\int\!\!{\rm d}^{D}x\overline{\psi}(x)\big( \gamma^\mu\partial_\mu+ \ii m_{2n+1}\gamma^{2n+1}+ \ii m_\star\gamma_\star\big)\psi(x),
\end{equation}
 where SO($2n+2$)$\,\,\mapsto\,\,$SO($2n+1$). By comparing this dimensionally-reduced  action to that  of Eq.~\eqref{eq:Euc_action}, we see that the sigma fields  play  a similar role to the mass terms when they are homogeneous.  The main difference is that we have more freedom in the definition of parity, and these two mass terms can open a gap in the parity-symmetric case.  For odd spacetime dimensions, parity must be understood as a transformation that reverses only an odd number of the spatial directions, for example 
\beq
\begin{split}
\mathcal{P}\psi(\tau,\boldsymbol{x})&=\gamma^{2n}\psi(\tau,x_1,\cdots x_{2n-1},-x_{2n}),\\
\mathcal{P}\overline{\psi}(\tau,\boldsymbol{x})&=-\overline{\psi}(\tau,x_1,\cdots x_{2n-1},-x_{2n})\gamma^{2n},
\end{split}
\eeq
such that both mass terms~\eqref{eq:dim_reduced_action} are invariant under parity
\beq
\label{eq:parity}
\begin{split}
\overline{\psi}\gamma^{2n+1}\psi&\mapsto\overline{\psi}\gamma^{2n+1}\psi,
\\
\overline{\psi}\gamma_{\star}\psi&\mapsto\overline{\psi}\gamma_\star\psi.
\end{split}
\eeq
Such a transformation can be defined for any other spatial axis, or an combination of an odd number of them. 

We note again that this effective action~\eqref{eq:dim_reduced_action}  is SO($2n+1$) invariant in the reduced spacetime, but there is a higher SO($2n+3$) invariance  if one considers  rotations in the full spacetime with the two additional compactified dimensions. The important point is that, as a remnant of the compactified dimensions, the spinors inherit the dimensionality given by the corresponding representation of the higher-dimensional Clifford algebra (see Appendix~\ref{appA}). This enlarged number of spinor components and bigger  symmetry group  can play an important role once we introduce interactions. Indeed,   we can also consider introducing SO($2n+3$) invariant quartic interactions, which can be obtain by considering the transformations of fermionic bilinears built from the Clifford algebra elements~\eqref{eq:so_d_transformations_clifford}. In particular, we can add an SO($2n+3$)-invariant  4-Fermi term to the action
\beq
\label{eq:interactions}
S_{\rm int}=\!\!\int\!{\rm d}^{D}x\frac{g^2}{2}\!\left(-(\overline{\psi}\psi)^2-(\overline{\psi}\ii\gamma^a\psi)(\overline{\psi}\ii\gamma_a\psi)\right)\!\!.
\eeq
The first term corresponds to the so-called Gross-Neveu interaction~\cite{PhysRevD.10.3235}, and is a scalar under the SO($2n+3$) Lorentz transformations. The other $2n+3$ terms are quartic interactions corresponding to the so-called Thirring term~\cite{THIRRING195891}, which   is a vector-vector interaction that is also invariant under SO($2n+3$) Lorentz transformations. On the other hand, if one rewrites these terms as $(\overline{\psi}\ii\gamma^a\psi)^2=(\overline{\psi}\ii\gamma^\mu\psi)^2-(\overline{\psi}\gamma^{2n+1}\psi)^2-(\overline{\psi}\gamma_{\star}\psi)^2$, and reinterprets them from the perspective of the dimensionally-reduced spacetime, only the $(\overline{\psi}\gamma^\mu\psi)^2$ terms correspond to the squared magnitude of a vector under SO($2n+1$), whereas the two additional terms $(\overline{\psi}\ii\gamma^{2n+1}\psi)^2$ and $(\overline{\psi}\ii\gamma_{\star}\psi)^2$ are scalars. We emphasise that these additional terms are only allowed by the fact that we are working with a reducible representation of the gamma matrices, which are allowed by the larger number of spinor degrees of freedom in the enlarged spacetime. In the main text, we have referred to the action $S_{0}+S_{\rm int}$ in Eqs.~\eqref{eq:S_E_dim_reduced} and~\eqref{eq:interactions} for $n=1$ as our model of  Dirac matter with SO($2n+3$)=SO($5$) 4-Fermi interactions. We discuss possible lattice regularizations that allow us to discuss higher-order topological phases and competing symmetry-breaking condensates as one increases the strength of the quartic interactions. This regularization requires rewriting the dimensionally-reduced masses in Eq.~\eqref{eq:dim_reduced_action} in terms of twisted Wilson masses, as discussed in the main text.

\section{Standard Wilson fermions and  Chern insulators}
\label{app:Chern}

In this Appendix, we present the details for a standard Wilson discretization of the reducible Dirac QFT in Eq.~\eqref{eq:action_free}. As noted in the main text,  this regularization   amounts to the introduction of a momentum-dependent shift $m_1\mapsto \overline{m}_1\!(\boldsymbol{k})$ of one of the  masses in Eq.~\eqref{eq:free_action_naive}, whereas the other mass is set to zero $m_2\mapsto\overline{m}_2=0$. The Wilson mass  depends on a real parameter  $r$ as follows
 \begin{equation}
 \label{eq:wilson_mass}
   \overline{m}_1\!(\boldsymbol{k})=m_1+\frac{r}{a}\bigg(2-\cos(k_1a)-\cos(k_2a)\bigg),
\end{equation}
 which can  be understood as the consequence of a  finite-difference discretizations of terms  involving higher-order spatial derivatives~\cite{Wilson1977}. We note that we have used an overline in the function in order to differentiate this standard Wilson mass from the twisted Wilson mass in Eq.~\eqref{eq:twisted_mass} 
 
 For this regularised model in $D=2+1$ dimensions, one actually finds that it corresponds to two copies of the square-lattice version~\cite{PhysRevB.74.085308,PhysRevB.78.195424,PhysRevResearch.4.L042012,ZIEGLER2022168763} of Haldane's quantum anomalous Hall effect~\cite{PhysRevLett.61.2015}, leading to a Chern insulator. The full band structure consists of four energy bands with a two-fold  degeneracy $\epsilon_{q,\pm}(\boldsymbol{k})=\pm \epsilon(\boldsymbol{k})$ for $ q\in\{1,2\}$, where 
\beq
\label{bands_chern}
\epsilon(\boldsymbol{k})=\sqrt{\hat{\boldsymbol{k}}{\phantom{k}}^{\!\!\!\!\!2}+\overline{m}_1^2(\boldsymbol{k})},
\eeq  
The groundstate is then obtained by filling all the negative energy states $\ket{\rm gs}=\prod_{\boldsymbol{k}\in{\rm BZ}}\prod_{q=1,2}\ket{\epsilon_{q,-}(\boldsymbol{k})}$, and one clearly sees from the above dispersion  that the Wilson
term leads to a different mass for  each  Dirac point~\eqref{eq:Dirac_points}, namely
\beq
\label{eq:masses}
\overline{m}_1(\boldsymbol{k}_{{\rm D},\boldsymbol{\ell}})=m_1+\frac{r}{a}\bigg(2-(-1)^{\ell_1}-(-1)^{\ell_2}\bigg).
\eeq
For notational convenience, we can define a  mass matrix that contains the four Wilson masses
\beq
\label{eq:mass_matrix}
\overline{M}_{{\rm W},1}=\sum_{\boldsymbol{\ell}\in\mathbb{Z}_2\times\mathbb{Z}_2}\overline{m}_1(\boldsymbol{k}_{{\rm D},\boldsymbol{\ell}})\ket{\boldsymbol{\ell}}\!\bra{\boldsymbol{\ell}}:\hspace{2ex}\langle\boldsymbol{\ell}|\boldsymbol{\ell}'\rangle=\delta_{\boldsymbol{\ell},\boldsymbol{\ell}'}.
\eeq
As outlined above, by setting $m_1=0$, one sees that the spurious doublers become very heavy with a mass of the order of the lattice cutoff $\overline{m}_1\!(\boldsymbol{k}_{{\rm D},\boldsymbol{\ell}})\propto r/a, \forall\boldsymbol{\ell}\neq \boldsymbol{0}$. On the contrary,  the  fermion at the origin of the BZ remains massless; $\overline{m}_1\!(\boldsymbol{k}_{{\rm D},\boldsymbol{0}})=0$. Making a long-wavelength expansion  around this point $\boldsymbol{k}\mapsto \boldsymbol{k}_{{\rm D},\boldsymbol{0}}+\boldsymbol{k}$ for $|\boldsymbol{k}|\ll\Lambda_{\rm c}$ now yields a long-wavelength  action that coincides with Eq.~\eqref{eq:action_free} for $\overline{m}_2=0$. 
We remark that, although the lattice  discretization breaks explicitly the invariance under SO($3$) Lorentz transformations, 
one recovers it in the continuum  limit around $\boldsymbol{k}_{{\rm D},\boldsymbol{0}}$. 

Let us finally discuss the connection of the groundstate of this lattice QFT to standard first-order topological insulators, in particular, to the so-called Chern insulators. For the explicit choice of the gamma matrices~\eqref{eq:gamma_spatial}-\eqref{eq:gamma_masses}, there is a  block structure that can be exploited to find a  basis in which the problem reduces to a pair of decoupled  Chern insulators. Indeed, one can prove that the groundstate corresponding to the above Dirac sea can be characterised by a non-vanishing Chern number for the principal $U (1)$ bundle associated to the filled bands~\cite{nakahara_2017}. This topological invariant can be expressed as the  integral of  the Berry curvature   $\mathcal{F}^{ij}_{{\rm b},q}(\boldsymbol{k})=\partial_{k_j}\mathcal{A}^i_q(\boldsymbol{k})-\partial_{k_i}\mathcal{A}^j_q(\boldsymbol{k})$, where  the connection is $\mathcal{A}^i_q(\boldsymbol{k})=-\ii \bra{\epsilon_{q,-}(\boldsymbol{k})}\partial_{k_i}\ket{\epsilon_{q,-}(\boldsymbol{k})}$~\cite{doi:10.1098/rspa.1984.0023}. One can then show that, for our Wilson-fermion QFT, the Chern number is
\beq
\label{eq:Chern_number_def}
\mathsf{ Ch}=\frac{1}{{4\pi}}\sum_{q}\!\!\int\!{{\rm d}k_i\wedge{\rm d}k_j}\mathcal{F}^{ij}_{{\rm b},q}(\boldsymbol{k})=\frac{1}{\pi}\sum_{q}{\rm arg}\left\{{\rm Det}\big(\overline{M}_{{\rm W},1}\big)\right\},
\eeq
which thus attains a non-zero value when we have an odd number of Dirac points that carry a negative Wilson mass 
\beq
\label{eq:Chern_number}
\mathsf{ Ch}=\sum_{\boldsymbol{\ell}}(-1)^{\ell_1+\ell_2}{\rm sgn}\big(\tilde{m}_1\!(\boldsymbol{k}_{{\rm D},\boldsymbol{\ell}})\big).
\eeq
In light of Eq.~\eqref{eq:masses}, we thus see that, whenever $m_1a\in(-4r,-2r)\cup(-2r,0)$, there is a non-vanishing Chern number $\mathsf{ Ch}=\pm 2$, signalling that we have two copies of the standard Chern insulator $\mathsf{ Ch}=\pm 1$, each corresponding to the square-lattice version~\cite{PhysRevB.74.085308,PhysRevB.78.195424,PhysRevResearch.4.L042012,ZIEGLER2022168763} of Haldane's quantum anomalous Hall effect~\cite{PhysRevLett.61.2015}. Even if there is no net external magnetic field piercing the spatial lattice, the system displays a quantised Hall conductance that is related to the Chern number as in the standard quantum Hall effect~\cite{PhysRevLett.49.405}.  The bulk-boundary correspondence links these topological invariants to    the appearance of circulating edge   states 
localised at the spatial boundaries, which are in fact low-dimensional versions of Kaplan's domain-wall fermions in lattice field theories~\cite{KAPLAN1992342,GOLTERMAN1993219,PhysRevLett.108.181807}.

\section{ Calculation of the effective SO($5$) potential}
\label{app:eff_potential}

\begin{figure}[t]
	\centering
	\includegraphics[width=0.8\columnwidth]{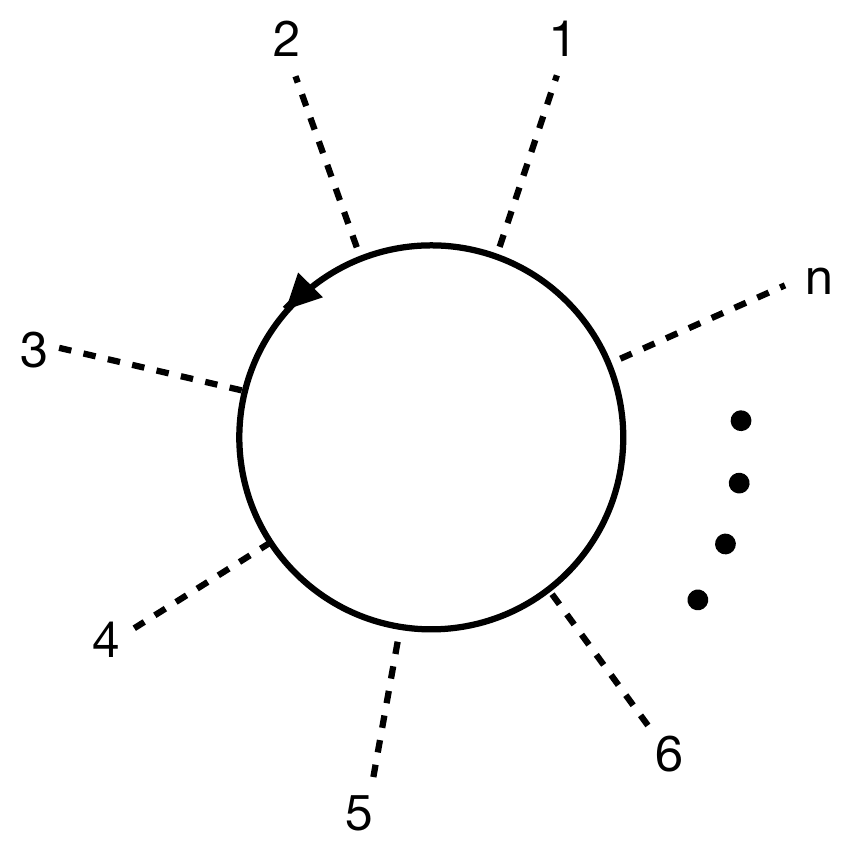}
	\caption{{\bf Large-$N$ Effective Potential:}  Example of a Feynman diagram with $n$ external zero-momentum auxiliary legs (dashed lines), and a single fermion loop (solid circle), yielding the main contribution to the effective potential $V_{\rm eff}$ in the large-$N$ limit.}
\label{fig:diag}
\end{figure}

Remarkably, all resummations required for the calculation of the effective potential $V_{\rm eff}(\boldsymbol{\Phi})$
are already encountered in the ``vanilla'' Gross-Neveu model in $2+1$ dimensions. We can rewrite this QFT in terms of an Euclidean Lagrangian containing just a single bosonic auxiliary field 
\begin{equation}
{\cal L}=\overline\psi\big(\ii\gamma^\mu p_\mu+m+\phi\big)\psi +
{N\over2g^2}\phi^2,
\end{equation}
where $\phi(x)$ will be latter identified with the various components of the completing condensates $\boldsymbol{\Phi}$ in our problem.
In the large-$N$ limit for a condensate $\Phi=\langle\phi\rangle$, $V_{\rm eff}$ then contains the sum of all 
diagrams with a single fermion loop and $n$ external $\phi$ legs~\cite{coleman_1985} (see Fig.~\ref{fig:diag})
which contribute with 
\begin{eqnarray}
{V_{\rm eff}(\Phi)\over N}&=&{\Sigma^2\over2g^2}+\sum_{n=1}^\infty{1\over
n}\int_p\mbox{tr}\left({{-\Phi}\over{\ii p^\mu\gamma_\mu+m}}\right)^n\nonumber\\
&=&{\Phi^2\over2g^2}+\sum_n{1\over n}\int_p\left({-\Phi\over{p^2+m^2}}\right)^n\mbox{tr}I_n
\label{eq:Vvanilla}
\end{eqnarray}
with
$I_n=(-ip^\mu\gamma_\mu+m)^n$ obeying the recurrence 
\begin{equation}
I_n=2mI_{n-1}-(p^2+m^2)I_{n-2}.
\label{eq:recvanilla}
\end{equation}
We now use $\mbox{tr}I_0=4$ and $\mbox{tr}I_1=4M$ to deduce the general result
\begin{equation}
\mbox{tr}I_n=4\sum_{k=0}^nA_{nk}m^k(p^2+m^2)^{{n-k}\over2},
\label{eq:In}
\end{equation}
where for $n$ even, $k$ is an even integer and the introduced matrix elements are
\begin{eqnarray}
A_{n0}&=&(-1)^{n\over2};\\
A_{nk}&=&{\cal
A}_k(-1)^{\textstyle{n\over2}}({\textstyle{n\over2}})^2(({\textstyle{n\over2}})^2-1^2)\ldots(({\textstyle{n\over2}})^2-({\textstyle{k\over2}}-1)^2);\nonumber
\end{eqnarray}
while for $n$ odd, $k$ is odd and
\begin{equation}
A_{nk}={1\over2}\left(A_{n+1,k+1}+A_{n-1,k+1}\right).
\label{eq:oddtoeven}
\end{equation}
The constant ${\cal A}_k$ in these expressions is defined such a way that one recovers $A_{nn}=2^{n-1}$.

Calculation of the sum in (\ref{eq:Vvanilla}) proceeds by considering a resummation of  three different cases.

\textit{ (a)  $n$ even, and $k=0$}: In this case, we find 
\begin{eqnarray}
\sum_{n=2,4,\ldots}{4(-1)^{n\over2}\over
n}&{\displaystyle\int_p}&\left({-\Phi\over\sqrt{p^2+m^2}}\right)^n=
\sum_q{2\over q}\int_p\left({-\Phi^2\over{p^2+q^2}}\right)^q\nonumber\\
&=&-2\int_p\log\left(1+{\Phi^2\over{p^2+m^2}}\right),
\label{eq:int1}
\end{eqnarray}
where we have made use of the index $q=n/2$.

\textit { (b) $n$ even and $k$ even:}
When the integer $k$ is allowed to be even, we reindex using $n=2q$ and $k=2\ell$ 
to find
\begin{equation}
\sum_{q\ell}{2A_{2q,2\ell}\over
m}\int_p\left({\Phi^2\over{p^2+m^2}}\right)^q\left({m^2\over{p^2+m^2}}\right)^\ell.
\end{equation}
We now use the identity
\begin{equation}
\sum_qA_{2q,2\ell}{z^q\over
q}={{2^{2\ell-1}}\over\ell}{z^\ell\over{(1+z)^{2\ell}}}
\end{equation}
to perform the following resummation 
\begin{equation}
\sum_\ell{1\over\ell}\int_p{{(2\Phi
m)^{2\ell}}\over{(p^2+m^2+\Phi^2)^{2\ell}}}=-\int_p\log\left(1-{{4\Phi^2m^2}\over{(p^2+m^2+\Phi^2)^2}}\right).
\label{eq:int2}
\end{equation}

\textit {(c) $n$ odd and $k$ odd:}
When both integers are odd, we use Eq.~(\ref{eq:oddtoeven}) to write ~the contribution as
\begin{equation}
\sum_{n,k=1,3,\dots}{4\over
n}\int_p\left({-\Phi\over\sqrt{p^2+m^2}}\right)^n\left(m\over\sqrt{p^2+m^2}\right)^kA_{nk}
\end{equation}
where  the matrix elements fulfill the following identity
\begin{equation}
\sum_{n {\rm odd}}(A_{n+1,k+1}+A_{n-1,k+1}){z^n\over n}={2^k\over
k}{z^k\over{(1+z^2)^k}}.
\end{equation}
Using these expressions, we can again resum on $n$ to find 
\begin{equation}
-\sum_{k {\rm odd}}\int_p{-2\over k}\left({{2\Phi
m}\over{p^2+m^2+\Phi^2}}\right)^k=-\int_p\log\left({{p^2+(m+\Phi)^2}\over{p^2+(m-\Phi)^2}}\right).
\label{eq:int3}
\end{equation}
Finally, after adding all the contributions in Eqs.~(\ref{eq:int1}), (\ref{eq:int2}) and
(\ref{eq:int3}) together,  we find   considerable simplifications such that the final result gets the following simple form 
\begin{equation}
{V_{\rm  eff}(\Phi)\over
N}={\Phi^2\over2g^2}-2\int_p\log\left({{p^2+(m+\Phi)^2}\over{p^2+m^2}}\right).
\label{eq:int5}
\end{equation}

Once we have this generic result, we need to consider the case with several competing condensation channels $\Phi\mapsto\boldsymbol{\Phi}=(\boldsymbol{A},\boldsymbol{\Sigma},\Pi)$, and a specific twisted Wilson mass regularization. If we still consider a continuum QFT, but  include the interaction term having full SO($5$) symmetry, the
expression for the effective potential generalises to 
\begin{eqnarray}
{V_{\rm eff}(\boldsymbol{A},\boldsymbol{\Sigma},\Pi)\over N}&=&{1\over
2g^2}(\boldsymbol{A}^2+\boldsymbol{\Sigma}^2+\Pi^2)\label{eq:Vefffull}\\
+
\sum_n{1\over
n}&{\displaystyle\int_p}&\mbox{tr}\left({{-\ii\boldsymbol{\gamma}\cdot\boldsymbol{
A}-\ii\gamma_3\Sigma_1-\ii\gamma_5\Sigma_2-\Pi}\over{\ii p^\mu\gamma_\mu+\ii m_1\gamma_3+\ii m_2\gamma_5}}\right)^n.\nonumber
\end{eqnarray}
Equations (\ref{eq:Vvanilla}) and (\ref{eq:recvanilla}) are replaced by 
\begin{equation}
{1\over
2g^2}(\boldsymbol{A}^2+\boldsymbol{\Sigma}^2+\Pi^2)+
\sum_n{1\over n}\int_p\left({1\over{p^2+m_1^2+m_2^2}}\right)^n\mbox{tr}I_n
\end{equation}
where we have introduced 
\begin{eqnarray}
I_n=&-&2(\boldsymbol{p}\cdot\boldsymbol{A}+m_1\Sigma_1+m_2\Sigma_2
)I_{n-1}\\&-&(\boldsymbol{A}^2+\boldsymbol{\Sigma}^2+\Pi^2)(p^2+m_1^2+m_2^2)I_{n-2}.\nonumber
\end{eqnarray}
It is now straightforward to repeat steps (\ref{eq:int1}-\ref{eq:int3}). After adding all the contributions, instead of Eq.~\eqref{eq:int5}, we find 
\beq
\frac{ V_{\rm eff}}{N}={1\over
2g^2}\boldsymbol{\Phi}^2-2\!\!\int_{p}\log\!\left(\frac{p_0^2+(\boldsymbol{p}+\boldsymbol{A})^2+\boldsymbol{m}^2(\boldsymbol{\Sigma})+\Pi^2}{p_0^2+\boldsymbol{p}^2+\boldsymbol{m}^2(\boldsymbol{0})}\right)\!\!.
\eeq
where we have introduced $\boldsymbol{m}(\boldsymbol{\Sigma})=(m_1+\Sigma_1,m_2+\Sigma_2)$.
At this point, in order to take into account the twisted-mass Wilson regularization, where we need to substitute 
$p=(p_0,\boldsymbol{p})\mapsto(k_0,\hat{\boldsymbol{k}})$,   $\boldsymbol{m}(\boldsymbol{\Sigma})\mapsto \boldsymbol{m}(\boldsymbol{k},\boldsymbol{\Sigma})$, and substitute the momentum integrals by the corresponding mode sums. In this way, we arrive at Eq.~\eqref{eq:quant_corr} of the main text. 



Finally we discuss incorporation of a further interaction between fermion charge densities mediated by an auxiliary $A_0$, whereupon Eqn.~(\ref{eq:Vefffull}) is supplemented by terms
\begin{equation}
    {A_0^2\over2\tilde{g}^{2}}+
    \sum_n{1\over n}{\displaystyle\int_p}\mbox{tr}\left({{-\ii\gamma_0A_0}\over{\ii p^\mu\gamma_\mu+\ii m_1\gamma_3+\ii m_2\gamma_5}}\right)^n.
\end{equation}
The choice $\tilde{g}^{2}=g^2$ yields full SO(6) symmetry of the  Euclidean 4-Fermi interactions. The algebra involving $\gamma_0$ is identical to that for $\boldsymbol{\gamma}$, and the one-loop contributions 
yield 
\begin{equation}
    -2\int_p\log
    {{[(p_0^2+A_0)^2+P^2]
    [(p_0^2-A_0)^2+P^2]
    }
    \over{(p^2+Q^2)^2}},
\end{equation}
where $Q^2=\boldsymbol{p}^2+\boldsymbol{m}^2(\boldsymbol{0})$, and  $P^2=(\boldsymbol{p}+\boldsymbol{A})^2+\boldsymbol{m}^2(\boldsymbol{\Sigma})+\Pi^2$.
Careful integration over $p_0$ with a UV cutoff $\Lambda$ show that all 
dependence on $A_0$ is O($\Lambda^{-1}$) and hence vanishes as $\Lambda\to\infty$.
As a consequence, the minimum of $V_{\rm eff}$ always lies at $A_0=0$ and it is therefore safe to ignore condensation in this channel at half-filling.

\bibliography{bibliography}

\begin{thebibliography}{231}%
\makeatletter
\providecommand \@ifxundefined [1]{%
 \@ifx{#1\undefined}
}%
\providecommand \@ifnum [1]{%
 \ifnum #1\expandafter \@firstoftwo
 \else \expandafter \@secondoftwo
 \fi
}%
\providecommand \@ifx [1]{%
 \ifx #1\expandafter \@firstoftwo
 \else \expandafter \@secondoftwo
 \fi
}%
\providecommand \natexlab [1]{#1}%
\providecommand \enquote  [1]{``#1''}%
\providecommand \bibnamefont  [1]{#1}%
\providecommand \bibfnamefont [1]{#1}%
\providecommand \citenamefont [1]{#1}%
\providecommand \href@noop [0]{\@secondoftwo}%
\providecommand \href [0]{\begingroup \@sanitize@url \@href}%
\providecommand \@href[1]{\@@startlink{#1}\@@href}%
\providecommand \@@href[1]{\endgroup#1\@@endlink}%
\providecommand \@sanitize@url [0]{\catcode `\\12\catcode `\$12\catcode
  `\&12\catcode `\#12\catcode `\^12\catcode `\_12\catcode `\%12\relax}%
\providecommand \@@startlink[1]{}%
\providecommand \@@endlink[0]{}%
\providecommand \url  [0]{\begingroup\@sanitize@url \@url }%
\providecommand \@url [1]{\endgroup\@href {#1}{\urlprefix }}%
\providecommand \urlprefix  [0]{URL }%
\providecommand \Eprint [0]{\href }%
\providecommand \doibase [0]{https://doi.org/}%
\providecommand \selectlanguage [0]{\@gobble}%
\providecommand \bibinfo  [0]{\@secondoftwo}%
\providecommand \bibfield  [0]{\@secondoftwo}%
\providecommand \translation [1]{[#1]}%
\providecommand \BibitemOpen [0]{}%
\providecommand \bibitemStop [0]{}%
\providecommand \bibitemNoStop [0]{.\EOS\space}%
\providecommand \EOS [0]{\spacefactor3000\relax}%
\providecommand \BibitemShut  [1]{\csname bibitem#1\endcsname}%
\let\auto@bib@innerbib\@empty
\bibitem [{\citenamefont {Peskin}\ and\ \citenamefont
  {Schroeder}(1995)}]{Peskin:1995ev}%
  \BibitemOpen
  \bibfield  {author} {\bibinfo {author} {\bibfnamefont {M.~E.}\ \bibnamefont
  {Peskin}}\ and\ \bibinfo {author} {\bibfnamefont {D.~V.}\ \bibnamefont
  {Schroeder}},\ }\href@noop {} {\emph {\bibinfo {title} {{An Introduction to
  quantum field theory}}}}\ (\bibinfo  {publisher} {Addison-Wesley},\ \bibinfo
  {address} {Reading, USA},\ \bibinfo {year} {1995})\BibitemShut {NoStop}%
\bibitem [{\citenamefont {Yang}\ and\ \citenamefont
  {Mills}(1954)}]{PhysRev.96.191}%
  \BibitemOpen
  \bibfield  {author} {\bibinfo {author} {\bibfnamefont {C.~N.}\ \bibnamefont
  {Yang}}\ and\ \bibinfo {author} {\bibfnamefont {R.~L.}\ \bibnamefont
  {Mills}},\ }\bibfield  {title} {\bibinfo {title} {Conservation of isotopic
  spin and isotopic gauge invariance},\ }\href
  {https://doi.org/10.1103/PhysRev.96.191} {\bibfield  {journal} {\bibinfo
  {journal} {Phys. Rev.}\ }\textbf {\bibinfo {volume} {96}},\ \bibinfo {pages}
  {191} (\bibinfo {year} {1954})}\BibitemShut {NoStop}%
\bibitem [{\citenamefont {Nambu}\ and\ \citenamefont
  {Jona-Lasinio}(1961{\natexlab{a}})}]{PhysRev.122.345}%
  \BibitemOpen
  \bibfield  {author} {\bibinfo {author} {\bibfnamefont {Y.}~\bibnamefont
  {Nambu}}\ and\ \bibinfo {author} {\bibfnamefont {G.}~\bibnamefont
  {Jona-Lasinio}},\ }\bibfield  {title} {\bibinfo {title} {Dynamical model of
  elementary particles based on an analogy with superconductivity. i},\ }\href
  {https://doi.org/10.1103/PhysRev.122.345} {\bibfield  {journal} {\bibinfo
  {journal} {Phys. Rev.}\ }\textbf {\bibinfo {volume} {122}},\ \bibinfo {pages}
  {345} (\bibinfo {year} {1961}{\natexlab{a}})}\BibitemShut {NoStop}%
\bibitem [{\citenamefont {Nambu}\ and\ \citenamefont
  {Jona-Lasinio}(1961{\natexlab{b}})}]{PhysRev.124.246}%
  \BibitemOpen
  \bibfield  {author} {\bibinfo {author} {\bibfnamefont {Y.}~\bibnamefont
  {Nambu}}\ and\ \bibinfo {author} {\bibfnamefont {G.}~\bibnamefont
  {Jona-Lasinio}},\ }\bibfield  {title} {\bibinfo {title} {Dynamical model of
  elementary particles based on an analogy with superconductivity. ii},\ }\href
  {https://doi.org/10.1103/PhysRev.124.246} {\bibfield  {journal} {\bibinfo
  {journal} {Phys. Rev.}\ }\textbf {\bibinfo {volume} {124}},\ \bibinfo {pages}
  {246} (\bibinfo {year} {1961}{\natexlab{b}})}\BibitemShut {NoStop}%
\bibitem [{\citenamefont {Klevansky}(1992)}]{RevModPhys.64.649}%
  \BibitemOpen
  \bibfield  {author} {\bibinfo {author} {\bibfnamefont {S.~P.}\ \bibnamefont
  {Klevansky}},\ }\bibfield  {title} {\bibinfo {title} {The
  nambu---jona-lasinio model of quantum chromodynamics},\ }\href
  {https://doi.org/10.1103/RevModPhys.64.649} {\bibfield  {journal} {\bibinfo
  {journal} {Rev. Mod. Phys.}\ }\textbf {\bibinfo {volume} {64}},\ \bibinfo
  {pages} {649} (\bibinfo {year} {1992})}\BibitemShut {NoStop}%
\bibitem [{\citenamefont {Hatsuda}\ and\ \citenamefont
  {Kunihiro}(1994)}]{HATSUDA1994221}%
  \BibitemOpen
  \bibfield  {author} {\bibinfo {author} {\bibfnamefont {T.}~\bibnamefont
  {Hatsuda}}\ and\ \bibinfo {author} {\bibfnamefont {T.}~\bibnamefont
  {Kunihiro}},\ }\bibfield  {title} {\bibinfo {title} {Qcd phenomenology based
  on a chiral effective lagrangian},\ }\href
  {https://doi.org/https://doi.org/10.1016/0370-1573(94)90022-1} {\bibfield
  {journal} {\bibinfo  {journal} {Physics Reports}\ }\textbf {\bibinfo {volume}
  {247}},\ \bibinfo {pages} {221} (\bibinfo {year} {1994})}\BibitemShut
  {NoStop}%
\bibitem [{\citenamefont {Gross}\ and\ \citenamefont
  {Neveu}(1974)}]{PhysRevD.10.3235}%
  \BibitemOpen
  \bibfield  {author} {\bibinfo {author} {\bibfnamefont {D.~J.}\ \bibnamefont
  {Gross}}\ and\ \bibinfo {author} {\bibfnamefont {A.}~\bibnamefont {Neveu}},\
  }\bibfield  {title} {\bibinfo {title} {Dynamical symmetry breaking in
  asymptotically free field theories},\ }\href
  {https://doi.org/10.1103/PhysRevD.10.3235} {\bibfield  {journal} {\bibinfo
  {journal} {Phys. Rev. D}\ }\textbf {\bibinfo {volume} {10}},\ \bibinfo
  {pages} {3235} (\bibinfo {year} {1974})}\BibitemShut {NoStop}%
\bibitem [{\citenamefont {Gross}()}]{doi:10.1142/9789814412674_0004}%
  \BibitemOpen
  \bibfield  {author} {\bibinfo {author} {\bibfnamefont {D.~J.}\ \bibnamefont
  {Gross}},\ }\bibinfo {title} {Applications of the renormalization group to
  high-energy physics},\ in\ \href {https://doi.org/10.1142/9789814412674_0004}
  {\emph {\bibinfo {booktitle} {Methods in Field Theory}}},\ pp.\ \bibinfo
  {pages} {141--250}\BibitemShut {NoStop}%
\bibitem [{\citenamefont {Zamolodchikov}\ and\ \citenamefont
  {Zamolodchikov}(1978)}]{ZAMOLODCHIKOV1978481}%
  \BibitemOpen
  \bibfield  {author} {\bibinfo {author} {\bibfnamefont {A.~B.}\ \bibnamefont
  {Zamolodchikov}}\ and\ \bibinfo {author} {\bibfnamefont {A.~B.}\ \bibnamefont
  {Zamolodchikov}},\ }\bibfield  {title} {\bibinfo {title} {Exact s matrix of
  gross-neveu “elementary” fermions},\ }\href
  {https://doi.org/https://doi.org/10.1016/0370-2693(78)90738-4} {\bibfield
  {journal} {\bibinfo  {journal} {Physics Letters B}\ }\textbf {\bibinfo
  {volume} {72}},\ \bibinfo {pages} {481} (\bibinfo {year} {1978})}\BibitemShut
  {NoStop}%
\bibitem [{\citenamefont {Witten}(1978)}]{WITTEN1978110}%
  \BibitemOpen
  \bibfield  {author} {\bibinfo {author} {\bibfnamefont {E.}~\bibnamefont
  {Witten}},\ }\bibfield  {title} {\bibinfo {title} {Chiral symmetry, the 1/n
  expansion and the su(n) thirring model},\ }\href
  {https://doi.org/https://doi.org/10.1016/0550-3213(78)90416-9} {\bibfield
  {journal} {\bibinfo  {journal} {Nuclear Physics B}\ }\textbf {\bibinfo
  {volume} {145}},\ \bibinfo {pages} {110} (\bibinfo {year}
  {1978})}\BibitemShut {NoStop}%
\bibitem [{\citenamefont {Berg}\ and\ \citenamefont
  {Weisz}(1978)}]{BERG1978205}%
  \BibitemOpen
  \bibfield  {author} {\bibinfo {author} {\bibfnamefont {B.}~\bibnamefont
  {Berg}}\ and\ \bibinfo {author} {\bibfnamefont {P.}~\bibnamefont {Weisz}},\
  }\bibfield  {title} {\bibinfo {title} {Exact s-matrix of the chiral invariant
  su(n) thirring model},\ }\href
  {https://doi.org/https://doi.org/10.1016/0550-3213(78)90438-8} {\bibfield
  {journal} {\bibinfo  {journal} {Nuclear Physics B}\ }\textbf {\bibinfo
  {volume} {146}},\ \bibinfo {pages} {205} (\bibinfo {year}
  {1978})}\BibitemShut {NoStop}%
\bibitem [{\citenamefont {Gross}\ and\ \citenamefont
  {Wilczek}(1973)}]{PhysRevLett.30.1343}%
  \BibitemOpen
  \bibfield  {author} {\bibinfo {author} {\bibfnamefont {D.~J.}\ \bibnamefont
  {Gross}}\ and\ \bibinfo {author} {\bibfnamefont {F.}~\bibnamefont
  {Wilczek}},\ }\bibfield  {title} {\bibinfo {title} {Ultraviolet behavior of
  non-abelian gauge theories},\ }\href
  {https://doi.org/10.1103/PhysRevLett.30.1343} {\bibfield  {journal} {\bibinfo
   {journal} {Phys. Rev. Lett.}\ }\textbf {\bibinfo {volume} {30}},\ \bibinfo
  {pages} {1343} (\bibinfo {year} {1973})}\BibitemShut {NoStop}%
\bibitem [{\citenamefont {Politzer}(1973)}]{PhysRevLett.30.1346}%
  \BibitemOpen
  \bibfield  {author} {\bibinfo {author} {\bibfnamefont {H.~D.}\ \bibnamefont
  {Politzer}},\ }\bibfield  {title} {\bibinfo {title} {Reliable perturbative
  results for strong interactions?},\ }\href
  {https://doi.org/10.1103/PhysRevLett.30.1346} {\bibfield  {journal} {\bibinfo
   {journal} {Phys. Rev. Lett.}\ }\textbf {\bibinfo {volume} {30}},\ \bibinfo
  {pages} {1346} (\bibinfo {year} {1973})}\BibitemShut {NoStop}%
\bibitem [{\citenamefont {Novoselov}\ \emph {et~al.}(2005)\citenamefont
  {Novoselov}, \citenamefont {Geim}, \citenamefont {Morozov}, \citenamefont
  {Jiang}, \citenamefont {Katsnelson}, \citenamefont {Grigorieva},
  \citenamefont {Dubonos},\ and\ \citenamefont {Firsov}}]{Novoselov2005}%
  \BibitemOpen
  \bibfield  {author} {\bibinfo {author} {\bibfnamefont {K.~S.}\ \bibnamefont
  {Novoselov}}, \bibinfo {author} {\bibfnamefont {A.~K.}\ \bibnamefont {Geim}},
  \bibinfo {author} {\bibfnamefont {S.~V.}\ \bibnamefont {Morozov}}, \bibinfo
  {author} {\bibfnamefont {D.}~\bibnamefont {Jiang}}, \bibinfo {author}
  {\bibfnamefont {M.~I.}\ \bibnamefont {Katsnelson}}, \bibinfo {author}
  {\bibfnamefont {I.~V.}\ \bibnamefont {Grigorieva}}, \bibinfo {author}
  {\bibfnamefont {S.~V.}\ \bibnamefont {Dubonos}},\ and\ \bibinfo {author}
  {\bibfnamefont {A.~A.}\ \bibnamefont {Firsov}},\ }\bibfield  {title}
  {\bibinfo {title} {Two-dimensional gas of massless dirac fermions in
  graphene},\ }\href {https://doi.org/10.1038/nature04233} {\bibfield
  {journal} {\bibinfo  {journal} {Nature}\ }\textbf {\bibinfo {volume} {438}},\
  \bibinfo {pages} {197} (\bibinfo {year} {2005})}\BibitemShut {NoStop}%
\bibitem [{\citenamefont {Zhang}\ \emph {et~al.}(2005)\citenamefont {Zhang},
  \citenamefont {Tan}, \citenamefont {Stormer},\ and\ \citenamefont
  {Kim}}]{Zhang2005}%
  \BibitemOpen
  \bibfield  {author} {\bibinfo {author} {\bibfnamefont {Y.}~\bibnamefont
  {Zhang}}, \bibinfo {author} {\bibfnamefont {Y.-W.}\ \bibnamefont {Tan}},
  \bibinfo {author} {\bibfnamefont {H.~L.}\ \bibnamefont {Stormer}},\ and\
  \bibinfo {author} {\bibfnamefont {P.}~\bibnamefont {Kim}},\ }\bibfield
  {title} {\bibinfo {title} {Experimental observation of the quantum hall
  effect and berry's phase in graphene},\ }\href
  {https://doi.org/10.1038/nature04235} {\bibfield  {journal} {\bibinfo
  {journal} {Nature}\ }\textbf {\bibinfo {volume} {438}},\ \bibinfo {pages}
  {201} (\bibinfo {year} {2005})}\BibitemShut {NoStop}%
\bibitem [{\citenamefont {Hsieh}\ \emph {et~al.}(2008)\citenamefont {Hsieh},
  \citenamefont {Qian}, \citenamefont {Wray}, \citenamefont {Xia},
  \citenamefont {Hor}, \citenamefont {Cava},\ and\ \citenamefont
  {Hasan}}]{Hsieh2008}%
  \BibitemOpen
  \bibfield  {author} {\bibinfo {author} {\bibfnamefont {D.}~\bibnamefont
  {Hsieh}}, \bibinfo {author} {\bibfnamefont {D.}~\bibnamefont {Qian}},
  \bibinfo {author} {\bibfnamefont {L.}~\bibnamefont {Wray}}, \bibinfo {author}
  {\bibfnamefont {Y.}~\bibnamefont {Xia}}, \bibinfo {author} {\bibfnamefont
  {Y.~S.}\ \bibnamefont {Hor}}, \bibinfo {author} {\bibfnamefont {R.~J.}\
  \bibnamefont {Cava}},\ and\ \bibinfo {author} {\bibfnamefont {M.~Z.}\
  \bibnamefont {Hasan}},\ }\bibfield  {title} {\bibinfo {title} {A topological
  dirac insulator in a quantum spin hall phase},\ }\href
  {https://doi.org/10.1038/nature06843} {\bibfield  {journal} {\bibinfo
  {journal} {Nature}\ }\textbf {\bibinfo {volume} {452}},\ \bibinfo {pages}
  {970} (\bibinfo {year} {2008})}\BibitemShut {NoStop}%
\bibitem [{\citenamefont {Hsieh}\ \emph {et~al.}(2009)\citenamefont {Hsieh},
  \citenamefont {Xia}, \citenamefont {Qian}, \citenamefont {Wray},
  \citenamefont {Dil}, \citenamefont {Meier}, \citenamefont {Osterwalder},
  \citenamefont {Patthey}, \citenamefont {Checkelsky}, \citenamefont {Ong},
  \citenamefont {Fedorov}, \citenamefont {Lin}, \citenamefont {Bansil},
  \citenamefont {Grauer}, \citenamefont {Hor}, \citenamefont {Cava},\ and\
  \citenamefont {Hasan}}]{Hsieh2009}%
  \BibitemOpen
  \bibfield  {author} {\bibinfo {author} {\bibfnamefont {D.}~\bibnamefont
  {Hsieh}}, \bibinfo {author} {\bibfnamefont {Y.}~\bibnamefont {Xia}}, \bibinfo
  {author} {\bibfnamefont {D.}~\bibnamefont {Qian}}, \bibinfo {author}
  {\bibfnamefont {L.}~\bibnamefont {Wray}}, \bibinfo {author} {\bibfnamefont
  {J.~H.}\ \bibnamefont {Dil}}, \bibinfo {author} {\bibfnamefont
  {F.}~\bibnamefont {Meier}}, \bibinfo {author} {\bibfnamefont
  {J.}~\bibnamefont {Osterwalder}}, \bibinfo {author} {\bibfnamefont
  {L.}~\bibnamefont {Patthey}}, \bibinfo {author} {\bibfnamefont {J.~G.}\
  \bibnamefont {Checkelsky}}, \bibinfo {author} {\bibfnamefont {N.~P.}\
  \bibnamefont {Ong}}, \bibinfo {author} {\bibfnamefont {A.~V.}\ \bibnamefont
  {Fedorov}}, \bibinfo {author} {\bibfnamefont {H.}~\bibnamefont {Lin}},
  \bibinfo {author} {\bibfnamefont {A.}~\bibnamefont {Bansil}}, \bibinfo
  {author} {\bibfnamefont {D.}~\bibnamefont {Grauer}}, \bibinfo {author}
  {\bibfnamefont {Y.~S.}\ \bibnamefont {Hor}}, \bibinfo {author} {\bibfnamefont
  {R.~J.}\ \bibnamefont {Cava}},\ and\ \bibinfo {author} {\bibfnamefont
  {M.~Z.}\ \bibnamefont {Hasan}},\ }\bibfield  {title} {\bibinfo {title} {A
  tunable topological insulator in the spin helical dirac transport regime},\
  }\href {https://doi.org/10.1038/nature08234} {\bibfield  {journal} {\bibinfo
  {journal} {Nature}\ }\textbf {\bibinfo {volume} {460}},\ \bibinfo {pages}
  {1101} (\bibinfo {year} {2009})}\BibitemShut {NoStop}%
\bibitem [{\citenamefont {Tarruell}\ \emph {et~al.}(2012)\citenamefont
  {Tarruell}, \citenamefont {Greif}, \citenamefont {Uehlinger}, \citenamefont
  {Jotzu},\ and\ \citenamefont {Esslinger}}]{Tarruell2012}%
  \BibitemOpen
  \bibfield  {author} {\bibinfo {author} {\bibfnamefont {L.}~\bibnamefont
  {Tarruell}}, \bibinfo {author} {\bibfnamefont {D.}~\bibnamefont {Greif}},
  \bibinfo {author} {\bibfnamefont {T.}~\bibnamefont {Uehlinger}}, \bibinfo
  {author} {\bibfnamefont {G.}~\bibnamefont {Jotzu}},\ and\ \bibinfo {author}
  {\bibfnamefont {T.}~\bibnamefont {Esslinger}},\ }\bibfield  {title} {\bibinfo
  {title} {Creating, moving and merging dirac points with a fermi gas in a
  tunable honeycomb lattice},\ }\href {https://doi.org/10.1038/nature10871}
  {\bibfield  {journal} {\bibinfo  {journal} {Nature}\ }\textbf {\bibinfo
  {volume} {483}},\ \bibinfo {pages} {302} (\bibinfo {year}
  {2012})}\BibitemShut {NoStop}%
\bibitem [{\citenamefont {Jotzu}\ \emph {et~al.}(2014)\citenamefont {Jotzu},
  \citenamefont {Messer}, \citenamefont {Desbuquois}, \citenamefont {Lebrat},
  \citenamefont {Uehlinger}, \citenamefont {Greif},\ and\ \citenamefont
  {Esslinger}}]{Jotzu2014}%
  \BibitemOpen
  \bibfield  {author} {\bibinfo {author} {\bibfnamefont {G.}~\bibnamefont
  {Jotzu}}, \bibinfo {author} {\bibfnamefont {M.}~\bibnamefont {Messer}},
  \bibinfo {author} {\bibfnamefont {R.}~\bibnamefont {Desbuquois}}, \bibinfo
  {author} {\bibfnamefont {M.}~\bibnamefont {Lebrat}}, \bibinfo {author}
  {\bibfnamefont {T.}~\bibnamefont {Uehlinger}}, \bibinfo {author}
  {\bibfnamefont {D.}~\bibnamefont {Greif}},\ and\ \bibinfo {author}
  {\bibfnamefont {T.}~\bibnamefont {Esslinger}},\ }\bibfield  {title} {\bibinfo
  {title} {Experimental realization of the topological haldane model with
  ultracold fermions},\ }\href {https://doi.org/10.1038/nature13915} {\bibfield
   {journal} {\bibinfo  {journal} {Nature}\ }\textbf {\bibinfo {volume}
  {515}},\ \bibinfo {pages} {237} (\bibinfo {year} {2014})}\BibitemShut
  {NoStop}%
\bibitem [{\citenamefont {Wehling}\ \emph {et~al.}(2014)\citenamefont
  {Wehling}, \citenamefont {Black-Schaffer},\ and\ \citenamefont
  {Balatsky}}]{doi:10.1080/00018732.2014.927109}%
  \BibitemOpen
  \bibfield  {author} {\bibinfo {author} {\bibfnamefont {T.}~\bibnamefont
  {Wehling}}, \bibinfo {author} {\bibfnamefont {A.}~\bibnamefont
  {Black-Schaffer}},\ and\ \bibinfo {author} {\bibfnamefont {A.}~\bibnamefont
  {Balatsky}},\ }\bibfield  {title} {\bibinfo {title} {Dirac materials},\
  }\href {https://doi.org/10.1080/00018732.2014.927109} {\bibfield  {journal}
  {\bibinfo  {journal} {Advances in Physics}\ }\textbf {\bibinfo {volume}
  {63}},\ \bibinfo {pages} {1} (\bibinfo {year} {2014})}\BibitemShut {NoStop}%
\bibitem [{\citenamefont {Gonz\'alez-Cuadra}\ \emph {et~al.}(2020)\citenamefont
  {Gonz\'alez-Cuadra}, \citenamefont {Dauphin}, \citenamefont {Aidelsburger},
  \citenamefont {Lewenstein},\ and\ \citenamefont {Bermudez}}]{Gonzalez_2020}%
  \BibitemOpen
  \bibfield  {author} {\bibinfo {author} {\bibfnamefont {D.}~\bibnamefont
  {Gonz\'alez-Cuadra}}, \bibinfo {author} {\bibfnamefont {A.}~\bibnamefont
  {Dauphin}}, \bibinfo {author} {\bibfnamefont {M.}~\bibnamefont
  {Aidelsburger}}, \bibinfo {author} {\bibfnamefont {M.}~\bibnamefont
  {Lewenstein}},\ and\ \bibinfo {author} {\bibfnamefont {A.}~\bibnamefont
  {Bermudez}},\ }\bibfield  {title} {\bibinfo {title} {Rotor jackiw-rebbi
  model: A cold-atom approach to chiral symmetry restoration and charge
  confinement},\ }\href {https://doi.org/10.1103/PRXQuantum.1.020321}
  {\bibfield  {journal} {\bibinfo  {journal} {PRX Quantum}\ }\textbf {\bibinfo
  {volume} {1}},\ \bibinfo {pages} {020321} (\bibinfo {year}
  {2020})}\BibitemShut {NoStop}%
\bibitem [{\citenamefont {Wilson}(1973)}]{PhysRevD.7.2911}%
  \BibitemOpen
  \bibfield  {author} {\bibinfo {author} {\bibfnamefont {K.~G.}\ \bibnamefont
  {Wilson}},\ }\bibfield  {title} {\bibinfo {title} {Quantum field - theory
  models in less than 4 dimensions},\ }\href
  {https://doi.org/10.1103/PhysRevD.7.2911} {\bibfield  {journal} {\bibinfo
  {journal} {Phys. Rev. D}\ }\textbf {\bibinfo {volume} {7}},\ \bibinfo {pages}
  {2911} (\bibinfo {year} {1973})}\BibitemShut {NoStop}%
\bibitem [{\citenamefont {Shizuya}(1980)}]{PhysRevD.21.2327}%
  \BibitemOpen
  \bibfield  {author} {\bibinfo {author} {\bibfnamefont {K.-i.}\ \bibnamefont
  {Shizuya}},\ }\bibfield  {title} {\bibinfo {title} {$\frac{1}{N}$ expansion
  and the theory of composite particles},\ }\href
  {https://doi.org/10.1103/PhysRevD.21.2327} {\bibfield  {journal} {\bibinfo
  {journal} {Phys. Rev. D}\ }\textbf {\bibinfo {volume} {21}},\ \bibinfo
  {pages} {2327} (\bibinfo {year} {1980})}\BibitemShut {NoStop}%
\bibitem [{\citenamefont {Hands}\ \emph {et~al.}(1993)\citenamefont {Hands},
  \citenamefont {Kocic},\ and\ \citenamefont {Kogut}}]{HANDS199329}%
  \BibitemOpen
  \bibfield  {author} {\bibinfo {author} {\bibfnamefont {S.}~\bibnamefont
  {Hands}}, \bibinfo {author} {\bibfnamefont {A.}~\bibnamefont {Kocic}},\ and\
  \bibinfo {author} {\bibfnamefont {J.}~\bibnamefont {Kogut}},\ }\bibfield
  {title} {\bibinfo {title} {Four-fermi theories in fewer than four
  dimensions},\ }\href {https://doi.org/https://doi.org/10.1006/aphy.1993.1039}
  {\bibfield  {journal} {\bibinfo  {journal} {Annals of Physics}\ }\textbf
  {\bibinfo {volume} {224}},\ \bibinfo {pages} {29} (\bibinfo {year}
  {1993})}\BibitemShut {NoStop}%
\bibitem [{\citenamefont {Rosenstein}\ \emph {et~al.}(1989)\citenamefont
  {Rosenstein}, \citenamefont {Warr},\ and\ \citenamefont
  {Park}}]{PhysRevLett.62.1433}%
  \BibitemOpen
  \bibfield  {author} {\bibinfo {author} {\bibfnamefont {B.}~\bibnamefont
  {Rosenstein}}, \bibinfo {author} {\bibfnamefont {B.~J.}\ \bibnamefont
  {Warr}},\ and\ \bibinfo {author} {\bibfnamefont {S.~H.}\ \bibnamefont
  {Park}},\ }\bibfield  {title} {\bibinfo {title} {Four-fermion theory is
  renormalizable in 2+1 dimensions},\ }\href
  {https://doi.org/10.1103/PhysRevLett.62.1433} {\bibfield  {journal} {\bibinfo
   {journal} {Phys. Rev. Lett.}\ }\textbf {\bibinfo {volume} {62}},\ \bibinfo
  {pages} {1433} (\bibinfo {year} {1989})}\BibitemShut {NoStop}%
\bibitem [{\citenamefont {Rosenstein}\ \emph {et~al.}(1991)\citenamefont
  {Rosenstein}, \citenamefont {Warr},\ and\ \citenamefont
  {Park}}]{ROSENSTEIN199159}%
  \BibitemOpen
  \bibfield  {author} {\bibinfo {author} {\bibfnamefont {B.}~\bibnamefont
  {Rosenstein}}, \bibinfo {author} {\bibfnamefont {B.~J.}\ \bibnamefont
  {Warr}},\ and\ \bibinfo {author} {\bibfnamefont {S.~H.}\ \bibnamefont
  {Park}},\ }\bibfield  {title} {\bibinfo {title} {Dynamical symmetry breaking
  in four-fermion interaction models},\ }\href
  {https://doi.org/https://doi.org/10.1016/0370-1573(91)90129-A} {\bibfield
  {journal} {\bibinfo  {journal} {Physics Reports}\ }\textbf {\bibinfo {volume}
  {205}},\ \bibinfo {pages} {59} (\bibinfo {year} {1991})}\BibitemShut
  {NoStop}%
\bibitem [{\citenamefont {Semenoff}\ and\ \citenamefont
  {Wijewardhana}(1989)}]{PhysRevLett.63.2633}%
  \BibitemOpen
  \bibfield  {author} {\bibinfo {author} {\bibfnamefont {G.~W.}\ \bibnamefont
  {Semenoff}}\ and\ \bibinfo {author} {\bibfnamefont {L.~C.~R.}\ \bibnamefont
  {Wijewardhana}},\ }\bibfield  {title} {\bibinfo {title} {Dynamical mass
  generation in 3d four-fermion theory},\ }\href
  {https://doi.org/10.1103/PhysRevLett.63.2633} {\bibfield  {journal} {\bibinfo
   {journal} {Phys. Rev. Lett.}\ }\textbf {\bibinfo {volume} {63}},\ \bibinfo
  {pages} {2633} (\bibinfo {year} {1989})}\BibitemShut {NoStop}%
\bibitem [{\citenamefont {Boyack}\ \emph {et~al.}(2021)\citenamefont {Boyack},
  \citenamefont {Yerzhakov},\ and\ \citenamefont {Maciejko}}]{Boyack2021}%
  \BibitemOpen
  \bibfield  {author} {\bibinfo {author} {\bibfnamefont {R.}~\bibnamefont
  {Boyack}}, \bibinfo {author} {\bibfnamefont {H.}~\bibnamefont {Yerzhakov}},\
  and\ \bibinfo {author} {\bibfnamefont {J.}~\bibnamefont {Maciejko}},\
  }\bibfield  {title} {\bibinfo {title} {Quantum phase transitions in dirac
  fermion systems},\ }\href {https://doi.org/10.1140/epjs/s11734-021-00069-1}
  {\bibfield  {journal} {\bibinfo  {journal} {The European Physical Journal
  Special Topics}\ }\textbf {\bibinfo {volume} {230}},\ \bibinfo {pages} {979}
  (\bibinfo {year} {2021})}\BibitemShut {NoStop}%
\bibitem [{\citenamefont {Herbut}(2023)}]{herbut2023wilsonfisher}%
  \BibitemOpen
  \bibfield  {author} {\bibinfo {author} {\bibfnamefont {I.~F.}\ \bibnamefont
  {Herbut}},\ }\href@noop {} {\bibinfo {title} {Wilson-fisher fixed points in
  presence of dirac fermions}} (\bibinfo {year} {2023}),\ \Eprint
  {https://arxiv.org/abs/2304.07654} {arXiv:2304.07654 [cond-mat.str-el]}
  \BibitemShut {NoStop}%
\bibitem [{\citenamefont {Bloch}\ \emph {et~al.}(2008)\citenamefont {Bloch},
  \citenamefont {Dalibard},\ and\ \citenamefont {Zwerger}}]{Bloch_2008}%
  \BibitemOpen
  \bibfield  {author} {\bibinfo {author} {\bibfnamefont {I.}~\bibnamefont
  {Bloch}}, \bibinfo {author} {\bibfnamefont {J.}~\bibnamefont {Dalibard}},\
  and\ \bibinfo {author} {\bibfnamefont {W.}~\bibnamefont {Zwerger}},\
  }\bibfield  {title} {\bibinfo {title} {Many-body physics with ultracold
  gases},\ }\href {https://doi.org/10.1103/revmodphys.80.885} {\bibfield
  {journal} {\bibinfo  {journal} {Rev. Mod. Phys.}\ }\textbf {\bibinfo {volume}
  {80}},\ \bibinfo {pages} {885} (\bibinfo {year} {2008})}\BibitemShut
  {NoStop}%
\bibitem [{\citenamefont {Feynman}(1982)}]{Feynman_1982}%
  \BibitemOpen
  \bibfield  {author} {\bibinfo {author} {\bibfnamefont {R.~P.}\ \bibnamefont
  {Feynman}},\ }\bibfield  {title} {\bibinfo {title} {Simulating physics with
  computers},\ }\href {https://doi.org/10.1007/bf02650179} {\bibfield
  {journal} {\bibinfo  {journal} {Int. J. Theor. Phys.}\ }\textbf {\bibinfo
  {volume} {21}},\ \bibinfo {pages} {467} (\bibinfo {year} {1982})}\BibitemShut
  {NoStop}%
\bibitem [{\citenamefont {Cirac}\ and\ \citenamefont
  {Zoller}(2012)}]{Cirac2012}%
  \BibitemOpen
  \bibfield  {author} {\bibinfo {author} {\bibfnamefont {J.~I.}\ \bibnamefont
  {Cirac}}\ and\ \bibinfo {author} {\bibfnamefont {P.}~\bibnamefont {Zoller}},\
  }\bibfield  {title} {\bibinfo {title} {Goals and opportunities in quantum
  simulation},\ }\href {https://doi.org/10.1038/nphys2275} {\bibfield
  {journal} {\bibinfo  {journal} {Nature Physics}\ }\textbf {\bibinfo {volume}
  {8}},\ \bibinfo {pages} {264} (\bibinfo {year} {2012})}\BibitemShut {NoStop}%
\bibitem [{\citenamefont {Bloch}\ \emph {et~al.}(2012)\citenamefont {Bloch},
  \citenamefont {Dalibard},\ and\ \citenamefont {Nascimb{\`e}ne}}]{Bloch2012}%
  \BibitemOpen
  \bibfield  {author} {\bibinfo {author} {\bibfnamefont {I.}~\bibnamefont
  {Bloch}}, \bibinfo {author} {\bibfnamefont {J.}~\bibnamefont {Dalibard}},\
  and\ \bibinfo {author} {\bibfnamefont {S.}~\bibnamefont {Nascimb{\`e}ne}},\
  }\bibfield  {title} {\bibinfo {title} {Quantum simulations with ultracold
  quantum gases},\ }\href {https://doi.org/10.1038/nphys2259} {\bibfield
  {journal} {\bibinfo  {journal} {Nature Physics}\ }\textbf {\bibinfo {volume}
  {8}},\ \bibinfo {pages} {267} (\bibinfo {year} {2012})}\BibitemShut {NoStop}%
\bibitem [{\citenamefont {Blatt}\ and\ \citenamefont {Roos}(2012)}]{Blatt2012}%
  \BibitemOpen
  \bibfield  {author} {\bibinfo {author} {\bibfnamefont {R.}~\bibnamefont
  {Blatt}}\ and\ \bibinfo {author} {\bibfnamefont {C.~F.}\ \bibnamefont
  {Roos}},\ }\bibfield  {title} {\bibinfo {title} {Quantum simulations with
  trapped ions},\ }\href {https://doi.org/10.1038/nphys2252} {\bibfield
  {journal} {\bibinfo  {journal} {Nature Physics}\ }\textbf {\bibinfo {volume}
  {8}},\ \bibinfo {pages} {277} (\bibinfo {year} {2012})}\BibitemShut {NoStop}%
\bibitem [{\citenamefont {Altman}\ \emph {et~al.}(2021)\citenamefont {Altman},
  \citenamefont {Brown}, \citenamefont {Carleo}, \citenamefont {Carr},
  \citenamefont {Demler}, \citenamefont {Chin}, \citenamefont {DeMarco},
  \citenamefont {Economou}, \citenamefont {Eriksson}, \citenamefont {Fu},
  \citenamefont {Greiner}, \citenamefont {Hazzard}, \citenamefont {Hulet},
  \citenamefont {Koll\'ar}, \citenamefont {Lev}, \citenamefont {Lukin},
  \citenamefont {Ma}, \citenamefont {Mi}, \citenamefont {Misra}, \citenamefont
  {Monroe}, \citenamefont {Murch}, \citenamefont {Nazario}, \citenamefont {Ni},
  \citenamefont {Potter}, \citenamefont {Roushan}, \citenamefont {Saffman},
  \citenamefont {Schleier-Smith}, \citenamefont {Siddiqi}, \citenamefont
  {Simmonds}, \citenamefont {Singh}, \citenamefont {Spielman}, \citenamefont
  {Temme}, \citenamefont {Weiss}, \citenamefont {Vu\ifmmode \check{c}\else
  \v{c}\fi{}kovi\ifmmode~\acute{c}\else \'{c}\fi{}}, \citenamefont
  {Vuleti\ifmmode~\acute{c}\else \'{c}\fi{}}, \citenamefont {Ye},\ and\
  \citenamefont {Zwierlein}}]{PRXQuantum.2.017003}%
  \BibitemOpen
  \bibfield  {author} {\bibinfo {author} {\bibfnamefont {E.}~\bibnamefont
  {Altman}}, \bibinfo {author} {\bibfnamefont {K.~R.}\ \bibnamefont {Brown}},
  \bibinfo {author} {\bibfnamefont {G.}~\bibnamefont {Carleo}}, \bibinfo
  {author} {\bibfnamefont {L.~D.}\ \bibnamefont {Carr}}, \bibinfo {author}
  {\bibfnamefont {E.}~\bibnamefont {Demler}}, \bibinfo {author} {\bibfnamefont
  {C.}~\bibnamefont {Chin}}, \bibinfo {author} {\bibfnamefont {B.}~\bibnamefont
  {DeMarco}}, \bibinfo {author} {\bibfnamefont {S.~E.}\ \bibnamefont
  {Economou}}, \bibinfo {author} {\bibfnamefont {M.~A.}\ \bibnamefont
  {Eriksson}}, \bibinfo {author} {\bibfnamefont {K.-M.~C.}\ \bibnamefont {Fu}},
  \bibinfo {author} {\bibfnamefont {M.}~\bibnamefont {Greiner}}, \bibinfo
  {author} {\bibfnamefont {K.~R.}\ \bibnamefont {Hazzard}}, \bibinfo {author}
  {\bibfnamefont {R.~G.}\ \bibnamefont {Hulet}}, \bibinfo {author}
  {\bibfnamefont {A.~J.}\ \bibnamefont {Koll\'ar}}, \bibinfo {author}
  {\bibfnamefont {B.~L.}\ \bibnamefont {Lev}}, \bibinfo {author} {\bibfnamefont
  {M.~D.}\ \bibnamefont {Lukin}}, \bibinfo {author} {\bibfnamefont
  {R.}~\bibnamefont {Ma}}, \bibinfo {author} {\bibfnamefont {X.}~\bibnamefont
  {Mi}}, \bibinfo {author} {\bibfnamefont {S.}~\bibnamefont {Misra}}, \bibinfo
  {author} {\bibfnamefont {C.}~\bibnamefont {Monroe}}, \bibinfo {author}
  {\bibfnamefont {K.}~\bibnamefont {Murch}}, \bibinfo {author} {\bibfnamefont
  {Z.}~\bibnamefont {Nazario}}, \bibinfo {author} {\bibfnamefont {K.-K.}\
  \bibnamefont {Ni}}, \bibinfo {author} {\bibfnamefont {A.~C.}\ \bibnamefont
  {Potter}}, \bibinfo {author} {\bibfnamefont {P.}~\bibnamefont {Roushan}},
  \bibinfo {author} {\bibfnamefont {M.}~\bibnamefont {Saffman}}, \bibinfo
  {author} {\bibfnamefont {M.}~\bibnamefont {Schleier-Smith}}, \bibinfo
  {author} {\bibfnamefont {I.}~\bibnamefont {Siddiqi}}, \bibinfo {author}
  {\bibfnamefont {R.}~\bibnamefont {Simmonds}}, \bibinfo {author}
  {\bibfnamefont {M.}~\bibnamefont {Singh}}, \bibinfo {author} {\bibfnamefont
  {I.}~\bibnamefont {Spielman}}, \bibinfo {author} {\bibfnamefont
  {K.}~\bibnamefont {Temme}}, \bibinfo {author} {\bibfnamefont {D.~S.}\
  \bibnamefont {Weiss}}, \bibinfo {author} {\bibfnamefont {J.}~\bibnamefont
  {Vu\ifmmode \check{c}\else \v{c}\fi{}kovi\ifmmode~\acute{c}\else
  \'{c}\fi{}}}, \bibinfo {author} {\bibfnamefont {V.}~\bibnamefont
  {Vuleti\ifmmode~\acute{c}\else \'{c}\fi{}}}, \bibinfo {author} {\bibfnamefont
  {J.}~\bibnamefont {Ye}},\ and\ \bibinfo {author} {\bibfnamefont
  {M.}~\bibnamefont {Zwierlein}},\ }\bibfield  {title} {\bibinfo {title}
  {Quantum simulators: Architectures and opportunities},\ }\href
  {https://doi.org/10.1103/PRXQuantum.2.017003} {\bibfield  {journal} {\bibinfo
   {journal} {PRX Quantum}\ }\textbf {\bibinfo {volume} {2}},\ \bibinfo {pages}
  {017003} (\bibinfo {year} {2021})}\BibitemShut {NoStop}%
\bibitem [{\citenamefont {Troyer}\ and\ \citenamefont
  {Wiese}(2005)}]{PhysRevLett.94.170201}%
  \BibitemOpen
  \bibfield  {author} {\bibinfo {author} {\bibfnamefont {M.}~\bibnamefont
  {Troyer}}\ and\ \bibinfo {author} {\bibfnamefont {U.-J.}\ \bibnamefont
  {Wiese}},\ }\bibfield  {title} {\bibinfo {title} {Computational complexity
  and fundamental limitations to fermionic quantum monte carlo simulations},\
  }\href {https://doi.org/10.1103/PhysRevLett.94.170201} {\bibfield  {journal}
  {\bibinfo  {journal} {Phys. Rev. Lett.}\ }\textbf {\bibinfo {volume} {94}},\
  \bibinfo {pages} {170201} (\bibinfo {year} {2005})}\BibitemShut {NoStop}%
\bibitem [{\citenamefont {Nagata}(2022)}]{NAGATA2022103991}%
  \BibitemOpen
  \bibfield  {author} {\bibinfo {author} {\bibfnamefont {K.}~\bibnamefont
  {Nagata}},\ }\bibfield  {title} {\bibinfo {title} {Finite-density lattice qcd
  and sign problem: Current status and open problems},\ }\href
  {https://doi.org/https://doi.org/10.1016/j.ppnp.2022.103991} {\bibfield
  {journal} {\bibinfo  {journal} {Progress in Particle and Nuclear Physics}\
  }\textbf {\bibinfo {volume} {127}},\ \bibinfo {pages} {103991} (\bibinfo
  {year} {2022})}\BibitemShut {NoStop}%
\bibitem [{\citenamefont {Wu}\ \emph {et~al.}(2003)\citenamefont {Wu},
  \citenamefont {Hu},\ and\ \citenamefont {Zhang}}]{PhysRevLett.91.186402}%
  \BibitemOpen
  \bibfield  {author} {\bibinfo {author} {\bibfnamefont {C.}~\bibnamefont
  {Wu}}, \bibinfo {author} {\bibfnamefont {J.-p.}\ \bibnamefont {Hu}},\ and\
  \bibinfo {author} {\bibfnamefont {S.-c.}\ \bibnamefont {Zhang}},\ }\bibfield
  {title} {\bibinfo {title} {Exact so(5) symmetry in the spin-$3/2$ fermionic
  system},\ }\href {https://doi.org/10.1103/PhysRevLett.91.186402} {\bibfield
  {journal} {\bibinfo  {journal} {Phys. Rev. Lett.}\ }\textbf {\bibinfo
  {volume} {91}},\ \bibinfo {pages} {186402} (\bibinfo {year}
  {2003})}\BibitemShut {NoStop}%
\bibitem [{\citenamefont {Wu}(2006)}]{doi:10.1142/S0217984906012213}%
  \BibitemOpen
  \bibfield  {author} {\bibinfo {author} {\bibfnamefont {C.}~\bibnamefont
  {Wu}},\ }\bibfield  {title} {\bibinfo {title} {Hidden symmetry and quantum
  phases in spin-3/2 cold atomic systems},\ }\href
  {https://doi.org/10.1142/S0217984906012213} {\bibfield  {journal} {\bibinfo
  {journal} {Modern Physics Letters B}\ }\textbf {\bibinfo {volume} {20}},\
  \bibinfo {pages} {1707} (\bibinfo {year} {2006})}\BibitemShut {NoStop}%
\bibitem [{\citenamefont {Hubbard}(1963)}]{PRSLSA_276_238}%
  \BibitemOpen
  \bibfield  {author} {\bibinfo {author} {\bibfnamefont {J.}~\bibnamefont
  {Hubbard}},\ }\bibfield  {title} {\bibinfo {title} {Electron correlations in
  narrow energy bands},\ }\href {https://doi.org/10.1098/rspa.1963.0204}
  {\bibfield  {journal} {\bibinfo  {journal} {Proc. R. Soc. London, Ser. A}\
  }\textbf {\bibinfo {volume} {276}},\ \bibinfo {pages} {238} (\bibinfo {year}
  {1963})}\BibitemShut {NoStop}%
\bibitem [{\citenamefont {Eckert}\ \emph {et~al.}(2007)\citenamefont {Eckert},
  \citenamefont {Zawitkowski}, \citenamefont {Leskinen}, \citenamefont
  {Sanpera},\ and\ \citenamefont {Lewenstein}}]{Eckert_2007}%
  \BibitemOpen
  \bibfield  {author} {\bibinfo {author} {\bibfnamefont {K.}~\bibnamefont
  {Eckert}}, \bibinfo {author} {\bibfnamefont {L.}~\bibnamefont {Zawitkowski}},
  \bibinfo {author} {\bibfnamefont {M.~J.}\ \bibnamefont {Leskinen}}, \bibinfo
  {author} {\bibfnamefont {A.}~\bibnamefont {Sanpera}},\ and\ \bibinfo {author}
  {\bibfnamefont {M.}~\bibnamefont {Lewenstein}},\ }\bibfield  {title}
  {\bibinfo {title} {Ultracold atomic bose and fermi spinor gases in optical
  lattices},\ }\href {https://doi.org/10.1088/1367-2630/9/5/133} {\bibfield
  {journal} {\bibinfo  {journal} {New Journal of Physics}\ }\textbf {\bibinfo
  {volume} {9}},\ \bibinfo {pages} {133} (\bibinfo {year} {2007})}\BibitemShut
  {NoStop}%
\bibitem [{\citenamefont {Szirmai}\ and\ \citenamefont
  {Lewenstein}(2011)}]{Szirmai_2011}%
  \BibitemOpen
  \bibfield  {author} {\bibinfo {author} {\bibfnamefont {E.}~\bibnamefont
  {Szirmai}}\ and\ \bibinfo {author} {\bibfnamefont {M.}~\bibnamefont
  {Lewenstein}},\ }\bibfield  {title} {\bibinfo {title} {Exotic magnetic orders
  for high-spin ultracold fermions},\ }\href
  {https://doi.org/10.1209/0295-5075/93/66005} {\bibfield  {journal} {\bibinfo
  {journal} {Europhysics Letters}\ }\textbf {\bibinfo {volume} {93}},\ \bibinfo
  {pages} {66005} (\bibinfo {year} {2011})}\BibitemShut {NoStop}%
\bibitem [{\citenamefont {Lecheminant}\ \emph {et~al.}(2005)\citenamefont
  {Lecheminant}, \citenamefont {Boulat},\ and\ \citenamefont
  {Azaria}}]{PhysRevLett.95.240402}%
  \BibitemOpen
  \bibfield  {author} {\bibinfo {author} {\bibfnamefont {P.}~\bibnamefont
  {Lecheminant}}, \bibinfo {author} {\bibfnamefont {E.}~\bibnamefont
  {Boulat}},\ and\ \bibinfo {author} {\bibfnamefont {P.}~\bibnamefont
  {Azaria}},\ }\bibfield  {title} {\bibinfo {title} {Confinement and
  superfluidity in one-dimensional degenerate fermionic cold atoms},\ }\href
  {https://doi.org/10.1103/PhysRevLett.95.240402} {\bibfield  {journal}
  {\bibinfo  {journal} {Phys. Rev. Lett.}\ }\textbf {\bibinfo {volume} {95}},\
  \bibinfo {pages} {240402} (\bibinfo {year} {2005})}\BibitemShut {NoStop}%
\bibitem [{\citenamefont {Wu}(2005)}]{PhysRevLett.95.266404}%
  \BibitemOpen
  \bibfield  {author} {\bibinfo {author} {\bibfnamefont {C.}~\bibnamefont
  {Wu}},\ }\bibfield  {title} {\bibinfo {title} {Competing orders in
  one-dimensional spin-$3/2$ fermionic systems},\ }\href
  {https://doi.org/10.1103/PhysRevLett.95.266404} {\bibfield  {journal}
  {\bibinfo  {journal} {Phys. Rev. Lett.}\ }\textbf {\bibinfo {volume} {95}},\
  \bibinfo {pages} {266404} (\bibinfo {year} {2005})}\BibitemShut {NoStop}%
\bibitem [{\citenamefont {Controzzi}\ and\ \citenamefont
  {Tsvelik}(2006)}]{PhysRevLett.96.097205}%
  \BibitemOpen
  \bibfield  {author} {\bibinfo {author} {\bibfnamefont {D.}~\bibnamefont
  {Controzzi}}\ and\ \bibinfo {author} {\bibfnamefont {A.~M.}\ \bibnamefont
  {Tsvelik}},\ }\bibfield  {title} {\bibinfo {title} {Exactly solvable model
  for isospin $s=3/2$ fermionic atoms on an optical lattice},\ }\href
  {https://doi.org/10.1103/PhysRevLett.96.097205} {\bibfield  {journal}
  {\bibinfo  {journal} {Phys. Rev. Lett.}\ }\textbf {\bibinfo {volume} {96}},\
  \bibinfo {pages} {097205} (\bibinfo {year} {2006})}\BibitemShut {NoStop}%
\bibitem [{\citenamefont {Rodr\'{\i}guez}\ \emph {et~al.}(2010)\citenamefont
  {Rodr\'{\i}guez}, \citenamefont {Arg\"uelles}, \citenamefont
  {Colom\'e-Tatch\'e}, \citenamefont {Vekua},\ and\ \citenamefont
  {Santos}}]{PhysRevLett.105.050402}%
  \BibitemOpen
  \bibfield  {author} {\bibinfo {author} {\bibfnamefont {K.}~\bibnamefont
  {Rodr\'{\i}guez}}, \bibinfo {author} {\bibfnamefont {A.}~\bibnamefont
  {Arg\"uelles}}, \bibinfo {author} {\bibfnamefont {M.}~\bibnamefont
  {Colom\'e-Tatch\'e}}, \bibinfo {author} {\bibfnamefont {T.}~\bibnamefont
  {Vekua}},\ and\ \bibinfo {author} {\bibfnamefont {L.}~\bibnamefont
  {Santos}},\ }\bibfield  {title} {\bibinfo {title} {Mott-insulator phases of
  spin-$3/2$ fermions in the presence of quadratic zeeman coupling},\ }\href
  {https://doi.org/10.1103/PhysRevLett.105.050402} {\bibfield  {journal}
  {\bibinfo  {journal} {Phys. Rev. Lett.}\ }\textbf {\bibinfo {volume} {105}},\
  \bibinfo {pages} {050402} (\bibinfo {year} {2010})}\BibitemShut {NoStop}%
\bibitem [{\citenamefont {Zhang}(1997)}]{doi:10.1126/science.275.5303.1089}%
  \BibitemOpen
  \bibfield  {author} {\bibinfo {author} {\bibfnamefont {S.-C.}\ \bibnamefont
  {Zhang}},\ }\bibfield  {title} {\bibinfo {title} {A unified theory based on
  $so(5)$ symmetry of superconductivity and antiferromagnetism},\ }\href
  {https://doi.org/10.1126/science.275.5303.1089} {\bibfield  {journal}
  {\bibinfo  {journal} {Science}\ }\textbf {\bibinfo {volume} {275}},\ \bibinfo
  {pages} {1089} (\bibinfo {year} {1997})}\BibitemShut {NoStop}%
\bibitem [{\citenamefont {Demler}\ \emph {et~al.}(2004)\citenamefont {Demler},
  \citenamefont {Hanke},\ and\ \citenamefont {Zhang}}]{RevModPhys.76.909}%
  \BibitemOpen
  \bibfield  {author} {\bibinfo {author} {\bibfnamefont {E.}~\bibnamefont
  {Demler}}, \bibinfo {author} {\bibfnamefont {W.}~\bibnamefont {Hanke}},\ and\
  \bibinfo {author} {\bibfnamefont {S.-C.}\ \bibnamefont {Zhang}},\ }\bibfield
  {title} {\bibinfo {title} {$\mathit{SO}(5)$ theory of antiferromagnetism and
  superconductivity},\ }\href {https://doi.org/10.1103/RevModPhys.76.909}
  {\bibfield  {journal} {\bibinfo  {journal} {Rev. Mod. Phys.}\ }\textbf
  {\bibinfo {volume} {76}},\ \bibinfo {pages} {909} (\bibinfo {year}
  {2004})}\BibitemShut {NoStop}%
\bibitem [{\citenamefont {Schindler}\ \emph {et~al.}(2018)\citenamefont
  {Schindler}, \citenamefont {Cook}, \citenamefont {Vergniory}, \citenamefont
  {Wang}, \citenamefont {Parkin}, \citenamefont {Bernevig},\ and\ \citenamefont
  {Neupert}}]{doi:10.1126/sciadv.aat0346}%
  \BibitemOpen
  \bibfield  {author} {\bibinfo {author} {\bibfnamefont {F.}~\bibnamefont
  {Schindler}}, \bibinfo {author} {\bibfnamefont {A.~M.}\ \bibnamefont {Cook}},
  \bibinfo {author} {\bibfnamefont {M.~G.}\ \bibnamefont {Vergniory}}, \bibinfo
  {author} {\bibfnamefont {Z.}~\bibnamefont {Wang}}, \bibinfo {author}
  {\bibfnamefont {S.~S.~P.}\ \bibnamefont {Parkin}}, \bibinfo {author}
  {\bibfnamefont {B.~A.}\ \bibnamefont {Bernevig}},\ and\ \bibinfo {author}
  {\bibfnamefont {T.}~\bibnamefont {Neupert}},\ }\bibfield  {title} {\bibinfo
  {title} {Higher-order topological insulators},\ }\href
  {https://doi.org/10.1126/sciadv.aat0346} {\bibfield  {journal} {\bibinfo
  {journal} {Science Advances}\ }\textbf {\bibinfo {volume} {4}},\ \bibinfo
  {pages} {eaat0346} (\bibinfo {year} {2018})}\BibitemShut {NoStop}%
\bibitem [{\citenamefont {Benalcazar}\ \emph
  {et~al.}(2017{\natexlab{a}})\citenamefont {Benalcazar}, \citenamefont
  {Bernevig},\ and\ \citenamefont {Hughes}}]{doi:10.1126/science.aah6442}%
  \BibitemOpen
  \bibfield  {author} {\bibinfo {author} {\bibfnamefont {W.~A.}\ \bibnamefont
  {Benalcazar}}, \bibinfo {author} {\bibfnamefont {B.~A.}\ \bibnamefont
  {Bernevig}},\ and\ \bibinfo {author} {\bibfnamefont {T.~L.}\ \bibnamefont
  {Hughes}},\ }\bibfield  {title} {\bibinfo {title} {Quantized electric
  multipole insulators},\ }\href {https://doi.org/10.1126/science.aah6442}
  {\bibfield  {journal} {\bibinfo  {journal} {Science}\ }\textbf {\bibinfo
  {volume} {357}},\ \bibinfo {pages} {61} (\bibinfo {year}
  {2017}{\natexlab{a}})}\BibitemShut {NoStop}%
\bibitem [{\citenamefont {Benalcazar}\ \emph
  {et~al.}(2017{\natexlab{b}})\citenamefont {Benalcazar}, \citenamefont
  {Bernevig},\ and\ \citenamefont {Hughes}}]{PhysRevB.96.245115}%
  \BibitemOpen
  \bibfield  {author} {\bibinfo {author} {\bibfnamefont {W.~A.}\ \bibnamefont
  {Benalcazar}}, \bibinfo {author} {\bibfnamefont {B.~A.}\ \bibnamefont
  {Bernevig}},\ and\ \bibinfo {author} {\bibfnamefont {T.~L.}\ \bibnamefont
  {Hughes}},\ }\bibfield  {title} {\bibinfo {title} {Electric multipole
  moments, topological multipole moment pumping, and chiral hinge states in
  crystalline insulators},\ }\href {https://doi.org/10.1103/PhysRevB.96.245115}
  {\bibfield  {journal} {\bibinfo  {journal} {Phys. Rev. B}\ }\textbf {\bibinfo
  {volume} {96}},\ \bibinfo {pages} {245115} (\bibinfo {year}
  {2017}{\natexlab{b}})}\BibitemShut {NoStop}%
\bibitem [{\citenamefont {Geier}\ \emph {et~al.}(2018)\citenamefont {Geier},
  \citenamefont {Trifunovic}, \citenamefont {Hoskam},\ and\ \citenamefont
  {Brouwer}}]{PhysRevB.97.205135}%
  \BibitemOpen
  \bibfield  {author} {\bibinfo {author} {\bibfnamefont {M.}~\bibnamefont
  {Geier}}, \bibinfo {author} {\bibfnamefont {L.}~\bibnamefont {Trifunovic}},
  \bibinfo {author} {\bibfnamefont {M.}~\bibnamefont {Hoskam}},\ and\ \bibinfo
  {author} {\bibfnamefont {P.~W.}\ \bibnamefont {Brouwer}},\ }\bibfield
  {title} {\bibinfo {title} {Second-order topological insulators and
  superconductors with an order-two crystalline symmetry},\ }\href
  {https://doi.org/10.1103/PhysRevB.97.205135} {\bibfield  {journal} {\bibinfo
  {journal} {Phys. Rev. B}\ }\textbf {\bibinfo {volume} {97}},\ \bibinfo
  {pages} {205135} (\bibinfo {year} {2018})}\BibitemShut {NoStop}%
\bibitem [{\citenamefont {You}\ \emph {et~al.}(2018)\citenamefont {You},
  \citenamefont {Devakul}, \citenamefont {Burnell},\ and\ \citenamefont
  {Neupert}}]{You_2018}%
  \BibitemOpen
  \bibfield  {author} {\bibinfo {author} {\bibfnamefont {Y.}~\bibnamefont
  {You}}, \bibinfo {author} {\bibfnamefont {T.}~\bibnamefont {Devakul}},
  \bibinfo {author} {\bibfnamefont {F.~J.}\ \bibnamefont {Burnell}},\ and\
  \bibinfo {author} {\bibfnamefont {T.}~\bibnamefont {Neupert}},\ }\bibfield
  {title} {\bibinfo {title} {Higher-order symmetry-protected topological states
  for interacting bosons and fermions},\ }\href
  {https://doi.org/10.1103/PhysRevB.98.235102} {\bibfield  {journal} {\bibinfo
  {journal} {Phys. Rev. B}\ }\textbf {\bibinfo {volume} {98}},\ \bibinfo
  {pages} {235102} (\bibinfo {year} {2018})}\BibitemShut {NoStop}%
\bibitem [{\citenamefont {Dubinkin}\ and\ \citenamefont
  {Hughes}(2019)}]{Dubinkin_2019}%
  \BibitemOpen
  \bibfield  {author} {\bibinfo {author} {\bibfnamefont {O.}~\bibnamefont
  {Dubinkin}}\ and\ \bibinfo {author} {\bibfnamefont {T.~L.}\ \bibnamefont
  {Hughes}},\ }\bibfield  {title} {\bibinfo {title} {Higher-order bosonic
  topological phases in spin models},\ }\href
  {https://doi.org/10.1103/PhysRevB.99.235132} {\bibfield  {journal} {\bibinfo
  {journal} {Phys. Rev. B}\ }\textbf {\bibinfo {volume} {99}},\ \bibinfo
  {pages} {235132} (\bibinfo {year} {2019})}\BibitemShut {NoStop}%
\bibitem [{\citenamefont {Kudo}\ \emph
  {et~al.}(2019{\natexlab{a}})\citenamefont {Kudo}, \citenamefont {Yoshida},\
  and\ \citenamefont {Hatsugai}}]{Kudo_2019}%
  \BibitemOpen
  \bibfield  {author} {\bibinfo {author} {\bibfnamefont {K.}~\bibnamefont
  {Kudo}}, \bibinfo {author} {\bibfnamefont {T.}~\bibnamefont {Yoshida}},\ and\
  \bibinfo {author} {\bibfnamefont {Y.}~\bibnamefont {Hatsugai}},\ }\bibfield
  {title} {\bibinfo {title} {Higher-order topological mott insulators},\ }\href
  {https://doi.org/10.1103/PhysRevLett.123.196402} {\bibfield  {journal}
  {\bibinfo  {journal} {Phys. Rev. Lett.}\ }\textbf {\bibinfo {volume} {123}},\
  \bibinfo {pages} {196402} (\bibinfo {year} {2019}{\natexlab{a}})}\BibitemShut
  {NoStop}%
\bibitem [{\citenamefont {Laubscher}\ \emph {et~al.}(2019)\citenamefont
  {Laubscher}, \citenamefont {Loss},\ and\ \citenamefont
  {Klinovaja}}]{Laubscher_2019}%
  \BibitemOpen
  \bibfield  {author} {\bibinfo {author} {\bibfnamefont {K.}~\bibnamefont
  {Laubscher}}, \bibinfo {author} {\bibfnamefont {D.}~\bibnamefont {Loss}},\
  and\ \bibinfo {author} {\bibfnamefont {J.}~\bibnamefont {Klinovaja}},\
  }\bibfield  {title} {\bibinfo {title} {Fractional topological
  superconductivity and parafermion corner states},\ }\href
  {https://doi.org/10.1103/PhysRevResearch.1.032017} {\bibfield  {journal}
  {\bibinfo  {journal} {Phys. Rev. Research}\ }\textbf {\bibinfo {volume}
  {1}},\ \bibinfo {pages} {032017} (\bibinfo {year} {2019})}\BibitemShut
  {NoStop}%
\bibitem [{\citenamefont {Laubscher}\ \emph {et~al.}(2020)\citenamefont
  {Laubscher}, \citenamefont {Loss},\ and\ \citenamefont
  {Klinovaja}}]{Laubscher_2020}%
  \BibitemOpen
  \bibfield  {author} {\bibinfo {author} {\bibfnamefont {K.}~\bibnamefont
  {Laubscher}}, \bibinfo {author} {\bibfnamefont {D.}~\bibnamefont {Loss}},\
  and\ \bibinfo {author} {\bibfnamefont {J.}~\bibnamefont {Klinovaja}},\
  }\bibfield  {title} {\bibinfo {title} {Majorana and parafermion corner states
  from two coupled sheets of bilayer graphene},\ }\href
  {https://doi.org/10.1103/PhysRevResearch.2.013330} {\bibfield  {journal}
  {\bibinfo  {journal} {Phys. Rev. Research}\ }\textbf {\bibinfo {volume}
  {2}},\ \bibinfo {pages} {013330} (\bibinfo {year} {2020})}\BibitemShut
  {NoStop}%
\bibitem [{\citenamefont {Sil}\ and\ \citenamefont {Ghosh}(2020)}]{Sil_2020}%
  \BibitemOpen
  \bibfield  {author} {\bibinfo {author} {\bibfnamefont {A.}~\bibnamefont
  {Sil}}\ and\ \bibinfo {author} {\bibfnamefont {A.~K.}\ \bibnamefont
  {Ghosh}},\ }\bibfield  {title} {\bibinfo {title} {First and second order
  topological phases on ferromagnetic breathing kagome lattice},\ }\href
  {https://doi.org/10.1088/1361-648x/ab6f8b} {\bibfield  {journal} {\bibinfo
  {journal} {Journal of Physics: Condensed Matter}\ }\textbf {\bibinfo {volume}
  {32}},\ \bibinfo {pages} {205601} (\bibinfo {year} {2020})}\BibitemShut
  {NoStop}%
\bibitem [{\citenamefont {Rasmussen}\ and\ \citenamefont
  {Lu}(2020)}]{Rasmussen_2020}%
  \BibitemOpen
  \bibfield  {author} {\bibinfo {author} {\bibfnamefont {A.}~\bibnamefont
  {Rasmussen}}\ and\ \bibinfo {author} {\bibfnamefont {Y.-M.}\ \bibnamefont
  {Lu}},\ }\bibfield  {title} {\bibinfo {title} {Classification and
  construction of higher-order symmetry-protected topological phases of
  interacting bosons},\ }\href {https://doi.org/10.1103/PhysRevB.101.085137}
  {\bibfield  {journal} {\bibinfo  {journal} {Phys. Rev. B}\ }\textbf {\bibinfo
  {volume} {101}},\ \bibinfo {pages} {085137} (\bibinfo {year}
  {2020})}\BibitemShut {NoStop}%
\bibitem [{\citenamefont {Bibo}\ \emph {et~al.}(2020)\citenamefont {Bibo},
  \citenamefont {Lovas}, \citenamefont {You}, \citenamefont {Grusdt},\ and\
  \citenamefont {Pollmann}}]{bibo2020}%
  \BibitemOpen
  \bibfield  {author} {\bibinfo {author} {\bibfnamefont {J.}~\bibnamefont
  {Bibo}}, \bibinfo {author} {\bibfnamefont {I.}~\bibnamefont {Lovas}},
  \bibinfo {author} {\bibfnamefont {Y.}~\bibnamefont {You}}, \bibinfo {author}
  {\bibfnamefont {F.}~\bibnamefont {Grusdt}},\ and\ \bibinfo {author}
  {\bibfnamefont {F.}~\bibnamefont {Pollmann}},\ }\bibfield  {title} {\bibinfo
  {title} {Fractional corner charges in a two-dimensional superlattice
  bose-hubbard model},\ }\href {https://doi.org/10.1103/PhysRevB.102.041126}
  {\bibfield  {journal} {\bibinfo  {journal} {Phys. Rev. B}\ }\textbf {\bibinfo
  {volume} {102}},\ \bibinfo {pages} {041126} (\bibinfo {year}
  {2020})}\BibitemShut {NoStop}%
\bibitem [{\citenamefont {Peng}\ \emph {et~al.}(2021)\citenamefont {Peng},
  \citenamefont {Zhang},\ and\ \citenamefont {Lu}}]{Peng_2021}%
  \BibitemOpen
  \bibfield  {author} {\bibinfo {author} {\bibfnamefont {C.}~\bibnamefont
  {Peng}}, \bibinfo {author} {\bibfnamefont {L.}~\bibnamefont {Zhang}},\ and\
  \bibinfo {author} {\bibfnamefont {Z.-Y.}\ \bibnamefont {Lu}},\ }\bibfield
  {title} {\bibinfo {title} {Deconfined quantum phase transition of a
  higher-order symmetry-protected topological state},\ }\href
  {https://doi.org/10.1103/PhysRevB.104.075112} {\bibfield  {journal} {\bibinfo
   {journal} {Phys. Rev. B}\ }\textbf {\bibinfo {volume} {104}},\ \bibinfo
  {pages} {075112} (\bibinfo {year} {2021})}\BibitemShut {NoStop}%
\bibitem [{\citenamefont {Guo}\ \emph {et~al.}(2021)\citenamefont {Guo},
  \citenamefont {Sun}, \citenamefont {Zhu}, \citenamefont {Li}, \citenamefont
  {Guo},\ and\ \citenamefont {Feng}}]{Guo_2022}%
  \BibitemOpen
  \bibfield  {author} {\bibinfo {author} {\bibfnamefont {J.}~\bibnamefont
  {Guo}}, \bibinfo {author} {\bibfnamefont {J.}~\bibnamefont {Sun}}, \bibinfo
  {author} {\bibfnamefont {X.}~\bibnamefont {Zhu}}, \bibinfo {author}
  {\bibfnamefont {C.-A.}\ \bibnamefont {Li}}, \bibinfo {author} {\bibfnamefont
  {H.}~\bibnamefont {Guo}},\ and\ \bibinfo {author} {\bibfnamefont
  {S.}~\bibnamefont {Feng}},\ }\bibfield  {title} {\bibinfo {title} {Quantum
  monte carlo study of topological phases on a spin analogue of
  benalcazar--bernevig--hughes model},\ }\href
  {https://doi.org/10.1088/1361-648X/ac30b4} {\bibfield  {journal} {\bibinfo
  {journal} {Journal of Physics: Condensed Matter}\ }\textbf {\bibinfo {volume}
  {34}},\ \bibinfo {pages} {035603} (\bibinfo {year} {2021})}\BibitemShut
  {NoStop}%
\bibitem [{\citenamefont {Hackenbroich}\ \emph {et~al.}(2021)\citenamefont
  {Hackenbroich}, \citenamefont {Hudomal}, \citenamefont {Schuch},
  \citenamefont {Bernevig},\ and\ \citenamefont
  {Regnault}}]{Hackenbroich_2021}%
  \BibitemOpen
  \bibfield  {author} {\bibinfo {author} {\bibfnamefont {A.}~\bibnamefont
  {Hackenbroich}}, \bibinfo {author} {\bibfnamefont {A.}~\bibnamefont
  {Hudomal}}, \bibinfo {author} {\bibfnamefont {N.}~\bibnamefont {Schuch}},
  \bibinfo {author} {\bibfnamefont {B.~A.}\ \bibnamefont {Bernevig}},\ and\
  \bibinfo {author} {\bibfnamefont {N.}~\bibnamefont {Regnault}},\ }\bibfield
  {title} {\bibinfo {title} {Fractional chiral hinge insulator},\ }\href
  {https://doi.org/10.1103/PhysRevB.103.L161110} {\bibfield  {journal}
  {\bibinfo  {journal} {Phys. Rev. B}\ }\textbf {\bibinfo {volume} {103}},\
  \bibinfo {pages} {L161110} (\bibinfo {year} {2021})}\BibitemShut {NoStop}%
\bibitem [{\citenamefont {Otsuka}\ \emph {et~al.}(2021)\citenamefont {Otsuka},
  \citenamefont {Yoshida}, \citenamefont {Kudo}, \citenamefont {Yunoki},\ and\
  \citenamefont {Hatsugai}}]{Otsuka_2021}%
  \BibitemOpen
  \bibfield  {author} {\bibinfo {author} {\bibfnamefont {Y.}~\bibnamefont
  {Otsuka}}, \bibinfo {author} {\bibfnamefont {T.}~\bibnamefont {Yoshida}},
  \bibinfo {author} {\bibfnamefont {K.}~\bibnamefont {Kudo}}, \bibinfo {author}
  {\bibfnamefont {S.}~\bibnamefont {Yunoki}},\ and\ \bibinfo {author}
  {\bibfnamefont {Y.}~\bibnamefont {Hatsugai}},\ }\bibfield  {title} {\bibinfo
  {title} {Higher-order topological mott insulator on the pyrochlore lattice},\
  }\href {https://doi.org/10.1038/s41598-021-99213-z} {\bibfield  {journal}
  {\bibinfo  {journal} {Scientific Reports}\ }\textbf {\bibinfo {volume}
  {11}},\ \bibinfo {pages} {20270} (\bibinfo {year} {2021})}\BibitemShut
  {NoStop}%
\bibitem [{\citenamefont {Gonz\'alez-Cuadra}(2022)}]{gonzalez2022}%
  \BibitemOpen
  \bibfield  {author} {\bibinfo {author} {\bibfnamefont {D.}~\bibnamefont
  {Gonz\'alez-Cuadra}},\ }\bibfield  {title} {\bibinfo {title} {Higher-order
  topological quantum paramagnets},\ }\href
  {https://doi.org/10.1103/PhysRevB.105.L020403} {\bibfield  {journal}
  {\bibinfo  {journal} {Phys. Rev. B}\ }\textbf {\bibinfo {volume} {105}},\
  \bibinfo {pages} {L020403} (\bibinfo {year} {2022})}\BibitemShut {NoStop}%
\bibitem [{\citenamefont {Li}\ \emph {et~al.}(2022)\citenamefont {Li},
  \citenamefont {Wu}, \citenamefont {Luo}, \citenamefont {Huang},\ and\
  \citenamefont {Chang}}]{Li_2022}%
  \BibitemOpen
  \bibfield  {author} {\bibinfo {author} {\bibfnamefont {Y.-M.}\ \bibnamefont
  {Li}}, \bibinfo {author} {\bibfnamefont {Y.-J.}\ \bibnamefont {Wu}}, \bibinfo
  {author} {\bibfnamefont {X.-W.}\ \bibnamefont {Luo}}, \bibinfo {author}
  {\bibfnamefont {Y.}~\bibnamefont {Huang}},\ and\ \bibinfo {author}
  {\bibfnamefont {K.}~\bibnamefont {Chang}},\ }\bibfield  {title} {\bibinfo
  {title} {Higher-order topological phases of magnons protected by magnetic
  crystalline symmetries},\ }\href
  {https://doi.org/10.1103/PhysRevB.106.054403} {\bibfield  {journal} {\bibinfo
   {journal} {Phys. Rev. B}\ }\textbf {\bibinfo {volume} {106}},\ \bibinfo
  {pages} {054403} (\bibinfo {year} {2022})}\BibitemShut {NoStop}%
\bibitem [{\citenamefont {Montorsi}\ \emph {et~al.}(2022)\citenamefont
  {Montorsi}, \citenamefont {Bhattacharya}, \citenamefont {Gonz\'alez-Cuadra},
  \citenamefont {Lewenstein}, \citenamefont {Palumbo},\ and\ \citenamefont
  {Barbiero}}]{Montorsi_2022}%
  \BibitemOpen
  \bibfield  {author} {\bibinfo {author} {\bibfnamefont {A.}~\bibnamefont
  {Montorsi}}, \bibinfo {author} {\bibfnamefont {U.}~\bibnamefont
  {Bhattacharya}}, \bibinfo {author} {\bibfnamefont {D.}~\bibnamefont
  {Gonz\'alez-Cuadra}}, \bibinfo {author} {\bibfnamefont {M.}~\bibnamefont
  {Lewenstein}}, \bibinfo {author} {\bibfnamefont {G.}~\bibnamefont
  {Palumbo}},\ and\ \bibinfo {author} {\bibfnamefont {L.}~\bibnamefont
  {Barbiero}},\ }\bibfield  {title} {\bibinfo {title} {Interacting second-order
  topological insulators in one-dimensional fermions with correlated hopping},\
  }\href {https://doi.org/10.1103/PhysRevB.106.L241115} {\bibfield  {journal}
  {\bibinfo  {journal} {Phys. Rev. B}\ }\textbf {\bibinfo {volume} {106}},\
  \bibinfo {pages} {L241115} (\bibinfo {year} {2022})}\BibitemShut {NoStop}%
\bibitem [{\citenamefont {Wienand}\ \emph {et~al.}(2022)\citenamefont
  {Wienand}, \citenamefont {Horn}, \citenamefont {Aidelsburger}, \citenamefont
  {Bibo},\ and\ \citenamefont {Grusdt}}]{Wienand_2022}%
  \BibitemOpen
  \bibfield  {author} {\bibinfo {author} {\bibfnamefont {J.~F.}\ \bibnamefont
  {Wienand}}, \bibinfo {author} {\bibfnamefont {F.}~\bibnamefont {Horn}},
  \bibinfo {author} {\bibfnamefont {M.}~\bibnamefont {Aidelsburger}}, \bibinfo
  {author} {\bibfnamefont {J.}~\bibnamefont {Bibo}},\ and\ \bibinfo {author}
  {\bibfnamefont {F.}~\bibnamefont {Grusdt}},\ }\bibfield  {title} {\bibinfo
  {title} {Thouless pumps and bulk-boundary correspondence in higher-order
  symmetry-protected topological phases},\ }\href
  {https://doi.org/10.1103/PhysRevLett.128.246602} {\bibfield  {journal}
  {\bibinfo  {journal} {Phys. Rev. Lett.}\ }\textbf {\bibinfo {volume} {128}},\
  \bibinfo {pages} {246602} (\bibinfo {year} {2022})}\BibitemShut {NoStop}%
\bibitem [{\citenamefont {Aksenov}\ \emph {et~al.}(2023)\citenamefont
  {Aksenov}, \citenamefont {Fedoseev}, \citenamefont {Shustin},\ and\
  \citenamefont {Zlotnikov}}]{Aksenov_2023}%
  \BibitemOpen
  \bibfield  {author} {\bibinfo {author} {\bibfnamefont {S.~V.}\ \bibnamefont
  {Aksenov}}, \bibinfo {author} {\bibfnamefont {A.~D.}\ \bibnamefont
  {Fedoseev}}, \bibinfo {author} {\bibfnamefont {M.~S.}\ \bibnamefont
  {Shustin}},\ and\ \bibinfo {author} {\bibfnamefont {A.~O.}\ \bibnamefont
  {Zlotnikov}},\ }\bibfield  {title} {\bibinfo {title} {Effect of local coulomb
  interaction on majorana corner modes: Weak and strong correlation limits},\
  }\href {https://doi.org/10.1103/PhysRevB.107.125401} {\bibfield  {journal}
  {\bibinfo  {journal} {Phys. Rev. B}\ }\textbf {\bibinfo {volume} {107}},\
  \bibinfo {pages} {125401} (\bibinfo {year} {2023})}\BibitemShut {NoStop}%
\bibitem [{\citenamefont {Fraxanet}\ \emph {et~al.}(2023)\citenamefont
  {Fraxanet}, \citenamefont {Dauphin}, \citenamefont {Lewenstein},
  \citenamefont {Barbiero},\ and\ \citenamefont
  {González-Cuadra}}]{Fraxanet_2023}%
  \BibitemOpen
  \bibfield  {author} {\bibinfo {author} {\bibfnamefont {J.}~\bibnamefont
  {Fraxanet}}, \bibinfo {author} {\bibfnamefont {A.}~\bibnamefont {Dauphin}},
  \bibinfo {author} {\bibfnamefont {M.}~\bibnamefont {Lewenstein}}, \bibinfo
  {author} {\bibfnamefont {L.}~\bibnamefont {Barbiero}},\ and\ \bibinfo
  {author} {\bibfnamefont {D.}~\bibnamefont {González-Cuadra}},\ }\href@noop
  {} {\bibinfo {title} {Higher-order topological peierls insulator in a
  two-dimensional atom-cavity system}} (\bibinfo {year} {2023}),\ \Eprint
  {https://arxiv.org/abs/2305.03409} {arXiv:2305.03409 [cond-mat.quant-gas]}
  \BibitemShut {NoStop}%
\bibitem [{\citenamefont {Hohenadler}\ and\ \citenamefont
  {Assaad}(2013)}]{Hohenadler_2013}%
  \BibitemOpen
  \bibfield  {author} {\bibinfo {author} {\bibfnamefont {M.}~\bibnamefont
  {Hohenadler}}\ and\ \bibinfo {author} {\bibfnamefont {F.~F.}\ \bibnamefont
  {Assaad}},\ }\bibfield  {title} {\bibinfo {title} {Correlation effects in
  two-dimensional topological insulators},\ }\href
  {https://doi.org/10.1088/0953-8984/25/14/143201} {\bibfield  {journal}
  {\bibinfo  {journal} {Journal of Physics: Condensed Matter}\ }\textbf
  {\bibinfo {volume} {25}},\ \bibinfo {pages} {143201} (\bibinfo {year}
  {2013})}\BibitemShut {NoStop}%
\bibitem [{\citenamefont {Bergholtz}\ and\ \citenamefont
  {Liu}(2013)}]{doi:10.1142/S021797921330017X}%
  \BibitemOpen
  \bibfield  {author} {\bibinfo {author} {\bibfnamefont {E.~J.}\ \bibnamefont
  {Bergholtz}}\ and\ \bibinfo {author} {\bibfnamefont {Z.}~\bibnamefont
  {Liu}},\ }\bibfield  {title} {\bibinfo {title} {Topological flat band models
  and fractional chern insulators},\ }\href
  {https://doi.org/10.1142/S021797921330017X} {\bibfield  {journal} {\bibinfo
  {journal} {International Journal of Modern Physics B}\ }\textbf {\bibinfo
  {volume} {27}},\ \bibinfo {pages} {1330017} (\bibinfo {year}
  {2013})}\BibitemShut {NoStop}%
\bibitem [{\citenamefont {Neupert}\ \emph {et~al.}(2015)\citenamefont
  {Neupert}, \citenamefont {Chamon}, \citenamefont {Iadecola}, \citenamefont
  {Santos},\ and\ \citenamefont {Mudry}}]{Neupert_2015}%
  \BibitemOpen
  \bibfield  {author} {\bibinfo {author} {\bibfnamefont {T.}~\bibnamefont
  {Neupert}}, \bibinfo {author} {\bibfnamefont {C.}~\bibnamefont {Chamon}},
  \bibinfo {author} {\bibfnamefont {T.}~\bibnamefont {Iadecola}}, \bibinfo
  {author} {\bibfnamefont {L.~H.}\ \bibnamefont {Santos}},\ and\ \bibinfo
  {author} {\bibfnamefont {C.}~\bibnamefont {Mudry}},\ }\bibfield  {title}
  {\bibinfo {title} {Fractional (chern and topological) insulators},\ }\href
  {https://doi.org/10.1088/0031-8949/2015/t164/014005} {\bibfield  {journal}
  {\bibinfo  {journal} {Physica Scripta}\ }\textbf {\bibinfo {volume} {T164}},\
  \bibinfo {pages} {014005} (\bibinfo {year} {2015})}\BibitemShut {NoStop}%
\bibitem [{\citenamefont {Rachel}(2018)}]{Rachel_2018}%
  \BibitemOpen
  \bibfield  {author} {\bibinfo {author} {\bibfnamefont {S.}~\bibnamefont
  {Rachel}},\ }\bibfield  {title} {\bibinfo {title} {Interacting topological
  insulators: a review},\ }\href {https://doi.org/10.1088/1361-6633/aad6a6}
  {\bibfield  {journal} {\bibinfo  {journal} {Reports on Progress in Physics}\
  }\textbf {\bibinfo {volume} {81}},\ \bibinfo {pages} {116501} (\bibinfo
  {year} {2018})}\BibitemShut {NoStop}%
\bibitem [{\citenamefont {Negele}\ and\ \citenamefont
  {Orland}(2019)}]{negele_orland_2019}%
  \BibitemOpen
  \bibfield  {author} {\bibinfo {author} {\bibfnamefont {J.~W.}\ \bibnamefont
  {Negele}}\ and\ \bibinfo {author} {\bibfnamefont {H.}~\bibnamefont
  {Orland}},\ }\href@noop {} {\emph {\bibinfo {title} {Quantum many-particle
  systems}}}\ (\bibinfo  {publisher} {CRC Press Taylor and Francis Group},\
  \bibinfo {year} {2019})\BibitemShut {NoStop}%
\bibitem [{\citenamefont {Ramond}(2001)}]{0201304503}%
  \BibitemOpen
  \bibfield  {author} {\bibinfo {author} {\bibfnamefont {P.}~\bibnamefont
  {Ramond}},\ }\href
  {http://www.amazon.com/Field-Theory-Modern-Frontiers-Physics/dp/0201304503%3FSubscriptionId%3D13CT5CVB80YFWJEPWS02%26tag%3Dws%26linkCode%3Dxm2%26camp%3D2025%26creative%3D165953%26creativeASIN%3D0201304503}
  {\emph {\bibinfo {title} {Field Theory : A Modern Primer (Frontiers in
  Physics Series, Vol 74)}}},\ \bibinfo {edition} {2nd}\ ed.\ (\bibinfo
  {publisher} {Westview Press},\ \bibinfo {year} {2001})\BibitemShut {NoStop}%
\bibitem [{\citenamefont {Proeyen}(2016)}]{vanproeyen2016tools}%
  \BibitemOpen
  \bibfield  {author} {\bibinfo {author} {\bibfnamefont {A.~V.}\ \bibnamefont
  {Proeyen}},\ }\href@noop {} {\bibinfo {title} {Tools for supersymmetry}}
  (\bibinfo {year} {2016}),\ \Eprint {https://arxiv.org/abs/hep-th/9910030}
  {arXiv:hep-th/9910030 [hep-th]} \BibitemShut {NoStop}%
\bibitem [{\citenamefont {Freedman}\ and\ \citenamefont
  {Van~Proeyen}(2012)}]{Freedman:2012zz}%
  \BibitemOpen
  \bibfield  {author} {\bibinfo {author} {\bibfnamefont {D.~Z.}\ \bibnamefont
  {Freedman}}\ and\ \bibinfo {author} {\bibfnamefont {A.}~\bibnamefont
  {Van~Proeyen}},\ }\href@noop {} {\emph {\bibinfo {title} {{Supergravity}}}}\
  (\bibinfo  {publisher} {Cambridge Univ. Press},\ \bibinfo {address}
  {Cambridge, UK},\ \bibinfo {year} {2012})\BibitemShut {NoStop}%
\bibitem [{\citenamefont {Bailin}\ and\ \citenamefont
  {Love}(1987)}]{D_Bailin_1987}%
  \BibitemOpen
  \bibfield  {author} {\bibinfo {author} {\bibfnamefont {D.}~\bibnamefont
  {Bailin}}\ and\ \bibinfo {author} {\bibfnamefont {A.}~\bibnamefont {Love}},\
  }\bibfield  {title} {\bibinfo {title} {Kaluza-klein theories},\ }\href
  {https://doi.org/10.1088/0034-4885/50/9/001} {\bibfield  {journal} {\bibinfo
  {journal} {Reports on Progress in Physics}\ }\textbf {\bibinfo {volume}
  {50}},\ \bibinfo {pages} {1087} (\bibinfo {year} {1987})}\BibitemShut
  {NoStop}%
\bibitem [{\citenamefont {Ryu}\ \emph {et~al.}(2010)\citenamefont {Ryu},
  \citenamefont {Schnyder}, \citenamefont {Furusaki},\ and\ \citenamefont
  {Ludwig}}]{Ryu_2010}%
  \BibitemOpen
  \bibfield  {author} {\bibinfo {author} {\bibfnamefont {S.}~\bibnamefont
  {Ryu}}, \bibinfo {author} {\bibfnamefont {A.~P.}\ \bibnamefont {Schnyder}},
  \bibinfo {author} {\bibfnamefont {A.}~\bibnamefont {Furusaki}},\ and\
  \bibinfo {author} {\bibfnamefont {A.~W.~W.}\ \bibnamefont {Ludwig}},\
  }\bibfield  {title} {\bibinfo {title} {Topological insulators and
  superconductors: tenfold way and dimensional hierarchy},\ }\href
  {https://doi.org/10.1088/1367-2630/12/6/065010} {\bibfield  {journal}
  {\bibinfo  {journal} {New Journal of Physics}\ }\textbf {\bibinfo {volume}
  {12}},\ \bibinfo {pages} {065010} (\bibinfo {year} {2010})}\BibitemShut
  {NoStop}%
\bibitem [{\citenamefont {Thirring}(1958)}]{THIRRING195891}%
  \BibitemOpen
  \bibfield  {author} {\bibinfo {author} {\bibfnamefont {W.~E.}\ \bibnamefont
  {Thirring}},\ }\bibfield  {title} {\bibinfo {title} {A soluble relativistic
  field theory},\ }\href
  {https://doi.org/https://doi.org/10.1016/0003-4916(58)90015-0} {\bibfield
  {journal} {\bibinfo  {journal} {Annals of Physics}\ }\textbf {\bibinfo
  {volume} {3}},\ \bibinfo {pages} {91 } (\bibinfo {year} {1958})}\BibitemShut
  {NoStop}%
\bibitem [{\citenamefont {Coleman}(1975)}]{PhysRevD.11.2088}%
  \BibitemOpen
  \bibfield  {author} {\bibinfo {author} {\bibfnamefont {S.}~\bibnamefont
  {Coleman}},\ }\bibfield  {title} {\bibinfo {title} {Quantum sine-gordon
  equation as the massive thirring model},\ }\href
  {https://doi.org/10.1103/PhysRevD.11.2088} {\bibfield  {journal} {\bibinfo
  {journal} {Phys. Rev. D}\ }\textbf {\bibinfo {volume} {11}},\ \bibinfo
  {pages} {2088} (\bibinfo {year} {1975})}\BibitemShut {NoStop}%
\bibitem [{\citenamefont {Hands}(1997)}]{hep-lat/9706018}%
  \BibitemOpen
  \bibfield  {author} {\bibinfo {author} {\bibfnamefont {S.}~\bibnamefont
  {Hands}},\ }\href@noop {} {\bibinfo {title} {Fixed point four-fermi
  theories}} (\bibinfo {year} {1997}),\ \Eprint
  {https://arxiv.org/abs/arXiv:hep-lat/9706018} {arXiv:hep-lat/9706018}
  \BibitemShut {NoStop}%
\bibitem [{\citenamefont {Wilson}(1977)}]{Wilson1977}%
  \BibitemOpen
  \bibfield  {author} {\bibinfo {author} {\bibfnamefont {K.~G.}\ \bibnamefont
  {Wilson}},\ }\bibfield  {title} {\bibinfo {title} {Quarks and strings on a
  lattice},\ }in\ \href {https://doi.org/10.1007/978-1-4613-4208-3_6} {\emph
  {\bibinfo {booktitle} {New Phenomena in Subnuclear Physics}}}\ (\bibinfo
  {publisher} {Springer {US}},\ \bibinfo {year} {1977})\ pp.\ \bibinfo {pages}
  {69--142}\BibitemShut {NoStop}%
\bibitem [{\citenamefont {Hasan}\ and\ \citenamefont
  {Kane}(2010)}]{RevModPhys.82.3045}%
  \BibitemOpen
  \bibfield  {author} {\bibinfo {author} {\bibfnamefont {M.~Z.}\ \bibnamefont
  {Hasan}}\ and\ \bibinfo {author} {\bibfnamefont {C.~L.}\ \bibnamefont
  {Kane}},\ }\bibfield  {title} {\bibinfo {title} {Colloquium: Topological
  insulators},\ }\href {https://doi.org/10.1103/RevModPhys.82.3045} {\bibfield
  {journal} {\bibinfo  {journal} {Rev. Mod. Phys.}\ }\textbf {\bibinfo {volume}
  {82}},\ \bibinfo {pages} {3045} (\bibinfo {year} {2010})}\BibitemShut
  {NoStop}%
\bibitem [{\citenamefont {Qi}\ and\ \citenamefont
  {Zhang}(2011)}]{RevModPhys.83.1057}%
  \BibitemOpen
  \bibfield  {author} {\bibinfo {author} {\bibfnamefont {X.-L.}\ \bibnamefont
  {Qi}}\ and\ \bibinfo {author} {\bibfnamefont {S.-C.}\ \bibnamefont {Zhang}},\
  }\bibfield  {title} {\bibinfo {title} {Topological insulators and
  superconductors},\ }\href {https://doi.org/10.1103/RevModPhys.83.1057}
  {\bibfield  {journal} {\bibinfo  {journal} {Rev. Mod. Phys.}\ }\textbf
  {\bibinfo {volume} {83}},\ \bibinfo {pages} {1057} (\bibinfo {year}
  {2011})}\BibitemShut {NoStop}%
\bibitem [{\citenamefont {Nielsen}\ and\ \citenamefont
  {Ninomiya}(1981{\natexlab{a}})}]{NIELSEN198120}%
  \BibitemOpen
  \bibfield  {author} {\bibinfo {author} {\bibfnamefont {H.}~\bibnamefont
  {Nielsen}}\ and\ \bibinfo {author} {\bibfnamefont {M.}~\bibnamefont
  {Ninomiya}},\ }\bibfield  {title} {\bibinfo {title} {Absence of neutrinos on
  a lattice: (i). proof by homotopy theory},\ }\href
  {https://doi.org/https://doi.org/10.1016/0550-3213(81)90361-8} {\bibfield
  {journal} {\bibinfo  {journal} {Nuclear Physics B}\ }\textbf {\bibinfo
  {volume} {185}},\ \bibinfo {pages} {20 } (\bibinfo {year}
  {1981}{\natexlab{a}})}\BibitemShut {NoStop}%
\bibitem [{\citenamefont {Nielsen}\ and\ \citenamefont
  {Ninomiya}(1981{\natexlab{b}})}]{NIELSEN1981173}%
  \BibitemOpen
  \bibfield  {author} {\bibinfo {author} {\bibfnamefont {H.}~\bibnamefont
  {Nielsen}}\ and\ \bibinfo {author} {\bibfnamefont {M.}~\bibnamefont
  {Ninomiya}},\ }\bibfield  {title} {\bibinfo {title} {Absence of neutrinos on
  a lattice: (ii). intuitive topological proof},\ }\href
  {https://doi.org/https://doi.org/10.1016/0550-3213(81)90524-1} {\bibfield
  {journal} {\bibinfo  {journal} {Nuclear Physics B}\ }\textbf {\bibinfo
  {volume} {193}},\ \bibinfo {pages} {173 } (\bibinfo {year}
  {1981}{\natexlab{b}})}\BibitemShut {NoStop}%
\bibitem [{\citenamefont {Gattringer}\ and\ \citenamefont
  {Lang}(2010)}]{gattringer_lang_2010}%
  \BibitemOpen
  \bibfield  {author} {\bibinfo {author} {\bibfnamefont {C.}~\bibnamefont
  {Gattringer}}\ and\ \bibinfo {author} {\bibfnamefont {C.~B.}\ \bibnamefont
  {Lang}},\ }\href@noop {} {\emph {\bibinfo {title} {Quantum chromodynamics on
  the lattice: an introductory presentation}}}\ (\bibinfo  {publisher}
  {Springer},\ \bibinfo {year} {2010})\BibitemShut {NoStop}%
\bibitem [{\citenamefont {Haldane}(1988)}]{PhysRevLett.61.2015}%
  \BibitemOpen
  \bibfield  {author} {\bibinfo {author} {\bibfnamefont {F.~D.~M.}\
  \bibnamefont {Haldane}},\ }\bibfield  {title} {\bibinfo {title} {Model for a
  quantum hall effect without landau levels: Condensed-matter realization of
  the "parity anomaly"},\ }\href {https://doi.org/10.1103/PhysRevLett.61.2015}
  {\bibfield  {journal} {\bibinfo  {journal} {Phys. Rev. Lett.}\ }\textbf
  {\bibinfo {volume} {61}},\ \bibinfo {pages} {2015} (\bibinfo {year}
  {1988})}\BibitemShut {NoStop}%
\bibitem [{\citenamefont {Qi}\ \emph {et~al.}(2006)\citenamefont {Qi},
  \citenamefont {Wu},\ and\ \citenamefont {Zhang}}]{PhysRevB.74.085308}%
  \BibitemOpen
  \bibfield  {author} {\bibinfo {author} {\bibfnamefont {X.-L.}\ \bibnamefont
  {Qi}}, \bibinfo {author} {\bibfnamefont {Y.-S.}\ \bibnamefont {Wu}},\ and\
  \bibinfo {author} {\bibfnamefont {S.-C.}\ \bibnamefont {Zhang}},\ }\bibfield
  {title} {\bibinfo {title} {Topological quantization of the spin hall effect
  in two-dimensional paramagnetic semiconductors},\ }\href
  {https://doi.org/10.1103/PhysRevB.74.085308} {\bibfield  {journal} {\bibinfo
  {journal} {Phys. Rev. B}\ }\textbf {\bibinfo {volume} {74}},\ \bibinfo
  {pages} {085308} (\bibinfo {year} {2006})}\BibitemShut {NoStop}%
\bibitem [{\citenamefont {Qi}\ \emph {et~al.}(2008)\citenamefont {Qi},
  \citenamefont {Hughes},\ and\ \citenamefont {Zhang}}]{PhysRevB.78.195424}%
  \BibitemOpen
  \bibfield  {author} {\bibinfo {author} {\bibfnamefont {X.-L.}\ \bibnamefont
  {Qi}}, \bibinfo {author} {\bibfnamefont {T.~L.}\ \bibnamefont {Hughes}},\
  and\ \bibinfo {author} {\bibfnamefont {S.-C.}\ \bibnamefont {Zhang}},\
  }\bibfield  {title} {\bibinfo {title} {Topological field theory of
  time-reversal invariant insulators},\ }\href
  {https://doi.org/10.1103/PhysRevB.78.195424} {\bibfield  {journal} {\bibinfo
  {journal} {Phys. Rev. B}\ }\textbf {\bibinfo {volume} {78}},\ \bibinfo
  {pages} {195424} (\bibinfo {year} {2008})}\BibitemShut {NoStop}%
\bibitem [{\citenamefont {Ziegler}\ \emph
  {et~al.}(2022{\natexlab{a}})\citenamefont {Ziegler}, \citenamefont {Tirrito},
  \citenamefont {Lewenstein}, \citenamefont {Hands},\ and\ \citenamefont
  {Bermudez}}]{PhysRevResearch.4.L042012}%
  \BibitemOpen
  \bibfield  {author} {\bibinfo {author} {\bibfnamefont {L.}~\bibnamefont
  {Ziegler}}, \bibinfo {author} {\bibfnamefont {E.}~\bibnamefont {Tirrito}},
  \bibinfo {author} {\bibfnamefont {M.}~\bibnamefont {Lewenstein}}, \bibinfo
  {author} {\bibfnamefont {S.}~\bibnamefont {Hands}},\ and\ \bibinfo {author}
  {\bibfnamefont {A.}~\bibnamefont {Bermudez}},\ }\bibfield  {title} {\bibinfo
  {title} {Correlated chern insulators in two-dimensional raman lattices: A
  cold-atom regularization of strongly coupled four-fermi field theories},\
  }\href {https://doi.org/10.1103/PhysRevResearch.4.L042012} {\bibfield
  {journal} {\bibinfo  {journal} {Phys. Rev. Res.}\ }\textbf {\bibinfo {volume}
  {4}},\ \bibinfo {pages} {L042012} (\bibinfo {year}
  {2022}{\natexlab{a}})}\BibitemShut {NoStop}%
\bibitem [{\citenamefont {Ziegler}\ \emph
  {et~al.}(2022{\natexlab{b}})\citenamefont {Ziegler}, \citenamefont {Tirrito},
  \citenamefont {Lewenstein}, \citenamefont {Hands},\ and\ \citenamefont
  {Bermudez}}]{ZIEGLER2022168763}%
  \BibitemOpen
  \bibfield  {author} {\bibinfo {author} {\bibfnamefont {L.}~\bibnamefont
  {Ziegler}}, \bibinfo {author} {\bibfnamefont {E.}~\bibnamefont {Tirrito}},
  \bibinfo {author} {\bibfnamefont {M.}~\bibnamefont {Lewenstein}}, \bibinfo
  {author} {\bibfnamefont {S.}~\bibnamefont {Hands}},\ and\ \bibinfo {author}
  {\bibfnamefont {A.}~\bibnamefont {Bermudez}},\ }\bibfield  {title} {\bibinfo
  {title} {Large-n chern insulators: Lattice field theory and quantum
  simulation approaches to correlation effects in the quantum anomalous hall
  effect},\ }\href {https://doi.org/https://doi.org/10.1016/j.aop.2022.168763}
  {\bibfield  {journal} {\bibinfo  {journal} {Annals of Physics}\ }\textbf
  {\bibinfo {volume} {439}},\ \bibinfo {pages} {168763} (\bibinfo {year}
  {2022}{\natexlab{b}})}\BibitemShut {NoStop}%
\bibitem [{\citenamefont {Frezzotti}\ \emph
  {et~al.}(2001{\natexlab{a}})\citenamefont {Frezzotti}, \citenamefont
  {Grassi}, \citenamefont {Sint},\ and\ \citenamefont
  {Weisz}}]{ALPHA_Collaboration_2001}%
  \BibitemOpen
  \bibfield  {author} {\bibinfo {author} {\bibfnamefont {R.}~\bibnamefont
  {Frezzotti}}, \bibinfo {author} {\bibfnamefont {P.~A.}\ \bibnamefont
  {Grassi}}, \bibinfo {author} {\bibfnamefont {S.}~\bibnamefont {Sint}},\ and\
  \bibinfo {author} {\bibfnamefont {P.}~\bibnamefont {Weisz}},\ }\bibfield
  {title} {\bibinfo {title} {Lattice qcd with a chirally twisted mass term},\
  }\href {https://doi.org/10.1088/1126-6708/2001/08/058} {\bibfield  {journal}
  {\bibinfo  {journal} {Journal of High Energy Physics}\ }\textbf {\bibinfo
  {volume} {2001}},\ \bibinfo {pages} {058} (\bibinfo {year}
  {2001}{\natexlab{a}})}\BibitemShut {NoStop}%
\bibitem [{\citenamefont {Frezzotti}\ \emph
  {et~al.}(2001{\natexlab{b}})\citenamefont {Frezzotti}, \citenamefont {Sint},\
  and\ \citenamefont {Weisz}}]{Roberto_Frezzotti_2001}%
  \BibitemOpen
  \bibfield  {author} {\bibinfo {author} {\bibfnamefont {R.}~\bibnamefont
  {Frezzotti}}, \bibinfo {author} {\bibfnamefont {S.}~\bibnamefont {Sint}},\
  and\ \bibinfo {author} {\bibfnamefont {P.}~\bibnamefont {Weisz}},\ }\bibfield
   {title} {\bibinfo {title} {O(a) improved twisted mass lattice qcd},\ }\href
  {https://doi.org/10.1088/1126-6708/2001/07/048} {\bibfield  {journal}
  {\bibinfo  {journal} {Journal of High Energy Physics}\ }\textbf {\bibinfo
  {volume} {2001}},\ \bibinfo {pages} {048} (\bibinfo {year}
  {2001}{\natexlab{b}})}\BibitemShut {NoStop}%
\bibitem [{\citenamefont {Pena}\ \emph {et~al.}(2004)\citenamefont {Pena},
  \citenamefont {Sint},\ and\ \citenamefont {Vladikas}}]{Carlos_Pena_2004}%
  \BibitemOpen
  \bibfield  {author} {\bibinfo {author} {\bibfnamefont {C.}~\bibnamefont
  {Pena}}, \bibinfo {author} {\bibfnamefont {S.}~\bibnamefont {Sint}},\ and\
  \bibinfo {author} {\bibfnamefont {A.}~\bibnamefont {Vladikas}},\ }\bibfield
  {title} {\bibinfo {title} {Twisted mass qcd and lattice approaches to the i =
  1/2 rule},\ }\href {https://doi.org/10.1088/1126-6708/2004/09/069} {\bibfield
   {journal} {\bibinfo  {journal} {Journal of High Energy Physics}\ }\textbf
  {\bibinfo {volume} {2004}},\ \bibinfo {pages} {069} (\bibinfo {year}
  {2004})}\BibitemShut {NoStop}%
\bibitem [{\citenamefont {Frezzotti}\ and\ \citenamefont
  {Rossi}(2004)}]{Frezzotti_2004}%
  \BibitemOpen
  \bibfield  {author} {\bibinfo {author} {\bibfnamefont {R.}~\bibnamefont
  {Frezzotti}}\ and\ \bibinfo {author} {\bibfnamefont {G.~C.}\ \bibnamefont
  {Rossi}},\ }\bibfield  {title} {\bibinfo {title} {Chirally improving wilson
  fermions 1. o(a) improvement},\ }\href
  {https://doi.org/10.1088/1126-6708/2004/08/007} {\bibfield  {journal}
  {\bibinfo  {journal} {Journal of High Energy Physics}\ }\textbf {\bibinfo
  {volume} {2004}},\ \bibinfo {pages} {007} (\bibinfo {year}
  {2004})}\BibitemShut {NoStop}%
\bibitem [{\citenamefont {Alexandrou}\ \emph {et~al.}(2008)\citenamefont
  {Alexandrou}, \citenamefont {Baron}, \citenamefont {Blossier}, \citenamefont
  {Brinet}, \citenamefont {Carbonell}, \citenamefont {Dimopoulos},
  \citenamefont {Drach}, \citenamefont {Farchioni}, \citenamefont {Frezzotti},
  \citenamefont {Guichon}, \citenamefont {Herdoiza}, \citenamefont {Jansen},
  \citenamefont {Korzec}, \citenamefont {Koutsou}, \citenamefont {Liu},
  \citenamefont {Michael}, \citenamefont {P\`ene}, \citenamefont {Shindler},
  \citenamefont {Urbach},\ and\ \citenamefont {Wenger}}]{PhysRevD.78.014509}%
  \BibitemOpen
  \bibfield  {author} {\bibinfo {author} {\bibfnamefont {C.}~\bibnamefont
  {Alexandrou}}, \bibinfo {author} {\bibfnamefont {R.}~\bibnamefont {Baron}},
  \bibinfo {author} {\bibfnamefont {B.}~\bibnamefont {Blossier}}, \bibinfo
  {author} {\bibfnamefont {M.}~\bibnamefont {Brinet}}, \bibinfo {author}
  {\bibfnamefont {J.}~\bibnamefont {Carbonell}}, \bibinfo {author}
  {\bibfnamefont {P.}~\bibnamefont {Dimopoulos}}, \bibinfo {author}
  {\bibfnamefont {V.}~\bibnamefont {Drach}}, \bibinfo {author} {\bibfnamefont
  {F.}~\bibnamefont {Farchioni}}, \bibinfo {author} {\bibfnamefont
  {R.}~\bibnamefont {Frezzotti}}, \bibinfo {author} {\bibfnamefont
  {P.}~\bibnamefont {Guichon}}, \bibinfo {author} {\bibfnamefont
  {G.}~\bibnamefont {Herdoiza}}, \bibinfo {author} {\bibfnamefont
  {K.}~\bibnamefont {Jansen}}, \bibinfo {author} {\bibfnamefont
  {T.}~\bibnamefont {Korzec}}, \bibinfo {author} {\bibfnamefont
  {G.}~\bibnamefont {Koutsou}}, \bibinfo {author} {\bibfnamefont
  {Z.}~\bibnamefont {Liu}}, \bibinfo {author} {\bibfnamefont {C.}~\bibnamefont
  {Michael}}, \bibinfo {author} {\bibfnamefont {O.}~\bibnamefont {P\`ene}},
  \bibinfo {author} {\bibfnamefont {A.}~\bibnamefont {Shindler}}, \bibinfo
  {author} {\bibfnamefont {C.}~\bibnamefont {Urbach}},\ and\ \bibinfo {author}
  {\bibfnamefont {U.}~\bibnamefont {Wenger}},\ }\bibfield  {title} {\bibinfo
  {title} {Light baryon masses with dynamical twisted mass fermions},\ }\href
  {https://doi.org/10.1103/PhysRevD.78.014509} {\bibfield  {journal} {\bibinfo
  {journal} {Phys. Rev. D}\ }\textbf {\bibinfo {volume} {78}},\ \bibinfo
  {pages} {014509} (\bibinfo {year} {2008})}\BibitemShut {NoStop}%
\bibitem [{\citenamefont {Abdel-Rehim}\ \emph {et~al.}(2017)\citenamefont
  {Abdel-Rehim}, \citenamefont {Alexandrou}, \citenamefont {Burger},
  \citenamefont {Constantinou}, \citenamefont {Dimopoulos}, \citenamefont
  {Frezzotti}, \citenamefont {Hadjiyiannakou}, \citenamefont {Helmes},
  \citenamefont {Jansen}, \citenamefont {Jost}, \citenamefont {Kallidonis},
  \citenamefont {Knippschild}, \citenamefont {Kostrzewa}, \citenamefont
  {Koutsou}, \citenamefont {Liu}, \citenamefont {Mangin-Brinet}, \citenamefont
  {Ottnad}, \citenamefont {Petschlies}, \citenamefont {Pientka}, \citenamefont
  {Rossi}, \citenamefont {Urbach}, \citenamefont {Wenger},\ and\ \citenamefont
  {Werner}}]{PhysRevD.95.094515}%
  \BibitemOpen
  \bibfield  {author} {\bibinfo {author} {\bibfnamefont {A.}~\bibnamefont
  {Abdel-Rehim}}, \bibinfo {author} {\bibfnamefont {C.}~\bibnamefont
  {Alexandrou}}, \bibinfo {author} {\bibfnamefont {F.}~\bibnamefont {Burger}},
  \bibinfo {author} {\bibfnamefont {M.}~\bibnamefont {Constantinou}}, \bibinfo
  {author} {\bibfnamefont {P.}~\bibnamefont {Dimopoulos}}, \bibinfo {author}
  {\bibfnamefont {R.}~\bibnamefont {Frezzotti}}, \bibinfo {author}
  {\bibfnamefont {K.}~\bibnamefont {Hadjiyiannakou}}, \bibinfo {author}
  {\bibfnamefont {C.}~\bibnamefont {Helmes}}, \bibinfo {author} {\bibfnamefont
  {K.}~\bibnamefont {Jansen}}, \bibinfo {author} {\bibfnamefont
  {C.}~\bibnamefont {Jost}}, \bibinfo {author} {\bibfnamefont {C.}~\bibnamefont
  {Kallidonis}}, \bibinfo {author} {\bibfnamefont {B.}~\bibnamefont
  {Knippschild}}, \bibinfo {author} {\bibfnamefont {B.}~\bibnamefont
  {Kostrzewa}}, \bibinfo {author} {\bibfnamefont {G.}~\bibnamefont {Koutsou}},
  \bibinfo {author} {\bibfnamefont {L.}~\bibnamefont {Liu}}, \bibinfo {author}
  {\bibfnamefont {M.}~\bibnamefont {Mangin-Brinet}}, \bibinfo {author}
  {\bibfnamefont {K.}~\bibnamefont {Ottnad}}, \bibinfo {author} {\bibfnamefont
  {M.}~\bibnamefont {Petschlies}}, \bibinfo {author} {\bibfnamefont
  {G.}~\bibnamefont {Pientka}}, \bibinfo {author} {\bibfnamefont {G.~C.}\
  \bibnamefont {Rossi}}, \bibinfo {author} {\bibfnamefont {C.}~\bibnamefont
  {Urbach}}, \bibinfo {author} {\bibfnamefont {U.}~\bibnamefont {Wenger}},\
  and\ \bibinfo {author} {\bibfnamefont {M.}~\bibnamefont {Werner}},\
  }\bibfield  {title} {\bibinfo {title} {First physics results at the physical
  pion mass from ${N}_{f}=2$ wilson twisted mass fermions at maximal twist},\
  }\href {https://doi.org/10.1103/PhysRevD.95.094515} {\bibfield  {journal}
  {\bibinfo  {journal} {Phys. Rev. D}\ }\textbf {\bibinfo {volume} {95}},\
  \bibinfo {pages} {094515} (\bibinfo {year} {2017})}\BibitemShut {NoStop}%
\bibitem [{\citenamefont {Alexandrou}\ \emph {et~al.}(2021)\citenamefont
  {Alexandrou}, \citenamefont {Bacchio}, \citenamefont {Bergner}, \citenamefont
  {Finkenrath}, \citenamefont {Gasbarro}, \citenamefont {Hadjiyiannakou},
  \citenamefont {Jansen}, \citenamefont {Kostrzewa}, \citenamefont {Ottnad},
  \citenamefont {Petschlies}, \citenamefont {Pittler}, \citenamefont
  {Steffens}, \citenamefont {Urbach},\ and\ \citenamefont
  {Wenger}}]{PhysRevLett.127.252001}%
  \BibitemOpen
  \bibfield  {author} {\bibinfo {author} {\bibfnamefont {C.}~\bibnamefont
  {Alexandrou}}, \bibinfo {author} {\bibfnamefont {S.}~\bibnamefont {Bacchio}},
  \bibinfo {author} {\bibfnamefont {G.}~\bibnamefont {Bergner}}, \bibinfo
  {author} {\bibfnamefont {J.}~\bibnamefont {Finkenrath}}, \bibinfo {author}
  {\bibfnamefont {A.}~\bibnamefont {Gasbarro}}, \bibinfo {author}
  {\bibfnamefont {K.}~\bibnamefont {Hadjiyiannakou}}, \bibinfo {author}
  {\bibfnamefont {K.}~\bibnamefont {Jansen}}, \bibinfo {author} {\bibfnamefont
  {B.}~\bibnamefont {Kostrzewa}}, \bibinfo {author} {\bibfnamefont
  {K.}~\bibnamefont {Ottnad}}, \bibinfo {author} {\bibfnamefont
  {M.}~\bibnamefont {Petschlies}}, \bibinfo {author} {\bibfnamefont
  {F.}~\bibnamefont {Pittler}}, \bibinfo {author} {\bibfnamefont
  {F.}~\bibnamefont {Steffens}}, \bibinfo {author} {\bibfnamefont
  {C.}~\bibnamefont {Urbach}},\ and\ \bibinfo {author} {\bibfnamefont
  {U.}~\bibnamefont {Wenger}} (\bibinfo {collaboration} {Extended Twisted Mass
  Collaboration}),\ }\bibfield  {title} {\bibinfo {title} {Quark and gluon
  momentum fractions in the pion from ${N}_{f}=2+1+1$ lattice qcd},\ }\href
  {https://doi.org/10.1103/PhysRevLett.127.252001} {\bibfield  {journal}
  {\bibinfo  {journal} {Phys. Rev. Lett.}\ }\textbf {\bibinfo {volume} {127}},\
  \bibinfo {pages} {252001} (\bibinfo {year} {2021})}\BibitemShut {NoStop}%
\bibitem [{\citenamefont {Alexandrou}\ \emph {et~al.}(2023)\citenamefont
  {Alexandrou}, \citenamefont {Bacchio}, \citenamefont {De~Santis},
  \citenamefont {Dimopoulos}, \citenamefont {Finkenrath}, \citenamefont
  {Frezzotti}, \citenamefont {Gagliardi}, \citenamefont {Garofalo},
  \citenamefont {Hadjiyiannakou}, \citenamefont {Kostrzewa}, \citenamefont
  {Jansen}, \citenamefont {Lubicz}, \citenamefont {Petschlies}, \citenamefont
  {Sanfilippo}, \citenamefont {Simula}, \citenamefont {Tantalo}, \citenamefont
  {Urbach},\ and\ \citenamefont {Wenger}}]{PhysRevLett.130.241901}%
  \BibitemOpen
  \bibfield  {author} {\bibinfo {author} {\bibfnamefont {C.}~\bibnamefont
  {Alexandrou}}, \bibinfo {author} {\bibfnamefont {S.}~\bibnamefont {Bacchio}},
  \bibinfo {author} {\bibfnamefont {A.}~\bibnamefont {De~Santis}}, \bibinfo
  {author} {\bibfnamefont {P.}~\bibnamefont {Dimopoulos}}, \bibinfo {author}
  {\bibfnamefont {J.}~\bibnamefont {Finkenrath}}, \bibinfo {author}
  {\bibfnamefont {R.}~\bibnamefont {Frezzotti}}, \bibinfo {author}
  {\bibfnamefont {G.}~\bibnamefont {Gagliardi}}, \bibinfo {author}
  {\bibfnamefont {M.}~\bibnamefont {Garofalo}}, \bibinfo {author}
  {\bibfnamefont {K.}~\bibnamefont {Hadjiyiannakou}}, \bibinfo {author}
  {\bibfnamefont {B.}~\bibnamefont {Kostrzewa}}, \bibinfo {author}
  {\bibfnamefont {K.}~\bibnamefont {Jansen}}, \bibinfo {author} {\bibfnamefont
  {V.}~\bibnamefont {Lubicz}}, \bibinfo {author} {\bibfnamefont
  {M.}~\bibnamefont {Petschlies}}, \bibinfo {author} {\bibfnamefont
  {F.}~\bibnamefont {Sanfilippo}}, \bibinfo {author} {\bibfnamefont
  {S.}~\bibnamefont {Simula}}, \bibinfo {author} {\bibfnamefont
  {N.}~\bibnamefont {Tantalo}}, \bibinfo {author} {\bibfnamefont
  {C.}~\bibnamefont {Urbach}},\ and\ \bibinfo {author} {\bibfnamefont
  {U.}~\bibnamefont {Wenger}} (\bibinfo {collaboration} {Extended Twisted Mass
  Collaboration (ETMC)}),\ }\bibfield  {title} {\bibinfo {title} {Probing the
  energy-smeared $r$ ratio using lattice qcd},\ }\href
  {https://doi.org/10.1103/PhysRevLett.130.241901} {\bibfield  {journal}
  {\bibinfo  {journal} {Phys. Rev. Lett.}\ }\textbf {\bibinfo {volume} {130}},\
  \bibinfo {pages} {241901} (\bibinfo {year} {2023})}\BibitemShut {NoStop}%
\bibitem [{\citenamefont {Saez}\ \emph {et~al.}(2023)\citenamefont {Saez},
  \citenamefont {Conigli}, \citenamefont {Frison}, \citenamefont {Herdoiza},
  \citenamefont {Pena},\ and\ \citenamefont {Ugarrio}}]{Saez:2022ptd}%
  \BibitemOpen
  \bibfield  {author} {\bibinfo {author} {\bibfnamefont {A.}~\bibnamefont
  {Saez}}, \bibinfo {author} {\bibfnamefont {A.}~\bibnamefont {Conigli}},
  \bibinfo {author} {\bibfnamefont {J.}~\bibnamefont {Frison}}, \bibinfo
  {author} {\bibfnamefont {G.}~\bibnamefont {Herdoiza}}, \bibinfo {author}
  {\bibfnamefont {C.}~\bibnamefont {Pena}},\ and\ \bibinfo {author}
  {\bibfnamefont {J.}~\bibnamefont {Ugarrio}},\ }\bibfield  {title} {\bibinfo
  {title} {{Scale Setting from a Mixed Action with Twisted Mass Valence
  Quarks}},\ }\href {https://doi.org/10.22323/1.430.0357} {\bibfield  {journal}
  {\bibinfo  {journal} {PoS}\ }\textbf {\bibinfo {volume} {LATTICE2022}},\
  \bibinfo {pages} {357} (\bibinfo {year} {2023})}\BibitemShut {NoStop}%
\bibitem [{\citenamefont {Shindler}(2008)}]{SHINDLER200837}%
  \BibitemOpen
  \bibfield  {author} {\bibinfo {author} {\bibfnamefont {A.}~\bibnamefont
  {Shindler}},\ }\bibfield  {title} {\bibinfo {title} {Twisted mass lattice
  qcd},\ }\href {https://doi.org/https://doi.org/10.1016/j.physrep.2008.03.001}
  {\bibfield  {journal} {\bibinfo  {journal} {Physics Reports}\ }\textbf
  {\bibinfo {volume} {461}},\ \bibinfo {pages} {37} (\bibinfo {year}
  {2008})}\BibitemShut {NoStop}%
\bibitem [{\citenamefont {Peccei}\ and\ \citenamefont
  {Quinn}(1977)}]{PhysRevLett.38.1440}%
  \BibitemOpen
  \bibfield  {author} {\bibinfo {author} {\bibfnamefont {R.~D.}\ \bibnamefont
  {Peccei}}\ and\ \bibinfo {author} {\bibfnamefont {H.~R.}\ \bibnamefont
  {Quinn}},\ }\bibfield  {title} {\bibinfo {title} {$\mathrm{CP}$ conservation
  in the presence of pseudoparticles},\ }\href
  {https://doi.org/10.1103/PhysRevLett.38.1440} {\bibfield  {journal} {\bibinfo
   {journal} {Phys. Rev. Lett.}\ }\textbf {\bibinfo {volume} {38}},\ \bibinfo
  {pages} {1440} (\bibinfo {year} {1977})}\BibitemShut {NoStop}%
\bibitem [{\citenamefont {Wilczek}(1978)}]{PhysRevLett.40.279}%
  \BibitemOpen
  \bibfield  {author} {\bibinfo {author} {\bibfnamefont {F.}~\bibnamefont
  {Wilczek}},\ }\bibfield  {title} {\bibinfo {title} {Problem of strong $p$ and
  $t$ invariance in the presence of instantons},\ }\href
  {https://doi.org/10.1103/PhysRevLett.40.279} {\bibfield  {journal} {\bibinfo
  {journal} {Phys. Rev. Lett.}\ }\textbf {\bibinfo {volume} {40}},\ \bibinfo
  {pages} {279} (\bibinfo {year} {1978})}\BibitemShut {NoStop}%
\bibitem [{\citenamefont {Weinberg}(1978)}]{PhysRevLett.40.223}%
  \BibitemOpen
  \bibfield  {author} {\bibinfo {author} {\bibfnamefont {S.}~\bibnamefont
  {Weinberg}},\ }\bibfield  {title} {\bibinfo {title} {A new light boson?},\
  }\href {https://doi.org/10.1103/PhysRevLett.40.223} {\bibfield  {journal}
  {\bibinfo  {journal} {Phys. Rev. Lett.}\ }\textbf {\bibinfo {volume} {40}},\
  \bibinfo {pages} {223} (\bibinfo {year} {1978})}\BibitemShut {NoStop}%
\bibitem [{\citenamefont {Bermudez}\ \emph {et~al.}(2010)\citenamefont
  {Bermudez}, \citenamefont {Mazza}, \citenamefont {Rizzi}, \citenamefont
  {Goldman}, \citenamefont {Lewenstein},\ and\ \citenamefont
  {Martin-Delgado}}]{PhysRevLett.105.190404}%
  \BibitemOpen
  \bibfield  {author} {\bibinfo {author} {\bibfnamefont {A.}~\bibnamefont
  {Bermudez}}, \bibinfo {author} {\bibfnamefont {L.}~\bibnamefont {Mazza}},
  \bibinfo {author} {\bibfnamefont {M.}~\bibnamefont {Rizzi}}, \bibinfo
  {author} {\bibfnamefont {N.}~\bibnamefont {Goldman}}, \bibinfo {author}
  {\bibfnamefont {M.}~\bibnamefont {Lewenstein}},\ and\ \bibinfo {author}
  {\bibfnamefont {M.~A.}\ \bibnamefont {Martin-Delgado}},\ }\bibfield  {title}
  {\bibinfo {title} {Wilson fermions and axion electrodynamics in optical
  lattices},\ }\href {https://doi.org/10.1103/PhysRevLett.105.190404}
  {\bibfield  {journal} {\bibinfo  {journal} {Phys. Rev. Lett.}\ }\textbf
  {\bibinfo {volume} {105}},\ \bibinfo {pages} {190404} (\bibinfo {year}
  {2010})}\BibitemShut {NoStop}%
\bibitem [{\citenamefont {Wilczek}(1987)}]{PhysRevLett.58.1799}%
  \BibitemOpen
  \bibfield  {author} {\bibinfo {author} {\bibfnamefont {F.}~\bibnamefont
  {Wilczek}},\ }\bibfield  {title} {\bibinfo {title} {Two applications of axion
  electrodynamics},\ }\href {https://doi.org/10.1103/PhysRevLett.58.1799}
  {\bibfield  {journal} {\bibinfo  {journal} {Phys. Rev. Lett.}\ }\textbf
  {\bibinfo {volume} {58}},\ \bibinfo {pages} {1799} (\bibinfo {year}
  {1987})}\BibitemShut {NoStop}%
\bibitem [{\citenamefont {Kaplan}(1992)}]{KAPLAN1992342}%
  \BibitemOpen
  \bibfield  {author} {\bibinfo {author} {\bibfnamefont {D.~B.}\ \bibnamefont
  {Kaplan}},\ }\bibfield  {title} {\bibinfo {title} {A method for simulating
  chiral fermions on the lattice},\ }\href
  {https://doi.org/https://doi.org/10.1016/0370-2693(92)91112-M} {\bibfield
  {journal} {\bibinfo  {journal} {Physics Letters B}\ }\textbf {\bibinfo
  {volume} {288}},\ \bibinfo {pages} {342 } (\bibinfo {year}
  {1992})}\BibitemShut {NoStop}%
\bibitem [{\citenamefont {Golterman}\ \emph {et~al.}(1993)\citenamefont
  {Golterman}, \citenamefont {Jansen},\ and\ \citenamefont
  {Kaplan}}]{GOLTERMAN1993219}%
  \BibitemOpen
  \bibfield  {author} {\bibinfo {author} {\bibfnamefont {M.~F.}\ \bibnamefont
  {Golterman}}, \bibinfo {author} {\bibfnamefont {K.}~\bibnamefont {Jansen}},\
  and\ \bibinfo {author} {\bibfnamefont {D.~B.}\ \bibnamefont {Kaplan}},\
  }\bibfield  {title} {\bibinfo {title} {Chern-simons currents and chiral
  fermions on the lattice},\ }\href
  {https://doi.org/https://doi.org/10.1016/0370-2693(93)90692-B} {\bibfield
  {journal} {\bibinfo  {journal} {Physics Letters B}\ }\textbf {\bibinfo
  {volume} {301}},\ \bibinfo {pages} {219 } (\bibinfo {year}
  {1993})}\BibitemShut {NoStop}%
\bibitem [{\citenamefont {Kaplan}\ and\ \citenamefont
  {Sun}(2012)}]{PhysRevLett.108.181807}%
  \BibitemOpen
  \bibfield  {author} {\bibinfo {author} {\bibfnamefont {D.~B.}\ \bibnamefont
  {Kaplan}}\ and\ \bibinfo {author} {\bibfnamefont {S.}~\bibnamefont {Sun}},\
  }\bibfield  {title} {\bibinfo {title} {Spacetime as a topological insulator:
  Mechanism for the origin of the fermion generations},\ }\href
  {https://doi.org/10.1103/PhysRevLett.108.181807} {\bibfield  {journal}
  {\bibinfo  {journal} {Phys. Rev. Lett.}\ }\textbf {\bibinfo {volume} {108}},\
  \bibinfo {pages} {181807} (\bibinfo {year} {2012})}\BibitemShut {NoStop}%
\bibitem [{\citenamefont {Symanzik}(1983)}]{SYMANZIK1983187}%
  \BibitemOpen
  \bibfield  {author} {\bibinfo {author} {\bibfnamefont {K.}~\bibnamefont
  {Symanzik}},\ }\bibfield  {title} {\bibinfo {title} {Continuum limit and
  improved action in lattice theories: (i). principles and theory},\ }\href
  {https://doi.org/https://doi.org/10.1016/0550-3213(83)90468-6} {\bibfield
  {journal} {\bibinfo  {journal} {Nuclear Physics B}\ }\textbf {\bibinfo
  {volume} {226}},\ \bibinfo {pages} {187} (\bibinfo {year}
  {1983})}\BibitemShut {NoStop}%
\bibitem [{\citenamefont {Lüscher}\ and\ \citenamefont
  {Weisz}(1985)}]{LUSCHER1985250}%
  \BibitemOpen
  \bibfield  {author} {\bibinfo {author} {\bibfnamefont {M.}~\bibnamefont
  {Lüscher}}\ and\ \bibinfo {author} {\bibfnamefont {P.}~\bibnamefont
  {Weisz}},\ }\bibfield  {title} {\bibinfo {title} {Computation of the action
  for on-shell improved lattice gauge theories at weak coupling},\ }\href
  {https://doi.org/https://doi.org/10.1016/0370-2693(85)90966-9} {\bibfield
  {journal} {\bibinfo  {journal} {Physics Letters B}\ }\textbf {\bibinfo
  {volume} {158}},\ \bibinfo {pages} {250} (\bibinfo {year}
  {1985})}\BibitemShut {NoStop}%
\bibitem [{\citenamefont {Jansen}\ \emph {et~al.}(1996)\citenamefont {Jansen},
  \citenamefont {Liu}, \citenamefont {Lüscher}, \citenamefont {Simma},
  \citenamefont {Sint}, \citenamefont {Sommer}, \citenamefont {Weisz},\ and\
  \citenamefont {Wolff}}]{JANSEN1996275}%
  \BibitemOpen
  \bibfield  {author} {\bibinfo {author} {\bibfnamefont {K.}~\bibnamefont
  {Jansen}}, \bibinfo {author} {\bibfnamefont {C.}~\bibnamefont {Liu}},
  \bibinfo {author} {\bibfnamefont {M.}~\bibnamefont {Lüscher}}, \bibinfo
  {author} {\bibfnamefont {H.}~\bibnamefont {Simma}}, \bibinfo {author}
  {\bibfnamefont {S.}~\bibnamefont {Sint}}, \bibinfo {author} {\bibfnamefont
  {R.}~\bibnamefont {Sommer}}, \bibinfo {author} {\bibfnamefont
  {P.}~\bibnamefont {Weisz}},\ and\ \bibinfo {author} {\bibfnamefont
  {U.}~\bibnamefont {Wolff}},\ }\bibfield  {title} {\bibinfo {title}
  {Non-perturbative renormalization of lattice qcd at all scales},\ }\href
  {https://doi.org/https://doi.org/10.1016/0370-2693(96)00075-5} {\bibfield
  {journal} {\bibinfo  {journal} {Physics Letters B}\ }\textbf {\bibinfo
  {volume} {372}},\ \bibinfo {pages} {275} (\bibinfo {year}
  {1996})}\BibitemShut {NoStop}%
\bibitem [{\citenamefont {Jansen}\ and\ \citenamefont
  {Sommer}(1998)}]{JANSEN1998185}%
  \BibitemOpen
  \bibfield  {author} {\bibinfo {author} {\bibfnamefont {K.}~\bibnamefont
  {Jansen}}\ and\ \bibinfo {author} {\bibfnamefont {R.}~\bibnamefont
  {Sommer}},\ }\bibfield  {title} {\bibinfo {title} {O(a) improvement of
  lattice qcd with two flavors of wilson quarks},\ }\href
  {https://doi.org/https://doi.org/10.1016/S0550-3213(98)00396-4} {\bibfield
  {journal} {\bibinfo  {journal} {Nuclear Physics B}\ }\textbf {\bibinfo
  {volume} {530}},\ \bibinfo {pages} {185} (\bibinfo {year}
  {1998})}\BibitemShut {NoStop}%
\bibitem [{\citenamefont {{Della Morte}}\ \emph {et~al.}(2005)\citenamefont
  {{Della Morte}}, \citenamefont {Hoffmann}, \citenamefont {Knechtli},
  \citenamefont {Rolf}, \citenamefont {Sommer}, \citenamefont {Wetzorke},\ and\
  \citenamefont {Wolff}}]{DELLAMORTE2005117}%
  \BibitemOpen
  \bibfield  {author} {\bibinfo {author} {\bibfnamefont {M.}~\bibnamefont
  {{Della Morte}}}, \bibinfo {author} {\bibfnamefont {R.}~\bibnamefont
  {Hoffmann}}, \bibinfo {author} {\bibfnamefont {F.}~\bibnamefont {Knechtli}},
  \bibinfo {author} {\bibfnamefont {J.}~\bibnamefont {Rolf}}, \bibinfo {author}
  {\bibfnamefont {R.}~\bibnamefont {Sommer}}, \bibinfo {author} {\bibfnamefont
  {I.}~\bibnamefont {Wetzorke}},\ and\ \bibinfo {author} {\bibfnamefont
  {U.}~\bibnamefont {Wolff}},\ }\bibfield  {title} {\bibinfo {title}
  {Non-perturbative quark mass renormalization in two-flavor qcd},\ }\href
  {https://doi.org/https://doi.org/10.1016/j.nuclphysb.2005.09.028} {\bibfield
  {journal} {\bibinfo  {journal} {Nuclear Physics B}\ }\textbf {\bibinfo
  {volume} {729}},\ \bibinfo {pages} {117} (\bibinfo {year}
  {2005})}\BibitemShut {NoStop}%
\bibitem [{\citenamefont {Trifunovic}\ and\ \citenamefont
  {Brouwer}(2021)}]{https://doi.org/10.1002/pssb.202000090}%
  \BibitemOpen
  \bibfield  {author} {\bibinfo {author} {\bibfnamefont {L.}~\bibnamefont
  {Trifunovic}}\ and\ \bibinfo {author} {\bibfnamefont {P.~W.}\ \bibnamefont
  {Brouwer}},\ }\bibfield  {title} {\bibinfo {title} {Higher-order topological
  band structures},\ }\href
  {https://doi.org/https://doi.org/10.1002/pssb.202000090} {\bibfield
  {journal} {\bibinfo  {journal} {physica status solidi (b)}\ }\textbf
  {\bibinfo {volume} {258}},\ \bibinfo {pages} {2000090} (\bibinfo {year}
  {2021})}\BibitemShut {NoStop}%
\bibitem [{\citenamefont {Kogut}\ and\ \citenamefont
  {Susskind}(1975)}]{PhysRevD.11.395}%
  \BibitemOpen
  \bibfield  {author} {\bibinfo {author} {\bibfnamefont {J.}~\bibnamefont
  {Kogut}}\ and\ \bibinfo {author} {\bibfnamefont {L.}~\bibnamefont
  {Susskind}},\ }\bibfield  {title} {\bibinfo {title} {Hamiltonian formulation
  of wilson's lattice gauge theories},\ }\href
  {https://doi.org/10.1103/PhysRevD.11.395} {\bibfield  {journal} {\bibinfo
  {journal} {Phys. Rev. D}\ }\textbf {\bibinfo {volume} {11}},\ \bibinfo
  {pages} {395} (\bibinfo {year} {1975})}\BibitemShut {NoStop}%
\bibitem [{\citenamefont {Vidal}\ \emph {et~al.}(1998)\citenamefont {Vidal},
  \citenamefont {Mosseri},\ and\ \citenamefont {Dou\ifmmode~\mbox{\c{c}}\else
  \c{c}\fi{}ot}}]{PhysRevLett.81.5888}%
  \BibitemOpen
  \bibfield  {author} {\bibinfo {author} {\bibfnamefont {J.}~\bibnamefont
  {Vidal}}, \bibinfo {author} {\bibfnamefont {R.}~\bibnamefont {Mosseri}},\
  and\ \bibinfo {author} {\bibfnamefont {B.}~\bibnamefont
  {Dou\ifmmode~\mbox{\c{c}}\else \c{c}\fi{}ot}},\ }\bibfield  {title} {\bibinfo
  {title} {Aharonov-bohm cages in two-dimensional structures},\ }\href
  {https://doi.org/10.1103/PhysRevLett.81.5888} {\bibfield  {journal} {\bibinfo
   {journal} {Phys. Rev. Lett.}\ }\textbf {\bibinfo {volume} {81}},\ \bibinfo
  {pages} {5888} (\bibinfo {year} {1998})}\BibitemShut {NoStop}%
\bibitem [{\citenamefont {Creutz}(1999)}]{PhysRevLett.83.2636}%
  \BibitemOpen
  \bibfield  {author} {\bibinfo {author} {\bibfnamefont {M.}~\bibnamefont
  {Creutz}},\ }\bibfield  {title} {\bibinfo {title} {End states, ladder
  compounds, and domain-wall fermions},\ }\href
  {https://doi.org/10.1103/PhysRevLett.83.2636} {\bibfield  {journal} {\bibinfo
   {journal} {Phys. Rev. Lett.}\ }\textbf {\bibinfo {volume} {83}},\ \bibinfo
  {pages} {2636} (\bibinfo {year} {1999})}\BibitemShut {NoStop}%
\bibitem [{\citenamefont {Creutz}(2001)}]{RevModPhys.73.119}%
  \BibitemOpen
  \bibfield  {author} {\bibinfo {author} {\bibfnamefont {M.}~\bibnamefont
  {Creutz}},\ }\bibfield  {title} {\bibinfo {title} {Aspects of chiral symmetry
  and the lattice},\ }\href {https://doi.org/10.1103/RevModPhys.73.119}
  {\bibfield  {journal} {\bibinfo  {journal} {Rev. Mod. Phys.}\ }\textbf
  {\bibinfo {volume} {73}},\ \bibinfo {pages} {119} (\bibinfo {year}
  {2001})}\BibitemShut {NoStop}%
\bibitem [{\citenamefont {Bermudez}\ \emph {et~al.}(2009)\citenamefont
  {Bermudez}, \citenamefont {Patan\`e}, \citenamefont {Amico},\ and\
  \citenamefont {Martin-Delgado}}]{PhysRevLett.102.135702}%
  \BibitemOpen
  \bibfield  {author} {\bibinfo {author} {\bibfnamefont {A.}~\bibnamefont
  {Bermudez}}, \bibinfo {author} {\bibfnamefont {D.}~\bibnamefont {Patan\`e}},
  \bibinfo {author} {\bibfnamefont {L.}~\bibnamefont {Amico}},\ and\ \bibinfo
  {author} {\bibfnamefont {M.~A.}\ \bibnamefont {Martin-Delgado}},\ }\bibfield
  {title} {\bibinfo {title} {Topology-induced anomalous defect production by
  crossing a quantum critical point},\ }\href
  {https://doi.org/10.1103/PhysRevLett.102.135702} {\bibfield  {journal}
  {\bibinfo  {journal} {Phys. Rev. Lett.}\ }\textbf {\bibinfo {volume} {102}},\
  \bibinfo {pages} {135702} (\bibinfo {year} {2009})}\BibitemShut {NoStop}%
\bibitem [{\citenamefont {Viyuela}\ \emph {et~al.}(2012)\citenamefont
  {Viyuela}, \citenamefont {Rivas},\ and\ \citenamefont
  {Martin-Delgado}}]{PhysRevB.86.155140}%
  \BibitemOpen
  \bibfield  {author} {\bibinfo {author} {\bibfnamefont {O.}~\bibnamefont
  {Viyuela}}, \bibinfo {author} {\bibfnamefont {A.}~\bibnamefont {Rivas}},\
  and\ \bibinfo {author} {\bibfnamefont {M.~A.}\ \bibnamefont
  {Martin-Delgado}},\ }\bibfield  {title} {\bibinfo {title} {Thermal
  instability of protected end states in a one-dimensional topological
  insulator},\ }\href {https://doi.org/10.1103/PhysRevB.86.155140} {\bibfield
  {journal} {\bibinfo  {journal} {Phys. Rev. B}\ }\textbf {\bibinfo {volume}
  {86}},\ \bibinfo {pages} {155140} (\bibinfo {year} {2012})}\BibitemShut
  {NoStop}%
\bibitem [{\citenamefont {Viyuela}\ \emph {et~al.}(2014)\citenamefont
  {Viyuela}, \citenamefont {Rivas},\ and\ \citenamefont
  {Martin-Delgado}}]{PhysRevLett.112.130401}%
  \BibitemOpen
  \bibfield  {author} {\bibinfo {author} {\bibfnamefont {O.}~\bibnamefont
  {Viyuela}}, \bibinfo {author} {\bibfnamefont {A.}~\bibnamefont {Rivas}},\
  and\ \bibinfo {author} {\bibfnamefont {M.~A.}\ \bibnamefont
  {Martin-Delgado}},\ }\bibfield  {title} {\bibinfo {title} {Uhlmann phase as a
  topological measure for one-dimensional fermion systems},\ }\href
  {https://doi.org/10.1103/PhysRevLett.112.130401} {\bibfield  {journal}
  {\bibinfo  {journal} {Phys. Rev. Lett.}\ }\textbf {\bibinfo {volume} {112}},\
  \bibinfo {pages} {130401} (\bibinfo {year} {2014})}\BibitemShut {NoStop}%
\bibitem [{\citenamefont {Sun}\ and\ \citenamefont
  {Lim}(2017)}]{PhysRevB.96.035139}%
  \BibitemOpen
  \bibfield  {author} {\bibinfo {author} {\bibfnamefont {N.}~\bibnamefont
  {Sun}}\ and\ \bibinfo {author} {\bibfnamefont {L.-K.}\ \bibnamefont {Lim}},\
  }\bibfield  {title} {\bibinfo {title} {Quantum charge pumps with topological
  phases in a creutz ladder},\ }\href
  {https://doi.org/10.1103/PhysRevB.96.035139} {\bibfield  {journal} {\bibinfo
  {journal} {Phys. Rev. B}\ }\textbf {\bibinfo {volume} {96}},\ \bibinfo
  {pages} {035139} (\bibinfo {year} {2017})}\BibitemShut {NoStop}%
\bibitem [{\citenamefont {Jafari}\ \emph {et~al.}(2019)\citenamefont {Jafari},
  \citenamefont {Johannesson}, \citenamefont {Langari},\ and\ \citenamefont
  {Martin-Delgado}}]{PhysRevB.99.054302}%
  \BibitemOpen
  \bibfield  {author} {\bibinfo {author} {\bibfnamefont {R.}~\bibnamefont
  {Jafari}}, \bibinfo {author} {\bibfnamefont {H.}~\bibnamefont {Johannesson}},
  \bibinfo {author} {\bibfnamefont {A.}~\bibnamefont {Langari}},\ and\ \bibinfo
  {author} {\bibfnamefont {M.~A.}\ \bibnamefont {Martin-Delgado}},\ }\bibfield
  {title} {\bibinfo {title} {Quench dynamics and zero-energy modes: The case of
  the creutz model},\ }\href {https://doi.org/10.1103/PhysRevB.99.054302}
  {\bibfield  {journal} {\bibinfo  {journal} {Phys. Rev. B}\ }\textbf {\bibinfo
  {volume} {99}},\ \bibinfo {pages} {054302} (\bibinfo {year}
  {2019})}\BibitemShut {NoStop}%
\bibitem [{\citenamefont {J\"unemann}\ \emph {et~al.}(2017)\citenamefont
  {J\"unemann}, \citenamefont {Piga}, \citenamefont {Ran}, \citenamefont
  {Lewenstein}, \citenamefont {Rizzi},\ and\ \citenamefont
  {Bermudez}}]{Junemann_2017}%
  \BibitemOpen
  \bibfield  {author} {\bibinfo {author} {\bibfnamefont {J.}~\bibnamefont
  {J\"unemann}}, \bibinfo {author} {\bibfnamefont {A.}~\bibnamefont {Piga}},
  \bibinfo {author} {\bibfnamefont {S.-J.}\ \bibnamefont {Ran}}, \bibinfo
  {author} {\bibfnamefont {M.}~\bibnamefont {Lewenstein}}, \bibinfo {author}
  {\bibfnamefont {M.}~\bibnamefont {Rizzi}},\ and\ \bibinfo {author}
  {\bibfnamefont {A.}~\bibnamefont {Bermudez}},\ }\bibfield  {title} {\bibinfo
  {title} {Exploring interacting topological insulators with ultracold atoms:
  The synthetic creutz-hubbard model},\ }\href
  {https://doi.org/10.1103/PhysRevX.7.031057} {\bibfield  {journal} {\bibinfo
  {journal} {Phys. Rev. X}\ }\textbf {\bibinfo {volume} {7}},\ \bibinfo {pages}
  {031057} (\bibinfo {year} {2017})}\BibitemShut {NoStop}%
\bibitem [{\citenamefont {Bermudez}\ \emph {et~al.}(2018)\citenamefont
  {Bermudez}, \citenamefont {Tirrito}, \citenamefont {Rizzi}, \citenamefont
  {Lewenstein},\ and\ \citenamefont {Hands}}]{BERMUDEZ2018149}%
  \BibitemOpen
  \bibfield  {author} {\bibinfo {author} {\bibfnamefont {A.}~\bibnamefont
  {Bermudez}}, \bibinfo {author} {\bibfnamefont {E.}~\bibnamefont {Tirrito}},
  \bibinfo {author} {\bibfnamefont {M.}~\bibnamefont {Rizzi}}, \bibinfo
  {author} {\bibfnamefont {M.}~\bibnamefont {Lewenstein}},\ and\ \bibinfo
  {author} {\bibfnamefont {S.}~\bibnamefont {Hands}},\ }\bibfield  {title}
  {\bibinfo {title} {Gross–neveu–wilson model and correlated
  symmetry-protected topological phases},\ }\href
  {https://doi.org/https://doi.org/10.1016/j.aop.2018.10.007} {\bibfield
  {journal} {\bibinfo  {journal} {Annals of Physics}\ }\textbf {\bibinfo
  {volume} {399}},\ \bibinfo {pages} {149 } (\bibinfo {year}
  {2018})}\BibitemShut {NoStop}%
\bibitem [{\citenamefont {Tirrito}\ \emph {et~al.}(2019)\citenamefont
  {Tirrito}, \citenamefont {Rizzi}, \citenamefont {Sierra}, \citenamefont
  {Lewenstein},\ and\ \citenamefont {Bermudez}}]{PhysRevB.99.125106}%
  \BibitemOpen
  \bibfield  {author} {\bibinfo {author} {\bibfnamefont {E.}~\bibnamefont
  {Tirrito}}, \bibinfo {author} {\bibfnamefont {M.}~\bibnamefont {Rizzi}},
  \bibinfo {author} {\bibfnamefont {G.}~\bibnamefont {Sierra}}, \bibinfo
  {author} {\bibfnamefont {M.}~\bibnamefont {Lewenstein}},\ and\ \bibinfo
  {author} {\bibfnamefont {A.}~\bibnamefont {Bermudez}},\ }\bibfield  {title}
  {\bibinfo {title} {Renormalization group flows for wilson-hubbard matter and
  the topological hamiltonian},\ }\href
  {https://doi.org/10.1103/PhysRevB.99.125106} {\bibfield  {journal} {\bibinfo
  {journal} {Phys. Rev. B}\ }\textbf {\bibinfo {volume} {99}},\ \bibinfo
  {pages} {125106} (\bibinfo {year} {2019})}\BibitemShut {NoStop}%
\bibitem [{\citenamefont {Tirrito}\ \emph {et~al.}(2022)\citenamefont
  {Tirrito}, \citenamefont {Lewenstein},\ and\ \citenamefont
  {Bermudez}}]{PhysRevB.106.045147}%
  \BibitemOpen
  \bibfield  {author} {\bibinfo {author} {\bibfnamefont {E.}~\bibnamefont
  {Tirrito}}, \bibinfo {author} {\bibfnamefont {M.}~\bibnamefont
  {Lewenstein}},\ and\ \bibinfo {author} {\bibfnamefont {A.}~\bibnamefont
  {Bermudez}},\ }\bibfield  {title} {\bibinfo {title} {Topological chiral
  currents in the gross-neveu model extension},\ }\href
  {https://doi.org/10.1103/PhysRevB.106.045147} {\bibfield  {journal} {\bibinfo
   {journal} {Phys. Rev. B}\ }\textbf {\bibinfo {volume} {106}},\ \bibinfo
  {pages} {045147} (\bibinfo {year} {2022})}\BibitemShut {NoStop}%
\bibitem [{\citenamefont {Cirac}\ \emph {et~al.}(2010)\citenamefont {Cirac},
  \citenamefont {Maraner},\ and\ \citenamefont
  {Pachos}}]{PhysRevLett.105.190403}%
  \BibitemOpen
  \bibfield  {author} {\bibinfo {author} {\bibfnamefont {J.~I.}\ \bibnamefont
  {Cirac}}, \bibinfo {author} {\bibfnamefont {P.}~\bibnamefont {Maraner}},\
  and\ \bibinfo {author} {\bibfnamefont {J.~K.}\ \bibnamefont {Pachos}},\
  }\bibfield  {title} {\bibinfo {title} {Cold atom simulation of interacting
  relativistic quantum field theories},\ }\href
  {https://doi.org/10.1103/PhysRevLett.105.190403} {\bibfield  {journal}
  {\bibinfo  {journal} {Phys. Rev. Lett.}\ }\textbf {\bibinfo {volume} {105}},\
  \bibinfo {pages} {190403} (\bibinfo {year} {2010})}\BibitemShut {NoStop}%
\bibitem [{\citenamefont
  {Wiese}(2013)}]{https://doi.org/10.1002/andp.201300104}%
  \BibitemOpen
  \bibfield  {author} {\bibinfo {author} {\bibfnamefont {U.-J.}\ \bibnamefont
  {Wiese}},\ }\bibfield  {title} {\bibinfo {title} {Ultracold quantum gases and
  lattice systems: quantum simulation of lattice gauge theories},\ }\href
  {https://doi.org/https://doi.org/10.1002/andp.201300104} {\bibfield
  {journal} {\bibinfo  {journal} {Annalen der Physik}\ }\textbf {\bibinfo
  {volume} {525}},\ \bibinfo {pages} {777} (\bibinfo {year}
  {2013})}\BibitemShut {NoStop}%
\bibitem [{\citenamefont {Zohar}\ \emph {et~al.}(2015)\citenamefont {Zohar},
  \citenamefont {Cirac},\ and\ \citenamefont {Reznik}}]{Zohar_2016}%
  \BibitemOpen
  \bibfield  {author} {\bibinfo {author} {\bibfnamefont {E.}~\bibnamefont
  {Zohar}}, \bibinfo {author} {\bibfnamefont {J.~I.}\ \bibnamefont {Cirac}},\
  and\ \bibinfo {author} {\bibfnamefont {B.}~\bibnamefont {Reznik}},\
  }\bibfield  {title} {\bibinfo {title} {Quantum simulations of lattice gauge
  theories using ultracold atoms in optical lattices},\ }\href
  {https://doi.org/10.1088/0034-4885/79/1/014401} {\bibfield  {journal}
  {\bibinfo  {journal} {Reports on Progress in Physics}\ }\textbf {\bibinfo
  {volume} {79}},\ \bibinfo {pages} {014401} (\bibinfo {year}
  {2015})}\BibitemShut {NoStop}%
\bibitem [{\citenamefont {Dalmonte}\ and\ \citenamefont
  {Montangero}(2016)}]{doi:10.1080/00107514.2016.1151199}%
  \BibitemOpen
  \bibfield  {author} {\bibinfo {author} {\bibfnamefont {M.}~\bibnamefont
  {Dalmonte}}\ and\ \bibinfo {author} {\bibfnamefont {S.}~\bibnamefont
  {Montangero}},\ }\bibfield  {title} {\bibinfo {title} {Lattice gauge theory
  simulations in the quantum information era},\ }\href
  {https://doi.org/10.1080/00107514.2016.1151199} {\bibfield  {journal}
  {\bibinfo  {journal} {Contemporary Physics}\ }\textbf {\bibinfo {volume}
  {57}},\ \bibinfo {pages} {388} (\bibinfo {year} {2016})}\BibitemShut
  {NoStop}%
\bibitem [{\citenamefont {Ba{\~{n}}uls}\ \emph {et~al.}(2020)\citenamefont
  {Ba{\~{n}}uls}, \citenamefont {Blatt}, \citenamefont {Catani}, \citenamefont
  {Celi}, \citenamefont {Cirac}, \citenamefont {Dalmonte}, \citenamefont
  {Fallani}, \citenamefont {Jansen}, \citenamefont {Lewenstein}, \citenamefont
  {Montangero}, \citenamefont {Muschik}, \citenamefont {Reznik}, \citenamefont
  {Rico}, \citenamefont {Tagliacozzo}, \citenamefont {Van~Acoleyen},
  \citenamefont {Verstraete}, \citenamefont {Wiese}, \citenamefont {Wingate},
  \citenamefont {Zakrzewski},\ and\ \citenamefont {Zoller}}]{Banuls2020}%
  \BibitemOpen
  \bibfield  {author} {\bibinfo {author} {\bibfnamefont {M.~C.}\ \bibnamefont
  {Ba{\~{n}}uls}}, \bibinfo {author} {\bibfnamefont {R.}~\bibnamefont {Blatt}},
  \bibinfo {author} {\bibfnamefont {J.}~\bibnamefont {Catani}}, \bibinfo
  {author} {\bibfnamefont {A.}~\bibnamefont {Celi}}, \bibinfo {author}
  {\bibfnamefont {J.~I.}\ \bibnamefont {Cirac}}, \bibinfo {author}
  {\bibfnamefont {M.}~\bibnamefont {Dalmonte}}, \bibinfo {author}
  {\bibfnamefont {L.}~\bibnamefont {Fallani}}, \bibinfo {author} {\bibfnamefont
  {K.}~\bibnamefont {Jansen}}, \bibinfo {author} {\bibfnamefont
  {M.}~\bibnamefont {Lewenstein}}, \bibinfo {author} {\bibfnamefont
  {S.}~\bibnamefont {Montangero}}, \bibinfo {author} {\bibfnamefont {C.~A.}\
  \bibnamefont {Muschik}}, \bibinfo {author} {\bibfnamefont {B.}~\bibnamefont
  {Reznik}}, \bibinfo {author} {\bibfnamefont {E.}~\bibnamefont {Rico}},
  \bibinfo {author} {\bibfnamefont {L.}~\bibnamefont {Tagliacozzo}}, \bibinfo
  {author} {\bibfnamefont {K.}~\bibnamefont {Van~Acoleyen}}, \bibinfo {author}
  {\bibfnamefont {F.}~\bibnamefont {Verstraete}}, \bibinfo {author}
  {\bibfnamefont {U.-J.}\ \bibnamefont {Wiese}}, \bibinfo {author}
  {\bibfnamefont {M.}~\bibnamefont {Wingate}}, \bibinfo {author} {\bibfnamefont
  {J.}~\bibnamefont {Zakrzewski}},\ and\ \bibinfo {author} {\bibfnamefont
  {P.}~\bibnamefont {Zoller}},\ }\bibfield  {title} {\bibinfo {title}
  {Simulating lattice gauge theories within quantum technologies},\ }\href
  {https://doi.org/10.1140/epjd/e2020-100571-8} {\bibfield  {journal} {\bibinfo
   {journal} {The European Physical Journal D}\ }\textbf {\bibinfo {volume}
  {74}},\ \bibinfo {pages} {165} (\bibinfo {year} {2020})}\BibitemShut
  {NoStop}%
\bibitem [{\citenamefont {Ba{\~{n}}uls}\ and\ \citenamefont
  {Cichy}(2020)}]{Carmen_Banuls_2020}%
  \BibitemOpen
  \bibfield  {author} {\bibinfo {author} {\bibfnamefont {M.~C.}\ \bibnamefont
  {Ba{\~{n}}uls}}\ and\ \bibinfo {author} {\bibfnamefont {K.}~\bibnamefont
  {Cichy}},\ }\bibfield  {title} {\bibinfo {title} {Review on novel methods for
  lattice gauge theories},\ }\href {https://doi.org/10.1088/1361-6633/ab6311}
  {\bibfield  {journal} {\bibinfo  {journal} {Reports on Progress in Physics}\
  }\textbf {\bibinfo {volume} {83}},\ \bibinfo {pages} {024401} (\bibinfo
  {year} {2020})}\BibitemShut {NoStop}%
\bibitem [{\citenamefont {Aidelsburger}\ \emph {et~al.}(2022)\citenamefont
  {Aidelsburger}, \citenamefont {Barbiero}, \citenamefont {Bermudez},
  \citenamefont {Chanda}, \citenamefont {Dauphin}, \citenamefont
  {González-Cuadra}, \citenamefont {Grzybowski}, \citenamefont {Hands},
  \citenamefont {Jendrzejewski}, \citenamefont {Jünemann}, \citenamefont
  {Juzeliūnas}, \citenamefont {Kasper}, \citenamefont {Piga}, \citenamefont
  {Ran}, \citenamefont {Rizzi}, \citenamefont {Sierra}, \citenamefont
  {Tagliacozzo}, \citenamefont {Tirrito}, \citenamefont {Zache}, \citenamefont
  {Zakrzewski}, \citenamefont {Zohar},\ and\ \citenamefont
  {Lewenstein}}]{doi:10.1098/rsta.2021.0064}%
  \BibitemOpen
  \bibfield  {author} {\bibinfo {author} {\bibfnamefont {M.}~\bibnamefont
  {Aidelsburger}}, \bibinfo {author} {\bibfnamefont {L.}~\bibnamefont
  {Barbiero}}, \bibinfo {author} {\bibfnamefont {A.}~\bibnamefont {Bermudez}},
  \bibinfo {author} {\bibfnamefont {T.}~\bibnamefont {Chanda}}, \bibinfo
  {author} {\bibfnamefont {A.}~\bibnamefont {Dauphin}}, \bibinfo {author}
  {\bibfnamefont {D.}~\bibnamefont {González-Cuadra}}, \bibinfo {author}
  {\bibfnamefont {P.~R.}\ \bibnamefont {Grzybowski}}, \bibinfo {author}
  {\bibfnamefont {S.}~\bibnamefont {Hands}}, \bibinfo {author} {\bibfnamefont
  {F.}~\bibnamefont {Jendrzejewski}}, \bibinfo {author} {\bibfnamefont
  {J.}~\bibnamefont {Jünemann}}, \bibinfo {author} {\bibfnamefont
  {G.}~\bibnamefont {Juzeliūnas}}, \bibinfo {author} {\bibfnamefont
  {V.}~\bibnamefont {Kasper}}, \bibinfo {author} {\bibfnamefont
  {A.}~\bibnamefont {Piga}}, \bibinfo {author} {\bibfnamefont {S.-J.}\
  \bibnamefont {Ran}}, \bibinfo {author} {\bibfnamefont {M.}~\bibnamefont
  {Rizzi}}, \bibinfo {author} {\bibfnamefont {G.}~\bibnamefont {Sierra}},
  \bibinfo {author} {\bibfnamefont {L.}~\bibnamefont {Tagliacozzo}}, \bibinfo
  {author} {\bibfnamefont {E.}~\bibnamefont {Tirrito}}, \bibinfo {author}
  {\bibfnamefont {T.~V.}\ \bibnamefont {Zache}}, \bibinfo {author}
  {\bibfnamefont {J.}~\bibnamefont {Zakrzewski}}, \bibinfo {author}
  {\bibfnamefont {E.}~\bibnamefont {Zohar}},\ and\ \bibinfo {author}
  {\bibfnamefont {M.}~\bibnamefont {Lewenstein}},\ }\bibfield  {title}
  {\bibinfo {title} {Cold atoms meet lattice gauge theory},\ }\href
  {https://doi.org/10.1098/rsta.2021.0064} {\bibfield  {journal} {\bibinfo
  {journal} {Philosophical Transactions of the Royal Society A: Mathematical,
  Physical and Engineering Sciences}\ }\textbf {\bibinfo {volume} {380}},\
  \bibinfo {pages} {20210064} (\bibinfo {year} {2022})}\BibitemShut {NoStop}%
\bibitem [{\citenamefont {Klco}\ \emph {et~al.}(2022)\citenamefont {Klco},
  \citenamefont {Roggero},\ and\ \citenamefont {Savage}}]{Klco_2022}%
  \BibitemOpen
  \bibfield  {author} {\bibinfo {author} {\bibfnamefont {N.}~\bibnamefont
  {Klco}}, \bibinfo {author} {\bibfnamefont {A.}~\bibnamefont {Roggero}},\ and\
  \bibinfo {author} {\bibfnamefont {M.~J.}\ \bibnamefont {Savage}},\ }\bibfield
   {title} {\bibinfo {title} {Standard model physics and the digital quantum
  revolution: thoughts about the interface},\ }\href
  {https://doi.org/10.1088/1361-6633/ac58a4} {\bibfield  {journal} {\bibinfo
  {journal} {Reports on Progress in Physics}\ }\textbf {\bibinfo {volume}
  {85}},\ \bibinfo {pages} {064301} (\bibinfo {year} {2022})}\BibitemShut
  {NoStop}%
\bibitem [{\citenamefont {Bauer}\ \emph {et~al.}(2022)\citenamefont {Bauer},
  \citenamefont {Davoudi}, \citenamefont {Balantekin}, \citenamefont
  {Bhattacharya}, \citenamefont {Carena}, \citenamefont {de~Jong},
  \citenamefont {Draper}, \citenamefont {El-Khadra}, \citenamefont {Gemelke},
  \citenamefont {Hanada}, \citenamefont {Kharzeev}, \citenamefont {Lamm},
  \citenamefont {Li}, \citenamefont {Liu}, \citenamefont {Lukin}, \citenamefont
  {Meurice}, \citenamefont {Monroe}, \citenamefont {Nachman}, \citenamefont
  {Pagano}, \citenamefont {Preskill}, \citenamefont {Rinaldi}, \citenamefont
  {Roggero}, \citenamefont {Santiago}, \citenamefont {Savage}, \citenamefont
  {Siddiqi}, \citenamefont {Siopsis}, \citenamefont {Van~Zanten}, \citenamefont
  {Wiebe}, \citenamefont {Yamauchi}, \citenamefont {Yeter-Aydeniz},\ and\
  \citenamefont {Zorzetti}}]{https://doi.org/10.48550/arxiv.2204.03381}%
  \BibitemOpen
  \bibfield  {author} {\bibinfo {author} {\bibfnamefont {C.~W.}\ \bibnamefont
  {Bauer}}, \bibinfo {author} {\bibfnamefont {Z.}~\bibnamefont {Davoudi}},
  \bibinfo {author} {\bibfnamefont {A.~B.}\ \bibnamefont {Balantekin}},
  \bibinfo {author} {\bibfnamefont {T.}~\bibnamefont {Bhattacharya}}, \bibinfo
  {author} {\bibfnamefont {M.}~\bibnamefont {Carena}}, \bibinfo {author}
  {\bibfnamefont {W.~A.}\ \bibnamefont {de~Jong}}, \bibinfo {author}
  {\bibfnamefont {P.}~\bibnamefont {Draper}}, \bibinfo {author} {\bibfnamefont
  {A.}~\bibnamefont {El-Khadra}}, \bibinfo {author} {\bibfnamefont
  {N.}~\bibnamefont {Gemelke}}, \bibinfo {author} {\bibfnamefont
  {M.}~\bibnamefont {Hanada}}, \bibinfo {author} {\bibfnamefont
  {D.}~\bibnamefont {Kharzeev}}, \bibinfo {author} {\bibfnamefont
  {H.}~\bibnamefont {Lamm}}, \bibinfo {author} {\bibfnamefont {Y.-Y.}\
  \bibnamefont {Li}}, \bibinfo {author} {\bibfnamefont {J.}~\bibnamefont
  {Liu}}, \bibinfo {author} {\bibfnamefont {M.}~\bibnamefont {Lukin}}, \bibinfo
  {author} {\bibfnamefont {Y.}~\bibnamefont {Meurice}}, \bibinfo {author}
  {\bibfnamefont {C.}~\bibnamefont {Monroe}}, \bibinfo {author} {\bibfnamefont
  {B.}~\bibnamefont {Nachman}}, \bibinfo {author} {\bibfnamefont
  {G.}~\bibnamefont {Pagano}}, \bibinfo {author} {\bibfnamefont
  {J.}~\bibnamefont {Preskill}}, \bibinfo {author} {\bibfnamefont
  {E.}~\bibnamefont {Rinaldi}}, \bibinfo {author} {\bibfnamefont
  {A.}~\bibnamefont {Roggero}}, \bibinfo {author} {\bibfnamefont {D.~I.}\
  \bibnamefont {Santiago}}, \bibinfo {author} {\bibfnamefont {M.~J.}\
  \bibnamefont {Savage}}, \bibinfo {author} {\bibfnamefont {I.}~\bibnamefont
  {Siddiqi}}, \bibinfo {author} {\bibfnamefont {G.}~\bibnamefont {Siopsis}},
  \bibinfo {author} {\bibfnamefont {D.}~\bibnamefont {Van~Zanten}}, \bibinfo
  {author} {\bibfnamefont {N.}~\bibnamefont {Wiebe}}, \bibinfo {author}
  {\bibfnamefont {Y.}~\bibnamefont {Yamauchi}}, \bibinfo {author}
  {\bibfnamefont {K.}~\bibnamefont {Yeter-Aydeniz}},\ and\ \bibinfo {author}
  {\bibfnamefont {S.}~\bibnamefont {Zorzetti}},\ }\href
  {https://doi.org/10.48550/ARXIV.2204.03381} {\bibinfo {title} {Quantum
  simulation for high energy physics}} (\bibinfo {year} {2022})\BibitemShut
  {NoStop}%
\bibitem [{\citenamefont {Meglio}\ \emph {et~al.}(2023)\citenamefont {Meglio},
  \citenamefont {Jansen}, \citenamefont {Tavernelli}, \citenamefont
  {Alexandrou}, \citenamefont {Arunachalam}, \citenamefont {Bauer},
  \citenamefont {Borras}, \citenamefont {Carrazza}, \citenamefont {Crippa},
  \citenamefont {Croft}, \citenamefont {de~Putter}, \citenamefont {Delgado},
  \citenamefont {Dunjko}, \citenamefont {Egger}, \citenamefont
  {Fernandez-Combarro}, \citenamefont {Fuchs}, \citenamefont {Funcke},
  \citenamefont {Gonzalez-Cuadra}, \citenamefont {Grossi}, \citenamefont
  {Halimeh}, \citenamefont {Holmes}, \citenamefont {Kuhn}, \citenamefont
  {Lacroix}, \citenamefont {Lewis}, \citenamefont {Lucchesi}, \citenamefont
  {Martinez}, \citenamefont {Meloni}, \citenamefont {Mezzacapo}, \citenamefont
  {Montangero}, \citenamefont {Nagano}, \citenamefont {Radescu}, \citenamefont
  {Ortega}, \citenamefont {Roggero}, \citenamefont {Schuhmacher}, \citenamefont
  {Seixas}, \citenamefont {Silvi}, \citenamefont {Spentzouris}, \citenamefont
  {Tacchino}, \citenamefont {Temme}, \citenamefont {Terashi}, \citenamefont
  {Tura}, \citenamefont {Tuysuz}, \citenamefont {Vallecorsa}, \citenamefont
  {Wiese}, \citenamefont {Yoo},\ and\ \citenamefont {Zhang}}]{dimeglio_2023}%
  \BibitemOpen
  \bibfield  {author} {\bibinfo {author} {\bibfnamefont {A.~D.}\ \bibnamefont
  {Meglio}}, \bibinfo {author} {\bibfnamefont {K.}~\bibnamefont {Jansen}},
  \bibinfo {author} {\bibfnamefont {I.}~\bibnamefont {Tavernelli}}, \bibinfo
  {author} {\bibfnamefont {C.}~\bibnamefont {Alexandrou}}, \bibinfo {author}
  {\bibfnamefont {S.}~\bibnamefont {Arunachalam}}, \bibinfo {author}
  {\bibfnamefont {C.~W.}\ \bibnamefont {Bauer}}, \bibinfo {author}
  {\bibfnamefont {K.}~\bibnamefont {Borras}}, \bibinfo {author} {\bibfnamefont
  {S.}~\bibnamefont {Carrazza}}, \bibinfo {author} {\bibfnamefont
  {A.}~\bibnamefont {Crippa}}, \bibinfo {author} {\bibfnamefont
  {V.}~\bibnamefont {Croft}}, \bibinfo {author} {\bibfnamefont
  {R.}~\bibnamefont {de~Putter}}, \bibinfo {author} {\bibfnamefont
  {A.}~\bibnamefont {Delgado}}, \bibinfo {author} {\bibfnamefont
  {V.}~\bibnamefont {Dunjko}}, \bibinfo {author} {\bibfnamefont {D.~J.}\
  \bibnamefont {Egger}}, \bibinfo {author} {\bibfnamefont {E.}~\bibnamefont
  {Fernandez-Combarro}}, \bibinfo {author} {\bibfnamefont {E.}~\bibnamefont
  {Fuchs}}, \bibinfo {author} {\bibfnamefont {L.}~\bibnamefont {Funcke}},
  \bibinfo {author} {\bibfnamefont {D.}~\bibnamefont {Gonzalez-Cuadra}},
  \bibinfo {author} {\bibfnamefont {M.}~\bibnamefont {Grossi}}, \bibinfo
  {author} {\bibfnamefont {J.~C.}\ \bibnamefont {Halimeh}}, \bibinfo {author}
  {\bibfnamefont {Z.}~\bibnamefont {Holmes}}, \bibinfo {author} {\bibfnamefont
  {S.}~\bibnamefont {Kuhn}}, \bibinfo {author} {\bibfnamefont {D.}~\bibnamefont
  {Lacroix}}, \bibinfo {author} {\bibfnamefont {R.}~\bibnamefont {Lewis}},
  \bibinfo {author} {\bibfnamefont {D.}~\bibnamefont {Lucchesi}}, \bibinfo
  {author} {\bibfnamefont {M.~L.}\ \bibnamefont {Martinez}}, \bibinfo {author}
  {\bibfnamefont {F.}~\bibnamefont {Meloni}}, \bibinfo {author} {\bibfnamefont
  {A.}~\bibnamefont {Mezzacapo}}, \bibinfo {author} {\bibfnamefont
  {S.}~\bibnamefont {Montangero}}, \bibinfo {author} {\bibfnamefont
  {L.}~\bibnamefont {Nagano}}, \bibinfo {author} {\bibfnamefont
  {V.}~\bibnamefont {Radescu}}, \bibinfo {author} {\bibfnamefont {E.~R.}\
  \bibnamefont {Ortega}}, \bibinfo {author} {\bibfnamefont {A.}~\bibnamefont
  {Roggero}}, \bibinfo {author} {\bibfnamefont {J.}~\bibnamefont
  {Schuhmacher}}, \bibinfo {author} {\bibfnamefont {J.}~\bibnamefont {Seixas}},
  \bibinfo {author} {\bibfnamefont {P.}~\bibnamefont {Silvi}}, \bibinfo
  {author} {\bibfnamefont {P.}~\bibnamefont {Spentzouris}}, \bibinfo {author}
  {\bibfnamefont {F.}~\bibnamefont {Tacchino}}, \bibinfo {author}
  {\bibfnamefont {K.}~\bibnamefont {Temme}}, \bibinfo {author} {\bibfnamefont
  {K.}~\bibnamefont {Terashi}}, \bibinfo {author} {\bibfnamefont
  {J.}~\bibnamefont {Tura}}, \bibinfo {author} {\bibfnamefont {C.}~\bibnamefont
  {Tuysuz}}, \bibinfo {author} {\bibfnamefont {S.}~\bibnamefont {Vallecorsa}},
  \bibinfo {author} {\bibfnamefont {U.-J.}\ \bibnamefont {Wiese}}, \bibinfo
  {author} {\bibfnamefont {S.}~\bibnamefont {Yoo}},\ and\ \bibinfo {author}
  {\bibfnamefont {J.}~\bibnamefont {Zhang}},\ }\href@noop {} {\bibinfo {title}
  {Quantum computing for high-energy physics: State of the art and challenges.
  summary of the qc4hep working group}} (\bibinfo {year} {2023}),\ \Eprint
  {https://arxiv.org/abs/2307.03236} {arXiv:2307.03236 [quant-ph]} \BibitemShut
  {NoStop}%
\bibitem [{\citenamefont {Gerritsma}\ \emph {et~al.}(2010)\citenamefont
  {Gerritsma}, \citenamefont {Kirchmair}, \citenamefont {Z{\"a}hringer},
  \citenamefont {Solano}, \citenamefont {Blatt},\ and\ \citenamefont
  {Roos}}]{Gerritsma2010}%
  \BibitemOpen
  \bibfield  {author} {\bibinfo {author} {\bibfnamefont {R.}~\bibnamefont
  {Gerritsma}}, \bibinfo {author} {\bibfnamefont {G.}~\bibnamefont
  {Kirchmair}}, \bibinfo {author} {\bibfnamefont {F.}~\bibnamefont
  {Z{\"a}hringer}}, \bibinfo {author} {\bibfnamefont {E.}~\bibnamefont
  {Solano}}, \bibinfo {author} {\bibfnamefont {R.}~\bibnamefont {Blatt}},\ and\
  \bibinfo {author} {\bibfnamefont {C.~F.}\ \bibnamefont {Roos}},\ }\bibfield
  {title} {\bibinfo {title} {Quantum simulation of the dirac equation},\ }\href
  {https://doi.org/10.1038/nature08688} {\bibfield  {journal} {\bibinfo
  {journal} {Nature}\ }\textbf {\bibinfo {volume} {463}},\ \bibinfo {pages}
  {68} (\bibinfo {year} {2010})}\BibitemShut {NoStop}%
\bibitem [{\citenamefont {Gerritsma}\ \emph {et~al.}(2011)\citenamefont
  {Gerritsma}, \citenamefont {Lanyon}, \citenamefont {Kirchmair}, \citenamefont
  {Z\"ahringer}, \citenamefont {Hempel}, \citenamefont {Casanova},
  \citenamefont {Garc\'{\i}a-Ripoll}, \citenamefont {Solano}, \citenamefont
  {Blatt},\ and\ \citenamefont {Roos}}]{PhysRevLett.106.060503}%
  \BibitemOpen
  \bibfield  {author} {\bibinfo {author} {\bibfnamefont {R.}~\bibnamefont
  {Gerritsma}}, \bibinfo {author} {\bibfnamefont {B.~P.}\ \bibnamefont
  {Lanyon}}, \bibinfo {author} {\bibfnamefont {G.}~\bibnamefont {Kirchmair}},
  \bibinfo {author} {\bibfnamefont {F.}~\bibnamefont {Z\"ahringer}}, \bibinfo
  {author} {\bibfnamefont {C.}~\bibnamefont {Hempel}}, \bibinfo {author}
  {\bibfnamefont {J.}~\bibnamefont {Casanova}}, \bibinfo {author}
  {\bibfnamefont {J.~J.}\ \bibnamefont {Garc\'{\i}a-Ripoll}}, \bibinfo {author}
  {\bibfnamefont {E.}~\bibnamefont {Solano}}, \bibinfo {author} {\bibfnamefont
  {R.}~\bibnamefont {Blatt}},\ and\ \bibinfo {author} {\bibfnamefont {C.~F.}\
  \bibnamefont {Roos}},\ }\bibfield  {title} {\bibinfo {title} {Quantum
  simulation of the klein paradox with trapped ions},\ }\href
  {https://doi.org/10.1103/PhysRevLett.106.060503} {\bibfield  {journal}
  {\bibinfo  {journal} {Phys. Rev. Lett.}\ }\textbf {\bibinfo {volume} {106}},\
  \bibinfo {pages} {060503} (\bibinfo {year} {2011})}\BibitemShut {NoStop}%
\bibitem [{\citenamefont {Salger}\ \emph {et~al.}(2011)\citenamefont {Salger},
  \citenamefont {Grossert}, \citenamefont {Kling},\ and\ \citenamefont
  {Weitz}}]{PhysRevLett.107.240401}%
  \BibitemOpen
  \bibfield  {author} {\bibinfo {author} {\bibfnamefont {T.}~\bibnamefont
  {Salger}}, \bibinfo {author} {\bibfnamefont {C.}~\bibnamefont {Grossert}},
  \bibinfo {author} {\bibfnamefont {S.}~\bibnamefont {Kling}},\ and\ \bibinfo
  {author} {\bibfnamefont {M.}~\bibnamefont {Weitz}},\ }\bibfield  {title}
  {\bibinfo {title} {Klein tunneling of a quasirelativistic bose-einstein
  condensate in an optical lattice},\ }\href
  {https://doi.org/10.1103/PhysRevLett.107.240401} {\bibfield  {journal}
  {\bibinfo  {journal} {Phys. Rev. Lett.}\ }\textbf {\bibinfo {volume} {107}},\
  \bibinfo {pages} {240401} (\bibinfo {year} {2011})}\BibitemShut {NoStop}%
\bibitem [{\citenamefont {LeBlanc}\ \emph {et~al.}(2013)\citenamefont
  {LeBlanc}, \citenamefont {Beeler}, \citenamefont {Jiménez-García},
  \citenamefont {Perry}, \citenamefont {Sugawa}, \citenamefont {Williams},\
  and\ \citenamefont {Spielman}}]{LeBlanc_2013}%
  \BibitemOpen
  \bibfield  {author} {\bibinfo {author} {\bibfnamefont {L.~J.}\ \bibnamefont
  {LeBlanc}}, \bibinfo {author} {\bibfnamefont {M.~C.}\ \bibnamefont {Beeler}},
  \bibinfo {author} {\bibfnamefont {K.}~\bibnamefont {Jiménez-García}},
  \bibinfo {author} {\bibfnamefont {A.~R.}\ \bibnamefont {Perry}}, \bibinfo
  {author} {\bibfnamefont {S.}~\bibnamefont {Sugawa}}, \bibinfo {author}
  {\bibfnamefont {R.~A.}\ \bibnamefont {Williams}},\ and\ \bibinfo {author}
  {\bibfnamefont {I.~B.}\ \bibnamefont {Spielman}},\ }\bibfield  {title}
  {\bibinfo {title} {Direct observation of zitterbewegung in a bose–einstein
  condensate},\ }\href {https://doi.org/10.1088/1367-2630/15/7/073011}
  {\bibfield  {journal} {\bibinfo  {journal} {New Journal of Physics}\ }\textbf
  {\bibinfo {volume} {15}},\ \bibinfo {pages} {073011} (\bibinfo {year}
  {2013})}\BibitemShut {NoStop}%
\bibitem [{\citenamefont {Uehlinger}\ \emph
  {et~al.}(2013{\natexlab{a}})\citenamefont {Uehlinger}, \citenamefont {Greif},
  \citenamefont {Jotzu}, \citenamefont {Tarruell}, \citenamefont {Esslinger},
  \citenamefont {Wang},\ and\ \citenamefont {Troyer}}]{Uehlinger2013}%
  \BibitemOpen
  \bibfield  {author} {\bibinfo {author} {\bibfnamefont {T.}~\bibnamefont
  {Uehlinger}}, \bibinfo {author} {\bibfnamefont {D.}~\bibnamefont {Greif}},
  \bibinfo {author} {\bibfnamefont {G.}~\bibnamefont {Jotzu}}, \bibinfo
  {author} {\bibfnamefont {L.}~\bibnamefont {Tarruell}}, \bibinfo {author}
  {\bibfnamefont {T.}~\bibnamefont {Esslinger}}, \bibinfo {author}
  {\bibfnamefont {L.}~\bibnamefont {Wang}},\ and\ \bibinfo {author}
  {\bibfnamefont {M.}~\bibnamefont {Troyer}},\ }\bibfield  {title} {\bibinfo
  {title} {Double transfer through dirac points in a tunable honeycomb optical
  lattice},\ }\href {https://doi.org/10.1140/epjst/e2013-01761-y} {\bibfield
  {journal} {\bibinfo  {journal} {The European Physical Journal Special
  Topics}\ }\textbf {\bibinfo {volume} {217}},\ \bibinfo {pages} {121}
  (\bibinfo {year} {2013}{\natexlab{a}})}\BibitemShut {NoStop}%
\bibitem [{\citenamefont {Uehlinger}\ \emph
  {et~al.}(2013{\natexlab{b}})\citenamefont {Uehlinger}, \citenamefont {Jotzu},
  \citenamefont {Messer}, \citenamefont {Greif}, \citenamefont {Hofstetter},
  \citenamefont {Bissbort},\ and\ \citenamefont
  {Esslinger}}]{PhysRevLett.111.185307}%
  \BibitemOpen
  \bibfield  {author} {\bibinfo {author} {\bibfnamefont {T.}~\bibnamefont
  {Uehlinger}}, \bibinfo {author} {\bibfnamefont {G.}~\bibnamefont {Jotzu}},
  \bibinfo {author} {\bibfnamefont {M.}~\bibnamefont {Messer}}, \bibinfo
  {author} {\bibfnamefont {D.}~\bibnamefont {Greif}}, \bibinfo {author}
  {\bibfnamefont {W.}~\bibnamefont {Hofstetter}}, \bibinfo {author}
  {\bibfnamefont {U.}~\bibnamefont {Bissbort}},\ and\ \bibinfo {author}
  {\bibfnamefont {T.}~\bibnamefont {Esslinger}},\ }\bibfield  {title} {\bibinfo
  {title} {Artificial graphene with tunable interactions},\ }\href
  {https://doi.org/10.1103/PhysRevLett.111.185307} {\bibfield  {journal}
  {\bibinfo  {journal} {Phys. Rev. Lett.}\ }\textbf {\bibinfo {volume} {111}},\
  \bibinfo {pages} {185307} (\bibinfo {year} {2013}{\natexlab{b}})}\BibitemShut
  {NoStop}%
\bibitem [{\citenamefont {Duca}\ \emph {et~al.}(2015)\citenamefont {Duca},
  \citenamefont {Li}, \citenamefont {Reitter}, \citenamefont {Bloch},
  \citenamefont {Schleier-Smith},\ and\ \citenamefont
  {Schneider}}]{doi:10.1126/science.1259052}%
  \BibitemOpen
  \bibfield  {author} {\bibinfo {author} {\bibfnamefont {L.}~\bibnamefont
  {Duca}}, \bibinfo {author} {\bibfnamefont {T.}~\bibnamefont {Li}}, \bibinfo
  {author} {\bibfnamefont {M.}~\bibnamefont {Reitter}}, \bibinfo {author}
  {\bibfnamefont {I.}~\bibnamefont {Bloch}}, \bibinfo {author} {\bibfnamefont
  {M.}~\bibnamefont {Schleier-Smith}},\ and\ \bibinfo {author} {\bibfnamefont
  {U.}~\bibnamefont {Schneider}},\ }\bibfield  {title} {\bibinfo {title} {An
  aharonov-bohm interferometer for determining bloch band topology},\ }\href
  {https://doi.org/10.1126/science.1259052} {\bibfield  {journal} {\bibinfo
  {journal} {Science}\ }\textbf {\bibinfo {volume} {347}},\ \bibinfo {pages}
  {288} (\bibinfo {year} {2015})},\ \Eprint
  {https://arxiv.org/abs/https://www.science.org/doi/pdf/10.1126/science.1259052}
  {https://www.science.org/doi/pdf/10.1126/science.1259052} \BibitemShut
  {NoStop}%
\bibitem [{\citenamefont {Li}\ \emph {et~al.}(2016)\citenamefont {Li},
  \citenamefont {Duca}, \citenamefont {Reitter}, \citenamefont {Grusdt},
  \citenamefont {Demler}, \citenamefont {Endres}, \citenamefont
  {Schleier-Smith}, \citenamefont {Bloch},\ and\ \citenamefont
  {Schneider}}]{doi:10.1126/science.aad5812}%
  \BibitemOpen
  \bibfield  {author} {\bibinfo {author} {\bibfnamefont {T.}~\bibnamefont
  {Li}}, \bibinfo {author} {\bibfnamefont {L.}~\bibnamefont {Duca}}, \bibinfo
  {author} {\bibfnamefont {M.}~\bibnamefont {Reitter}}, \bibinfo {author}
  {\bibfnamefont {F.}~\bibnamefont {Grusdt}}, \bibinfo {author} {\bibfnamefont
  {E.}~\bibnamefont {Demler}}, \bibinfo {author} {\bibfnamefont
  {M.}~\bibnamefont {Endres}}, \bibinfo {author} {\bibfnamefont
  {M.}~\bibnamefont {Schleier-Smith}}, \bibinfo {author} {\bibfnamefont
  {I.}~\bibnamefont {Bloch}},\ and\ \bibinfo {author} {\bibfnamefont
  {U.}~\bibnamefont {Schneider}},\ }\bibfield  {title} {\bibinfo {title} {Bloch
  state tomography using wilson lines},\ }\href
  {https://doi.org/10.1126/science.aad5812} {\bibfield  {journal} {\bibinfo
  {journal} {Science}\ }\textbf {\bibinfo {volume} {352}},\ \bibinfo {pages}
  {1094} (\bibinfo {year} {2016})}\BibitemShut {NoStop}%
\bibitem [{\citenamefont {Martinez}\ \emph {et~al.}(2016)\citenamefont
  {Martinez}, \citenamefont {Muschik}, \citenamefont {Schindler}, \citenamefont
  {Nigg}, \citenamefont {Erhard}, \citenamefont {Heyl}, \citenamefont {Hauke},
  \citenamefont {Dalmonte}, \citenamefont {Monz}, \citenamefont {Zoller},\ and\
  \citenamefont {Blatt}}]{Martinez2016}%
  \BibitemOpen
  \bibfield  {author} {\bibinfo {author} {\bibfnamefont {E.~A.}\ \bibnamefont
  {Martinez}}, \bibinfo {author} {\bibfnamefont {C.~A.}\ \bibnamefont
  {Muschik}}, \bibinfo {author} {\bibfnamefont {P.}~\bibnamefont {Schindler}},
  \bibinfo {author} {\bibfnamefont {D.}~\bibnamefont {Nigg}}, \bibinfo {author}
  {\bibfnamefont {A.}~\bibnamefont {Erhard}}, \bibinfo {author} {\bibfnamefont
  {M.}~\bibnamefont {Heyl}}, \bibinfo {author} {\bibfnamefont {P.}~\bibnamefont
  {Hauke}}, \bibinfo {author} {\bibfnamefont {M.}~\bibnamefont {Dalmonte}},
  \bibinfo {author} {\bibfnamefont {T.}~\bibnamefont {Monz}}, \bibinfo {author}
  {\bibfnamefont {P.}~\bibnamefont {Zoller}},\ and\ \bibinfo {author}
  {\bibfnamefont {R.}~\bibnamefont {Blatt}},\ }\bibfield  {title} {\bibinfo
  {title} {Real-time dynamics of lattice gauge theories with a few-qubit
  quantum computer},\ }\href {https://doi.org/10.1038/nature18318} {\bibfield
  {journal} {\bibinfo  {journal} {Nature}\ }\textbf {\bibinfo {volume} {534}},\
  \bibinfo {pages} {516} (\bibinfo {year} {2016})}\BibitemShut {NoStop}%
\bibitem [{\citenamefont {Dai}\ \emph {et~al.}(2017)\citenamefont {Dai},
  \citenamefont {Yang}, \citenamefont {Reingruber}, \citenamefont {Sun},
  \citenamefont {Xu}, \citenamefont {Chen}, \citenamefont {Yuan},\ and\
  \citenamefont {Pan}}]{Dai2017}%
  \BibitemOpen
  \bibfield  {author} {\bibinfo {author} {\bibfnamefont {H.-N.}\ \bibnamefont
  {Dai}}, \bibinfo {author} {\bibfnamefont {B.}~\bibnamefont {Yang}}, \bibinfo
  {author} {\bibfnamefont {A.}~\bibnamefont {Reingruber}}, \bibinfo {author}
  {\bibfnamefont {H.}~\bibnamefont {Sun}}, \bibinfo {author} {\bibfnamefont
  {X.-F.}\ \bibnamefont {Xu}}, \bibinfo {author} {\bibfnamefont {Y.-A.}\
  \bibnamefont {Chen}}, \bibinfo {author} {\bibfnamefont {Z.-S.}\ \bibnamefont
  {Yuan}},\ and\ \bibinfo {author} {\bibfnamefont {J.-W.}\ \bibnamefont
  {Pan}},\ }\bibfield  {title} {\bibinfo {title} {Four-body ring-exchange
  interactions and anyonic statistics within a minimal toric-code
  hamiltonian},\ }\href {https://doi.org/10.1038/nphys4243} {\bibfield
  {journal} {\bibinfo  {journal} {Nature Physics}\ }\textbf {\bibinfo {volume}
  {13}},\ \bibinfo {pages} {1195} (\bibinfo {year} {2017})}\BibitemShut
  {NoStop}%
\bibitem [{\citenamefont {Klco}\ \emph {et~al.}(2018)\citenamefont {Klco},
  \citenamefont {Dumitrescu}, \citenamefont {McCaskey}, \citenamefont {Morris},
  \citenamefont {Pooser}, \citenamefont {Sanz}, \citenamefont {Solano},
  \citenamefont {Lougovski},\ and\ \citenamefont
  {Savage}}]{PhysRevA.98.032331}%
  \BibitemOpen
  \bibfield  {author} {\bibinfo {author} {\bibfnamefont {N.}~\bibnamefont
  {Klco}}, \bibinfo {author} {\bibfnamefont {E.~F.}\ \bibnamefont
  {Dumitrescu}}, \bibinfo {author} {\bibfnamefont {A.~J.}\ \bibnamefont
  {McCaskey}}, \bibinfo {author} {\bibfnamefont {T.~D.}\ \bibnamefont
  {Morris}}, \bibinfo {author} {\bibfnamefont {R.~C.}\ \bibnamefont {Pooser}},
  \bibinfo {author} {\bibfnamefont {M.}~\bibnamefont {Sanz}}, \bibinfo {author}
  {\bibfnamefont {E.}~\bibnamefont {Solano}}, \bibinfo {author} {\bibfnamefont
  {P.}~\bibnamefont {Lougovski}},\ and\ \bibinfo {author} {\bibfnamefont
  {M.~J.}\ \bibnamefont {Savage}},\ }\bibfield  {title} {\bibinfo {title}
  {Quantum-classical computation of schwinger model dynamics using quantum
  computers},\ }\href {https://doi.org/10.1103/PhysRevA.98.032331} {\bibfield
  {journal} {\bibinfo  {journal} {Phys. Rev. A}\ }\textbf {\bibinfo {volume}
  {98}},\ \bibinfo {pages} {032331} (\bibinfo {year} {2018})}\BibitemShut
  {NoStop}%
\bibitem [{\citenamefont {Schweizer}\ \emph {et~al.}(2019)\citenamefont
  {Schweizer}, \citenamefont {Grusdt}, \citenamefont {Berngruber},
  \citenamefont {Barbiero}, \citenamefont {Demler}, \citenamefont {Goldman},
  \citenamefont {Bloch},\ and\ \citenamefont {Aidelsburger}}]{Schweizer2019}%
  \BibitemOpen
  \bibfield  {author} {\bibinfo {author} {\bibfnamefont {C.}~\bibnamefont
  {Schweizer}}, \bibinfo {author} {\bibfnamefont {F.}~\bibnamefont {Grusdt}},
  \bibinfo {author} {\bibfnamefont {M.}~\bibnamefont {Berngruber}}, \bibinfo
  {author} {\bibfnamefont {L.}~\bibnamefont {Barbiero}}, \bibinfo {author}
  {\bibfnamefont {E.}~\bibnamefont {Demler}}, \bibinfo {author} {\bibfnamefont
  {N.}~\bibnamefont {Goldman}}, \bibinfo {author} {\bibfnamefont
  {I.}~\bibnamefont {Bloch}},\ and\ \bibinfo {author} {\bibfnamefont
  {M.}~\bibnamefont {Aidelsburger}},\ }\bibfield  {title} {\bibinfo {title}
  {Floquet approach to z2 lattice gauge theories with ultracold atoms in
  optical lattices},\ }\href {https://doi.org/10.1038/s41567-019-0649-7}
  {\bibfield  {journal} {\bibinfo  {journal} {Nature Physics}\ }\textbf
  {\bibinfo {volume} {15}},\ \bibinfo {pages} {1168} (\bibinfo {year}
  {2019})}\BibitemShut {NoStop}%
\bibitem [{\citenamefont {Kokail}\ \emph {et~al.}(2019)\citenamefont {Kokail},
  \citenamefont {Maier}, \citenamefont {van Bijnen}, \citenamefont {Brydges},
  \citenamefont {Joshi}, \citenamefont {Jurcevic}, \citenamefont {Muschik},
  \citenamefont {Silvi}, \citenamefont {Blatt}, \citenamefont {Roos},\ and\
  \citenamefont {Zoller}}]{Kokail2019}%
  \BibitemOpen
  \bibfield  {author} {\bibinfo {author} {\bibfnamefont {C.}~\bibnamefont
  {Kokail}}, \bibinfo {author} {\bibfnamefont {C.}~\bibnamefont {Maier}},
  \bibinfo {author} {\bibfnamefont {R.}~\bibnamefont {van Bijnen}}, \bibinfo
  {author} {\bibfnamefont {T.}~\bibnamefont {Brydges}}, \bibinfo {author}
  {\bibfnamefont {M.~K.}\ \bibnamefont {Joshi}}, \bibinfo {author}
  {\bibfnamefont {P.}~\bibnamefont {Jurcevic}}, \bibinfo {author}
  {\bibfnamefont {C.~A.}\ \bibnamefont {Muschik}}, \bibinfo {author}
  {\bibfnamefont {P.}~\bibnamefont {Silvi}}, \bibinfo {author} {\bibfnamefont
  {R.}~\bibnamefont {Blatt}}, \bibinfo {author} {\bibfnamefont {C.~F.}\
  \bibnamefont {Roos}},\ and\ \bibinfo {author} {\bibfnamefont
  {P.}~\bibnamefont {Zoller}},\ }\bibfield  {title} {\bibinfo {title}
  {Self-verifying variational quantum simulation of lattice models},\ }\href
  {https://doi.org/10.1038/s41586-019-1177-4} {\bibfield  {journal} {\bibinfo
  {journal} {Nature}\ }\textbf {\bibinfo {volume} {569}},\ \bibinfo {pages}
  {355} (\bibinfo {year} {2019})}\BibitemShut {NoStop}%
\bibitem [{\citenamefont {Piñeiro}\ \emph {et~al.}(2019)\citenamefont
  {Piñeiro}, \citenamefont {Genkina}, \citenamefont {Lu},\ and\ \citenamefont
  {Spielman}}]{Pineiro_2019}%
  \BibitemOpen
  \bibfield  {author} {\bibinfo {author} {\bibfnamefont {A.~M.}\ \bibnamefont
  {Piñeiro}}, \bibinfo {author} {\bibfnamefont {D.}~\bibnamefont {Genkina}},
  \bibinfo {author} {\bibfnamefont {M.}~\bibnamefont {Lu}},\ and\ \bibinfo
  {author} {\bibfnamefont {I.~B.}\ \bibnamefont {Spielman}},\ }\bibfield
  {title} {\bibinfo {title} {Sauter–schwinger effect with a quantum gas},\
  }\href {https://doi.org/10.1088/1367-2630/ab3840} {\bibfield  {journal}
  {\bibinfo  {journal} {New Journal of Physics}\ }\textbf {\bibinfo {volume}
  {21}},\ \bibinfo {pages} {083035} (\bibinfo {year} {2019})}\BibitemShut
  {NoStop}%
\bibitem [{\citenamefont {Huerta~Alderete}\ \emph {et~al.}(2020)\citenamefont
  {Huerta~Alderete}, \citenamefont {Singh}, \citenamefont {Nguyen},
  \citenamefont {Zhu}, \citenamefont {Balu}, \citenamefont {Monroe},
  \citenamefont {Chandrashekar},\ and\ \citenamefont
  {Linke}}]{HuertaAlderete2020}%
  \BibitemOpen
  \bibfield  {author} {\bibinfo {author} {\bibfnamefont {C.}~\bibnamefont
  {Huerta~Alderete}}, \bibinfo {author} {\bibfnamefont {S.}~\bibnamefont
  {Singh}}, \bibinfo {author} {\bibfnamefont {N.~H.}\ \bibnamefont {Nguyen}},
  \bibinfo {author} {\bibfnamefont {D.}~\bibnamefont {Zhu}}, \bibinfo {author}
  {\bibfnamefont {R.}~\bibnamefont {Balu}}, \bibinfo {author} {\bibfnamefont
  {C.}~\bibnamefont {Monroe}}, \bibinfo {author} {\bibfnamefont {C.~M.}\
  \bibnamefont {Chandrashekar}},\ and\ \bibinfo {author} {\bibfnamefont
  {N.~M.}\ \bibnamefont {Linke}},\ }\bibfield  {title} {\bibinfo {title}
  {Quantum walks and dirac cellular automata on a programmable trapped-ion
  quantum computer},\ }\href {https://doi.org/10.1038/s41467-020-17519-4}
  {\bibfield  {journal} {\bibinfo  {journal} {Nature Communications}\ }\textbf
  {\bibinfo {volume} {11}},\ \bibinfo {pages} {3720} (\bibinfo {year}
  {2020})}\BibitemShut {NoStop}%
\bibitem [{\citenamefont {Liang}\ \emph {et~al.}(2023)\citenamefont {Liang},
  \citenamefont {Wei}, \citenamefont {Zhang}, \citenamefont {Wang},
  \citenamefont {Zhang}, \citenamefont {Wang}, \citenamefont {Qi},
  \citenamefont {Liu},\ and\ \citenamefont
  {Zhang}}]{PhysRevResearch.5.L012006}%
  \BibitemOpen
  \bibfield  {author} {\bibinfo {author} {\bibfnamefont {M.-C.}\ \bibnamefont
  {Liang}}, \bibinfo {author} {\bibfnamefont {Y.-D.}\ \bibnamefont {Wei}},
  \bibinfo {author} {\bibfnamefont {L.}~\bibnamefont {Zhang}}, \bibinfo
  {author} {\bibfnamefont {X.-J.}\ \bibnamefont {Wang}}, \bibinfo {author}
  {\bibfnamefont {H.}~\bibnamefont {Zhang}}, \bibinfo {author} {\bibfnamefont
  {W.-W.}\ \bibnamefont {Wang}}, \bibinfo {author} {\bibfnamefont
  {W.}~\bibnamefont {Qi}}, \bibinfo {author} {\bibfnamefont {X.-J.}\
  \bibnamefont {Liu}},\ and\ \bibinfo {author} {\bibfnamefont {X.}~\bibnamefont
  {Zhang}},\ }\bibfield  {title} {\bibinfo {title} {Realization of qi-wu-zhang
  model in spin-orbit-coupled ultracold fermions},\ }\href
  {https://doi.org/10.1103/PhysRevResearch.5.L012006} {\bibfield  {journal}
  {\bibinfo  {journal} {Phys. Rev. Res.}\ }\textbf {\bibinfo {volume} {5}},\
  \bibinfo {pages} {L012006} (\bibinfo {year} {2023})}\BibitemShut {NoStop}%
\bibitem [{\citenamefont {Surace}\ \emph {et~al.}(2020)\citenamefont {Surace},
  \citenamefont {Mazza}, \citenamefont {Giudici}, \citenamefont {Lerose},
  \citenamefont {Gambassi},\ and\ \citenamefont
  {Dalmonte}}]{PhysRevX.10.021041}%
  \BibitemOpen
  \bibfield  {author} {\bibinfo {author} {\bibfnamefont {F.~M.}\ \bibnamefont
  {Surace}}, \bibinfo {author} {\bibfnamefont {P.~P.}\ \bibnamefont {Mazza}},
  \bibinfo {author} {\bibfnamefont {G.}~\bibnamefont {Giudici}}, \bibinfo
  {author} {\bibfnamefont {A.}~\bibnamefont {Lerose}}, \bibinfo {author}
  {\bibfnamefont {A.}~\bibnamefont {Gambassi}},\ and\ \bibinfo {author}
  {\bibfnamefont {M.}~\bibnamefont {Dalmonte}},\ }\bibfield  {title} {\bibinfo
  {title} {Lattice gauge theories and string dynamics in rydberg atom quantum
  simulators},\ }\href {https://doi.org/10.1103/PhysRevX.10.021041} {\bibfield
  {journal} {\bibinfo  {journal} {Phys. Rev. X}\ }\textbf {\bibinfo {volume}
  {10}},\ \bibinfo {pages} {021041} (\bibinfo {year} {2020})}\BibitemShut
  {NoStop}%
\bibitem [{\citenamefont {Klco}\ \emph {et~al.}(2020)\citenamefont {Klco},
  \citenamefont {Savage},\ and\ \citenamefont {Stryker}}]{PhysRevD.101.074512}%
  \BibitemOpen
  \bibfield  {author} {\bibinfo {author} {\bibfnamefont {N.}~\bibnamefont
  {Klco}}, \bibinfo {author} {\bibfnamefont {M.~J.}\ \bibnamefont {Savage}},\
  and\ \bibinfo {author} {\bibfnamefont {J.~R.}\ \bibnamefont {Stryker}},\
  }\bibfield  {title} {\bibinfo {title} {Su(2) non-abelian gauge field theory
  in one dimension on digital quantum computers},\ }\href
  {https://doi.org/10.1103/PhysRevD.101.074512} {\bibfield  {journal} {\bibinfo
   {journal} {Phys. Rev. D}\ }\textbf {\bibinfo {volume} {101}},\ \bibinfo
  {pages} {074512} (\bibinfo {year} {2020})}\BibitemShut {NoStop}%
\bibitem [{\citenamefont {Mil}\ \emph {et~al.}(2020)\citenamefont {Mil},
  \citenamefont {Zache}, \citenamefont {Hegde}, \citenamefont {Xia},
  \citenamefont {Bhatt}, \citenamefont {Oberthaler}, \citenamefont {Hauke},
  \citenamefont {Berges},\ and\ \citenamefont {Jendrzejewski}}]{Mil1128}%
  \BibitemOpen
  \bibfield  {author} {\bibinfo {author} {\bibfnamefont {A.}~\bibnamefont
  {Mil}}, \bibinfo {author} {\bibfnamefont {T.~V.}\ \bibnamefont {Zache}},
  \bibinfo {author} {\bibfnamefont {A.}~\bibnamefont {Hegde}}, \bibinfo
  {author} {\bibfnamefont {A.}~\bibnamefont {Xia}}, \bibinfo {author}
  {\bibfnamefont {R.~P.}\ \bibnamefont {Bhatt}}, \bibinfo {author}
  {\bibfnamefont {M.~K.}\ \bibnamefont {Oberthaler}}, \bibinfo {author}
  {\bibfnamefont {P.}~\bibnamefont {Hauke}}, \bibinfo {author} {\bibfnamefont
  {J.}~\bibnamefont {Berges}},\ and\ \bibinfo {author} {\bibfnamefont
  {F.}~\bibnamefont {Jendrzejewski}},\ }\bibfield  {title} {\bibinfo {title} {A
  scalable realization of local u(1) gauge invariance in cold atomic
  mixtures},\ }\href {https://doi.org/10.1126/science.aaz5312} {\bibfield
  {journal} {\bibinfo  {journal} {Science}\ }\textbf {\bibinfo {volume}
  {367}},\ \bibinfo {pages} {1128} (\bibinfo {year} {2020})}\BibitemShut
  {NoStop}%
\bibitem [{\citenamefont {Yang}\ \emph {et~al.}(2020)\citenamefont {Yang},
  \citenamefont {Sun}, \citenamefont {Ott}, \citenamefont {Wang}, \citenamefont
  {Zache}, \citenamefont {Halimeh}, \citenamefont {Yuan}, \citenamefont
  {Hauke},\ and\ \citenamefont {Pan}}]{Yang2020}%
  \BibitemOpen
  \bibfield  {author} {\bibinfo {author} {\bibfnamefont {B.}~\bibnamefont
  {Yang}}, \bibinfo {author} {\bibfnamefont {H.}~\bibnamefont {Sun}}, \bibinfo
  {author} {\bibfnamefont {R.}~\bibnamefont {Ott}}, \bibinfo {author}
  {\bibfnamefont {H.-Y.}\ \bibnamefont {Wang}}, \bibinfo {author}
  {\bibfnamefont {T.~V.}\ \bibnamefont {Zache}}, \bibinfo {author}
  {\bibfnamefont {J.~C.}\ \bibnamefont {Halimeh}}, \bibinfo {author}
  {\bibfnamefont {Z.-S.}\ \bibnamefont {Yuan}}, \bibinfo {author}
  {\bibfnamefont {P.}~\bibnamefont {Hauke}},\ and\ \bibinfo {author}
  {\bibfnamefont {J.-W.}\ \bibnamefont {Pan}},\ }\bibfield  {title} {\bibinfo
  {title} {Observation of gauge invariance in a 71-site bose--hubbard quantum
  simulator},\ }\href {https://doi.org/10.1038/s41586-020-2910-8} {\bibfield
  {journal} {\bibinfo  {journal} {Nature}\ }\textbf {\bibinfo {volume} {587}},\
  \bibinfo {pages} {392} (\bibinfo {year} {2020})}\BibitemShut {NoStop}%
\bibitem [{\citenamefont {Zhou}\ \emph {et~al.}(2022)\citenamefont {Zhou},
  \citenamefont {Su}, \citenamefont {Halimeh}, \citenamefont {Ott},
  \citenamefont {Sun}, \citenamefont {Hauke}, \citenamefont {Yang},
  \citenamefont {Yuan}, \citenamefont {Berges},\ and\ \citenamefont
  {Pan}}]{doi:10.1126/science.abl6277}%
  \BibitemOpen
  \bibfield  {author} {\bibinfo {author} {\bibfnamefont {Z.-Y.}\ \bibnamefont
  {Zhou}}, \bibinfo {author} {\bibfnamefont {G.-X.}\ \bibnamefont {Su}},
  \bibinfo {author} {\bibfnamefont {J.~C.}\ \bibnamefont {Halimeh}}, \bibinfo
  {author} {\bibfnamefont {R.}~\bibnamefont {Ott}}, \bibinfo {author}
  {\bibfnamefont {H.}~\bibnamefont {Sun}}, \bibinfo {author} {\bibfnamefont
  {P.}~\bibnamefont {Hauke}}, \bibinfo {author} {\bibfnamefont
  {B.}~\bibnamefont {Yang}}, \bibinfo {author} {\bibfnamefont {Z.-S.}\
  \bibnamefont {Yuan}}, \bibinfo {author} {\bibfnamefont {J.}~\bibnamefont
  {Berges}},\ and\ \bibinfo {author} {\bibfnamefont {J.-W.}\ \bibnamefont
  {Pan}},\ }\bibfield  {title} {\bibinfo {title} {Thermalization dynamics of a
  gauge theory on a quantum simulator},\ }\href
  {https://doi.org/10.1126/science.abl6277} {\bibfield  {journal} {\bibinfo
  {journal} {Science}\ }\textbf {\bibinfo {volume} {377}},\ \bibinfo {pages}
  {311} (\bibinfo {year} {2022})}\BibitemShut {NoStop}%
\bibitem [{\citenamefont {Atas}\ \emph {et~al.}(2021)\citenamefont {Atas},
  \citenamefont {Zhang}, \citenamefont {Lewis}, \citenamefont {Jahanpour},
  \citenamefont {Haase},\ and\ \citenamefont {Muschik}}]{Atas2021}%
  \BibitemOpen
  \bibfield  {author} {\bibinfo {author} {\bibfnamefont {Y.~Y.}\ \bibnamefont
  {Atas}}, \bibinfo {author} {\bibfnamefont {J.}~\bibnamefont {Zhang}},
  \bibinfo {author} {\bibfnamefont {R.}~\bibnamefont {Lewis}}, \bibinfo
  {author} {\bibfnamefont {A.}~\bibnamefont {Jahanpour}}, \bibinfo {author}
  {\bibfnamefont {J.~F.}\ \bibnamefont {Haase}},\ and\ \bibinfo {author}
  {\bibfnamefont {C.~A.}\ \bibnamefont {Muschik}},\ }\bibfield  {title}
  {\bibinfo {title} {Su(2) hadrons on a quantum computer via a variational
  approach},\ }\href {https://doi.org/10.1038/s41467-021-26825-4} {\bibfield
  {journal} {\bibinfo  {journal} {Nature Communications}\ }\textbf {\bibinfo
  {volume} {12}},\ \bibinfo {pages} {6499} (\bibinfo {year}
  {2021})}\BibitemShut {NoStop}%
\bibitem [{\citenamefont {Bauer}\ \emph {et~al.}(2021)\citenamefont {Bauer},
  \citenamefont {Nachman},\ and\ \citenamefont
  {Freytsis}}]{PhysRevLett.127.212001}%
  \BibitemOpen
  \bibfield  {author} {\bibinfo {author} {\bibfnamefont {C.~W.}\ \bibnamefont
  {Bauer}}, \bibinfo {author} {\bibfnamefont {B.}~\bibnamefont {Nachman}},\
  and\ \bibinfo {author} {\bibfnamefont {M.}~\bibnamefont {Freytsis}},\
  }\bibfield  {title} {\bibinfo {title} {Simulating collider physics on quantum
  computers using effective field theories},\ }\href
  {https://doi.org/10.1103/PhysRevLett.127.212001} {\bibfield  {journal}
  {\bibinfo  {journal} {Phys. Rev. Lett.}\ }\textbf {\bibinfo {volume} {127}},\
  \bibinfo {pages} {212001} (\bibinfo {year} {2021})}\BibitemShut {NoStop}%
\bibitem [{\citenamefont {Nguyen}\ \emph {et~al.}(2022)\citenamefont {Nguyen},
  \citenamefont {Tran}, \citenamefont {Zhu}, \citenamefont {Green},
  \citenamefont {Alderete}, \citenamefont {Davoudi},\ and\ \citenamefont
  {Linke}}]{PRXQuantum.3.020324}%
  \BibitemOpen
  \bibfield  {author} {\bibinfo {author} {\bibfnamefont {N.~H.}\ \bibnamefont
  {Nguyen}}, \bibinfo {author} {\bibfnamefont {M.~C.}\ \bibnamefont {Tran}},
  \bibinfo {author} {\bibfnamefont {Y.}~\bibnamefont {Zhu}}, \bibinfo {author}
  {\bibfnamefont {A.~M.}\ \bibnamefont {Green}}, \bibinfo {author}
  {\bibfnamefont {C.~H.}\ \bibnamefont {Alderete}}, \bibinfo {author}
  {\bibfnamefont {Z.}~\bibnamefont {Davoudi}},\ and\ \bibinfo {author}
  {\bibfnamefont {N.~M.}\ \bibnamefont {Linke}},\ }\bibfield  {title} {\bibinfo
  {title} {Digital quantum simulation of the schwinger model and symmetry
  protection with trapped ions},\ }\href
  {https://doi.org/10.1103/PRXQuantum.3.020324} {\bibfield  {journal} {\bibinfo
   {journal} {PRX Quantum}\ }\textbf {\bibinfo {volume} {3}},\ \bibinfo {pages}
  {020324} (\bibinfo {year} {2022})}\BibitemShut {NoStop}%
\bibitem [{\citenamefont {Ciavarella}\ \emph {et~al.}(2021)\citenamefont
  {Ciavarella}, \citenamefont {Klco},\ and\ \citenamefont
  {Savage}}]{PhysRevD.103.094501}%
  \BibitemOpen
  \bibfield  {author} {\bibinfo {author} {\bibfnamefont {A.}~\bibnamefont
  {Ciavarella}}, \bibinfo {author} {\bibfnamefont {N.}~\bibnamefont {Klco}},\
  and\ \bibinfo {author} {\bibfnamefont {M.~J.}\ \bibnamefont {Savage}},\
  }\bibfield  {title} {\bibinfo {title} {Trailhead for quantum simulation of
  su(3) yang-mills lattice gauge theory in the local multiplet basis},\ }\href
  {https://doi.org/10.1103/PhysRevD.103.094501} {\bibfield  {journal} {\bibinfo
   {journal} {Phys. Rev. D}\ }\textbf {\bibinfo {volume} {103}},\ \bibinfo
  {pages} {094501} (\bibinfo {year} {2021})}\BibitemShut {NoStop}%
\bibitem [{\citenamefont {A~Rahman}\ \emph {et~al.}(2021)\citenamefont
  {A~Rahman}, \citenamefont {Lewis}, \citenamefont {Mendicelli},\ and\
  \citenamefont {Powell}}]{PhysRevD.104.034501}%
  \BibitemOpen
  \bibfield  {author} {\bibinfo {author} {\bibfnamefont {S.}~\bibnamefont
  {A~Rahman}}, \bibinfo {author} {\bibfnamefont {R.}~\bibnamefont {Lewis}},
  \bibinfo {author} {\bibfnamefont {E.}~\bibnamefont {Mendicelli}},\ and\
  \bibinfo {author} {\bibfnamefont {S.}~\bibnamefont {Powell}},\ }\bibfield
  {title} {\bibinfo {title} {Su(2) lattice gauge theory on a quantum
  annealer},\ }\href {https://doi.org/10.1103/PhysRevD.104.034501} {\bibfield
  {journal} {\bibinfo  {journal} {Phys. Rev. D}\ }\textbf {\bibinfo {volume}
  {104}},\ \bibinfo {pages} {034501} (\bibinfo {year} {2021})}\BibitemShut
  {NoStop}%
\bibitem [{\citenamefont {Ciavarella}\ and\ \citenamefont
  {Chernyshev}(2022)}]{PhysRevD.105.074504}%
  \BibitemOpen
  \bibfield  {author} {\bibinfo {author} {\bibfnamefont {A.~N.}\ \bibnamefont
  {Ciavarella}}\ and\ \bibinfo {author} {\bibfnamefont {I.~A.}\ \bibnamefont
  {Chernyshev}},\ }\bibfield  {title} {\bibinfo {title} {Preparation of the
  su(3) lattice yang-mills vacuum with variational quantum methods},\ }\href
  {https://doi.org/10.1103/PhysRevD.105.074504} {\bibfield  {journal} {\bibinfo
   {journal} {Phys. Rev. D}\ }\textbf {\bibinfo {volume} {105}},\ \bibinfo
  {pages} {074504} (\bibinfo {year} {2022})}\BibitemShut {NoStop}%
\bibitem [{\citenamefont {Lu}\ \emph {et~al.}(2022)\citenamefont {Lu},
  \citenamefont {Reid}, \citenamefont {Fritsch}, \citenamefont {Pi\~neiro},\
  and\ \citenamefont {Spielman}}]{PhysRevLett.129.040402}%
  \BibitemOpen
  \bibfield  {author} {\bibinfo {author} {\bibfnamefont {M.}~\bibnamefont
  {Lu}}, \bibinfo {author} {\bibfnamefont {G.~H.}\ \bibnamefont {Reid}},
  \bibinfo {author} {\bibfnamefont {A.~R.}\ \bibnamefont {Fritsch}}, \bibinfo
  {author} {\bibfnamefont {A.~M.}\ \bibnamefont {Pi\~neiro}},\ and\ \bibinfo
  {author} {\bibfnamefont {I.~B.}\ \bibnamefont {Spielman}},\ }\bibfield
  {title} {\bibinfo {title} {Floquet engineering topological dirac bands},\
  }\href {https://doi.org/10.1103/PhysRevLett.129.040402} {\bibfield  {journal}
  {\bibinfo  {journal} {Phys. Rev. Lett.}\ }\textbf {\bibinfo {volume} {129}},\
  \bibinfo {pages} {040402} (\bibinfo {year} {2022})}\BibitemShut {NoStop}%
\bibitem [{\citenamefont {Wang}\ \emph {et~al.}(2022)\citenamefont {Wang},
  \citenamefont {Ge}, \citenamefont {Xiang}, \citenamefont {Song},
  \citenamefont {Huang}, \citenamefont {Song}, \citenamefont {Guo},
  \citenamefont {Su}, \citenamefont {Xu}, \citenamefont {Zheng},\ and\
  \citenamefont {Fan}}]{PhysRevResearch.4.L022060}%
  \BibitemOpen
  \bibfield  {author} {\bibinfo {author} {\bibfnamefont {Z.}~\bibnamefont
  {Wang}}, \bibinfo {author} {\bibfnamefont {Z.-Y.}\ \bibnamefont {Ge}},
  \bibinfo {author} {\bibfnamefont {Z.}~\bibnamefont {Xiang}}, \bibinfo
  {author} {\bibfnamefont {X.}~\bibnamefont {Song}}, \bibinfo {author}
  {\bibfnamefont {R.-Z.}\ \bibnamefont {Huang}}, \bibinfo {author}
  {\bibfnamefont {P.}~\bibnamefont {Song}}, \bibinfo {author} {\bibfnamefont
  {X.-Y.}\ \bibnamefont {Guo}}, \bibinfo {author} {\bibfnamefont
  {L.}~\bibnamefont {Su}}, \bibinfo {author} {\bibfnamefont {K.}~\bibnamefont
  {Xu}}, \bibinfo {author} {\bibfnamefont {D.}~\bibnamefont {Zheng}},\ and\
  \bibinfo {author} {\bibfnamefont {H.}~\bibnamefont {Fan}},\ }\bibfield
  {title} {\bibinfo {title} {Observation of emergent ${\mathbb{z}}_{2}$ gauge
  invariance in a superconducting circuit},\ }\href
  {https://doi.org/10.1103/PhysRevResearch.4.L022060} {\bibfield  {journal}
  {\bibinfo  {journal} {Phys. Rev. Research}\ }\textbf {\bibinfo {volume}
  {4}},\ \bibinfo {pages} {L022060} (\bibinfo {year} {2022})}\BibitemShut
  {NoStop}%
\bibitem [{\citenamefont {Atas}\ \emph {et~al.}(2022)\citenamefont {Atas},
  \citenamefont {Haase}, \citenamefont {Zhang}, \citenamefont {Wei},
  \citenamefont {Pfaendler}, \citenamefont {Lewis},\ and\ \citenamefont
  {Muschik}}]{https://doi.org/10.48550/arxiv.2207.03473}%
  \BibitemOpen
  \bibfield  {author} {\bibinfo {author} {\bibfnamefont {Y.~Y.}\ \bibnamefont
  {Atas}}, \bibinfo {author} {\bibfnamefont {J.~F.}\ \bibnamefont {Haase}},
  \bibinfo {author} {\bibfnamefont {J.}~\bibnamefont {Zhang}}, \bibinfo
  {author} {\bibfnamefont {V.}~\bibnamefont {Wei}}, \bibinfo {author}
  {\bibfnamefont {S.~M.~L.}\ \bibnamefont {Pfaendler}}, \bibinfo {author}
  {\bibfnamefont {R.}~\bibnamefont {Lewis}},\ and\ \bibinfo {author}
  {\bibfnamefont {C.~A.}\ \bibnamefont {Muschik}},\ }\href
  {https://doi.org/10.48550/ARXIV.2207.03473} {\bibinfo {title} {Real-time
  evolution of su(3) hadrons on a quantum computer}} (\bibinfo {year}
  {2022})\BibitemShut {NoStop}%
\bibitem [{\citenamefont {Farrell}\ \emph
  {et~al.}(2022{\natexlab{a}})\citenamefont {Farrell}, \citenamefont
  {Chernyshev}, \citenamefont {Powell}, \citenamefont {Zemlevskiy},
  \citenamefont {Illa},\ and\ \citenamefont
  {Savage}}]{https://doi.org/10.48550/arxiv.2207.01731}%
  \BibitemOpen
  \bibfield  {author} {\bibinfo {author} {\bibfnamefont {R.~C.}\ \bibnamefont
  {Farrell}}, \bibinfo {author} {\bibfnamefont {I.~A.}\ \bibnamefont
  {Chernyshev}}, \bibinfo {author} {\bibfnamefont {S.~J.~M.}\ \bibnamefont
  {Powell}}, \bibinfo {author} {\bibfnamefont {N.~A.}\ \bibnamefont
  {Zemlevskiy}}, \bibinfo {author} {\bibfnamefont {M.}~\bibnamefont {Illa}},\
  and\ \bibinfo {author} {\bibfnamefont {M.~J.}\ \bibnamefont {Savage}},\
  }\href {https://doi.org/10.48550/ARXIV.2207.01731} {\bibinfo {title}
  {Preparations for quantum simulations of quantum chromodynamics in 1+1
  dimensions: (i) axial gauge}} (\bibinfo {year}
  {2022}{\natexlab{a}})\BibitemShut {NoStop}%
\bibitem [{\citenamefont {Mildenberger}\ \emph {et~al.}(2022)\citenamefont
  {Mildenberger}, \citenamefont {Mruczkiewicz}, \citenamefont {Halimeh},
  \citenamefont {Jiang},\ and\ \citenamefont
  {Hauke}}]{mildenberger2022probing}%
  \BibitemOpen
  \bibfield  {author} {\bibinfo {author} {\bibfnamefont {J.}~\bibnamefont
  {Mildenberger}}, \bibinfo {author} {\bibfnamefont {W.}~\bibnamefont
  {Mruczkiewicz}}, \bibinfo {author} {\bibfnamefont {J.~C.}\ \bibnamefont
  {Halimeh}}, \bibinfo {author} {\bibfnamefont {Z.}~\bibnamefont {Jiang}},\
  and\ \bibinfo {author} {\bibfnamefont {P.}~\bibnamefont {Hauke}},\
  }\href@noop {} {\bibinfo {title} {Probing confinement in a $\mathbb{Z}_2$
  lattice gauge theory on a quantum computer}} (\bibinfo {year} {2022}),\
  \Eprint {https://arxiv.org/abs/2203.08905} {arXiv:2203.08905 [quant-ph]}
  \BibitemShut {NoStop}%
\bibitem [{\citenamefont {Farrell}\ \emph
  {et~al.}(2022{\natexlab{b}})\citenamefont {Farrell}, \citenamefont
  {Chernyshev}, \citenamefont {Powell}, \citenamefont {Zemlevskiy},
  \citenamefont {Illa},\ and\ \citenamefont
  {Savage}}]{https://doi.org/10.48550/arxiv.2209.10781}%
  \BibitemOpen
  \bibfield  {author} {\bibinfo {author} {\bibfnamefont {R.~C.}\ \bibnamefont
  {Farrell}}, \bibinfo {author} {\bibfnamefont {I.~A.}\ \bibnamefont
  {Chernyshev}}, \bibinfo {author} {\bibfnamefont {S.~J.~M.}\ \bibnamefont
  {Powell}}, \bibinfo {author} {\bibfnamefont {N.~A.}\ \bibnamefont
  {Zemlevskiy}}, \bibinfo {author} {\bibfnamefont {M.}~\bibnamefont {Illa}},\
  and\ \bibinfo {author} {\bibfnamefont {M.~J.}\ \bibnamefont {Savage}},\
  }\href {https://doi.org/10.48550/ARXIV.2209.10781} {\bibinfo {title}
  {Preparations for quantum simulations of quantum chromodynamics in 1+1
  dimensions: (ii) single-baryon $\beta$-decay in real time}} (\bibinfo {year}
  {2022}{\natexlab{b}})\BibitemShut {NoStop}%
\bibitem [{\citenamefont {Fr{\"o}lian}\ \emph {et~al.}(2022)\citenamefont
  {Fr{\"o}lian}, \citenamefont {Chisholm}, \citenamefont {Neri}, \citenamefont
  {Cabrera}, \citenamefont {Ramos}, \citenamefont {Celi},\ and\ \citenamefont
  {Tarruell}}]{Frolian2022}%
  \BibitemOpen
  \bibfield  {author} {\bibinfo {author} {\bibfnamefont {A.}~\bibnamefont
  {Fr{\"o}lian}}, \bibinfo {author} {\bibfnamefont {C.~S.}\ \bibnamefont
  {Chisholm}}, \bibinfo {author} {\bibfnamefont {E.}~\bibnamefont {Neri}},
  \bibinfo {author} {\bibfnamefont {C.~R.}\ \bibnamefont {Cabrera}}, \bibinfo
  {author} {\bibfnamefont {R.}~\bibnamefont {Ramos}}, \bibinfo {author}
  {\bibfnamefont {A.}~\bibnamefont {Celi}},\ and\ \bibinfo {author}
  {\bibfnamefont {L.}~\bibnamefont {Tarruell}},\ }\bibfield  {title} {\bibinfo
  {title} {Realizing a 1d topological gauge theory in an optically dressed
  bec},\ }\href {https://doi.org/10.1038/s41586-022-04943-3} {\bibfield
  {journal} {\bibinfo  {journal} {Nature}\ }\textbf {\bibinfo {volume} {608}},\
  \bibinfo {pages} {293} (\bibinfo {year} {2022})}\BibitemShut {NoStop}%
\bibitem [{\citenamefont {Su}\ \emph {et~al.}(2023)\citenamefont {Su},
  \citenamefont {Sun}, \citenamefont {Hudomal}, \citenamefont {Desaules},
  \citenamefont {Zhou}, \citenamefont {Yang}, \citenamefont {Halimeh},
  \citenamefont {Yuan}, \citenamefont {Papi\ifmmode~\acute{c}\else
  \'{c}\fi{}},\ and\ \citenamefont {Pan}}]{PhysRevResearch.5.023010}%
  \BibitemOpen
  \bibfield  {author} {\bibinfo {author} {\bibfnamefont {G.-X.}\ \bibnamefont
  {Su}}, \bibinfo {author} {\bibfnamefont {H.}~\bibnamefont {Sun}}, \bibinfo
  {author} {\bibfnamefont {A.}~\bibnamefont {Hudomal}}, \bibinfo {author}
  {\bibfnamefont {J.-Y.}\ \bibnamefont {Desaules}}, \bibinfo {author}
  {\bibfnamefont {Z.-Y.}\ \bibnamefont {Zhou}}, \bibinfo {author}
  {\bibfnamefont {B.}~\bibnamefont {Yang}}, \bibinfo {author} {\bibfnamefont
  {J.~C.}\ \bibnamefont {Halimeh}}, \bibinfo {author} {\bibfnamefont {Z.-S.}\
  \bibnamefont {Yuan}}, \bibinfo {author} {\bibfnamefont {Z.}~\bibnamefont
  {Papi\ifmmode~\acute{c}\else \'{c}\fi{}}},\ and\ \bibinfo {author}
  {\bibfnamefont {J.-W.}\ \bibnamefont {Pan}},\ }\bibfield  {title} {\bibinfo
  {title} {Observation of many-body scarring in a bose-hubbard quantum
  simulator},\ }\href {https://doi.org/10.1103/PhysRevResearch.5.023010}
  {\bibfield  {journal} {\bibinfo  {journal} {Phys. Rev. Res.}\ }\textbf
  {\bibinfo {volume} {5}},\ \bibinfo {pages} {023010} (\bibinfo {year}
  {2023})}\BibitemShut {NoStop}%
\bibitem [{\citenamefont {Charles}\ \emph {et~al.}(2023)\citenamefont
  {Charles}, \citenamefont {Gustafson}, \citenamefont {Hardt}, \citenamefont
  {Herren}, \citenamefont {Hogan}, \citenamefont {Lamm}, \citenamefont
  {Starecheski}, \citenamefont {de~Water},\ and\ \citenamefont
  {Wagman}}]{charles2023simulating}%
  \BibitemOpen
  \bibfield  {author} {\bibinfo {author} {\bibfnamefont {C.}~\bibnamefont
  {Charles}}, \bibinfo {author} {\bibfnamefont {E.~J.}\ \bibnamefont
  {Gustafson}}, \bibinfo {author} {\bibfnamefont {E.}~\bibnamefont {Hardt}},
  \bibinfo {author} {\bibfnamefont {F.}~\bibnamefont {Herren}}, \bibinfo
  {author} {\bibfnamefont {N.}~\bibnamefont {Hogan}}, \bibinfo {author}
  {\bibfnamefont {H.}~\bibnamefont {Lamm}}, \bibinfo {author} {\bibfnamefont
  {S.}~\bibnamefont {Starecheski}}, \bibinfo {author} {\bibfnamefont
  {R.~S.~V.}\ \bibnamefont {de~Water}},\ and\ \bibinfo {author} {\bibfnamefont
  {M.~L.}\ \bibnamefont {Wagman}},\ }\href@noop {} {\bibinfo {title}
  {Simulating $\mathbb{Z}_2$ lattice gauge theory on a quantum computer}}
  (\bibinfo {year} {2023}),\ \Eprint {https://arxiv.org/abs/2305.02361}
  {arXiv:2305.02361 [hep-lat]} \BibitemShut {NoStop}%
\bibitem [{\citenamefont {Zhang}\ \emph {et~al.}(2023)\citenamefont {Zhang},
  \citenamefont {Liu}, \citenamefont {Cheng}, \citenamefont {He}, \citenamefont
  {Wang}, \citenamefont {Wang}, \citenamefont {Zhu}, \citenamefont {Su},
  \citenamefont {Zhou}, \citenamefont {Zheng}, \citenamefont {Sun},
  \citenamefont {Yang}, \citenamefont {Hauke}, \citenamefont {Zheng},
  \citenamefont {Halimeh}, \citenamefont {Yuan},\ and\ \citenamefont
  {Pan}}]{zhang2023observation}%
  \BibitemOpen
  \bibfield  {author} {\bibinfo {author} {\bibfnamefont {W.-Y.}\ \bibnamefont
  {Zhang}}, \bibinfo {author} {\bibfnamefont {Y.}~\bibnamefont {Liu}}, \bibinfo
  {author} {\bibfnamefont {Y.}~\bibnamefont {Cheng}}, \bibinfo {author}
  {\bibfnamefont {M.-G.}\ \bibnamefont {He}}, \bibinfo {author} {\bibfnamefont
  {H.-Y.}\ \bibnamefont {Wang}}, \bibinfo {author} {\bibfnamefont {T.-Y.}\
  \bibnamefont {Wang}}, \bibinfo {author} {\bibfnamefont {Z.-H.}\ \bibnamefont
  {Zhu}}, \bibinfo {author} {\bibfnamefont {G.-X.}\ \bibnamefont {Su}},
  \bibinfo {author} {\bibfnamefont {Z.-Y.}\ \bibnamefont {Zhou}}, \bibinfo
  {author} {\bibfnamefont {Y.-G.}\ \bibnamefont {Zheng}}, \bibinfo {author}
  {\bibfnamefont {H.}~\bibnamefont {Sun}}, \bibinfo {author} {\bibfnamefont
  {B.}~\bibnamefont {Yang}}, \bibinfo {author} {\bibfnamefont {P.}~\bibnamefont
  {Hauke}}, \bibinfo {author} {\bibfnamefont {W.}~\bibnamefont {Zheng}},
  \bibinfo {author} {\bibfnamefont {J.~C.}\ \bibnamefont {Halimeh}}, \bibinfo
  {author} {\bibfnamefont {Z.-S.}\ \bibnamefont {Yuan}},\ and\ \bibinfo
  {author} {\bibfnamefont {J.-W.}\ \bibnamefont {Pan}},\ }\href@noop {}
  {\bibinfo {title} {Observation of microscopic confinement dynamics by a
  tunable topological $\theta$-angle}} (\bibinfo {year} {2023}),\ \Eprint
  {https://arxiv.org/abs/2306.11794} {arXiv:2306.11794 [cond-mat.quant-gas]}
  \BibitemShut {NoStop}%
\bibitem [{\citenamefont {Wilson}(1974)}]{PhysRevD.10.2445}%
  \BibitemOpen
  \bibfield  {author} {\bibinfo {author} {\bibfnamefont {K.~G.}\ \bibnamefont
  {Wilson}},\ }\bibfield  {title} {\bibinfo {title} {Confinement of quarks},\
  }\href {https://doi.org/10.1103/PhysRevD.10.2445} {\bibfield  {journal}
  {\bibinfo  {journal} {Phys. Rev. D}\ }\textbf {\bibinfo {volume} {10}},\
  \bibinfo {pages} {2445} (\bibinfo {year} {1974})}\BibitemShut {NoStop}%
\bibitem [{\citenamefont {Jaksch}\ and\ \citenamefont
  {Zoller}(2005)}]{Jaksch_2005}%
  \BibitemOpen
  \bibfield  {author} {\bibinfo {author} {\bibfnamefont {D.}~\bibnamefont
  {Jaksch}}\ and\ \bibinfo {author} {\bibfnamefont {P.}~\bibnamefont
  {Zoller}},\ }\bibfield  {title} {\bibinfo {title} {The cold atom hubbard
  toolbox},\ }\href {https://doi.org/https://doi.org/10.1016/j.aop.2004.09.010}
  {\bibfield  {journal} {\bibinfo  {journal} {Annals of Physics}\ }\textbf
  {\bibinfo {volume} {315}},\ \bibinfo {pages} {52} (\bibinfo {year} {2005})},\
  \bibinfo {note} {special Issue}\BibitemShut {NoStop}%
\bibitem [{\citenamefont {Lewenstein}\ \emph {et~al.}(2007)\citenamefont
  {Lewenstein}, \citenamefont {Sanpera}, \citenamefont {Ahufinger},
  \citenamefont {Damski}, \citenamefont {Sen(De)},\ and\ \citenamefont
  {Sen}}]{Lewenstein_2007}%
  \BibitemOpen
  \bibfield  {author} {\bibinfo {author} {\bibfnamefont {M.}~\bibnamefont
  {Lewenstein}}, \bibinfo {author} {\bibfnamefont {A.}~\bibnamefont {Sanpera}},
  \bibinfo {author} {\bibfnamefont {V.}~\bibnamefont {Ahufinger}}, \bibinfo
  {author} {\bibfnamefont {B.}~\bibnamefont {Damski}}, \bibinfo {author}
  {\bibfnamefont {A.}~\bibnamefont {Sen(De)}},\ and\ \bibinfo {author}
  {\bibfnamefont {U.}~\bibnamefont {Sen}},\ }\bibfield  {title} {\bibinfo
  {title} {Ultracold atomic gases in optical lattices: mimicking condensed
  matter physics and beyond},\ }\href
  {https://doi.org/10.1080/00018730701223200} {\bibfield  {journal} {\bibinfo
  {journal} {Advances in Physics}\ }\textbf {\bibinfo {volume} {56}},\ \bibinfo
  {pages} {243} (\bibinfo {year} {2007})}\BibitemShut {NoStop}%
\bibitem [{\citenamefont {Gross}\ and\ \citenamefont
  {Bloch}(2017)}]{Gross_2017}%
  \BibitemOpen
  \bibfield  {author} {\bibinfo {author} {\bibfnamefont {C.}~\bibnamefont
  {Gross}}\ and\ \bibinfo {author} {\bibfnamefont {I.}~\bibnamefont {Bloch}},\
  }\bibfield  {title} {\bibinfo {title} {Quantum simulations with ultracold
  atoms in optical lattices},\ }\href {https://doi.org/10.1126/science.aal3837}
  {\bibfield  {journal} {\bibinfo  {journal} {Science}\ }\textbf {\bibinfo
  {volume} {357}},\ \bibinfo {pages} {995} (\bibinfo {year}
  {2017})}\BibitemShut {NoStop}%
\bibitem [{\citenamefont {Chin}\ \emph {et~al.}(2010)\citenamefont {Chin},
  \citenamefont {Grimm}, \citenamefont {Julienne},\ and\ \citenamefont
  {Tiesinga}}]{RevModPhys.82.1225}%
  \BibitemOpen
  \bibfield  {author} {\bibinfo {author} {\bibfnamefont {C.}~\bibnamefont
  {Chin}}, \bibinfo {author} {\bibfnamefont {R.}~\bibnamefont {Grimm}},
  \bibinfo {author} {\bibfnamefont {P.}~\bibnamefont {Julienne}},\ and\
  \bibinfo {author} {\bibfnamefont {E.}~\bibnamefont {Tiesinga}},\ }\bibfield
  {title} {\bibinfo {title} {Feshbach resonances in ultracold gases},\ }\href
  {https://doi.org/10.1103/RevModPhys.82.1225} {\bibfield  {journal} {\bibinfo
  {journal} {Rev. Mod. Phys.}\ }\textbf {\bibinfo {volume} {82}},\ \bibinfo
  {pages} {1225} (\bibinfo {year} {2010})}\BibitemShut {NoStop}%
\bibitem [{\citenamefont {Cazalilla}\ and\ \citenamefont
  {Rey}(2014)}]{Cazalilla_2014}%
  \BibitemOpen
  \bibfield  {author} {\bibinfo {author} {\bibfnamefont {M.~A.}\ \bibnamefont
  {Cazalilla}}\ and\ \bibinfo {author} {\bibfnamefont {A.~M.}\ \bibnamefont
  {Rey}},\ }\bibfield  {title} {\bibinfo {title} {Ultracold fermi gases with
  emergent su(n) symmetry},\ }\href
  {https://doi.org/10.1088/0034-4885/77/12/124401} {\bibfield  {journal}
  {\bibinfo  {journal} {Reports on Progress in Physics}\ }\textbf {\bibinfo
  {volume} {77}},\ \bibinfo {pages} {124401} (\bibinfo {year}
  {2014})}\BibitemShut {NoStop}%
\bibitem [{\citenamefont {Capponi}\ \emph {et~al.}(2016)\citenamefont
  {Capponi}, \citenamefont {Lecheminant},\ and\ \citenamefont
  {Totsuka}}]{CAPPONI201650}%
  \BibitemOpen
  \bibfield  {author} {\bibinfo {author} {\bibfnamefont {S.}~\bibnamefont
  {Capponi}}, \bibinfo {author} {\bibfnamefont {P.}~\bibnamefont
  {Lecheminant}},\ and\ \bibinfo {author} {\bibfnamefont {K.}~\bibnamefont
  {Totsuka}},\ }\bibfield  {title} {\bibinfo {title} {Phases of one-dimensional
  su(n) cold atomic fermi gases—from molecular luttinger liquids to
  topological phases},\ }\href
  {https://doi.org/https://doi.org/10.1016/j.aop.2016.01.011} {\bibfield
  {journal} {\bibinfo  {journal} {Annals of Physics}\ }\textbf {\bibinfo
  {volume} {367}},\ \bibinfo {pages} {50} (\bibinfo {year} {2016})}\BibitemShut
  {NoStop}%
\bibitem [{\citenamefont {Gorshkov}\ \emph {et~al.}(2010)\citenamefont
  {Gorshkov}, \citenamefont {Hermele}, \citenamefont {Gurarie}, \citenamefont
  {Xu}, \citenamefont {Julienne}, \citenamefont {Ye}, \citenamefont {Zoller},
  \citenamefont {Demler}, \citenamefont {Lukin},\ and\ \citenamefont
  {Rey}}]{Gorshkov2010}%
  \BibitemOpen
  \bibfield  {author} {\bibinfo {author} {\bibfnamefont {A.~V.}\ \bibnamefont
  {Gorshkov}}, \bibinfo {author} {\bibfnamefont {M.}~\bibnamefont {Hermele}},
  \bibinfo {author} {\bibfnamefont {V.}~\bibnamefont {Gurarie}}, \bibinfo
  {author} {\bibfnamefont {C.}~\bibnamefont {Xu}}, \bibinfo {author}
  {\bibfnamefont {P.~S.}\ \bibnamefont {Julienne}}, \bibinfo {author}
  {\bibfnamefont {J.}~\bibnamefont {Ye}}, \bibinfo {author} {\bibfnamefont
  {P.}~\bibnamefont {Zoller}}, \bibinfo {author} {\bibfnamefont
  {E.}~\bibnamefont {Demler}}, \bibinfo {author} {\bibfnamefont {M.~D.}\
  \bibnamefont {Lukin}},\ and\ \bibinfo {author} {\bibfnamefont {A.~M.}\
  \bibnamefont {Rey}},\ }\bibfield  {title} {\bibinfo {title} {Two-orbital s
  u(n) magnetism with ultracold alkaline-earth atoms},\ }\href
  {https://doi.org/10.1038/nphys1535} {\bibfield  {journal} {\bibinfo
  {journal} {Nature Physics}\ }\textbf {\bibinfo {volume} {6}},\ \bibinfo
  {pages} {289} (\bibinfo {year} {2010})}\BibitemShut {NoStop}%
\bibitem [{\citenamefont {Ho}(1998)}]{PhysRevLett.81.742}%
  \BibitemOpen
  \bibfield  {author} {\bibinfo {author} {\bibfnamefont {T.-L.}\ \bibnamefont
  {Ho}},\ }\bibfield  {title} {\bibinfo {title} {Spinor bose condensates in
  optical traps},\ }\href {https://doi.org/10.1103/PhysRevLett.81.742}
  {\bibfield  {journal} {\bibinfo  {journal} {Phys. Rev. Lett.}\ }\textbf
  {\bibinfo {volume} {81}},\ \bibinfo {pages} {742} (\bibinfo {year}
  {1998})}\BibitemShut {NoStop}%
\bibitem [{\citenamefont {Weber}\ \emph {et~al.}(2003)\citenamefont {Weber},
  \citenamefont {Herbig}, \citenamefont {Mark}, \citenamefont {N{\"a}gerl},\
  and\ \citenamefont {Grimm}}]{Weber2003}%
  \BibitemOpen
  \bibfield  {author} {\bibinfo {author} {\bibfnamefont {T.}~\bibnamefont
  {Weber}}, \bibinfo {author} {\bibfnamefont {J.}~\bibnamefont {Herbig}},
  \bibinfo {author} {\bibfnamefont {M.}~\bibnamefont {Mark}}, \bibinfo {author}
  {\bibfnamefont {H.-C.}\ \bibnamefont {N{\"a}gerl}},\ and\ \bibinfo {author}
  {\bibfnamefont {R.}~\bibnamefont {Grimm}},\ }\bibfield  {title} {\bibinfo
  {title} {Bose-einstein condensation of cesium},\ }\href
  {https://doi.org/10.1126/science.1079699} {\bibfield  {journal} {\bibinfo
  {journal} {Science}\ }\textbf {\bibinfo {volume} {299}},\ \bibinfo {pages}
  {232} (\bibinfo {year} {2003})}\BibitemShut {NoStop}%
\bibitem [{\citenamefont {Haller}\ \emph {et~al.}(2009)\citenamefont {Haller},
  \citenamefont {Gustavsson}, \citenamefont {Mark}, \citenamefont {Danzl},
  \citenamefont {Hart}, \citenamefont {Pupillo},\ and\ \citenamefont
  {Nägerl}}]{doi:10.1126/science.1175850}%
  \BibitemOpen
  \bibfield  {author} {\bibinfo {author} {\bibfnamefont {E.}~\bibnamefont
  {Haller}}, \bibinfo {author} {\bibfnamefont {M.}~\bibnamefont {Gustavsson}},
  \bibinfo {author} {\bibfnamefont {M.~J.}\ \bibnamefont {Mark}}, \bibinfo
  {author} {\bibfnamefont {J.~G.}\ \bibnamefont {Danzl}}, \bibinfo {author}
  {\bibfnamefont {R.}~\bibnamefont {Hart}}, \bibinfo {author} {\bibfnamefont
  {G.}~\bibnamefont {Pupillo}},\ and\ \bibinfo {author} {\bibfnamefont {H.-C.}\
  \bibnamefont {Nägerl}},\ }\bibfield  {title} {\bibinfo {title} {Realization
  of an excited, strongly correlated quantum gas phase},\ }\href
  {https://doi.org/10.1126/science.1175850} {\bibfield  {journal} {\bibinfo
  {journal} {Science}\ }\textbf {\bibinfo {volume} {325}},\ \bibinfo {pages}
  {1224} (\bibinfo {year} {2009})}\BibitemShut {NoStop}%
\bibitem [{\citenamefont {Liu}\ \emph {et~al.}(2014{\natexlab{a}})\citenamefont
  {Liu}, \citenamefont {Law},\ and\ \citenamefont
  {Ng}}]{PhysRevLett.112.086401}%
  \BibitemOpen
  \bibfield  {author} {\bibinfo {author} {\bibfnamefont {X.-J.}\ \bibnamefont
  {Liu}}, \bibinfo {author} {\bibfnamefont {K.~T.}\ \bibnamefont {Law}},\ and\
  \bibinfo {author} {\bibfnamefont {T.~K.}\ \bibnamefont {Ng}},\ }\bibfield
  {title} {\bibinfo {title} {Realization of 2d spin-orbit interaction and
  exotic topological orders in cold atoms},\ }\href
  {https://doi.org/10.1103/PhysRevLett.112.086401} {\bibfield  {journal}
  {\bibinfo  {journal} {Phys. Rev. Lett.}\ }\textbf {\bibinfo {volume} {112}},\
  \bibinfo {pages} {086401} (\bibinfo {year} {2014}{\natexlab{a}})}\BibitemShut
  {NoStop}%
\bibitem [{\citenamefont {Liu}\ \emph {et~al.}(2014{\natexlab{b}})\citenamefont
  {Liu}, \citenamefont {Law},\ and\ \citenamefont
  {Ng}}]{PhysRevLett.113.059901}%
  \BibitemOpen
  \bibfield  {author} {\bibinfo {author} {\bibfnamefont {X.-J.}\ \bibnamefont
  {Liu}}, \bibinfo {author} {\bibfnamefont {K.~T.}\ \bibnamefont {Law}},\ and\
  \bibinfo {author} {\bibfnamefont {T.~K.}\ \bibnamefont {Ng}},\ }\bibfield
  {title} {\bibinfo {title} {Erratum: Realization of 2d spin-orbit interaction
  and exotic topological orders in cold atoms [phys. rev. lett. 112, 086401
  (2014)]},\ }\href {https://doi.org/10.1103/PhysRevLett.113.059901} {\bibfield
   {journal} {\bibinfo  {journal} {Phys. Rev. Lett.}\ }\textbf {\bibinfo
  {volume} {113}},\ \bibinfo {pages} {059901} (\bibinfo {year}
  {2014}{\natexlab{b}})}\BibitemShut {NoStop}%
\bibitem [{\citenamefont {Wu}\ \emph {et~al.}(2016)\citenamefont {Wu},
  \citenamefont {Zhang}, \citenamefont {Sun}, \citenamefont {Xu}, \citenamefont
  {Wang}, \citenamefont {Ji}, \citenamefont {Deng}, \citenamefont {Chen},
  \citenamefont {Liu},\ and\ \citenamefont {Pan}}]{Wu83}%
  \BibitemOpen
  \bibfield  {author} {\bibinfo {author} {\bibfnamefont {Z.}~\bibnamefont
  {Wu}}, \bibinfo {author} {\bibfnamefont {L.}~\bibnamefont {Zhang}}, \bibinfo
  {author} {\bibfnamefont {W.}~\bibnamefont {Sun}}, \bibinfo {author}
  {\bibfnamefont {X.-T.}\ \bibnamefont {Xu}}, \bibinfo {author} {\bibfnamefont
  {B.-Z.}\ \bibnamefont {Wang}}, \bibinfo {author} {\bibfnamefont {S.-C.}\
  \bibnamefont {Ji}}, \bibinfo {author} {\bibfnamefont {Y.}~\bibnamefont
  {Deng}}, \bibinfo {author} {\bibfnamefont {S.}~\bibnamefont {Chen}}, \bibinfo
  {author} {\bibfnamefont {X.-J.}\ \bibnamefont {Liu}},\ and\ \bibinfo {author}
  {\bibfnamefont {J.-W.}\ \bibnamefont {Pan}},\ }\bibfield  {title} {\bibinfo
  {title} {Realization of two-dimensional spin-orbit coupling for bose-einstein
  condensates},\ }\href {https://doi.org/10.1126/science.aaf6689} {\bibfield
  {journal} {\bibinfo  {journal} {Science}\ }\textbf {\bibinfo {volume}
  {354}},\ \bibinfo {pages} {83} (\bibinfo {year} {2016})}\BibitemShut
  {NoStop}%
\bibitem [{\citenamefont {Sun}\ \emph {et~al.}(2018{\natexlab{a}})\citenamefont
  {Sun}, \citenamefont {Wang}, \citenamefont {Xu}, \citenamefont {Yi},
  \citenamefont {Zhang}, \citenamefont {Wu}, \citenamefont {Deng},
  \citenamefont {Liu}, \citenamefont {Chen},\ and\ \citenamefont
  {Pan}}]{PhysRevLett.121.150401}%
  \BibitemOpen
  \bibfield  {author} {\bibinfo {author} {\bibfnamefont {W.}~\bibnamefont
  {Sun}}, \bibinfo {author} {\bibfnamefont {B.-Z.}\ \bibnamefont {Wang}},
  \bibinfo {author} {\bibfnamefont {X.-T.}\ \bibnamefont {Xu}}, \bibinfo
  {author} {\bibfnamefont {C.-R.}\ \bibnamefont {Yi}}, \bibinfo {author}
  {\bibfnamefont {L.}~\bibnamefont {Zhang}}, \bibinfo {author} {\bibfnamefont
  {Z.}~\bibnamefont {Wu}}, \bibinfo {author} {\bibfnamefont {Y.}~\bibnamefont
  {Deng}}, \bibinfo {author} {\bibfnamefont {X.-J.}\ \bibnamefont {Liu}},
  \bibinfo {author} {\bibfnamefont {S.}~\bibnamefont {Chen}},\ and\ \bibinfo
  {author} {\bibfnamefont {J.-W.}\ \bibnamefont {Pan}},\ }\bibfield  {title}
  {\bibinfo {title} {Highly controllable and robust 2d spin-orbit coupling for
  quantum gases},\ }\href {https://doi.org/10.1103/PhysRevLett.121.150401}
  {\bibfield  {journal} {\bibinfo  {journal} {Phys. Rev. Lett.}\ }\textbf
  {\bibinfo {volume} {121}},\ \bibinfo {pages} {150401} (\bibinfo {year}
  {2018}{\natexlab{a}})}\BibitemShut {NoStop}%
\bibitem [{\citenamefont {Song}\ \emph {et~al.}(2018)\citenamefont {Song},
  \citenamefont {Zhang}, \citenamefont {He}, \citenamefont {Poon},
  \citenamefont {Hajiyev}, \citenamefont {Zhang}, \citenamefont {Liu},\ and\
  \citenamefont {Jo}}]{Songeaao4748}%
  \BibitemOpen
  \bibfield  {author} {\bibinfo {author} {\bibfnamefont {B.}~\bibnamefont
  {Song}}, \bibinfo {author} {\bibfnamefont {L.}~\bibnamefont {Zhang}},
  \bibinfo {author} {\bibfnamefont {C.}~\bibnamefont {He}}, \bibinfo {author}
  {\bibfnamefont {T.~F.~J.}\ \bibnamefont {Poon}}, \bibinfo {author}
  {\bibfnamefont {E.}~\bibnamefont {Hajiyev}}, \bibinfo {author} {\bibfnamefont
  {S.}~\bibnamefont {Zhang}}, \bibinfo {author} {\bibfnamefont {X.-J.}\
  \bibnamefont {Liu}},\ and\ \bibinfo {author} {\bibfnamefont {G.-B.}\
  \bibnamefont {Jo}},\ }\bibfield  {title} {\bibinfo {title} {Observation of
  symmetry-protected topological band with ultracold fermions},\ }\href
  {https://advances.sciencemag.org/content/4/2/eaao4748} {\bibfield  {journal}
  {\bibinfo  {journal} {Science Advances}\ }\textbf {\bibinfo {volume} {4}}
  (\bibinfo {year} {2018})}\BibitemShut {NoStop}%
\bibitem [{\citenamefont {Galitski}\ and\ \citenamefont
  {Spielman}(2013)}]{Galitski2013}%
  \BibitemOpen
  \bibfield  {author} {\bibinfo {author} {\bibfnamefont {V.}~\bibnamefont
  {Galitski}}\ and\ \bibinfo {author} {\bibfnamefont {I.~B.}\ \bibnamefont
  {Spielman}},\ }\bibfield  {title} {\bibinfo {title} {Spin--orbit coupling in
  quantum gases},\ }\href {https://doi.org/10.1038/nature11841} {\bibfield
  {journal} {\bibinfo  {journal} {Nature}\ }\textbf {\bibinfo {volume} {494}},\
  \bibinfo {pages} {49} (\bibinfo {year} {2013})}\BibitemShut {NoStop}%
\bibitem [{\citenamefont {Zhai}(2015)}]{zhai_2015}%
  \BibitemOpen
  \bibfield  {author} {\bibinfo {author} {\bibfnamefont {H.}~\bibnamefont
  {Zhai}},\ }\bibfield  {title} {\bibinfo {title} {Degenerate quantum gases
  with spin{\textendash}orbit coupling: a review},\ }\href
  {https://doi.org/10.1088/0034-4885/78/2/026001} {\bibfield  {journal}
  {\bibinfo  {journal} {Rep. Prog. Phys.}\ }\textbf {\bibinfo {volume} {78}},\
  \bibinfo {pages} {026001} (\bibinfo {year} {2015})}\BibitemShut {NoStop}%
\bibitem [{\citenamefont {Zhang}\ and\ \citenamefont {Liu}()}]{book_soc}%
  \BibitemOpen
  \bibfield  {author} {\bibinfo {author} {\bibfnamefont {L.}~\bibnamefont
  {Zhang}}\ and\ \bibinfo {author} {\bibfnamefont {X.-J.}\ \bibnamefont
  {Liu}},\ }\bibinfo {title} {Spin-orbit coupling and topological phases for
  ultracold atoms},\ in\ \href {https://doi.org/10.1142/9789813272538_0001}
  {\emph {\bibinfo {booktitle} {Synthetic Spin-Orbit Coupling in Cold
  Atoms}}},\ Chap.\ \bibinfo {chapter} {Chapter 1}, pp.\ \bibinfo {pages}
  {1--87}\BibitemShut {NoStop}%
\bibitem [{\citenamefont {Fulgado-Claudio}\ \emph {et~al.}(2023)\citenamefont
  {Fulgado-Claudio}, \citenamefont {Vel{\'{a}}zquez},\ and\ \citenamefont
  {Bermudez}}]{FulgadoClaudio2023fermionproduction}%
  \BibitemOpen
  \bibfield  {author} {\bibinfo {author} {\bibfnamefont {C.}~\bibnamefont
  {Fulgado-Claudio}}, \bibinfo {author} {\bibfnamefont {J.~M.~S.}\ \bibnamefont
  {Vel{\'{a}}zquez}},\ and\ \bibinfo {author} {\bibfnamefont {A.}~\bibnamefont
  {Bermudez}},\ }\bibfield  {title} {\bibinfo {title} {Fermion production at
  the boundary of an expanding universe: a cold-atom gravitational analogue},\
  }\href {https://doi.org/10.22331/q-2023-06-21-1042} {\bibfield  {journal}
  {\bibinfo  {journal} {{Quantum}}\ }\textbf {\bibinfo {volume} {7}},\ \bibinfo
  {pages} {1042} (\bibinfo {year} {2023})}\BibitemShut {NoStop}%
\bibitem [{\citenamefont {Struck}\ \emph {et~al.}(2012)\citenamefont {Struck},
  \citenamefont {\"Olschl\"ager}, \citenamefont {Weinberg}, \citenamefont
  {Hauke}, \citenamefont {Simonet}, \citenamefont {Eckardt}, \citenamefont
  {Lewenstein}, \citenamefont {Sengstock},\ and\ \citenamefont
  {Windpassinger}}]{Struck_2012}%
  \BibitemOpen
  \bibfield  {author} {\bibinfo {author} {\bibfnamefont {J.}~\bibnamefont
  {Struck}}, \bibinfo {author} {\bibfnamefont {C.}~\bibnamefont
  {\"Olschl\"ager}}, \bibinfo {author} {\bibfnamefont {M.}~\bibnamefont
  {Weinberg}}, \bibinfo {author} {\bibfnamefont {P.}~\bibnamefont {Hauke}},
  \bibinfo {author} {\bibfnamefont {J.}~\bibnamefont {Simonet}}, \bibinfo
  {author} {\bibfnamefont {A.}~\bibnamefont {Eckardt}}, \bibinfo {author}
  {\bibfnamefont {M.}~\bibnamefont {Lewenstein}}, \bibinfo {author}
  {\bibfnamefont {K.}~\bibnamefont {Sengstock}},\ and\ \bibinfo {author}
  {\bibfnamefont {P.}~\bibnamefont {Windpassinger}},\ }\bibfield  {title}
  {\bibinfo {title} {Tunable gauge potential for neutral and spinless particles
  in driven optical lattices},\ }\href
  {https://doi.org/10.1103/PhysRevLett.108.225304} {\bibfield  {journal}
  {\bibinfo  {journal} {Phys. Rev. Lett.}\ }\textbf {\bibinfo {volume} {108}},\
  \bibinfo {pages} {225304} (\bibinfo {year} {2012})}\BibitemShut {NoStop}%
\bibitem [{\citenamefont {Hayashi}(2018)}]{Hayashi2018}%
  \BibitemOpen
  \bibfield  {author} {\bibinfo {author} {\bibfnamefont {S.}~\bibnamefont
  {Hayashi}},\ }\bibfield  {title} {\bibinfo {title} {Topological invariants
  and corner states for hamiltonians on a three-dimensional lattice},\ }\href
  {https://doi.org/10.1007/s00220-018-3229-2} {\bibfield  {journal} {\bibinfo
  {journal} {Communications in Mathematical Physics}\ }\textbf {\bibinfo
  {volume} {364}},\ \bibinfo {pages} {343} (\bibinfo {year}
  {2018})}\BibitemShut {NoStop}%
\bibitem [{\citenamefont {Hayashi}(2019)}]{Hayashi2019}%
  \BibitemOpen
  \bibfield  {author} {\bibinfo {author} {\bibfnamefont {S.}~\bibnamefont
  {Hayashi}},\ }\bibfield  {title} {\bibinfo {title} {Toeplitz operators on
  concave corners and topologically protected corner states},\ }\href
  {https://doi.org/10.1007/s11005-019-01184-w} {\bibfield  {journal} {\bibinfo
  {journal} {Letters in Mathematical Physics}\ }\textbf {\bibinfo {volume}
  {109}},\ \bibinfo {pages} {2223} (\bibinfo {year} {2019})}\BibitemShut
  {NoStop}%
\bibitem [{\citenamefont {Nielsen}\ and\ \citenamefont
  {Chuang}(2000)}]{nielsen00}%
  \BibitemOpen
  \bibfield  {author} {\bibinfo {author} {\bibfnamefont {M.~A.}\ \bibnamefont
  {Nielsen}}\ and\ \bibinfo {author} {\bibfnamefont {I.~L.}\ \bibnamefont
  {Chuang}},\ }\href@noop {} {\emph {\bibinfo {title} {Quantum Computation and
  Quantum Information}}}\ (\bibinfo  {publisher} {Cambridge University Press},\
  \bibinfo {year} {2000})\BibitemShut {NoStop}%
\bibitem [{\citenamefont {Zak}(1989)}]{PhysRevLett.62.2747}%
  \BibitemOpen
  \bibfield  {author} {\bibinfo {author} {\bibfnamefont {J.}~\bibnamefont
  {Zak}},\ }\bibfield  {title} {\bibinfo {title} {Berry's phase for energy
  bands in solids},\ }\href {https://doi.org/10.1103/PhysRevLett.62.2747}
  {\bibfield  {journal} {\bibinfo  {journal} {Phys. Rev. Lett.}\ }\textbf
  {\bibinfo {volume} {62}},\ \bibinfo {pages} {2747} (\bibinfo {year}
  {1989})}\BibitemShut {NoStop}%
\bibitem [{\citenamefont {Bernevig}\ and\ \citenamefont
  {Hughes}(2013)}]{bernevig_hughes_2013}%
  \BibitemOpen
  \bibfield  {author} {\bibinfo {author} {\bibfnamefont {B.~A.}\ \bibnamefont
  {Bernevig}}\ and\ \bibinfo {author} {\bibfnamefont {T.~L.}\ \bibnamefont
  {Hughes}},\ }\href@noop {} {\emph {\bibinfo {title} {Topological insulators
  and topological superconductors}}}\ (\bibinfo  {publisher} {Princeton
  University Press},\ \bibinfo {year} {2013})\BibitemShut {NoStop}%
\bibitem [{\citenamefont {Coleman}(1985)}]{coleman_1985}%
  \BibitemOpen
  \bibfield  {author} {\bibinfo {author} {\bibfnamefont {S.}~\bibnamefont
  {Coleman}},\ }\href {https://doi.org/10.1017/CBO9780511565045} {\emph
  {\bibinfo {title} {Aspects of Symmetry: Selected Erice Lectures}}}\ (\bibinfo
   {publisher} {Cambridge University Press},\ \bibinfo {year}
  {1985})\BibitemShut {NoStop}%
\bibitem [{\citenamefont {Stratonovich}(1958)}]{startonovich}%
  \BibitemOpen
  \bibfield  {author} {\bibinfo {author} {\bibfnamefont {R.}~\bibnamefont
  {Stratonovich}},\ }\bibfield  {title} {\bibinfo {title} {On a method of
  calculating quantum distribution functions},\ }\href@noop {} {\bibfield
  {journal} {\bibinfo  {journal} {Soviet Physics, Doklady}\ }\textbf {\bibinfo
  {volume} {2}},\ \bibinfo {pages} {416} (\bibinfo {year} {1958})}\BibitemShut
  {NoStop}%
\bibitem [{\citenamefont {Hubbard}(1959)}]{PhysRevLett.3.77}%
  \BibitemOpen
  \bibfield  {author} {\bibinfo {author} {\bibfnamefont {J.}~\bibnamefont
  {Hubbard}},\ }\bibfield  {title} {\bibinfo {title} {Calculation of partition
  functions},\ }\href {https://doi.org/10.1103/PhysRevLett.3.77} {\bibfield
  {journal} {\bibinfo  {journal} {Phys. Rev. Lett.}\ }\textbf {\bibinfo
  {volume} {3}},\ \bibinfo {pages} {77} (\bibinfo {year} {1959})}\BibitemShut
  {NoStop}%
\bibitem [{\citenamefont {Ziegler}\ \emph
  {et~al.}(2022{\natexlab{c}})\citenamefont {Ziegler}, \citenamefont {Tirrito},
  \citenamefont {Lewenstein}, \citenamefont {Hands},\ and\ \citenamefont
  {Bermudez}}]{Ziegler_2022}%
  \BibitemOpen
  \bibfield  {author} {\bibinfo {author} {\bibfnamefont {L.}~\bibnamefont
  {Ziegler}}, \bibinfo {author} {\bibfnamefont {E.}~\bibnamefont {Tirrito}},
  \bibinfo {author} {\bibfnamefont {M.}~\bibnamefont {Lewenstein}}, \bibinfo
  {author} {\bibfnamefont {S.}~\bibnamefont {Hands}},\ and\ \bibinfo {author}
  {\bibfnamefont {A.}~\bibnamefont {Bermudez}},\ }\bibfield  {title} {\bibinfo
  {title} {Large-n chern insulators: Lattice field theory and quantum
  simulation approaches to correlation effects in the quantum anomalous hall
  effect},\ }\href {https://doi.org/https://doi.org/10.1016/j.aop.2022.168763}
  {\bibfield  {journal} {\bibinfo  {journal} {Annals of Physics}\ }\textbf
  {\bibinfo {volume} {439}},\ \bibinfo {pages} {168763} (\bibinfo {year}
  {2022}{\natexlab{c}})}\BibitemShut {NoStop}%
\bibitem [{\citenamefont {Aoki}(1984)}]{PhysRevD.30.2653}%
  \BibitemOpen
  \bibfield  {author} {\bibinfo {author} {\bibfnamefont {S.}~\bibnamefont
  {Aoki}},\ }\bibfield  {title} {\bibinfo {title} {New phase structure for
  lattice qcd with wilson fermions},\ }\href
  {https://doi.org/10.1103/PhysRevD.30.2653} {\bibfield  {journal} {\bibinfo
  {journal} {Phys. Rev. D}\ }\textbf {\bibinfo {volume} {30}},\ \bibinfo
  {pages} {2653} (\bibinfo {year} {1984})}\BibitemShut {NoStop}%
\bibitem [{\citenamefont {Eguchi}\ and\ \citenamefont
  {Nakayama}(1983)}]{Eguchi:1983gq}%
  \BibitemOpen
  \bibfield  {author} {\bibinfo {author} {\bibfnamefont {T.}~\bibnamefont
  {Eguchi}}\ and\ \bibinfo {author} {\bibfnamefont {R.}~\bibnamefont
  {Nakayama}},\ }\bibfield  {title} {\bibinfo {title} {{Wilson Lattice Fermion
  and the Recovery of Chiral Symmetry Near the Continuum Limit}},\ }\href
  {https://doi.org/10.1016/0370-2693(83)90024-2} {\bibfield  {journal}
  {\bibinfo  {journal} {Phys. Lett. B}\ }\textbf {\bibinfo {volume} {126}},\
  \bibinfo {pages} {89} (\bibinfo {year} {1983})}\BibitemShut {NoStop}%
\bibitem [{\citenamefont {Wang}\ \emph {et~al.}(2010)\citenamefont {Wang},
  \citenamefont {Qi},\ and\ \citenamefont {Zhang}}]{PhysRevLett.105.256803}%
  \BibitemOpen
  \bibfield  {author} {\bibinfo {author} {\bibfnamefont {Z.}~\bibnamefont
  {Wang}}, \bibinfo {author} {\bibfnamefont {X.-L.}\ \bibnamefont {Qi}},\ and\
  \bibinfo {author} {\bibfnamefont {S.-C.}\ \bibnamefont {Zhang}},\ }\bibfield
  {title} {\bibinfo {title} {Topological order parameters for interacting
  topological insulators},\ }\href
  {https://doi.org/10.1103/PhysRevLett.105.256803} {\bibfield  {journal}
  {\bibinfo  {journal} {Phys. Rev. Lett.}\ }\textbf {\bibinfo {volume} {105}},\
  \bibinfo {pages} {256803} (\bibinfo {year} {2010})}\BibitemShut {NoStop}%
\bibitem [{\citenamefont {Gurarie}(2011)}]{PhysRevB.83.085426}%
  \BibitemOpen
  \bibfield  {author} {\bibinfo {author} {\bibfnamefont {V.}~\bibnamefont
  {Gurarie}},\ }\bibfield  {title} {\bibinfo {title} {Single-particle green's
  functions and interacting topological insulators},\ }\href
  {https://doi.org/10.1103/PhysRevB.83.085426} {\bibfield  {journal} {\bibinfo
  {journal} {Phys. Rev. B}\ }\textbf {\bibinfo {volume} {83}},\ \bibinfo
  {pages} {085426} (\bibinfo {year} {2011})}\BibitemShut {NoStop}%
\bibitem [{\citenamefont {Wang}\ and\ \citenamefont
  {Zhang}(2012{\natexlab{a}})}]{PhysRevX.2.031008}%
  \BibitemOpen
  \bibfield  {author} {\bibinfo {author} {\bibfnamefont {Z.}~\bibnamefont
  {Wang}}\ and\ \bibinfo {author} {\bibfnamefont {S.-C.}\ \bibnamefont
  {Zhang}},\ }\bibfield  {title} {\bibinfo {title} {Simplified topological
  invariants for interacting insulators},\ }\href
  {https://doi.org/10.1103/PhysRevX.2.031008} {\bibfield  {journal} {\bibinfo
  {journal} {Phys. Rev. X}\ }\textbf {\bibinfo {volume} {2}},\ \bibinfo {pages}
  {031008} (\bibinfo {year} {2012}{\natexlab{a}})}\BibitemShut {NoStop}%
\bibitem [{\citenamefont {Wang}\ and\ \citenamefont
  {Zhang}(2012{\natexlab{b}})}]{PhysRevB.86.165116}%
  \BibitemOpen
  \bibfield  {author} {\bibinfo {author} {\bibfnamefont {Z.}~\bibnamefont
  {Wang}}\ and\ \bibinfo {author} {\bibfnamefont {S.-C.}\ \bibnamefont
  {Zhang}},\ }\bibfield  {title} {\bibinfo {title} {Strongly correlated
  topological superconductors and topological phase transitions via green's
  function},\ }\href {https://doi.org/10.1103/PhysRevB.86.165116} {\bibfield
  {journal} {\bibinfo  {journal} {Phys. Rev. B}\ }\textbf {\bibinfo {volume}
  {86}},\ \bibinfo {pages} {165116} (\bibinfo {year}
  {2012}{\natexlab{b}})}\BibitemShut {NoStop}%
\bibitem [{\citenamefont {Wang}\ and\ \citenamefont {Yan}(2013)}]{Wang_2013}%
  \BibitemOpen
  \bibfield  {author} {\bibinfo {author} {\bibfnamefont {Z.}~\bibnamefont
  {Wang}}\ and\ \bibinfo {author} {\bibfnamefont {B.}~\bibnamefont {Yan}},\
  }\bibfield  {title} {\bibinfo {title} {Topological hamiltonian as an exact
  tool for topological invariants},\ }\href
  {https://doi.org/10.1088/0953-8984/25/15/155601} {\bibfield  {journal}
  {\bibinfo  {journal} {Journal of Physics: Condensed Matter}\ }\textbf
  {\bibinfo {volume} {25}},\ \bibinfo {pages} {155601} (\bibinfo {year}
  {2013})}\BibitemShut {NoStop}%
\bibitem [{\citenamefont {Hatsugai}(2006)}]{doi:10.1143/JPSJ.75.123601}%
  \BibitemOpen
  \bibfield  {author} {\bibinfo {author} {\bibfnamefont {Y.}~\bibnamefont
  {Hatsugai}},\ }\bibfield  {title} {\bibinfo {title} {Quantized berry phases
  as a local order parameter of a quantum liquid},\ }\href
  {https://doi.org/10.1143/JPSJ.75.123601} {\bibfield  {journal} {\bibinfo
  {journal} {Journal of the Physical Society of Japan}\ }\textbf {\bibinfo
  {volume} {75}},\ \bibinfo {pages} {123601} (\bibinfo {year}
  {2006})}\BibitemShut {NoStop}%
\bibitem [{\citenamefont {Kudo}\ \emph
  {et~al.}(2019{\natexlab{b}})\citenamefont {Kudo}, \citenamefont {Watanabe},
  \citenamefont {Kariyado},\ and\ \citenamefont
  {Hatsugai}}]{PhysRevLett.122.146601}%
  \BibitemOpen
  \bibfield  {author} {\bibinfo {author} {\bibfnamefont {K.}~\bibnamefont
  {Kudo}}, \bibinfo {author} {\bibfnamefont {H.}~\bibnamefont {Watanabe}},
  \bibinfo {author} {\bibfnamefont {T.}~\bibnamefont {Kariyado}},\ and\
  \bibinfo {author} {\bibfnamefont {Y.}~\bibnamefont {Hatsugai}},\ }\bibfield
  {title} {\bibinfo {title} {Many-body chern number without integration},\
  }\href {https://doi.org/10.1103/PhysRevLett.122.146601} {\bibfield  {journal}
  {\bibinfo  {journal} {Phys. Rev. Lett.}\ }\textbf {\bibinfo {volume} {122}},\
  \bibinfo {pages} {146601} (\bibinfo {year} {2019}{\natexlab{b}})}\BibitemShut
  {NoStop}%
\bibitem [{\citenamefont {Sun}\ \emph {et~al.}(2018{\natexlab{b}})\citenamefont
  {Sun}, \citenamefont {Yi}, \citenamefont {Wang}, \citenamefont {Zhang},
  \citenamefont {Sanders}, \citenamefont {Xu}, \citenamefont {Wang},
  \citenamefont {Schmiedmayer}, \citenamefont {Deng}, \citenamefont {Liu},
  \citenamefont {Chen},\ and\ \citenamefont {Pan}}]{PhysRevLett.121.250403}%
  \BibitemOpen
  \bibfield  {author} {\bibinfo {author} {\bibfnamefont {W.}~\bibnamefont
  {Sun}}, \bibinfo {author} {\bibfnamefont {C.-R.}\ \bibnamefont {Yi}},
  \bibinfo {author} {\bibfnamefont {B.-Z.}\ \bibnamefont {Wang}}, \bibinfo
  {author} {\bibfnamefont {W.-W.}\ \bibnamefont {Zhang}}, \bibinfo {author}
  {\bibfnamefont {B.~C.}\ \bibnamefont {Sanders}}, \bibinfo {author}
  {\bibfnamefont {X.-T.}\ \bibnamefont {Xu}}, \bibinfo {author} {\bibfnamefont
  {Z.-Y.}\ \bibnamefont {Wang}}, \bibinfo {author} {\bibfnamefont
  {J.}~\bibnamefont {Schmiedmayer}}, \bibinfo {author} {\bibfnamefont
  {Y.}~\bibnamefont {Deng}}, \bibinfo {author} {\bibfnamefont {X.-J.}\
  \bibnamefont {Liu}}, \bibinfo {author} {\bibfnamefont {S.}~\bibnamefont
  {Chen}},\ and\ \bibinfo {author} {\bibfnamefont {J.-W.}\ \bibnamefont
  {Pan}},\ }\bibfield  {title} {\bibinfo {title} {Uncover topology by quantum
  quench dynamics},\ }\href {https://doi.org/10.1103/PhysRevLett.121.250403}
  {\bibfield  {journal} {\bibinfo  {journal} {Phys. Rev. Lett.}\ }\textbf
  {\bibinfo {volume} {121}},\ \bibinfo {pages} {250403} (\bibinfo {year}
  {2018}{\natexlab{b}})}\BibitemShut {NoStop}%
\bibitem [{\citenamefont {Jia}\ \emph {et~al.}(2023)\citenamefont {Jia},
  \citenamefont {Zhang}, \citenamefont {Zhang},\ and\ \citenamefont
  {Liu}}]{PhysRevB.107.125132}%
  \BibitemOpen
  \bibfield  {author} {\bibinfo {author} {\bibfnamefont {W.}~\bibnamefont
  {Jia}}, \bibinfo {author} {\bibfnamefont {L.}~\bibnamefont {Zhang}}, \bibinfo
  {author} {\bibfnamefont {L.}~\bibnamefont {Zhang}},\ and\ \bibinfo {author}
  {\bibfnamefont {X.-J.}\ \bibnamefont {Liu}},\ }\bibfield  {title} {\bibinfo
  {title} {Dynamical detection of mean-field topological phases in an
  interacting chern insulator},\ }\href
  {https://doi.org/10.1103/PhysRevB.107.125132} {\bibfield  {journal} {\bibinfo
   {journal} {Phys. Rev. B}\ }\textbf {\bibinfo {volume} {107}},\ \bibinfo
  {pages} {125132} (\bibinfo {year} {2023})}\BibitemShut {NoStop}%
\bibitem [{\citenamefont {Kuno}(2019)}]{PhysRevB.99.064105}%
  \BibitemOpen
  \bibfield  {author} {\bibinfo {author} {\bibfnamefont {Y.}~\bibnamefont
  {Kuno}},\ }\bibfield  {title} {\bibinfo {title} {Phase structure of the
  interacting su-schrieffer-heeger model and the relationship with the
  gross-neveu model on lattice},\ }\href
  {https://doi.org/10.1103/PhysRevB.99.064105} {\bibfield  {journal} {\bibinfo
  {journal} {Phys. Rev. B}\ }\textbf {\bibinfo {volume} {99}},\ \bibinfo
  {pages} {064105} (\bibinfo {year} {2019})}\BibitemShut {NoStop}%
\bibitem [{\citenamefont {Kaplan}\ and\ \citenamefont
  {Sen}(2020)}]{PhysRevLett.124.131601}%
  \BibitemOpen
  \bibfield  {author} {\bibinfo {author} {\bibfnamefont {D.~B.}\ \bibnamefont
  {Kaplan}}\ and\ \bibinfo {author} {\bibfnamefont {S.}~\bibnamefont {Sen}},\
  }\bibfield  {title} {\bibinfo {title} {Fractional quantum hall effect in a
  relativistic field theory},\ }\href
  {https://doi.org/10.1103/PhysRevLett.124.131601} {\bibfield  {journal}
  {\bibinfo  {journal} {Phys. Rev. Lett.}\ }\textbf {\bibinfo {volume} {124}},\
  \bibinfo {pages} {131601} (\bibinfo {year} {2020})}\BibitemShut {NoStop}%
\bibitem [{\citenamefont {Lenz}\ \emph {et~al.}(2020)\citenamefont {Lenz},
  \citenamefont {Pannullo}, \citenamefont {Wagner}, \citenamefont
  {Wellegehausen},\ and\ \citenamefont {Wipf}}]{PhysRevD.101.094512}%
  \BibitemOpen
  \bibfield  {author} {\bibinfo {author} {\bibfnamefont {J.}~\bibnamefont
  {Lenz}}, \bibinfo {author} {\bibfnamefont {L.}~\bibnamefont {Pannullo}},
  \bibinfo {author} {\bibfnamefont {M.}~\bibnamefont {Wagner}}, \bibinfo
  {author} {\bibfnamefont {B.}~\bibnamefont {Wellegehausen}},\ and\ \bibinfo
  {author} {\bibfnamefont {A.}~\bibnamefont {Wipf}},\ }\bibfield  {title}
  {\bibinfo {title} {Inhomogeneous phases in the gross-neveu model in $1+1$
  dimensions at finite number of flavors},\ }\href
  {https://doi.org/10.1103/PhysRevD.101.094512} {\bibfield  {journal} {\bibinfo
   {journal} {Phys. Rev. D}\ }\textbf {\bibinfo {volume} {101}},\ \bibinfo
  {pages} {094512} (\bibinfo {year} {2020})}\BibitemShut {NoStop}%
\bibitem [{\citenamefont {Ishikawa}\ \emph {et~al.}(2021)\citenamefont
  {Ishikawa}, \citenamefont {Nakayama},\ and\ \citenamefont
  {Suzuki}}]{PhysRevD.104.094515}%
  \BibitemOpen
  \bibfield  {author} {\bibinfo {author} {\bibfnamefont {T.}~\bibnamefont
  {Ishikawa}}, \bibinfo {author} {\bibfnamefont {K.}~\bibnamefont {Nakayama}},\
  and\ \bibinfo {author} {\bibfnamefont {K.}~\bibnamefont {Suzuki}},\
  }\bibfield  {title} {\bibinfo {title} {Kondo effect with wilson fermions},\
  }\href {https://doi.org/10.1103/PhysRevD.104.094515} {\bibfield  {journal}
  {\bibinfo  {journal} {Phys. Rev. D}\ }\textbf {\bibinfo {volume} {104}},\
  \bibinfo {pages} {094515} (\bibinfo {year} {2021})}\BibitemShut {NoStop}%
\bibitem [{\citenamefont {Asaduzzaman}\ \emph {et~al.}(2022)\citenamefont
  {Asaduzzaman}, \citenamefont {Toga}, \citenamefont {Catterall}, \citenamefont
  {Meurice},\ and\ \citenamefont {Sakai}}]{PhysRevD.106.114515}%
  \BibitemOpen
  \bibfield  {author} {\bibinfo {author} {\bibfnamefont {M.}~\bibnamefont
  {Asaduzzaman}}, \bibinfo {author} {\bibfnamefont {G.~C.}\ \bibnamefont
  {Toga}}, \bibinfo {author} {\bibfnamefont {S.}~\bibnamefont {Catterall}},
  \bibinfo {author} {\bibfnamefont {Y.}~\bibnamefont {Meurice}},\ and\ \bibinfo
  {author} {\bibfnamefont {R.}~\bibnamefont {Sakai}},\ }\bibfield  {title}
  {\bibinfo {title} {Quantum simulation of the $n$-flavor gross-neveu model},\
  }\href {https://doi.org/10.1103/PhysRevD.106.114515} {\bibfield  {journal}
  {\bibinfo  {journal} {Phys. Rev. D}\ }\textbf {\bibinfo {volume} {106}},\
  \bibinfo {pages} {114515} (\bibinfo {year} {2022})}\BibitemShut {NoStop}%
\bibitem [{\citenamefont {Roose}\ \emph {et~al.}(2021)\citenamefont {Roose},
  \citenamefont {Bultinck}, \citenamefont {Vanderstraeten}, \citenamefont
  {Verstraete}, \citenamefont {Van~Acoleyen},\ and\ \citenamefont
  {Haegeman}}]{Roose2021}%
  \BibitemOpen
  \bibfield  {author} {\bibinfo {author} {\bibfnamefont {G.}~\bibnamefont
  {Roose}}, \bibinfo {author} {\bibfnamefont {N.}~\bibnamefont {Bultinck}},
  \bibinfo {author} {\bibfnamefont {L.}~\bibnamefont {Vanderstraeten}},
  \bibinfo {author} {\bibfnamefont {F.}~\bibnamefont {Verstraete}}, \bibinfo
  {author} {\bibfnamefont {K.}~\bibnamefont {Van~Acoleyen}},\ and\ \bibinfo
  {author} {\bibfnamefont {J.}~\bibnamefont {Haegeman}},\ }\bibfield  {title}
  {\bibinfo {title} {Lattice regularisation and entanglement structure of the
  gross-neveu model},\ }\href {https://doi.org/10.1007/JHEP07(2021)207}
  {\bibfield  {journal} {\bibinfo  {journal} {Journal of High Energy Physics}\
  }\textbf {\bibinfo {volume} {2021}},\ \bibinfo {pages} {207} (\bibinfo {year}
  {2021})}\BibitemShut {NoStop}%
\bibitem [{\citenamefont {Czajka}\ \emph {et~al.}(2022)\citenamefont {Czajka},
  \citenamefont {Kang}, \citenamefont {Ma},\ and\ \citenamefont
  {Zhao}}]{Czajka2022}%
  \BibitemOpen
  \bibfield  {author} {\bibinfo {author} {\bibfnamefont {A.~M.}\ \bibnamefont
  {Czajka}}, \bibinfo {author} {\bibfnamefont {Z.-B.}\ \bibnamefont {Kang}},
  \bibinfo {author} {\bibfnamefont {H.}~\bibnamefont {Ma}},\ and\ \bibinfo
  {author} {\bibfnamefont {F.}~\bibnamefont {Zhao}},\ }\bibfield  {title}
  {\bibinfo {title} {Quantum simulation of chiral phase transitions},\ }\href
  {https://doi.org/10.1007/JHEP08(2022)209} {\bibfield  {journal} {\bibinfo
  {journal} {Journal of High Energy Physics}\ }\textbf {\bibinfo {volume}
  {2022}},\ \bibinfo {pages} {209} (\bibinfo {year} {2022})}\BibitemShut
  {NoStop}%
\bibitem [{\citenamefont {Sen}(2023)}]{PhysRevD.107.014509}%
  \BibitemOpen
  \bibfield  {author} {\bibinfo {author} {\bibfnamefont {S.}~\bibnamefont
  {Sen}},\ }\bibfield  {title} {\bibinfo {title} {Chiral fermions on lattice
  axion strings},\ }\href {https://doi.org/10.1103/PhysRevD.107.014509}
  {\bibfield  {journal} {\bibinfo  {journal} {Phys. Rev. D}\ }\textbf {\bibinfo
  {volume} {107}},\ \bibinfo {pages} {014509} (\bibinfo {year}
  {2023})}\BibitemShut {NoStop}%
\bibitem [{\citenamefont {Pannullo}\ \emph {et~al.}(2022)\citenamefont
  {Pannullo}, \citenamefont {Wagner},\ and\ \citenamefont
  {Winstel}}]{sym14020265}%
  \BibitemOpen
  \bibfield  {author} {\bibinfo {author} {\bibfnamefont {L.}~\bibnamefont
  {Pannullo}}, \bibinfo {author} {\bibfnamefont {M.}~\bibnamefont {Wagner}},\
  and\ \bibinfo {author} {\bibfnamefont {M.}~\bibnamefont {Winstel}},\
  }\bibfield  {title} {\bibinfo {title} {Inhomogeneous phases in the chirally
  imbalanced 2 + 1-dimensional gross-neveu model and their absence in the
  continuum limit},\ }\bibfield  {journal} {\bibinfo  {journal} {Symmetry}\
  }\textbf {\bibinfo {volume} {14}},\ \href
  {https://doi.org/10.3390/sym14020265} {10.3390/sym14020265} (\bibinfo {year}
  {2022})\BibitemShut {NoStop}%
\bibitem [{\citenamefont {Nakahara}(2017)}]{nakahara_2017}%
  \BibitemOpen
  \bibfield  {author} {\bibinfo {author} {\bibfnamefont {M.}~\bibnamefont
  {Nakahara}},\ }\href@noop {} {\emph {\bibinfo {title} {Geometry, Topology and
  Physics}}}\ (\bibinfo  {publisher} {CRC Press},\ \bibinfo {year}
  {2017})\BibitemShut {NoStop}%
\bibitem [{\citenamefont {Berry}(1984)}]{doi:10.1098/rspa.1984.0023}%
  \BibitemOpen
  \bibfield  {author} {\bibinfo {author} {\bibfnamefont {M.~V.}\ \bibnamefont
  {Berry}},\ }\bibfield  {title} {\bibinfo {title} {Quantal phase factors
  accompanying adiabatic changes},\ }\href
  {https://doi.org/10.1098/rspa.1984.0023} {\bibfield  {journal} {\bibinfo
  {journal} {Proceedings of the Royal Society of London. A. Mathematical and
  Physical Sciences}\ }\textbf {\bibinfo {volume} {392}},\ \bibinfo {pages}
  {45} (\bibinfo {year} {1984})}\BibitemShut {NoStop}%
\bibitem [{\citenamefont {Thouless}\ \emph {et~al.}(1982)\citenamefont
  {Thouless}, \citenamefont {Kohmoto}, \citenamefont {Nightingale},\ and\
  \citenamefont {den Nijs}}]{PhysRevLett.49.405}%
  \BibitemOpen
  \bibfield  {author} {\bibinfo {author} {\bibfnamefont {D.~J.}\ \bibnamefont
  {Thouless}}, \bibinfo {author} {\bibfnamefont {M.}~\bibnamefont {Kohmoto}},
  \bibinfo {author} {\bibfnamefont {M.~P.}\ \bibnamefont {Nightingale}},\ and\
  \bibinfo {author} {\bibfnamefont {M.}~\bibnamefont {den Nijs}},\ }\bibfield
  {title} {\bibinfo {title} {Quantized hall conductance in a two-dimensional
  periodic potential},\ }\href {https://doi.org/10.1103/PhysRevLett.49.405}
  {\bibfield  {journal} {\bibinfo  {journal} {Phys. Rev. Lett.}\ }\textbf
  {\bibinfo {volume} {49}},\ \bibinfo {pages} {405} (\bibinfo {year}
  {1982})}\BibitemShut {NoStop}%
\end{thebibliography}%

\end{document}